\documentclass{aa}
\usepackage{graphicx}
\usepackage[T1]{fontenc}
\usepackage{txfonts}
\usepackage[space]{grffile}
\usepackage{latexsym}
\usepackage{hyphenat}
\usepackage{textcomp}
\usepackage{longtable}
\usepackage{multirow,booktabs}
\usepackage{amsfonts,amsmath,amssymb}
\usepackage{url}
\usepackage[draft]{hyperref}
\hypersetup{colorlinks=false}
\usepackage[utf8]{inputenc}
\usepackage[english]{babel}
\usepackage[]{amsmath}
\usepackage{pdflscape}
\usepackage[]{subfigure}
\usepackage[]{color}
\usepackage[]{bm}
\usepackage[]{natbib}
\usepackage{makecell}
\usepackage{pifont}
\def\arcsec{\hbox{$^{\prime\prime}$}}

\begin{document}

   \title{Angular momentum profiles of Class~0 protostellar envelopes
   }

   \author{M. Gaudel
          \inst{1,2,3}
          \and A. J. Maury\inst{1,4}
          \and A. Belloche\inst{2}
          \and S. Maret\inst{5} 
          \and Ph. Andr\'e\inst{1}
          \and P. Hennebelle\inst{1}
          \and M. Galametz\inst{1}
          \and L. Testi\inst{6}
          \and S. Cabrit\inst{3}
          \and P. Palmeirim\inst{7}
          \and B. Ladjelate\inst{8}
          \and C. Codella\inst{9}
          \and L. Podio\inst{9}
          }

 \institute{Laboratoire AIM, CEA Saclay/DRF/IRFU, CNRS, Université Paris-Saclay, Université Paris Diderot, Sorbonne Paris Cité, F-91191 Gif-sur-Yvette, France\\
              \email{mathilde.gaudel@obspm.fr}
              \and Max-Planck-Institut f{\"u}r Radioastronomie, Auf dem H{\"u}gel 69, 53121 Bonn, Germany
              \and LERMA, Observatoire de Paris, PSL Research University, CNRS, Sorbonne Université, UPMC Université Paris 06, 75014 Paris, France
         \and Harvard-Smithsonian Center for Astrophysics, Cambridge, MA, USA
        \and Univ. Grenoble Alpes, CNRS, IPAG, 38000 Grenoble, France
        \and ESO, Karl Schwarzschild Strasse 2, 85748 Garching bei M\"unchen, Germany
        \and Instituto de Astrofísica e Ci\^{e}ncias do Espa\c{c}o, Universidade do Porto, CAUP, Rua das Estrelas, 4150-762, Porto, Portugal
        \and Instituto Radioastronomia Milim\'{e}trica, Av. Divina Pastora 7, Nucleo Central, 18012, Granada, Spain
       \and INAF - Osservatorio Astrofisico di Arcetri, Largo E. Fermi 5, 50125 Firenze, Italy
             }

   \date{Received 23 July 2019; accepted 24 January 2020}

% \abstract{}{}{}{}{} 
% 5 {} token are mandatory
 
  \abstract
  % context heading (optional)
  % {} leave it empty if necessary  
    {Understanding the initial properties of star forming material and how they affect the star formation process is a key question. The infalling gas must redistribute most of its initial angular momentum inherited from prestellar cores before reaching the central stellar embryo. Disk formation has been naturally considered as a possible solution to this "angular momentum problem".
    However, how the initial angular momentum of protostellar cores is distributed and evolves during the main accretion phase and the beginning of disk formation has largely remained unconstrained up to now.
    }
  % aims heading (mandatory)
   {In the framework of the IRAM CALYPSO survey, we obtained observations of the dense gas kinematics that we used to quantify the amount and distribution of specific angular momentum at all scales in collapsing-rotating Class~0 protostellar envelopes.}
  % methods heading (mandatory)
   {We used the high dynamic range C$^{18}$O (2$-$1) and N$_2$H$^+$ (1$-$0) datasets to produce centroid velocity maps and probe the rotational motions in the sample of 12 envelopes from scales $\sim$50 to $\sim$5000~au.}
  % results heading (mandatory)
   {We identify differential rotation motions at scales $\lesssim$1600~au in 11 out of the 12 protostellar envelopes of our sample by measuring the velocity gradient along the equatorial axis, which we fit with a power-law model $\mathrm{v} \propto r^{\alpha}$. This suggests that coherent motions dominate the kinematics in the inner protostellar envelopes. The radial distributions of specific angular momentum in the CALYPSO sample suggest the following two distinct regimes within protostellar envelopes: the specific angular momentum decreases as $j \propto r^{1.6 \pm 0.2}$ down to $\sim$1600~au and then tends to become relatively constant around $\sim$6 $\times$ 10$^{-4}$~km~s$^{-1}$~pc down to $\sim$50~au.}
  % conclusions heading (optional), leave it empty if necessary 
   {The values of specific angular momentum measured in the inner Class~0 envelopes suggest that material directly involved in the star formation process ($<$1600~au) has a specific angular momentum on the same order of magnitude as what is inferred in small T-Tauri disks. Thus, disk formation appears to be a direct consequence of angular momentum conservation during the collapse.
   Our analysis reveals a dispersion of the directions of velocity gradients at envelope scales $>$1600~au, suggesting that these gradients may not be directly related to rotational motions of the envelopes. We conclude that the specific angular momentum observed at these scales could find its origin in other mechanisms, such as core-forming motions (infall, turbulence), or trace an imprint of the initial conditions for the formation of protostellar cores.}


   \maketitle
%
%
%%%%%%%%%%%%%%%%%%%%%%%%%%%%%%%%%%%%
%
%________________________________________________________________
%
%%%%%%%%%%%%%%%%%%%%%%%%%%%%%%%%%%%%
%
%

\section{Introduction}
Stars form via the gravitational collapse of $\lesssim 0.1$\,pc dense cores, which are embedded within molecular clouds \citep{Andre00,Ward-Thompson07,DiFrancesco07}. Prestellar cores become unstable and collapse due to their own gravitational potential. One or several stellar embryos form in their center. This is the beginning of the main accretion phase called the protostellar phase.
Observations of the molecular line emission from large samples of cores in close star-forming regions revealed that velocity gradients are ubiquitous to prestellar structures at scales of 0.1$-$0.5~pc  \citep{Goodman93,Caselli02}. These were interpreted as slow rotation inherited from their formation process \citep{Goodman93,Caselli02, Ohashi99, Redman04, Williams06}. Assuming these gradients trace organized rotational motions, the observed velocities lead to a typical angular rotation velocity of $\Omega \sim$ 2~km~s$^{-1}$~pc$^{-1}$ and specific angular momentum values of $j=\rm{v} \times r \sim 10^{-3} - 10^{-1}$~km~s$^{-1}$~pc.

During the collapse, if the angular momentum of the parent prestellar cores is totally transferred to the stellar embryo, the gravitational force can not counteract the centrifugal force and the stellar embryo fragments prematurely before reaching the main sequence. This is the angular momentum problem for star formation \citep{Bodenheimer95}.
Although observational studies suggest a trend of decreasing specific angular momentum toward smaller core sizes of $j \propto r^{1.6}$, the $j$ measured in prestellar structures at scales of 10000~au ($\sim 10^{-3}$ km~s$^{-1}$~pc, \citealt{Caselli02}) is still typically three orders of magnitude higher than the one associated with the maximum rotational energy that a solar-type star can sustain ($j_\mathrm{break} \sim$ 10$^{18}$ cm$^{2}$~s$^{-1}$ $\sim$ 3 $\times$10$^{-6}$~km~s$^{-1}$~pc). The physical mechanisms responsible for the angular momentum redistribution before the matter is accreted by the central stellar object have still to be identified.

During the star formation process, disk formation is expected to be a consequence of angular momentum conservation during the collapse of rotating cores \citep{Cassen81, Terebey84}. From observational studies, disks are common in Class~II objects \citep{Andrews09, Isella09, Ricci10, Spezzi13, Pietu14, Cieza19}.
Thus, disk formation has been naturally considered as a possible solution to the angular momentum problem by redistributing the four orders of magnitude of $j$ measured from prestellar cores to the T-Tauri stars ($j \sim$2 $\times$10$^{-7}$~km~s$^{-1}$~pc, \citealt{Bouvier93}): the disk would store and evacuate the angular momentum of the matter by viscous friction \citep{Lynden-Bell74, Hartmann98, Najita18} or thanks to disk winds \citep{Blandford82,Pelletier92,Pudritz07} before the matter is accreted by the central stellar object. However, the spatial distribution of angular momentum during disk formation within star-forming structures at scales between the outer core radius and the stellar surface are still largely unconstrained.

Class~0 protostars are the first (proto)stellar objects observed after the collapse in prestellar cores \citep{Andre93, Andre00}. Due to their youth, most of their mass is still in the form of a dense, collapsing, reservoir envelope surrounding the central stellar embryo (M$_\mathrm{env} \gg$ M$_{\star}$). Thus, they are likely to retain the initial conditions inherited from prestellar cores, in particular regarding angular momentum. 
The young stellar embryo mass increases via the accretion of the gaseous and dusty envelope in a short timescale ($t<$10$^5$ yr, \citealt{Evans09,Maury11}). During this main accretion phase, most of the final stellar mass is accreted and, at the same time, the infalling gas must redistribute most of its initial angular momentum before reaching the central stellar embryo. Class~0 protostars are therefore key objects to understand the distribution of angular momentum of the material directly involved in the star formation process and constrain physical mechanisms responsible for the redistribution as disk formation.

Clear signatures of rotation \citep{Belloche02, Belloche04, Chen07} and infalling gas are generally detected in the envelopes of Class~0 protostars (see the review by \citealt{Ward-Thompson07}). 
Thanks to observations of the dense molecular gas emission, rotational motions where characterized in seven Class~0 or I protostellar envelopes at scales between 3500 and 10000~au \citep{Ohashi97,Belloche02,Chen07}. These envelopes exhibit an average angular momentum of $\sim$10$^{-3}$\,km~s$^{-1}$~pc at scales of $r<$5000~au, consistent with the $j$ measured in prestellar cores by \cite{Caselli02} ($\sim 10^{-3}$ km~s$^{-1}$~pc). These studies suggest an angular momentum that is constant with radius in Class~0 protostellar envelopes. These flat profiles are generally interpreted as the conservation of the angular momentum. 
From hydrodynamical simulations, the conservation of these typical values of angular momentum results in the formation of large rotationally supported disks with radii $>$100~au in a few thousand years \citep{Yorke99}. However, from observational studies, large disks are rare around Class~0 protostars (r$<$100~au, \citealt{Maury10, Kurono13, Yen13, Segura-Cox16, Maury18}).
Only recent numerical simulations including magnetohydrodynamics and non-ideal effects, such as ambipolar diffusion, Ohmic dissipation, or the Hall effect, allow small rotationally supported disks to be formed ($<$100~au; \citealt{Machida14, Tsukamoto15, Masson16}).

Very few studies have been able to produce resolved profiles of angular momentum to characterize the actual amount of angular momentum present at the smallest scales ($r \lesssim$1000~au) within protostellar envelopes. From interferometric observations, \cite{Yen15b} derive specific angular momentum values of $\sim$2 $\times$10$^{-4}$~km~s$^{-1}$~pc in seven Class~0 envelopes at scales of $r \sim$1000~au, values which are below the trend observed by \cite{Ohashi97}. These studies have put constraints on the angular momentum properties of Class~0 protostellar envelopes and suggest the material at $r \sim$1000~au must reduce its angular momentum by at least one order of magnitude from outer envelope to disk scales. \citet{Yen11,Yen17} show specific angular momentum profiles down to $\sim$350~au in two Class~0 protostellar envelopes: $j \sim$ 6 $\times$10$^{-4}$~km~s$^{-1}$~pc at $r \sim$1000~au and $j \lesssim$10$^{-4}$~km~s$^{-1}$~pc at $r \sim$350~au. In this case, conservation of angular momentum during rotating protostellar collapse might not be the dominant process leading to the formation of disks and stellar multiple systems. It is therefore crucial to obtain robust estimates of the angular momentum of the infalling material in protostellar envelopes during the main accretion phase by analyzing the kinematics from the outer regions of the envelope (10000~au, \citealt{Motte01}) to the protostellar disk ($<$50~au, \citealt{Maury18}).

%%%%%%%%%%%%%%%%%%%%%%%%%%%%%%%%%%%%
%
%________________________________________________________________
%
%%%%%%%%%%%%%%%%%%%%%%%%%%%%%%%%%%%%

\section{The CALYPSO survey}
The Continuum And Lines in Young ProtoStellar Objects (CALYPSO\footnote{See \url{http://irfu.cea.fr/Projets/Calypso/} and \url{http://www.iram-institute.org/EN/content-page-317-7-158-240-317-0.html}}) IRAM Large Program is a survey of 16 nearby Class 0 protostars (d$<$450~pc), carried out with the IRAM Plateau de Bure interferometer (PdBI) and IRAM 30-meter telescope (30m) at wavelengths of 1.29, 1.37, and 3.18~mm.
The CALYPSO sources are among the youngest known solar-type Class~0 objects \citep{Andre00} with envelope masses of $M_\mathrm{env} \sim $1.5~$M_{\odot}$ and internal luminosities of $L_\mathrm{bol} \sim$0.1$-$30~$L_{\odot}$ \citep{Maury18}.

The CALYPSO program allows us to study in detail the Class~0 envelope chemistry \citep{Maury14,  Anderl16, Simone17,Belloche20}, disk properties \citep{Maret14,  Maury18, Maret20} and protostellar jets (\citealt{Codella14, Santangelo15, Podio16, Lefevre17}; Anderl et al. subm.; Podio et al. in prep.).
One of the main goals of this large observing program is to understand how the circumstellar envelope is accreted onto the central protostellar object during the Class~0 phase, and ultimately tackle the angular momentum problem of star formation. This paper presents an analysis of envelope kinematics, for the 12 sources from the CALYPSO sample located at $d \leq$350~pc (see Table \ref{table:sample}) and discuss our results on the properties of the angular momentum in Class~0 protostellar envelopes.

We adopt the dust continuum peak at 1.3~mm (225~GHz) determined from the PdBI datasets by \cite{Maury18} as origin of the coordinate offsets of the protostellar envelopes (see Table \ref{table:sample}). We report for each source in Table \ref{table:sample} the outflow axis considered as the rotation axis and estimated by Podio \& CALYPSO (in prep.) from high-velocity emission of $^{12}$CO, SiO, and SO at scales $<$10\arcsec. We assume the equatorial axis of the protostellar envelopes, namely the intersection of the equatorial plane with the plane of the sky at the distance of the source, to be perpendicular to the rotation axis.
The SiO emission in the CALYPSO maps is very collimated, so the uncertainties on the direction of the rotation axis, and thus on the direction of the equatorial axis, are smaller than $\pm$10$^{\circ}$ (Podio \& CALYPSO, in prep.). For L1521-F, IRAM04191, and GF9-2, no collimated SiO jet is detected, thus, the uncertainties are a bit larger ($\pm$20$^{\circ}$). We also use estimates of the inclination of the equatorial plane with respect to the line of sight from the literature. These estimates, which come from geometric models that best reproduce the outflow kinematics observed in molecular emission, are highly uncertain since we do not have access to the 3D-structure of each source.

\begin{table*}[!ht]
\caption{Sample of CALYPSO Class~0 protostars considered for this analysis.}
\begin{center}
\resizebox{\hsize}{!}{\begin{tabular}{l c c c c c c c c l}
  \hline \hline
\hfill & \hfill & \hfill & \hfill & \hfill & \hfill & \hfill & \hfill & \hfill & \hfill \\
 Source~\tablefootmark{a} & R.A. (J2000) & DEC (J2000)~\tablefootmark{b} & $d$~\tablefootmark{c} & $L_\mathrm{int}$~\tablefootmark{d} & $M_\mathrm{env}$~\tablefootmark{e} &  $R_\mathrm{env}$~\tablefootmark{f}    & PA outflow~\tablefootmark{g} & $i$~\tablefootmark{h}  & Refs~\tablefootmark{i} \\
  & (h:m:s) & ($^{\circ}$:$\arcmin$:$\arcsec$) & (pc) & ($L_\mathrm{\odot}$) &  ($M_\mathrm{\odot}$)   &    (au)  & ($^{\circ}$) & ($^{\circ}$) &  \\
  \hfill & \hfill & \hfill & \hfill & \hfill & \hfill & \hfill & \hfill & \hfill  & \hfill \\
  \hline
  \hfill & \hfill & \hfill & \hfill & \hfill & \hfill & \hfill & \hfill & \hfill  & \hfill \\
    \multirow{7}{*}{\hfill} \\
 L1448-2A1 & 03:25:22.405 & 30:45:13.26 & \multirow{2}{*}{293} & \multirow{2}{*}{4.7} & \multirow{2}{*}{1.9}  &  \multirow{2}{*}{5300} &  \multirow{2}{*}{-63 (blue), +140 (red) \tablefootmark{$\star$}}   & \multirow{2}{*}{30 $\pm$ 10} & \multirow{2}{*}{1, 2, 3, 4} \\
 L1448-2A2 & 03:25:22.355 & 30:45:13.16 &  & & &  &   &   & \\
  \multirow{7}{*}{\hfill} \\
 L1448-NB1 & 03:25:36.378 & 30:45:14.77 & \multirow{2}{*}{293} & \multirow{2}{*}{3.9} & \multirow{2}{*}{4.8} & \multirow{2}{*}{9700}  &  \multirow{2}{*}{-80} & \multirow{2}{*}{30 $\pm$ 15} & \multirow{2}{*}{5, 6, 7, 4-8} \\
 L1448-NB2 & 03:25:36.315 & 30:45:15.15 & & &  & &  &  & \\
  \multirow{7}{*}{\hfill}  \\
 L1448-C & 03:25:38.875 & 30:44:05.33 & 293 & 10.9  & 2.0 & 7300  & -17 &  20 $\pm$ 5 & 9, 6, 7, 10-11\\
 \multirow{7}{*}{\hfill} \\
  IRAS2A & 03:28:55.570  & 31:14:37.07 & 293  & 47   & 7.9  & 10000 & +205 & 30 $\pm$ 15 & 12, 13, 7, 14-15\\
   \multirow{7}{*}{\hfill} \\
   SVS13-B & 03:29:03.078 & 31:15:51.74 & 293  & 3.1   &  2.8  & 2100  & +167 & 30 $\pm$ 20\tablefootmark{$\star$ $\star$} & 16, 17, 3, ...\\
  \multirow{7}{*}{\hfill} \\
  IRAS4A1 & 03:29:10.537 & 31:13:30.98 & \multirow{2}{*}{293} & \multirow{2}{*}{4.7} &  \multirow{2}{*}{12.3} &  \multirow{2}{*}{1700}   & \multirow{2}{*}{+180}  & \multirow{2}{*}{15 $\pm$ 10} &  \multirow{2}{*}{12, 6, 3, 18}\\
  IRAS4A2 & 03:29:10.432 &  31:13:32.12 &   &   &  &  &    &   & \\
  \multirow{7}{*}{\hfill} \\
 IRAS4B & 03:29:12.016 & 31:13:08.02 & 293  & 2.3   &  4.7  & 3200  & +167 & 20 $\pm$ 15 & 12, 6, 7, 19\\
 \multirow{7}{*}{\hfill} \\
  IRAM04191 & 04:21:56.899 & 15:29:46.11 & 140  & 0.05   &  0.5  & 14000 & +20 & 40 $\pm$ 10 &  20, 21, 7, 22\\
 \multirow{7}{*}{\hfill} \\
 L1521F & 04:28:38.941 & 26:51:35.14 & 140  & 0.035    &  0.7$-$2  & 4500  & +240 & 20 $\pm$ 20  & 23, 24, 3, 25\\
 \multirow{7}{*}{\hfill} \\
 L1527 & 04:39:53.875 & 26:03:09.66 & 140   & 0.9 & 1.2  & 17000  & +90 & 3 $\pm$ 5 & 26, 7, 7, 25 \\
 \multirow{7}{*}{\hfill} \\
 L1157 & 20:39:06.269 & 68:02:15.70 &  352  & 4.0 & 3.0  &  15800 & +163 & 10 $\pm$ 10 & 27, 7, 7, 28-29\\
  \multirow{7}{*}{\hfill} \\
  GF9-2 & 20:51:29.823 & 60:18:38.44 & 200  & 0.3 &  0.5  & 7000 & 0  & 30 $\pm$ 20 \tablefootmark{$\star$ $\star$} &  30, 31, 3, ...\\
  
 \hfill & \hfill & \hfill & \hfill  & \hfill &\hfill &\hfill  &\hfill &\hfill &\hfill \\
  \hline
  \end{tabular}}
\end{center}
\tablefoot{
\tablefoottext{a}{Name of the protostars with the multiple components resolved by the 1.3~mm continuum emission from PdBI observations \citep{Maury18}.}
\tablefoottext{b}{Coordinates of the continuum emission peak at 1.3~mm from \cite{Maury18}.}
\tablefoottext{c}{Distance assumed for the individual sources. We adopt a value of 140~pc for the Taurus distance estimated from a VLBA measurement \citep{Torres09}. The distances of Perseus and Cepheus are taken following recent Gaia parallax measurements that have determined a distance of (293 $\pm$ 20)~pc \citep{OrtizLeon18} and (352 $\pm$ 18)~pc \citep{Zucker19}, respectively. We adopt a value of 200~pc for the GF9-2 cloud distance \citep{Wiesemeyer97, Wiesemeyer98} but this distance is very uncertain and some studies estimated a higher distance between 440-470~pc (\citealt{Viotti69}, C. Zucker, priv. comm.) and 900~pc \citep{Reid16}.
}
\tablefoottext{d}{Internal luminosities which come from the analysis of \textit{Herschel} maps from the Gould Belt survey (HGBS, \citealt{Andre10} and Ladjelate et al. in prep.) and corrected by the assumed distance.}
\tablefoottext{e}{Envelope mass corrected by the assumed distance.}
\tablefoottext{f}{Outer radius of the individual protostellar envelope determined from dust continuum emission, corrected by the assumed distance. We adopt the radius from PdBI dust continuum emission \citep{Maury18} when we do not have any information on the 30m continuum from \cite{Motte01} and for IRAS4A which is known to be embedded into a compressing cloud \citep{Belloche06}.}
\tablefoottext{g}{Position angle of the blue lobe of the outflows estimated from CALYPSO PdBI $^{12}$CO and SiO emission maps (Podio \& CALYPSO, in prep.). PA is defined east from north. Sources indicated with} \tablefoottext{$\star$}{have an asymmetric outflow and the position angles of both lobes are reported. For IRAS2A, IRAS4A, and L1157, previous works done by \cite{Codella14-bis}, \cite{Santangelo15}, and \cite{Podio16}, respectively, show a detailed CALYPSO view of the jets. For L1521F, we use the PA estimated by \cite{Tokuda14, Tokuda16}.}
\tablefoottext{h}{Inclination angle of the equatorial plane with respect to the line of sight. Sources indicated with} \tablefoottext{$\star \star$}{have an inclination angle not well constrained, so we assumed a default value of (30 $\pm$ 20)$^{\circ}$.} 
\tablefoottext{i}{References for the protostar discovery paper, the envelope mass, the envelope radius and then the inclination are reported here.}
}
\tablebib{(1) \cite{Olinger99}; (2) \cite{Enoch09}; (3) \cite{Maury18}; (4) \cite{Tobin07}; (5) \cite{Curiel90}; (6) \cite{Sadavoy14}; (7) \cite{Motte01}; (8) \cite{Kwon06}; (9) \cite{Anglada89}; (10) \cite{Bachiller95}; (11) \cite{Girart01}; (12) \cite{Jennings87}; (13) \cite{Karska13}; (14) \cite{Codella04}; (15) \cite{Maret14}; (16) \cite{Grossman87}; (17) \cite{Chini97}; (18) \cite{Ching16}; (19) \cite{Desmurs09}; (20) \cite{Andre99}; (21) \cite{Andre00}; (22) \cite{Belloche02}; (23) \cite{Mizuno94}; (24) \cite{Tokuda16}; (25) \cite{Terebey09}; (26) \cite{Ladd91}; (27) \cite{Umemoto92}; (28) \cite{Gueth96}; (29) \cite{Bachiller01}; (30) \cite{Schneider79}; (31) \cite{Wiesemeyer97}.
}
\label{table:sample}
\end{table*}

\section{Observations and dataset reduction}

To probe the dense gas in our sample of protostellar envelopes, we use high spectral resolution observations of the emission of two molecular lines, C$^{18}$O (2$-$1) at 219.560~GHz and N$_{2}$H$^{+}$ (1$-$0) at 93.171~GHz. In this section, we describe the dataset\footnote{The datasets used in this paper are available at \url{http://www.iram.fr/ILPA/LP010/}.} properties exploited to characterize the kinematics of the envelopes at radii between $r\sim$50 and 5000~au from the central object. 

\subsection{Observations with the IRAM Plateau de Bure Interferometer}
Observations of the 12 protostellar envelopes considered here were carried out with the IRAM Plateau de Bure Interferometer (PdBI) between September 2010 and March 2013.
We used the 6-antenna array in two configurations (A and C), providing baselines ranging from 16 to 760~m, to carry out observations of the dust continuum emission and a dozen molecular lines, using three spectral setups (around 94~GHz, 219~GHz, and 231~GHz).
Gain and flux were calibrated using CLIC which is part of the GILDAS\footnote{See \url{http://www.iram.fr/IRAMFR/GILDAS/} for more information about the GILDAS software \citep{Pety05}.} software. 
The details of CALYPSO observations and the calibration carried out are presented in \citet{Maury18}. The phase self-calibration corrections derived from the continuum emission gain curves, described in \citet{Maury18}, were also applied to the line visibility dataset (for all sources in the restricted sample studied here except the faintest sources IRAM04191, L1521F, GF9-2, and L1448-2A).
Here, we focus on the C$^{18}$O (2$-$1) emission line at 219560.3190~MHz and the N$_{2}$H$^{+}$ (1$-$0) emission line at 93176.2595~MHz, observed with high spectral resolution (39~kHz channels, i.e., a spectral resolution of 0.05~km~s$^{-1}$ at 1.3~mm and 0.13~km~s$^{-1}$ at 3~mm).
The C$^{18}$O (2$-$1) maps were produced from the continuum-subtracted visibility tables using either (i) a robust weighting of 1 for the brightest sources to minimize the side-lobes, or (ii) a natural weighting for the faintest sources (IRAM04191, L1521F, GF9-2, and L1448-2A) to minimize the rms noise values. We resampled the spectral resolution to 0.2~km~s$^{-1}$ to improve the signal-to-noise ratio of compact emission. 
The N$_{2}$H$^{+}$ (1$-$0) maps were produced from the continuum-subtracted visibility tables using a natural weighting for all sources. In all cases, deconvolution was carried out using the Hogbom algorithm in the MAPPING program of the GILDAS software.

%
%
%%%%%%%%%%%%%%%%%%%%%%%%%%%%%%%%%%%%
%
%

\subsection{Short-spacing observations from the IRAM 30-meter telescope}
The short-spacing observations were obtained at the IRAM 30-meter telescope (30m) between November 2011 and November 2014. Details of the observations for each source are reported in Table \ref{table:temps-observations-30m}. We observed the C$^{18}$O (2$-$1) and N$_{2}$H$^{+}$ (1$-$0) lines using the heterodyne Eight MIxer Receiver (EMIR) in two atmospheric windows: E230 band at 1.3~mm and E090 band at 3~mm \citep{Carter12}. The Fast Fourier Transform Spectrometer (FTS) and the VErsatile SPectrometer Array (VESPA) were connected to the EMIR receiver in both cases. 
The FTS200 backend provided a large bandwidth (4~GHz) with a spectral resolution of 200~kHz (0.27~km~s$^{-1}$) for the C$^{18}$O line, while VESPA provided  high spectral resolution observations (20~kHz channel or 0.063~km~s$^{-1}$) of the N$_{2}$H$^{+}$ line.
We used the on-the-fly spectral line mapping, with the telescope beam moving at a constant angular velocity to sample regularly the region of interest (1$\arcmin$ $\times$ 1$\arcmin$ coverage for the C$^{18}$O emission and 2$\arcmin$ $\times$ 2$\arcmin$ coverage for the N$_{2}$H$^{+}$ emission).
The mean atmospheric opacity at 225 GHz was $\tau_{225} \sim 0.2$ during the observations at 1.3~mm and $\tau_{225} \sim 0.5$ during the observations at 3~mm.
The mean values of atmospheric opacity are reported for each source in Table \ref{table:temps-observations-30m}.
The telescope pointing was checked every 2$-$3~h on quasars close to the CALYPSO sources, and the telescope focus was corrected every 4$-$5~h using the planets available in the sky.
The single-dish dataset were reduced using the MIRA and CLASS programs of the GILDAS software following the standard steps: flagging of incorrect channels, temperature calibration, baseline subtraction, and gridding of individual spectra to produce regularly-sampled maps.

%
%
%%%%%%%%%%%%%%%%%%%%%%%%%%%%%%%%%%%%
%
%

\subsection{Combination of the PdBI and 30m data}
The IRAM PdBI observations are mostly sensitive to compact emission from the inner envelope.
Inversely, single-dish dataset contains information at envelope scales ($r \sim 5-40\arcsec$) but its angular resolution does not allow us to characterize the inner envelope emission at scales smaller than the beamwidth. 
To constrain the kinematics at all relevant scales of the envelope, one has to build high angular resolution dataset which recovers all emission of protostellar envelopes.
We merged the PdBI and the 30m datasets (hereafter PdBI+30m) for each tracer using the pseudo-visibility method\footnote{For details, see \url{http://www.iram.fr/IRAMFR/GILDAS/doc/pdf/map.pdf}.}: we generated pseudo-visibilities from the Fourier transformed 30m image data, which are then merged to the PdBI dataset in the MAPPING program of the GILDAS software. This process degrades the angular resolution of the PdBI dataset but recovers a large fraction of the extended emission. The spectral resolution of the combined PdBI+30m dataset is limited by the 30m dataset at 1.3~mm (0.27~km~s$^{-1}$) and the PdBI one at 3~mm (0.13~km~s$^{-1}$).

We produced the N$_{2}$H$^{+}$ (1$-$0) combined datacubes in such a way to have a synthesized beam size $<2\arcsec$ and a noise level $<$10~mJy~beam$^{-1}$. As the N$_{2}$H$^{+}$ emission traces preferably the outer protostellar envelope, we used a natural weighting to build the combined maps to minimize the noise rather than to maximize angular resolution.
The C$^{18}$O (2$-$1) combined maps were produced using a robust weighting scheme, in order to obtain synthesized beam sizes close to the PdBI ones, and to minimize the side-lobes. In all cases, the deconvolution was carried out using the Hogbom algorithm in MAPPING.

\subsection{Properties of the analyzed maps}
Following the procedure described above, we have obtained, for each source of the sample, a set of three cubes for each of the two molecular tracers C$^{18}$O (2$-$1) and N$_{2}$H$^{+}$ (1$-$0), probing the emission at different spatial scales (PdBI map, combined PdBI+30m map, and 30m map). 
In order to build maps with pixels that contain independent dataset and avoid oversampling, we inversed visibilities from the PdBI and the combined PdBI+30m datasets using only 4 pixels per synthesized beam, and we smoothed the resulting maps afterwards to obtain 2 pixels per element of resolution. 
The properties of the resulting maps are reported in Appendix \ref{properties-maps}. The spatial resolution of the molecular line emission maps is reported in Tables \ref{table:beams-c18o} and \ref{table:beams-n2hp}. 
The spatial extent of the molecular emission, the rms noise levels, and the integrated fluxes are reported in Tables \ref{table:details-obs-c18o} and \ref{table:details-obs-n2hp}.

%
%
%%%%%%%%%%%%%%%%%%%%%%%%%%%%%%%%%%%%
%
%

%
%
%__________________________________________________________________
%
%%%%%%%%%%%%%%%%%%%%%%%%%%%%%%%%%%%%

\begin{figure*}[!ht]
\centering
\includegraphics[scale=0.5,angle=0,trim=0cm 1.5cm 0cm 1.5cm,clip=true]{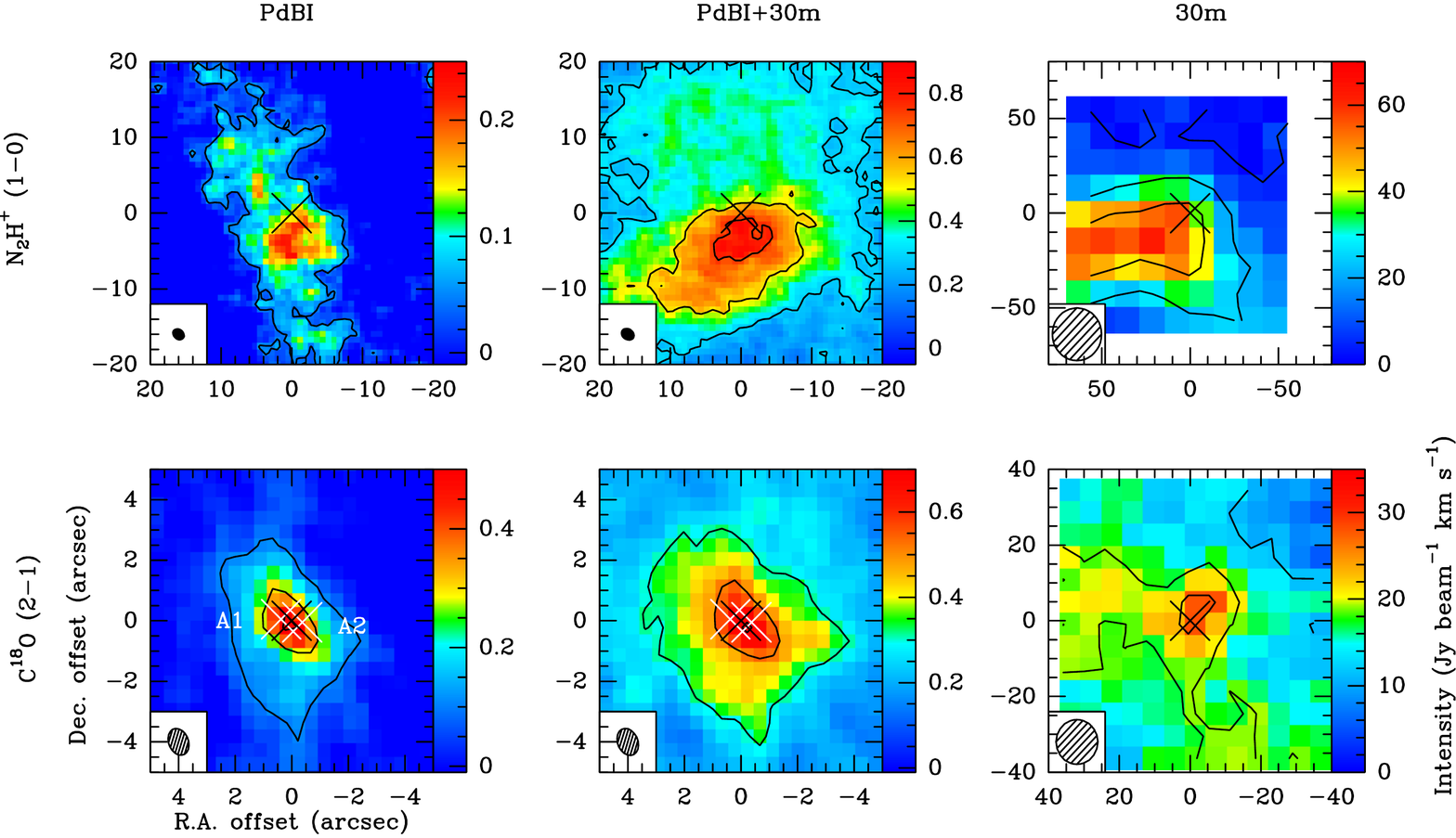}
\caption{Integrated intensity maps of N$_{2}$H$^{+}$ (1$-$0) (top) and C$^{18}$O (2$-$1) (bottom) emission from the PdBI (left), combined (middle), and 30m (right) datasets for L1448-2A.
The white crosses represent the positions of the binary system determined from the 1.3~mm dust continuum emission.
The black cross represents the middle position between the binary system. The clean beam is shown by an ellipse on the bottom left of each map.
The black lines represent the integrated intensity contours of each tracer starting at 5$\sigma$ and increasing in steps of 25$\sigma$ for N$_{2}$H$^{+}$ and 10$\sigma$ for C$^{18}$O (see Tables \ref{table:details-obs-c18o} and \ref{table:details-obs-n2hp}). Be careful, the spatial scales of the maps are not uniform in all panels.}
\label{fig:intensity-maps-L1448-2A}
\end{figure*}

\section{Envelope kinematics from high dynamic range datasets}

\subsection{Integrated intensity maps}

\begin{figure}[h]
\begin{center}
\includegraphics[scale=0.3,angle=0,trim=0.3cm 0cm 0.4cm 4cm,clip=true]{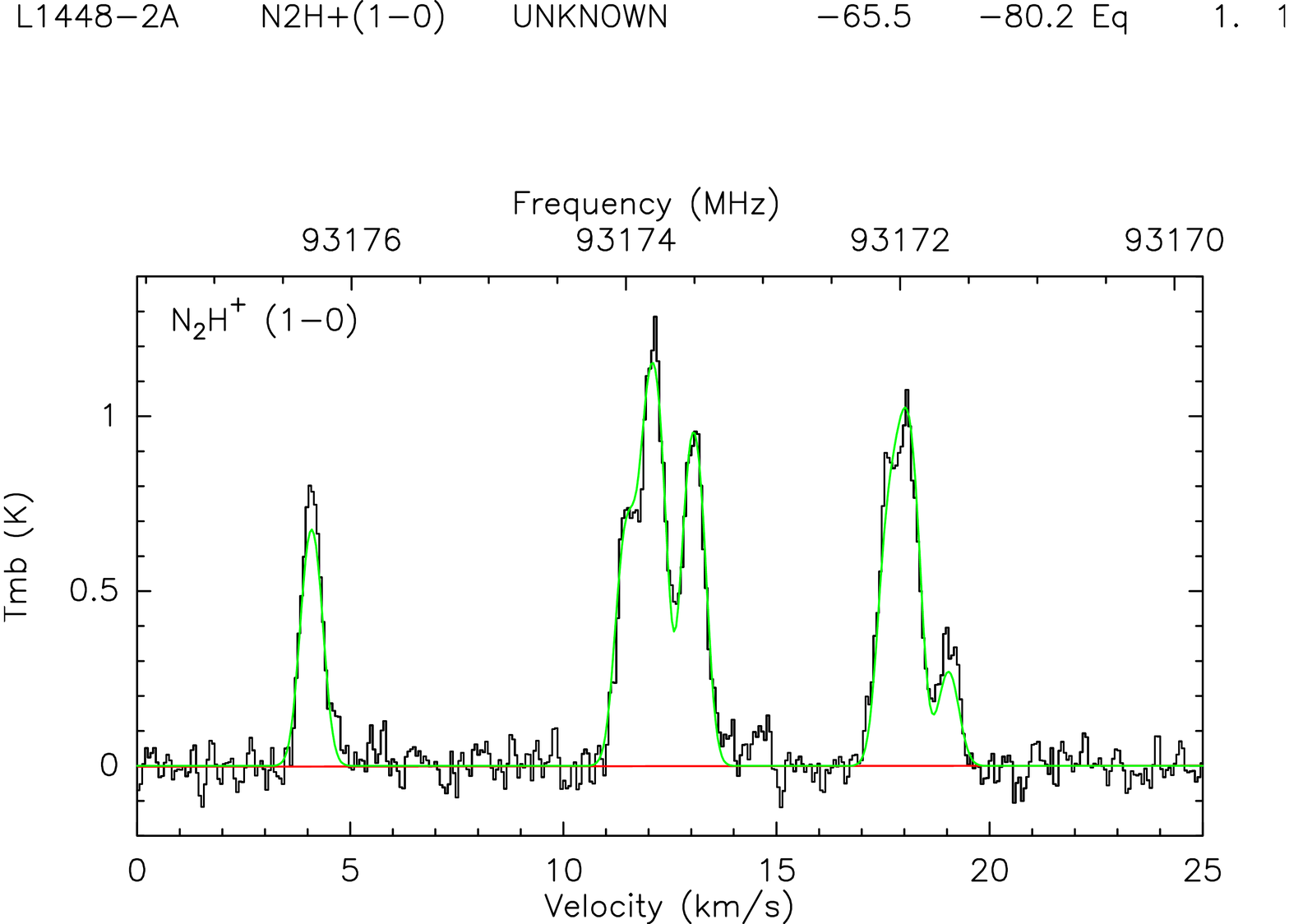}
\includegraphics[scale=0.3,angle=0,trim=0.3cm 0cm 0.4cm 4cm,clip=true]{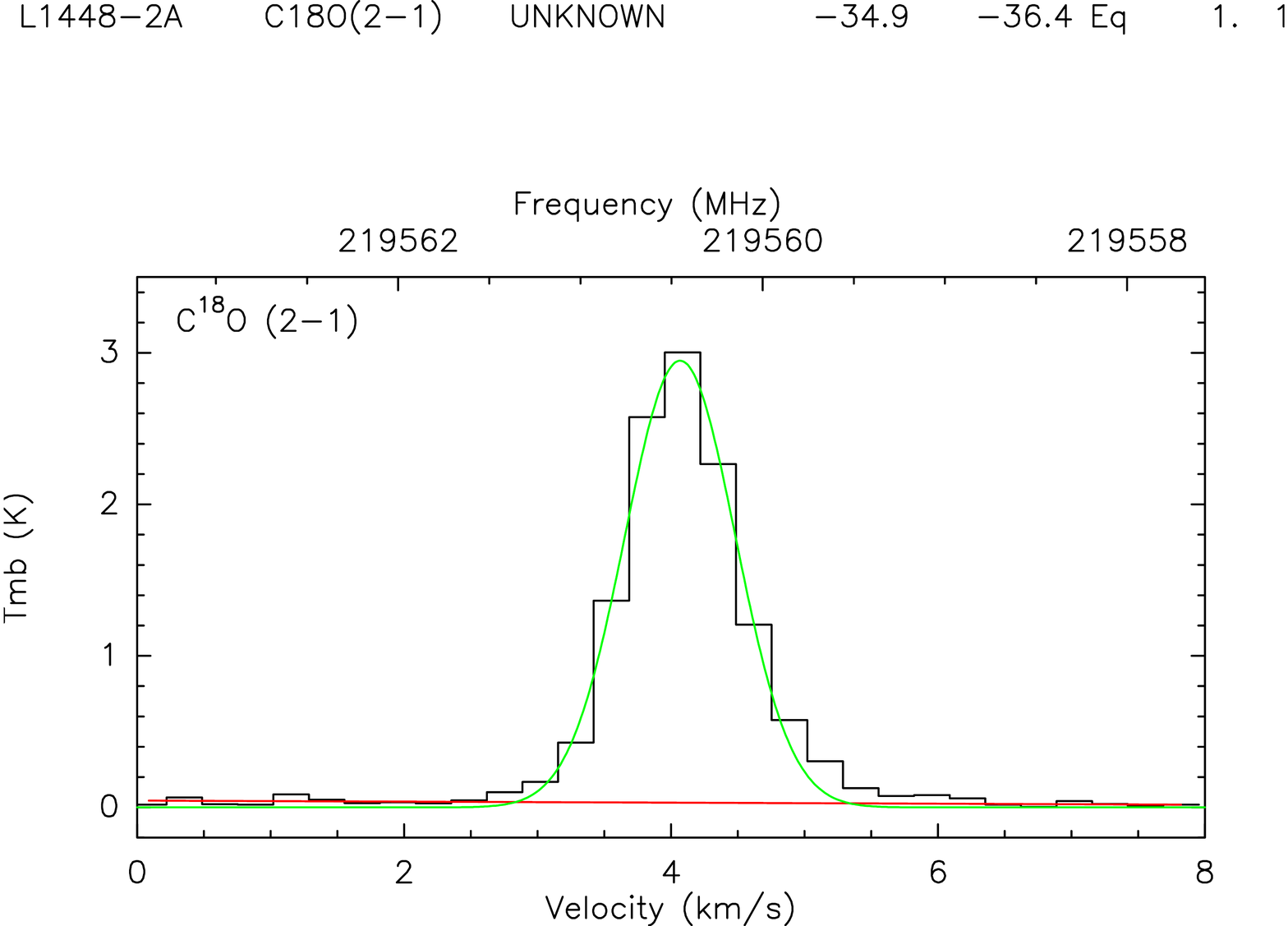}
\caption{Mean spectra of the N$_2$H$^{+}$ (top) and C$^{18}$O (bottom) molecular lines from the 30m datasets for L1448-2A. The best fits of the spectra, by a hyperfine structure and a Gaussian line profile models respectively, are represented in green solid lines. In the top panel, the velocity axis corresponds to the isolated HFS component $1_{01}-0_{12}$. The systemic velocity is estimated to be 4.10~km~s$^{-1}$ for this source (see Table \ref{table:vitesse-systemique-n2hp-30m}). }
\label{fig:spectra-30m-L1448C}
\end{center}
\end{figure}

To identify at which scales of the protostellar envelopes the different datasets are sensitive to, we produced integrated intensity maps by integrating spectra of each pixel for the molecular lines C$^{18}$O (2$-$1) and N$_{2}$H$^{+}$ (1$-$0) from the PdBI, combined, and 30m datasets for each source. 
For C$^{18}$O (2$-$1), we integrated each spectrum on a velocity range of $\pm$ 2.5~km~s$^{-1}$ around the velocity of the peak of the mean spectrum of each source. The 1$-$0 line of N$_{2}$H$^{+}$ has a hyperfine structure with seven components (see Fig. \ref{fig:spectra-30m-L1448C}). We integrated the N$_{2}$H$^{+}$ spectra over a range of 20~km~s$^{-1}$ encompassing the seven 
components. Figure~\ref{fig:intensity-maps-L1448-2A} shows as an example the integrated intensity maps obtained for L1448-2A. The integrated intensity maps of the other sources are provided in Appendix~\ref{sec:comments-indiv-sources}.

We used the integrated intensity maps to measure the average emission size of each tracer in each dataset above a 5$\sigma$ threshold.
The values reported in Tables \ref{table:details-obs-c18o} and \ref{table:details-obs-n2hp} are the average of two measurements: an intensity cut along the equatorial axis and circular averages at different radii around the intensity peak position of the source. Only pixels whose intensity is at least 5 times higher than the noise in the map are considered to build these intensity profiles. The FWHM of the adjustment by a Gaussian function allows us to determine the average emission size of the sources. For both tracers and for all sources in our sample, the emission is detected above 5$\sigma$ in an area larger in the combined datasets than in the PdBI datasets, and smaller than in the 30m ones (see Tables \ref{table:details-obs-c18o} and \ref{table:details-obs-n2hp}). Our three datasets are thus not sensitive to the same scales and allow us to probe different scales within the 12 sampled protostellar envelopes: the 30m datasets trace the outer envelope, the PdBI datasets the inner part and the combined ones the intermediate scales.

The C$^{18}$O and N$_{2}$H$^{+}$ molecules do not trace the same regions of the protostellar envelope either: \cite{Anderl16} report from an analysis of the CALYPSO survey that the N$_{2}$H$^{+}$ emission forms a ring around the central C18O emission in four sources.
Previous studies \citep{Bergin02,Maret02,Maret07,Anderl16} show that N$_{2}$H$^{+}$, which is abundant 
in the outer envelope, is chemically destroyed when the temperature in the envelope reaches the critical temperature ($T \gtrsim$20K) at which CO desorbs from dust ice mantles. Thus, while N$_{2}$H$^{+}$ can be used to probe the envelope kinematics at outer envelope scales, C$^{18}$O can be used as a complementary tracer of the gas kinematics at smaller radii where the embedded protostellar embryo heats the gas to higher temperatures. 

The C$^{18}$O emission is robustly detected ($>$5$\sigma$) in our PdBI observations for most sources, except for L1521F and IRAM04191 which are the lowest luminosity sources of our sample (see Table \ref{table:sample}), and for SVS13-B where the emission is dominated by its companion, the Class~I protostar SVS13-A. For most sources, the interferometric map obtained with the PdBI shows mostly compact emission ($r < 3 \arcsec$, see Table \ref{table:details-obs-c18o}). However, the C$^{18}$O emission from the 30m datasets shows more complex structures (see Appendix~\ref{sec:comments-indiv-sources}). Assuming that, under the hypothesis of spherical geometry, the emission from a protostellar envelope is compact ($r \lesssim$40$\arcsec$, i.e., $\lesssim$10000~au, see Table \ref{table:sample}) and stands out from the environment in which it is embedded, the 30m emission of L1448-2A, L1448-C, and IRAS4A comes mainly from the envelope.

The N$_{2}$H$^{+}$ emission is detected in our combined observations for all sources. In the four sources studied by \cite{Anderl16}, they do not detect the emission at the 1.3~mm continuum peak, but emission rings around the C$^{18}$O central emission. From Table \ref{table:details-obs-n2hp}, we noticed two types of emission morphologies based on the PdBI dataset: compact ($r <$7$\arcsec$, see Table \ref{table:details-obs-n2hp}) or filamentary ($r \geq$9$\arcsec$). In the same way as the C$^{18}$O emission, the N$_{2}$H$^{+}$ emission from the 30m datasets shows complex structures with radius $r \gtrsim$40$\arcsec$ for most sources, except for five sources (IRAM04191, L1521F, L1448-NB, L1448-C, and L1157) where the emission is consistent with the compact emission of the protostellar envelope.

The C$^{18}$O emission from the PdBI is not centered on the continuum peak for three sources in our sample: IRAS4A, L1448-NB, and L1448-2A (see Appendix~\ref{sec:comments-indiv-sources}). 
For each of these sources, the PdBI 1.3~mm dust continuum emission map resolves a close binary system ($<$600~au) with both components embedded in the same protostellar envelope (\citealt{Maury18}, see Table \ref{table:sample}). The origin of the coordinate offsets is chosen to be the main protostar, secondary protostar, and the middle of the binary system for IRAS4A, L1448-NB, and L1448-2A, respectively, to study the kinematics in a symmetrical way.

%
%
%%%%%%%%%%%%%%%%%%%%%%%%%%%%%%%%%%%%
%
%

\subsection{Velocity gradients in protostellar envelopes} \label{sec:velocity-maps} 
\begin{figure*}[ht]
\centering
\includegraphics[scale=0.5,angle=0,trim=0cm 1.5cm 0cm 1.5cm,clip=true]{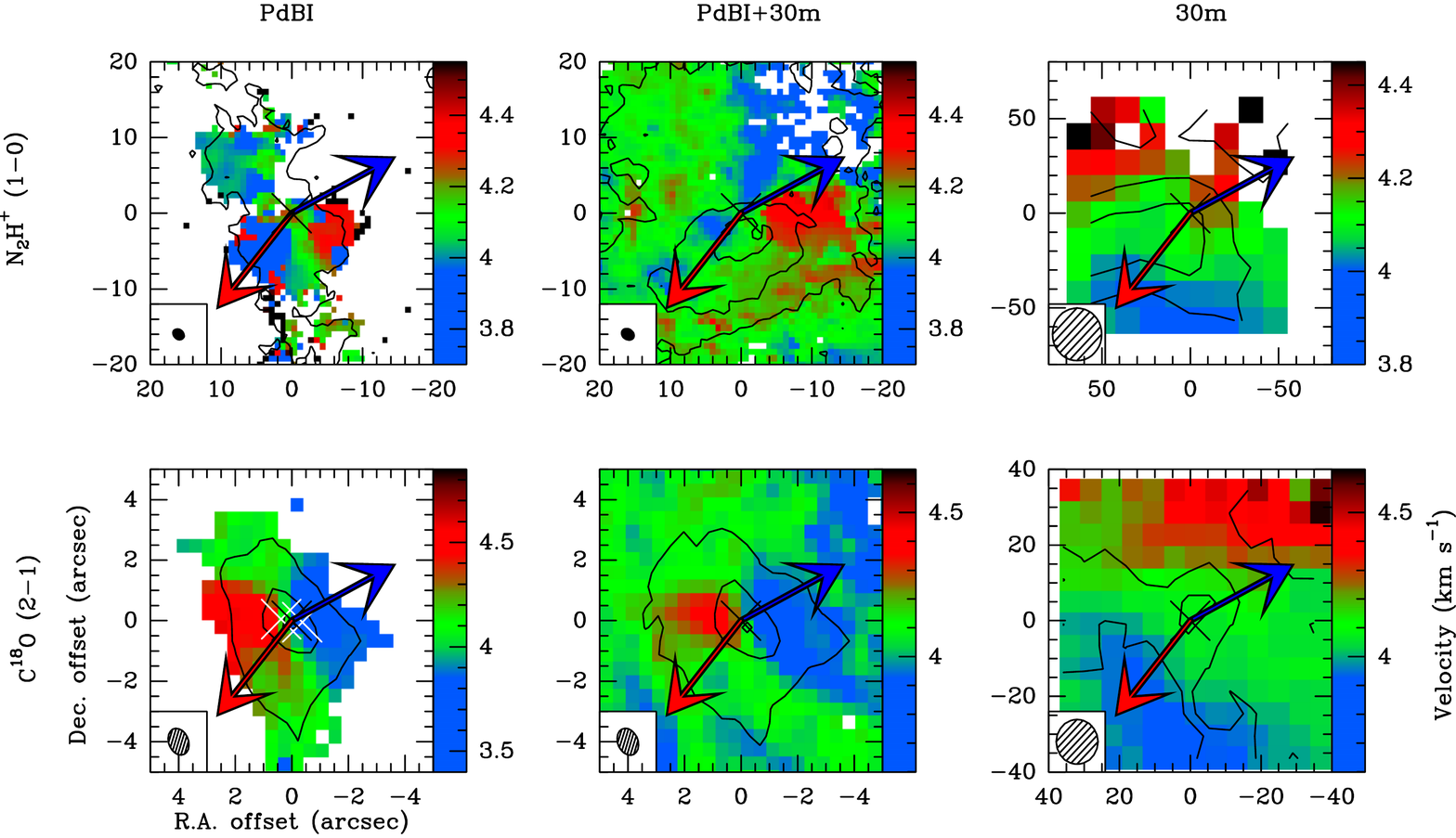}
\caption{Centroid velocity maps of N$_{2}$H$^{+}$ (1$-$0) (top) and C$^{18}$O (2$-$1) (bottom) emission from the PdBI (left), combined (middle), and 30m (right) datasets for L1448-2A. The blue and red solid arrows represent the directions of the blue- and red-shifted outflow lobes, respectively. The white crosses represent the positions of the binary system determined from the 1.3~mm dust continuum emission (see Table \ref{table:sample}).
The black cross represents the middle position between the binary system. The clean beam is shown by an ellipse on the bottom left. The integrated intensity contours in black are the same as in Fig. \ref{fig:intensity-maps-L1448-2A}.
}
\label{fig:velocity-maps-L1448-2A}
\end{figure*}

\begin{figure*}[!ht]
\centering
\includegraphics[scale=0.5,angle=0,trim=0cm 1.5cm 0cm 1.5cm,clip=true]{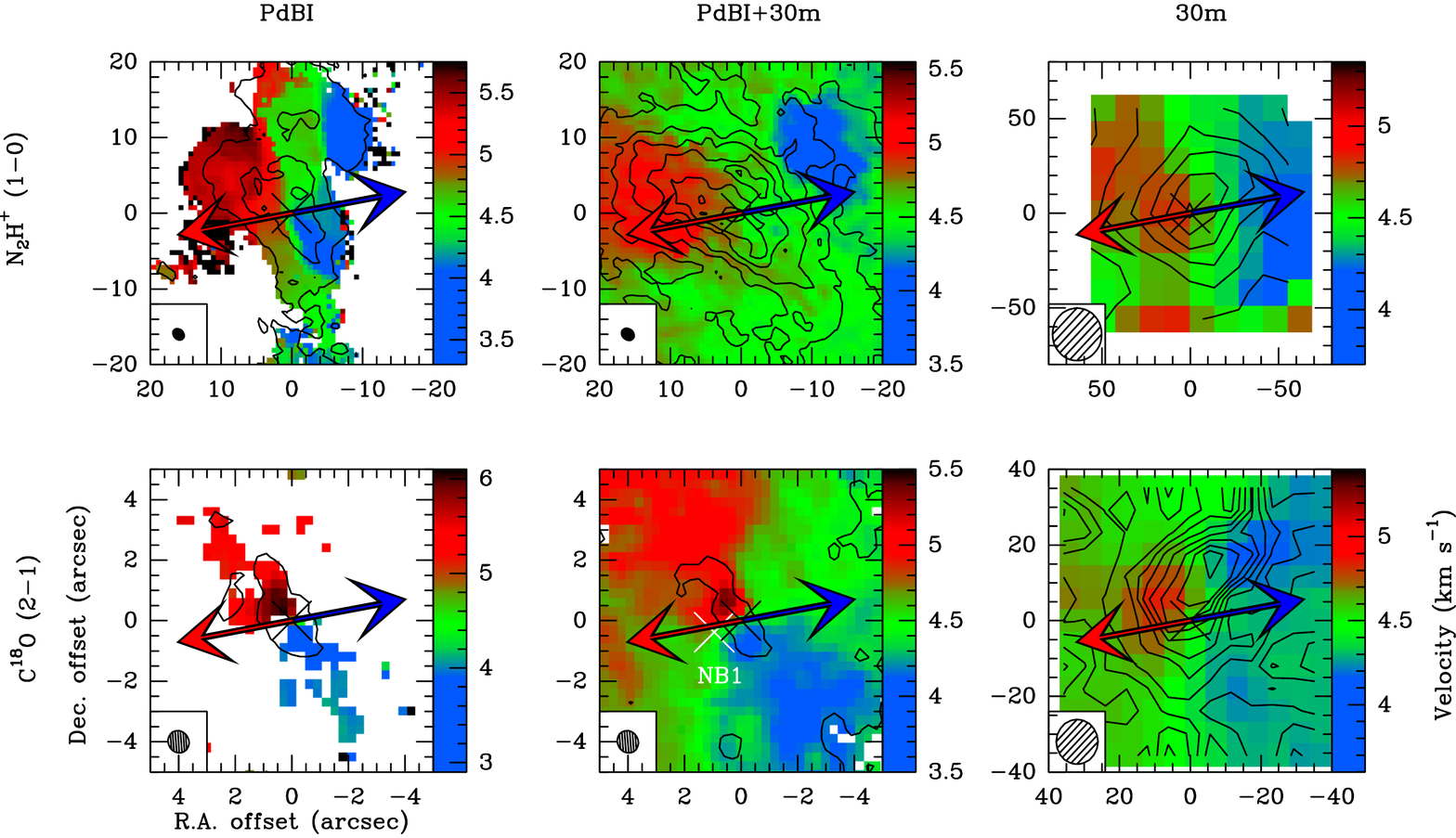}
\caption{Same as Figure \ref{fig:velocity-maps-L1448-2A}, but for L1448-NB. The white cross represents the position of main protostar L1448-NB1 determined from the 1.3~mm dust continuum emission (see Table \ref{table:sample}). The black cross represents the position of the secondary protostar L1448-NB2 of the multiple system.
}
\label{fig:velocity-maps-L1448NB}
\end{figure*}

\begin{figure*}[!ht]
\centering
\includegraphics[scale=0.5,angle=0,trim=0cm 1.5cm 0cm 1.5cm,clip=true]{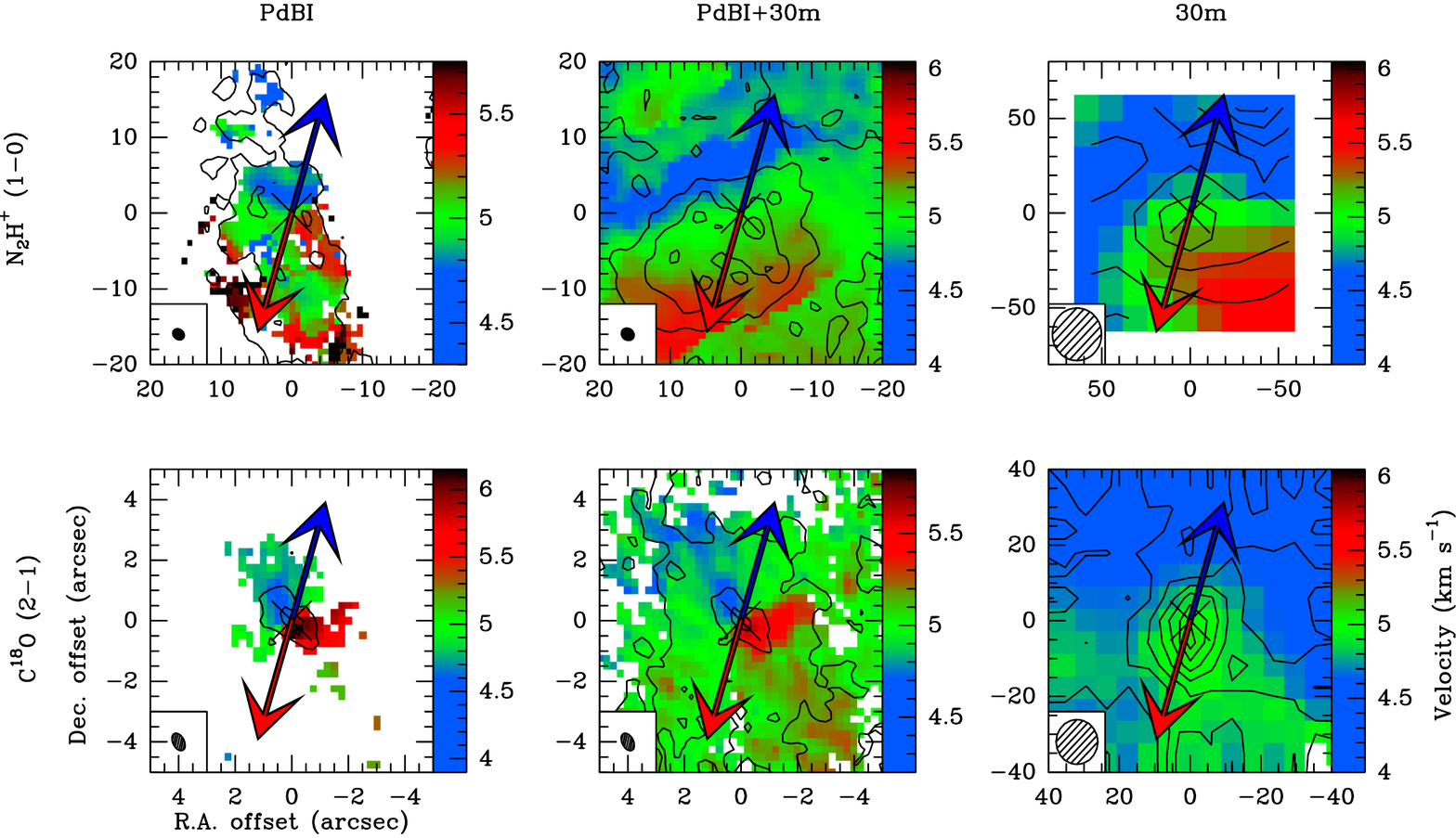}
\caption{Same as Figure \ref{fig:velocity-maps-L1448-2A}, but for L1448-C. 
}
\label{fig:velocity-maps-L1448C}
\end{figure*}

\begin{figure*}[!ht]
\centering
\includegraphics[scale=0.5,angle=0,trim=0cm 1.5cm 0cm 1.5cm,clip=true]{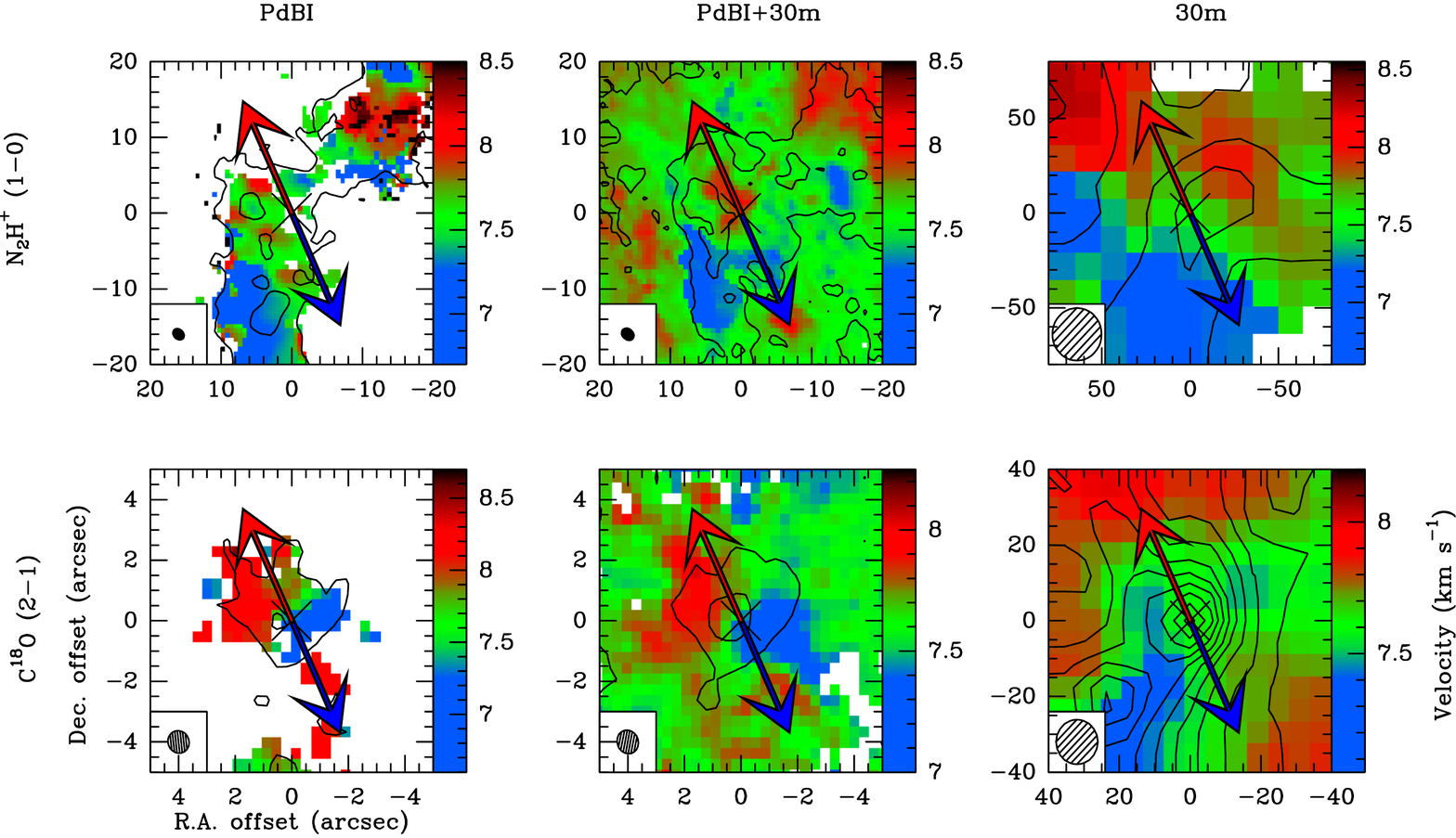}
\caption{Same as Figure \ref{fig:velocity-maps-L1448-2A}, but for IRAS2A. 
}
\label{fig:velocity-maps-IRAS2A}
\end{figure*}

\begin{figure*}[!ht]
\centering
\includegraphics[scale=0.5,angle=0,trim=0cm 1.5cm 0cm 1.5cm,clip=true]{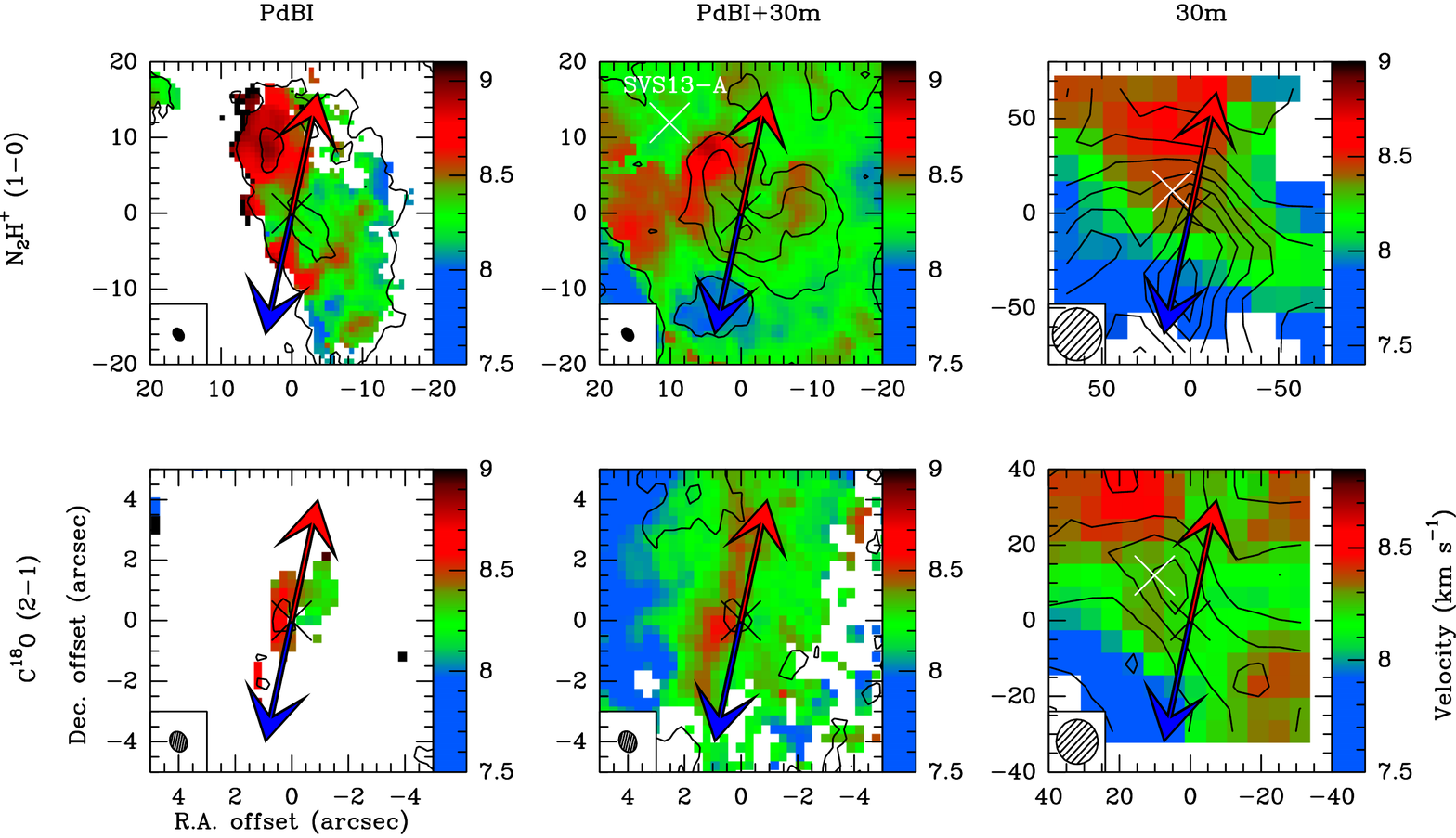}
\caption{Same as Figure \ref{fig:velocity-maps-L1448-2A}, but for SVS13-B. The white cross represents the position of the Class~I protostar SVS13-A determined from the 1.3~mm dust continuum emission \citep{Maury18}.
}
\label{fig:velocity-maps-SVS13B}
\end{figure*}

\begin{figure*}[!ht]
\centering
\includegraphics[scale=0.5,angle=0,trim=0cm 1.5cm 0cm 1.5cm,clip=true]{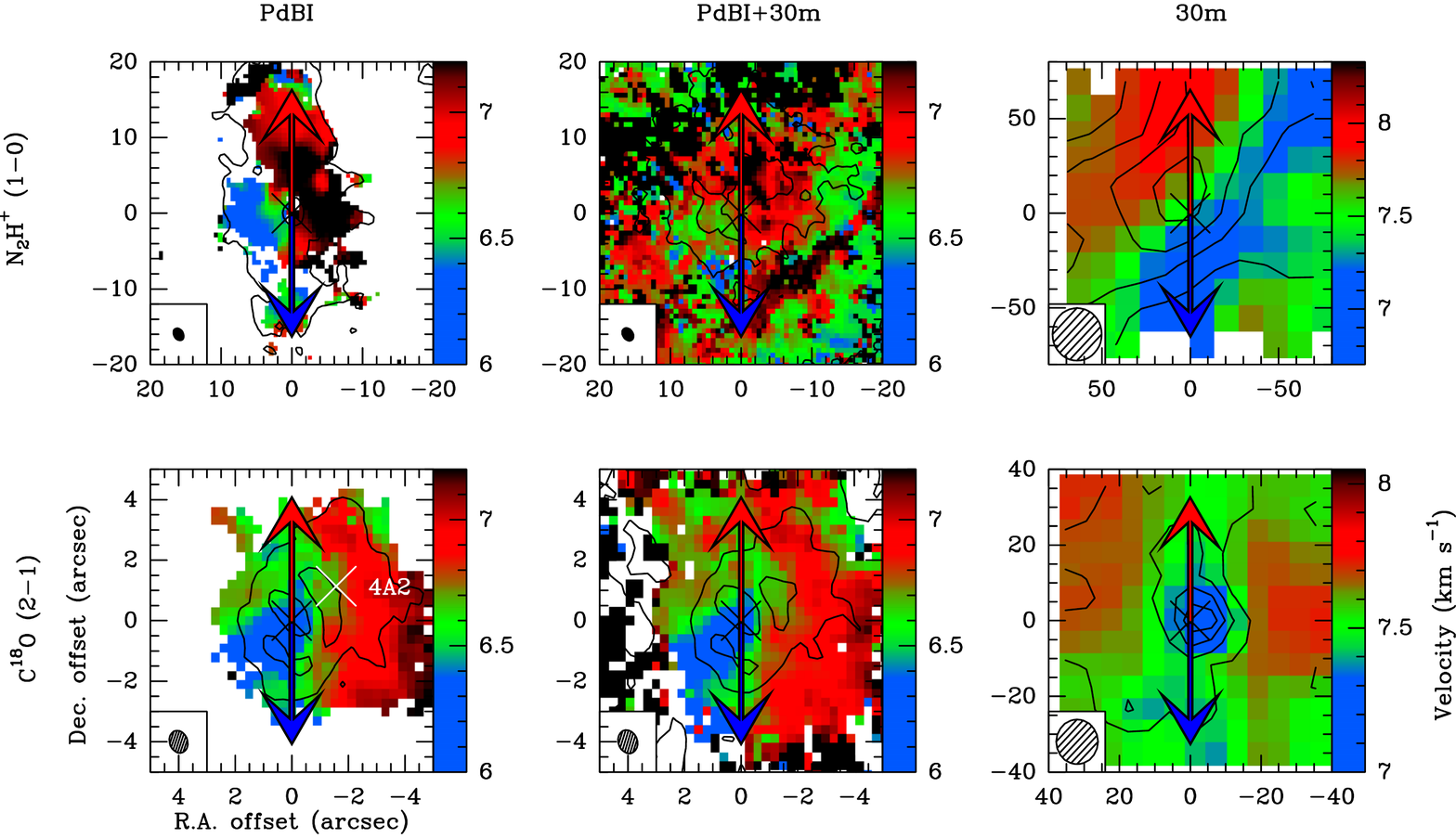}
\caption{Same as Figure \ref{fig:velocity-maps-L1448-2A}, but for IRAS4A. The white cross represents the position of secondary protostar IRAS4A2 determined from the 1.3~mm dust continuum emission (see Table \ref{table:sample}). The black cross represents the position of the main protostar IRAS4A1 of the multiple system. 
}
\label{fig:velocity-maps-IRAS4A}
\end{figure*}

\begin{figure*}[!ht]
\centering
\includegraphics[scale=0.5,angle=0,trim=0cm 1.5cm 0cm 1.5cm,clip=true]{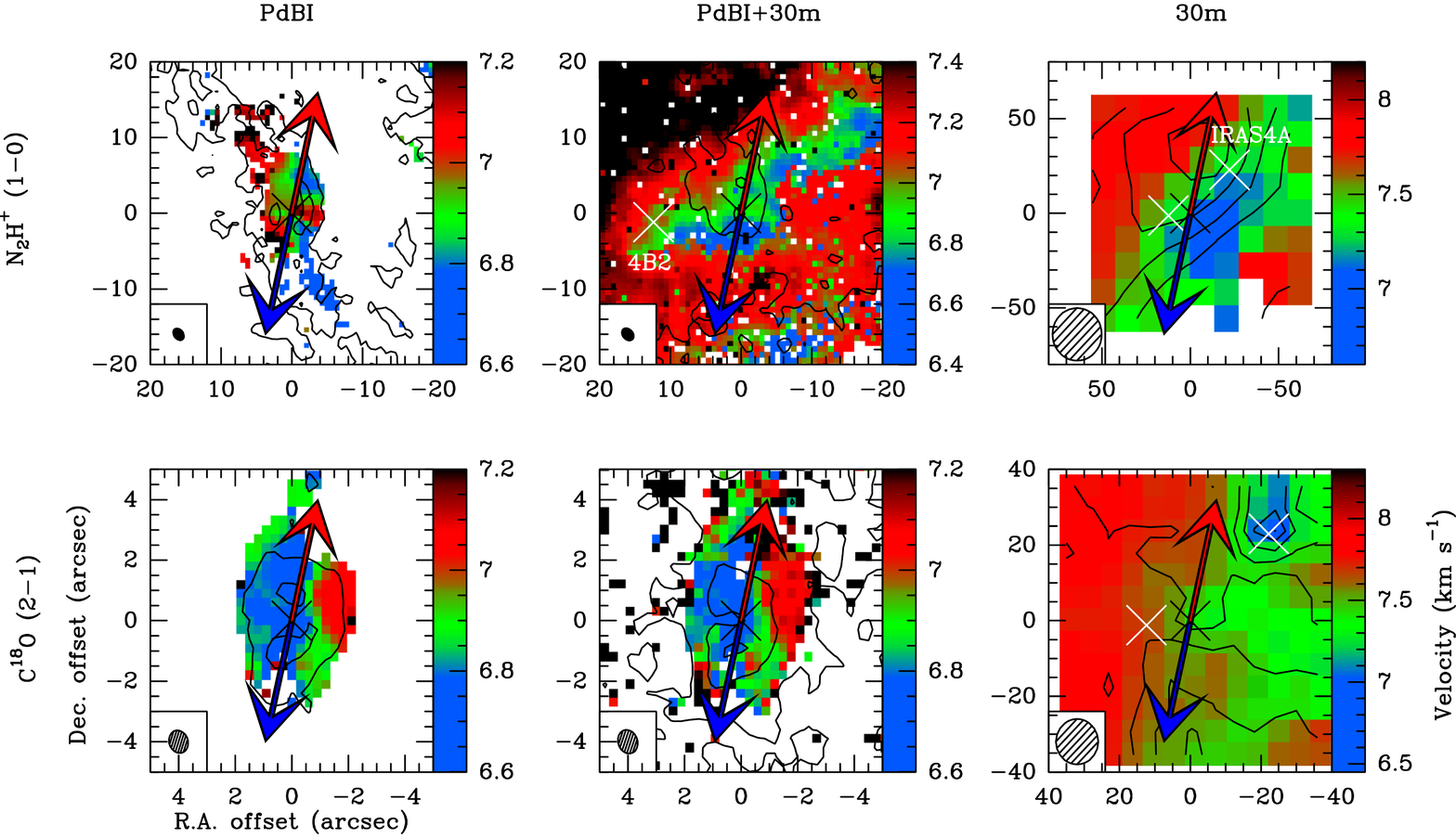}
\caption{Same as Figure \ref{fig:velocity-maps-L1448-2A}, but for IRAS4B. The white crosses represent the position of secondary protostar IRAS4B2 and the position of IRAS4A, respectively, determined from the 1.3~mm dust continuum emission (see Table \ref{table:sample}). The black cross represents the position of the secondary protostar L1448-NB2 of the multiple system.
}
\label{fig:velocity-maps-IRAS4B}
\end{figure*}

\begin{figure*}[!ht]
\centering
\includegraphics[scale=0.5,angle=0,trim=0cm 1.5cm 0cm 1.5cm,clip=true]{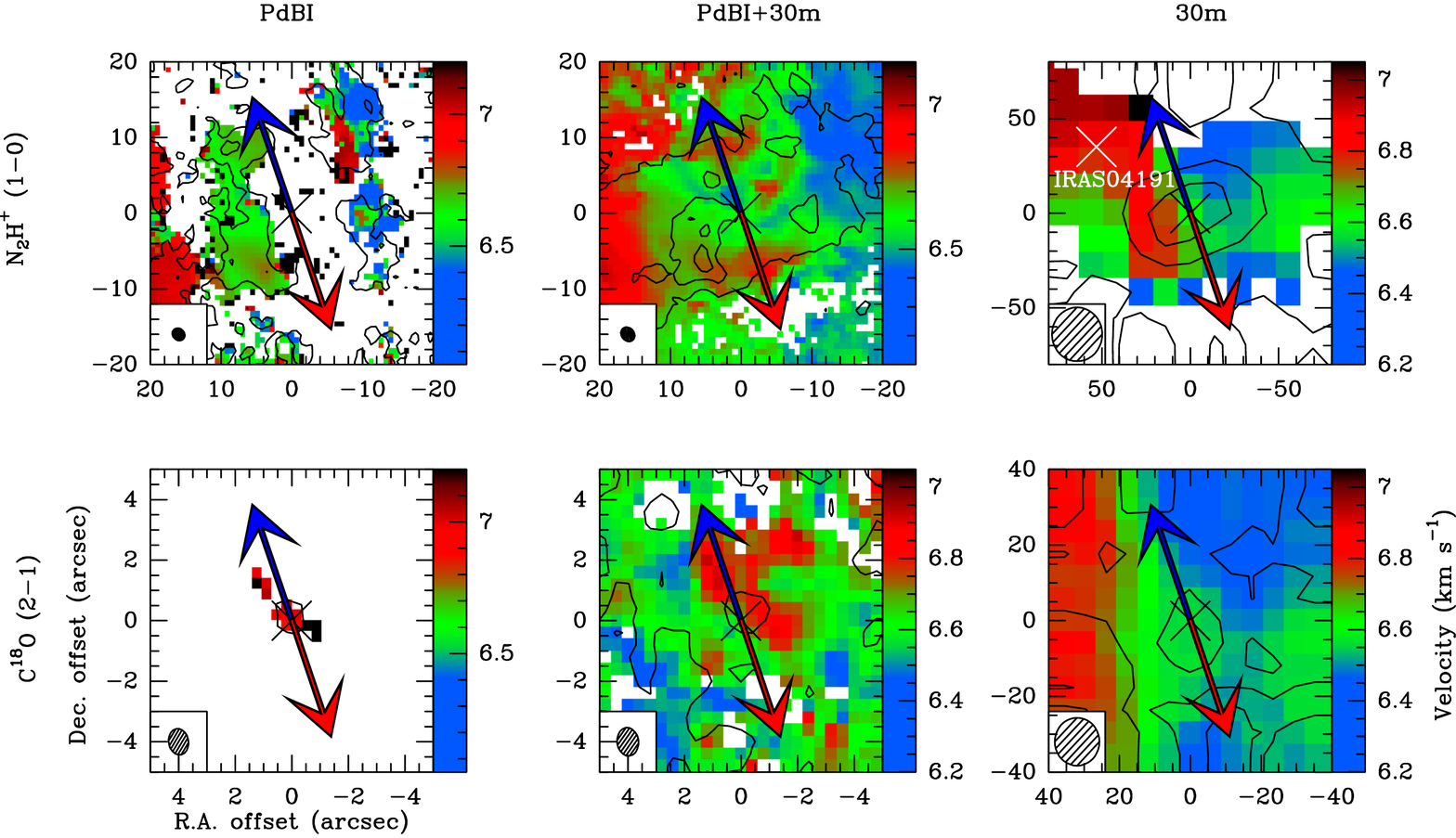}
\caption{Same as Figure \ref{fig:velocity-maps-L1448-2A}, but for IRAM04191. The white cross represents the position of the Class~I protostar IRAS04191. The black cross represents the position of IRAM04191 determined from the 1.3~mm dust continuum emission (see Table \ref{table:sample}).
}
\label{fig:velocity-maps-IRAM04191}
\end{figure*}

\begin{figure*}[!ht]
\centering
\includegraphics[scale=0.5,angle=0,trim=0cm 1.5cm 0cm 1.5cm,clip=true]{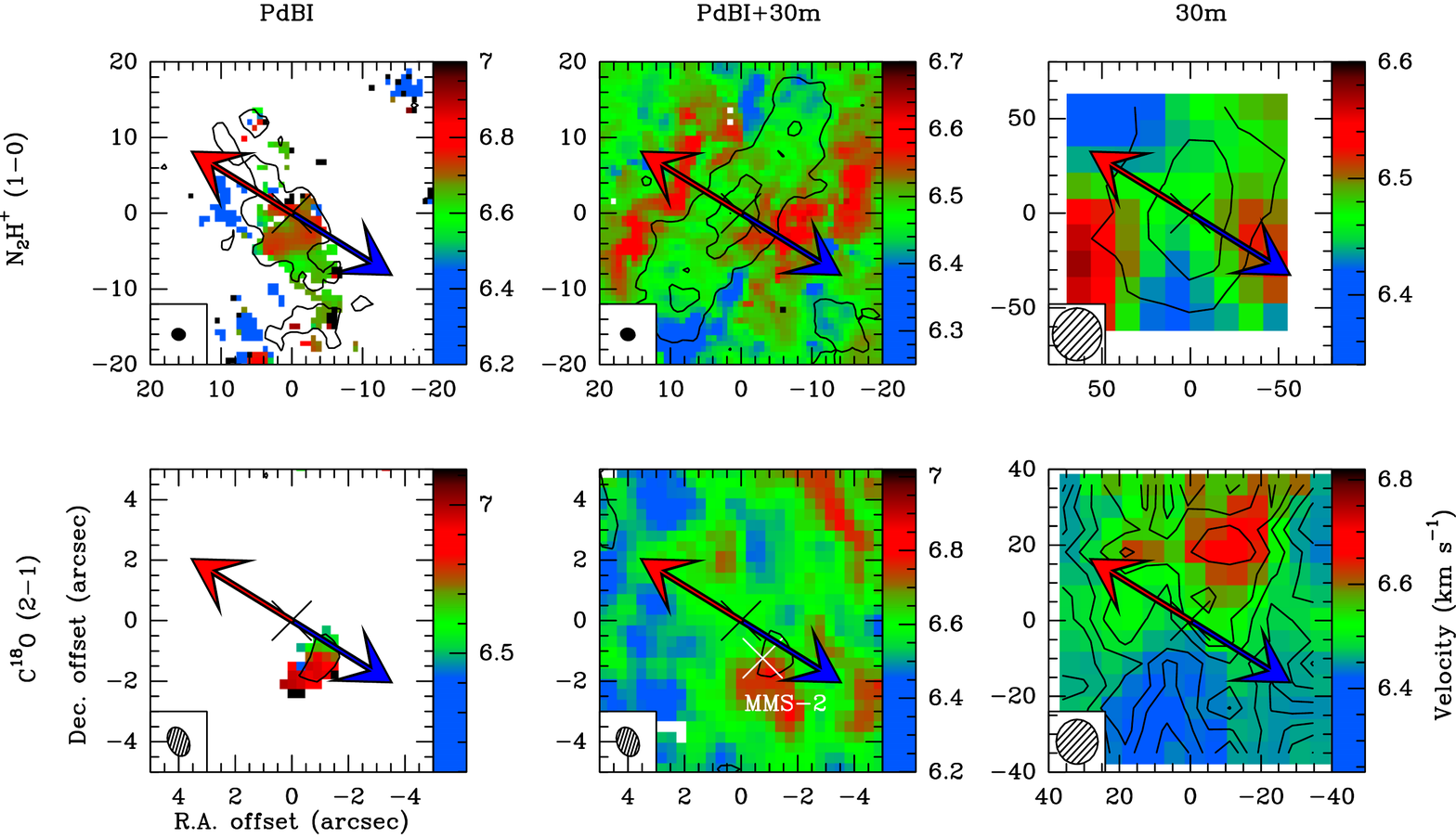}
\caption{Same as Figure \ref{fig:velocity-maps-L1448-2A} for L1521F. The white cross represents the position of the starless dense core MMS-2.
}
\label{fig:velocity-maps-L1521F}
\end{figure*}

\begin{figure*}[!ht]
\centering
\includegraphics[scale=0.5,angle=0,trim=0cm 1.5cm 0cm 1.5cm,clip=true]{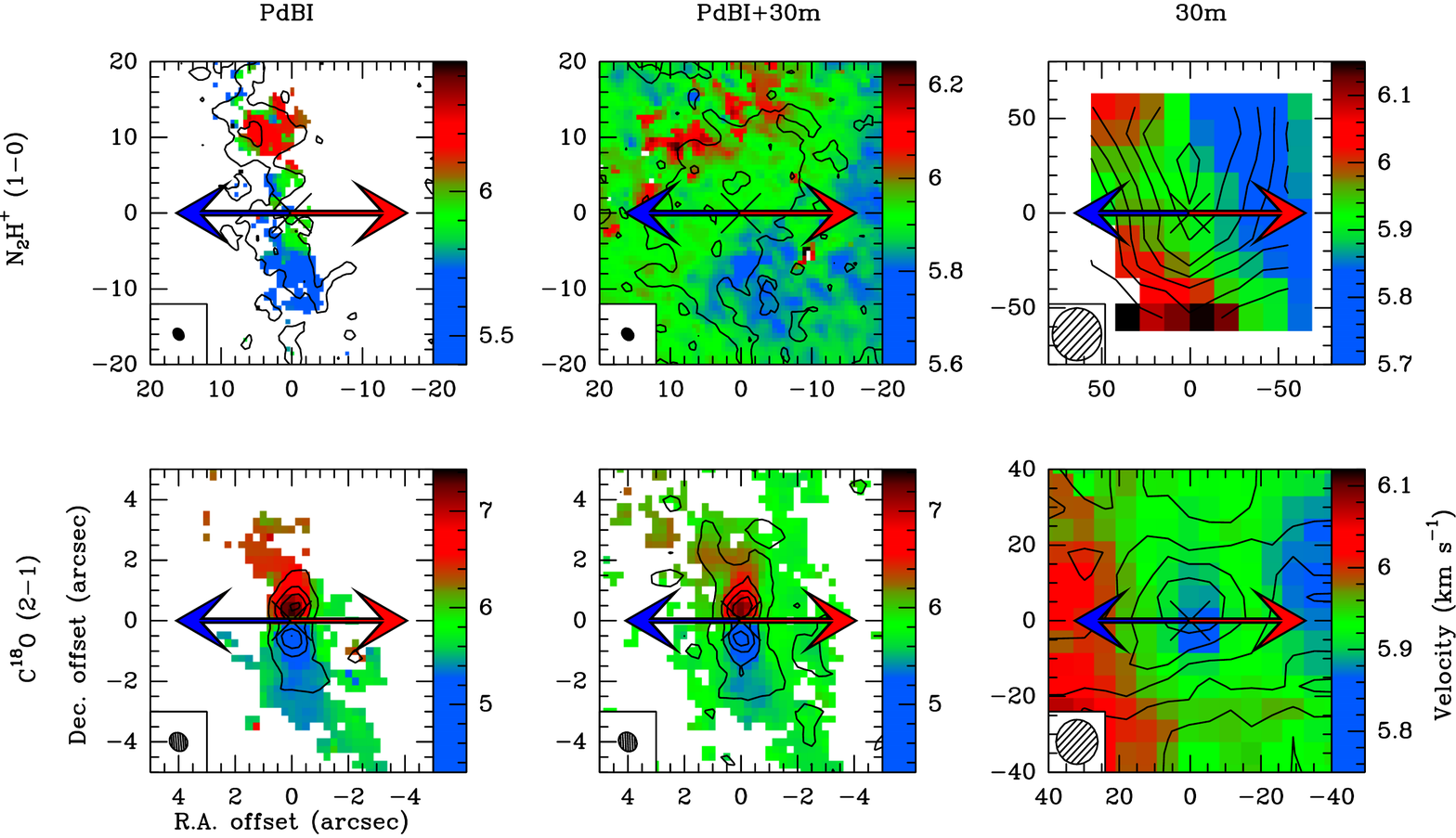}
\caption{Same as Figure \ref{fig:velocity-maps-L1448-2A}, but for L1527. 
}
\label{fig:velocity-maps-L1527}
\end{figure*}

\begin{figure*}[!ht]
\centering
\includegraphics[scale=0.5,angle=0,trim=0cm 1.5cm 0cm 1.5cm,clip=true]{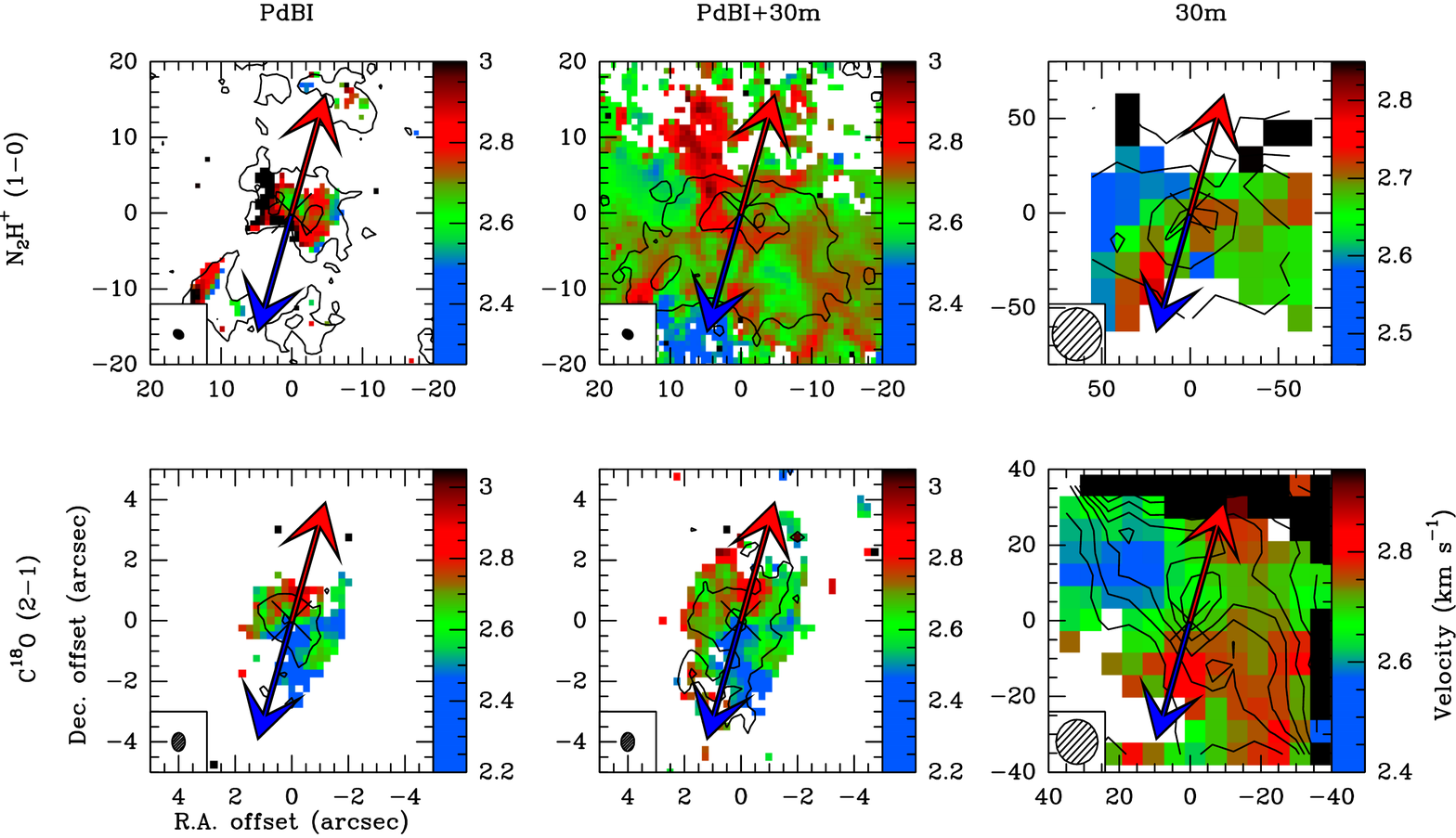}
\caption{Same as Figure \ref{fig:velocity-maps-L1448-2A}, but for L1157. 
}
\label{fig:velocity-maps-L1157}
\end{figure*}

\begin{figure*}[!ht]
\centering
\includegraphics[scale=0.5,angle=0,trim=0cm 1.5cm 0cm 1.5cm,clip=true]{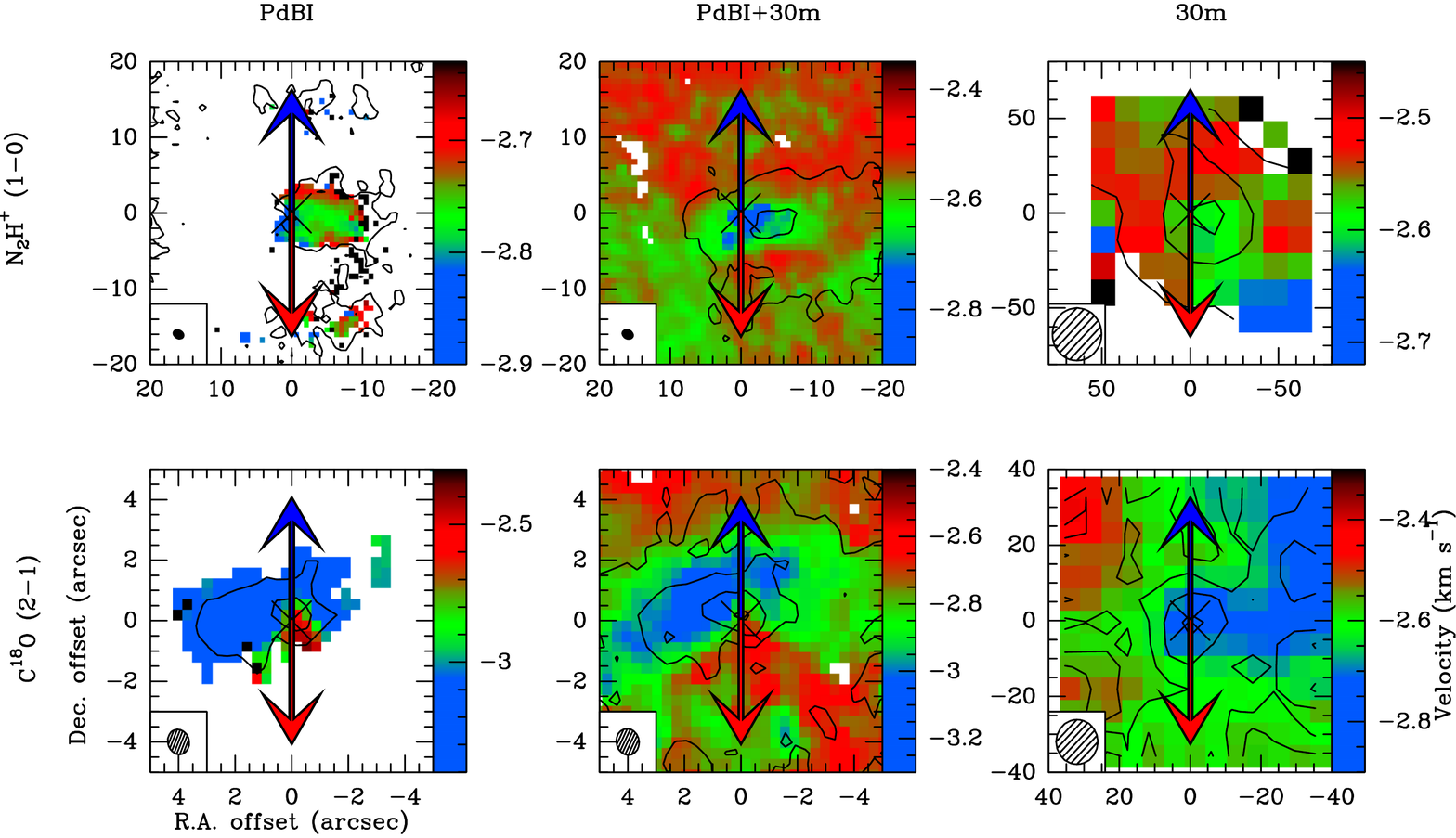}
\caption{Same as Figure \ref{fig:velocity-maps-L1448-2A}, but for GF9-2. 
}
\label{fig:velocity-maps-GF92}
\end{figure*}

To quantify centroid velocity variations at all scales of the protostellar envelopes, we produced centroid velocity maps of each Class~0 protostellar envelope by fitting all individual spectra (pixel by pixel) by line profile models in the CLASS program of the GILDAS software. We only considered the line intensity detected with a signal-to-noise ratio higher than 5.
We fit the spectra to be able to deal with multiple velocity components. Indeed, because protostellar envelopes are embedded in large-scale clouds, multiple velocity components can be expected on some lines of sight where both the protostellar envelope and the cloud emit. For example, \cite{Belloche06} find several velocity components in their 30m of the N$_2$H$^+$ emission of IRAS4A (see Appendix \ref{sec:comments-IRAS4A}). 
Except for IRAS4A and IRAS4B for which we fit two velocity components (see details in Appendix \ref{sec:comments-indiv-sources}), for most sources we used a Gaussian line profile to model the C$^{18}$O (2$-$1) emission, with the line intensity, full width at half maximum (FWHM), and centroid velocity let as free parameters (see Fig. \ref{fig:spectra-30m-L1448C}). In the case of N$_{2}$H$^{+}$ (1$-$0), we used a hyperfine structure (HFS) line profile to determine the FWHM and centroid velocity of the molecular line emission (see Fig. \ref{fig:spectra-30m-L1448C}). Figures~\ref{fig:velocity-maps-L1448-2A} to \ref{fig:velocity-maps-GF92} show the centroid velocity maps obtained for each source of the sample using the PdBI, combined, and 30m datasets for both the C$^{18}$O and N$_{2}$H$^{+}$ emission.


For most sources in our sample, these centroid velocity maps reveal organized velocity patterns with blue-shifted and red-shifted velocity components on both sides of the central stellar embryo, along the equatorial axis where such velocity gradients could be due to rotation of the envelopes.
The global kinematics in Class~0 envelopes is a complex combination of rotation, infall, and outflow motions. The observed velocities are projected on the line of sight and thus, are a mix of the various gas motions. Therefore, it is not straightforward to interpret a velocity gradient in terms of the underlying physical process producing it.
In order to have an indication of the origin of these gradients, we performed a least-square minimization of a linear velocity gradient model on the velocity maps following:
\begin{equation}
\rm{v}_\mathrm{grad}=\rm{v}_0 + a \Delta \alpha + b \Delta \beta, 
\label{eq:vgrad}
\end{equation}
with $\Delta \alpha$ and $\Delta \beta$ the offsets with respect to the central source \citep{Goodman93}.

This simple model provides an estimate of the reference velocity $\mathrm{v}_0$ called systemic velocity, the direction $\Theta$, and the amplitude $G$ of the mean velocity gradient. One would expect a mean gradient perpendicular to the outflow axis if the velocity gradient was due to rotational motions in an axisymmetric envelope. A mean gradient oriented along the outflow axis could be due to jets and outflows or infall in a flattened geometry. 
The gradients were fit on the region of the velocity maps shown in Fig. \ref{fig:velocity-maps-L1448-2A}, namely 10$\arcsec \times$ 10$\arcsec$ in the PdBI and combined datasets for the C$^{18}$O emission (lower left and central panels), 40$\arcsec \times$ 40$\arcsec$ for the N$_{2}$H$^{+}$ emission from the PdBI and combined datasets (upper left and central panels), and 80$\arcsec \times$ 80$\arcsec$ and 160$\arcsec \times$ 160$\arcsec$, respectively for the C$^{18}$O and N$_{2}$H$^{+}$ emission from the 30m datasets (right panels).
Table \ref{table:gradient-velocity-fit} reports for each source the significant mean velocity gradients detected with an amplitude higher than 2$\sigma$. No significant velocity gradient is observed for IRAM04191, L1521F, and SVS13-B in C$^{18}$O emission at scales of $r<$5\arcsec or for L1448-C and IRAS4B at $r>$30\arcsec (see Table \ref{table:gradient-velocity-fit}).

Seven of the 12 sources in our sample show a mean gradient in C$^{18}$O emission aligned with the equatorial axis ($\Delta \Theta <$30$^{\circ}$) which could trace rotational motions of the envelope at scales of $r<$5\arcsec. 
At similar scales, four sources (L1448-NB, L1521F, L1157, GF9-2) show gradients with intermediate orientation (30$^{\circ}< \Delta \Theta <$60$^{\circ}$). 
Finally, L1448-2A shows a mean gradient aligned to the outflow axis rather than the equatorial axis ($\Delta \Theta >$60$^{\circ}$). For these last five sources, the gradients observed could be due to a combination of rotation, ejection, and infall motions. 
For all sources, we noticed a systematic dispersion of the direction of the velocity gradient from inner to outer scales in the envelope (see Fig. \ref{fig:evolution-theta-plot}). We discuss in Sect. \ref{sec:velocity-gradient-1000au} whether this shift in direction of the velocity gradient is due to a transition from rotation-dominated inner envelope to collapse-dominated outer envelope at $r>$1500~au, or is due to the different molecular tracers used for this analysis.
In most sources, the gradient moves away from the equatorial axis as the scale increases. Only three sources (IRAM04191, L1521F, and L1527) show a gradient close to the equatorial axis with $\Delta \Theta <$30$^{\circ}$ at 2000~au in N$_{2}$H$^{+}$ emission from the combined dataset while four sources show a complex gradient and five sources have a $\Delta \Theta >$60$^{\circ}$.

\begin{sidewaystable*}
\centering
\caption{Estimation of the systemic velocity, the mean velocity gradient amplitude and its orientation from linear gradient fit of centroid velocity maps in C$^{18}$O and N$_{2}$H$^{+}$ emission from the PdBI, the combined, and the 30m datasets for the CALYPSO sample sources.}
\label{table:gradient-velocity-fit}
\resizebox{1\textwidth}{!}{\begin{tabular}{lc|cccc|cccc|cccc}
\hline \hline
 \multicolumn{10}{c}{\hfill}  \\
     Source                    &   Line         & \multicolumn{4}{c|}{PdBI}                           & \multicolumn{4}{c}{PdBI+30m}  &  \multicolumn{4}{c}{30m}                                           \\
\multicolumn{2}{c|}{\hfill} & \multicolumn{4}{c|}{\hfill} & \multicolumn{4}{c|}{\hfill} & \multicolumn{4}{c}{\hfill}  \\
      \hfill            & \hfill  & $G$ & v$_{0}$       & $\Theta$~\tablefootmark{a}     &  $\Delta \Theta$~\tablefootmark{b}  &    $G$ &  v$_{0}$ &  $\Theta$~\tablefootmark{a}           &       $\Delta \Theta$~\tablefootmark{b} & $G$ &      v$_{0}$       & $\Theta$~\tablefootmark{a}      &     $\Delta \Theta$~\tablefootmark{b}        \\
                         &       & (km~s$^{-1}$~pc$^{-1}$)     & (km~s$^{-1}$) & ($^{\circ}$) &($^{\circ}$) & (km~s$^{-1}$~pc$^{-1}$) & (km~s$^{-1}$) & ($^{\circ}$) &($^{\circ}$) &  (km~s$^{-1}$~pc$^{-1}$) &(km~s$^{-1}$) & ($^{\circ}$) & ($^{\circ}$)         \\
\hline
\multirow{2}{*}{L1448-2A}                    & C$^{18}$O (2-1)  &   118  $\pm$ 3            &  4.06 $\pm$ 0.03        &    107  $\pm$ 2       &       69      &              13 $\pm$ 2    &           4.05   $\pm$ 0.04          &   81 $\pm$ 1   & 43     &          6 $\pm$ 1     &    4.12  $\pm$ 0.02 &  -14 $\pm$ 2   & 52       \\
                         & N$_{2}$H$^{+}$ (1-0) &    19 $\pm$ 1        &    4.09 $\pm$ 0.03      &    -89 $\pm$ 2        & 53             &              2 $\pm$ 1      &     4.10 $\pm$ 0.04                    &   -177 $\pm$ 21 &  35  &          2 $\pm$ 1      &   4.16 $\pm$ 0.04      &  -8 $\pm$ 3    &  46 \\

\hline
\multirow{2}{*}{L1448-NB}                    & C$^{18}$O (2-1)  &   214  $\pm$ 1            &  4.71 $\pm$ 0.06        &    51  $\pm$ 1  &       41             &              75 $\pm$ 1    &           4.55   $\pm$ 0.05          &   50 $\pm$ 1     & 40    &            6 $\pm$ 1     &    4.45  $\pm$ 0.02 &  85 $\pm$ 4    &  75  \\
                         & N$_{2}$H$^{+}$ (1-0) &    108 $\pm$ 1        &    4.55 $\pm$ 0.02      &    74 $\pm$ 1           & 64          &           13 $\pm$ 1      &     4.57 $\pm$ 0.02                    &   100 $\pm$ 1  & 90  &       4 $\pm$ 1      &   4.51 $\pm$ 0.01      &  97 $\pm$ 1    &  87    \\

\hline
\multirow{2}{*}{L1448-C} &    C$^{18}$O (2-1)        &     218 $\pm$ 4          &   5.13 $\pm$ 0.05       &     -119 $\pm$ 2 &         12                     &  62 $\pm$ 1        &           5.03 $\pm$ 0.05              &     -138 $\pm$ 1 &            31       &       -- & -- & --        &   --         \\
                         &      N$_{2}$H$^{+}$ (1-0)      &     15 $\pm$ 2          &   4.92 $\pm$ 0.03       &           -175 $\pm$ 4           &     68      &   13 $\pm$ 1       &     4.97 $\pm$ 0.03                    &     -179 $\pm$ 1  &     72        &     7 $\pm$ 1 & 4.85 $\pm$ 0.01 & -152 $\pm$ 1       &    45        \\
                                                                     
\hline
\multirow{2}{*}{IRAS2A} &    C$^{18}$O (2-1)        &     150 $\pm$ 5         &   7.75 $\pm$ 0.08       &     101 $\pm$ 2  &        14                      &  17 $\pm$ 1       &           7.63 $\pm$ 0.06              &     97 $\pm$ 3        &     18       &        3 $\pm$ 1 & 7.66 $\pm$ 0.01 & -10 $\pm$ 1     &     55       \\
                         &      N$_{2}$H$^{+}$ (1-0)      &     11 $\pm$ 1          &   7.57 $\pm$ 0.04       &           -20 $\pm$ 6       &        45        &   4 $\pm$ 1       &     7.64 $\pm$ 0.03                    &     20 $\pm$ 4     &       85     &     5 $\pm$ 1     &     7.61 $\pm$ 0.01      &    -5 $\pm$ 1       &     60    \\
                                                 
\hline
\multirow{2}{*}{SVS13-B} &    C$^{18}$O (2-1)        &     --          &   --       &     --          &        --            &  28 $\pm$ 4        &           8.19 $\pm$ 0.08              &     89 $\pm$ 10 &          12       &       7 $\pm$ 1    &   8.15 $\pm$ 0.02    &    -33 $\pm$ 1      &     70        \\
                         &      N$_{2}$H$^{+}$ (1-0)      &     16 $\pm$ 1          &   8.48 $\pm$ 0.03       &           74 $\pm$ 1        &      3        &   5 $\pm$ 1       &     8.31 $\pm$ 0.02                    &     16 $\pm$ 4       &      61     &      7 $\pm$ 1     &     8.07 $\pm$ 0.02      &    -4 $\pm$ 1       &   81     \\

\hline
\multirow{2}{*}{IRAS4A} &    C$^{18}$O (2-1)        &     105 $\pm$ 1          &   6.57 $\pm$ 0.04       &     -79 $\pm$ 1         &         11            &  18 $\pm$ 4        &           6.72 $\pm$ 0.06              &     -80 $\pm$ 17 &         10         &      1.2 $\pm$ 0.4    &   7.57 $\pm$ 0.04    &    8 $\pm$ 13        &   82      \\
                         &      N$_{2}$H$^{+}$ (1-0)      &     43 $\pm$ 1          &   6.85 $\pm$ 0.05       &           -69 $\pm$ 1         &     21        &   7 $\pm$ 1       &     6.90 $\pm$ 0.06                    &     37 $\pm$ 2  &        53    &      3 $\pm$ 1     &     7.48 $\pm$ 0.02      &    51 $\pm$ 1       &     39       \\

\hline
\multirow{2}{*}{IRAS4B} &    C$^{18}$O (2-1)        &     65 $\pm$ 1          &   6.84 $\pm$ 0.03       &     -89 $\pm$ 1           &       14            &  52 $\pm$ 5        &           6.89 $\pm$ 0.06              &     -83 $\pm$ 8           &       20     &     --    &   --    &    --       &    --    \\
                         &      N$_{2}$H$^{+}$ (1-0)      &     25 $\pm$ 2          &   6.94 $\pm$ 0.04       &           96 $\pm$ 6       &       19         &   3 $\pm$ 1       &     7.06 $\pm$ 0.05                    &     -71 $\pm$ 14       &       32    &       3 $\pm$ 1     &     7.47 $\pm$ 0.02      &    51 $\pm$ 5       &     27     \\

\hline
\multirow{2}{*}{IRAM04191}                    & C$^{18}$O (2-1)  &  --             &   --         &      --                     &   -- &         25    $\pm$  2   &          6.62    $\pm$  0.07         &  -42  $\pm$ 5    &    28   &      6 $\pm$  1    &     6.59 $\pm$ 0.01 & 96 $\pm$ 1    & 14   \\
                         & N$_{2}$H$^{+}$ (1-0) &   11  $\pm$    2     &   6.76  $\pm$   0.05    &   109  $\pm$ 3             &   1    &  15 $\pm$   1    &    6.64  $\pm$      0.03               &  92   $\pm$      1 &       18     &          3    $\pm$ 1 &  6.61 $\pm$ 0.01 &  124 $\pm$ 1~\tablefootmark{$\star$}        &      14           \\

\hline
\multirow{2}{*}{L1521F}                    & C$^{18}$O (2-1)  &   --            &  --        &     --             &       --        &    16 $\pm$ 1    &             6.57 $\pm$ 0.04           &    -82 $\pm$ 2 &              52  &      2 $\pm$ 1     &      6.49 $\pm$ 0.01 & -8 $\pm$ 1   & 22 \\
                         & N$_{2}$H$^{+}$ (1-0) &     26 $\pm$ 2          &    6.67 $\pm$ 0.04      &    -76 $\pm$ 5          &     46      &             0.7 $\pm$ 0.1      &     6.48 $\pm$ 0.01                    &   -49 $\pm$ 2 &       19        &        0.4 $\pm$ 0.1 & 6.47 $\pm$ 0.01 & 6 $\pm$ 2       &    36    \\

\hline
\multirow{2}{*}{L1527}                    & C$^{18}$O (2-1)  &  171 $\pm$ 5            &  5.91 $\pm$ 0.06        &     -9 $\pm$ 3          &      9     &       66 $\pm$ 1    &             5.84 $\pm$ 0.06           &    22 $\pm$ 2        & 22    &          2 $\pm$ 1     &      5.94 $\pm$ 0.01 & 113 $\pm$ 1    &  67   \\
                         & N$_{2}$H$^{+}$ (1-0) &     17 $\pm$ 1          &    5.87 $\pm$ 0.03      &    9 $\pm$ 6  &              9           &   3 $\pm$ 1      &     5.91 $\pm$ 0.02                    &   26 $\pm$ 5  & 26   &       2 $\pm$ 1 & 5.91 $\pm$ 0.01 & 123 $\pm$ 4       &  57                  \\

\hline
\multirow{2}{*}{L1157} &    C$^{18}$O (2-1)        &     104 $\pm$ 3          &   2.61 $\pm$ 0.06       &    13 $\pm$ 2         &         60            &    68 $\pm$ 2       &           2.63 $\pm$ 0.06              &      35 $\pm$ 2      &       38     &      2 $\pm$ 1    &   2.70 $\pm$ 0.15    &    -125 $\pm$ 8       &    18    \\
                         &      N$_{2}$H$^{+}$ (1-0)      &     51 $\pm$ 2          &   2.82 $\pm$ 0.04       &           35 $\pm$ 2      &      38          &   0.8 $\pm$ 0.4       &     2.69 $\pm$ 0.02                    &     113 $\pm$ 65         &     40      &       1.0 $\pm$ 0.5     &     2.65 $\pm$ 0.04      &    -131 $\pm$ 35      &    24    \\  
\hline                        
\multirow{2}{*}{GF9-2}                    & C$^{18}$O (2-1)  &   126 $\pm$ 17            &   -3.01 $\pm$ 0.06         &    -154  $\pm$ 4         &    64       &                17  $\pm$  1    &          -2.83    $\pm$ 0.03          &   -133 $\pm$ 5         &  43   &    2 $\pm$ 1      &     -2.62 $\pm$ 0.01 & 102 $\pm$ 1  &  12 \\
                         & N$_{2}$H$^{+}$ (1-0) &   18  $\pm$   3      &   -2.80  $\pm$ 0.01       &   -123  $\pm$ 8  &         33           &            1.4  $\pm$ 0.1      &     -2.56 $\pm$  0.01                   &   -9 $\pm$ 5          & 81    &           0.5 $\pm$ 0.1 & -2.56 $\pm$ 0.01  & 40 $\pm$ 8         &    50       \\

\hline
\end{tabular}}
\tablefoot{ \tablefoottext{a}{Position angle of the redshifted lobe of the velocity gradient defined from north to east.}
\tablefoottext{b}{Absolute value, between 0$^{\circ}$ and 90$^{\circ}$, of the difference between the angle of the mean gradient and the angle of the equatorial axis. The equatorial axis is defined perpendicularly to the direction of the outflows (see Table \ref{table:sample}).}
\tablefoottext{$\star$}{For IRAM04191, the N$_{2}$H$^{+}$ emission in the 30m dataset was fit ignoring the pixels close to the Class~I protostar IRAS04191 in the field of view (see Appendix \ref{sec:comments-indiv-sources}).}}
\end{sidewaystable*}

%%%%%%%%%%%%%%%%%%%%%%%%%%%%%%%%%%%%
%
%________________________________________________________________
%
%%%%%%%%%%%%%%%%%%%%%%%%%%%%%%%%%%%%

\subsection{High dynamic range position-velocity diagrams to probe rotational motions} \label{subsec:diagram-PV-construction}
To investigate rotational motions and characterize the angular momentum properties in our sample of Class~0 protostellar envelopes, we build the position-velocity (PV$_\mathrm{rot}$) diagrams along the equatorial axis. We assumed the position angle of the equatorial axis as orthogonal to the jet axis reported in Table \ref{table:sample}. 
The choice of this equatorial axis allows us to maximize sensitivity to rotational motions and minimize potential contamination on the line of sight due to collapsing or outflowing gas \citep{Yen13}. The velocities reported in the PV$_\mathrm{rot}$ diagram are corrected for the inclination $i$ of the equatorial plane with respect to the line of sight (see Table \ref{table:sample}). 
We note that the correction for inclination is a multiplicative factor, thus if this inclination angle is not correctly estimated, the global observed shape is not distorted.

\begin{figure}[h]
\centering
\includegraphics[width=9cm]{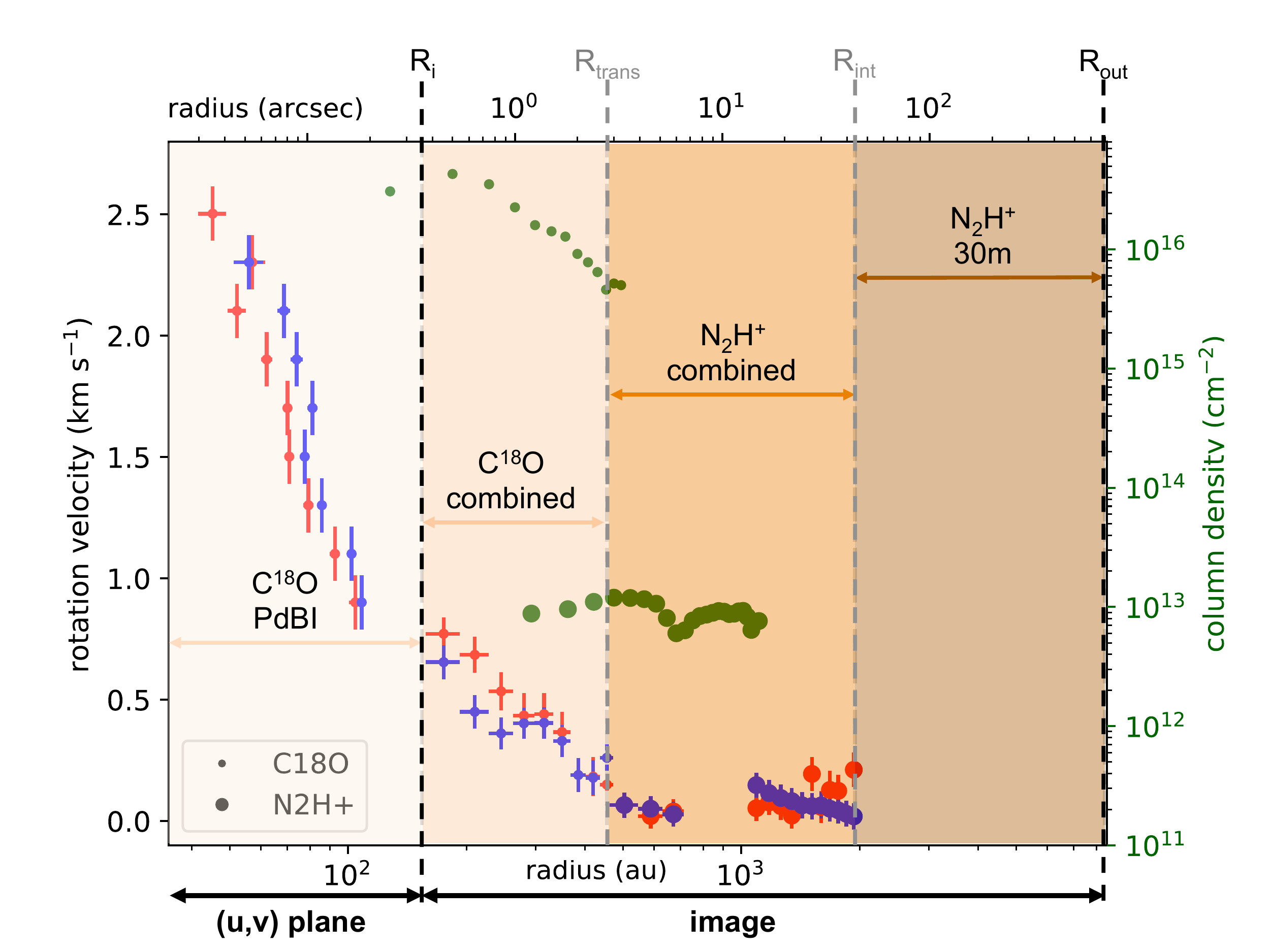}
\caption{Plot summarizing the combination of tracers and datasets used to build high dynamic range PV$_\mathrm{rot}$ diagrams in the L1527 envelope.
The transition radii between the different datasets (PdBI, combined, and 30m) and the two C$^{18}$O and N$_{2}$H$^{+}$ tracers represented by the dashed lines are given in Table \ref{table:radius-lines}. The green points show the column density profiles along the equatorial axis of C$^{18}$O and N$_{2}$H$^{+}$ estimated from the combined datasets (see Appendix \ref{sec:column-density}).
}
\label{fig:PV-diagram-construction}
\end{figure}

The analysis described in detail in Appendix \ref{details-diagram-PV-construction} allows us to build a PV$_\mathrm{rot}$ diagram with a high dynamic range from 50~au up to 5000~au for each source as follows (see the example of L1527 in Fig. \ref{fig:PV-diagram-construction}):\\
\indent $\bullet$ To constrain the PV$_\mathrm{rot}$ diagram at the smallest scales resolved by our dataset ($\sim$0.5$\arcsec$), we use the PdBI C$^{18}$O datasets that we analyze in the (u,v) plane to avoid imaging and deconvolution processes (see label "C$^{18}$O PdBI" in Fig. \ref{fig:PV-diagram-construction}). We only kept central emission positions in the channel maps at a position angle $<$|45$^{\circ}|$ with respect to the equatorial axis (see Appendix \ref{details-diagram-PV-construction}). \\
\indent $\bullet$ Since the C$^{18}$O extended emission is filtered out by the interferometer, we used the combined C$^{18}$O emission to populate the PV$_\mathrm{rot}$ diagram at the intermediate scales of the protostellar envelopes (see label "C$^{18}$O combined" in Fig. \ref{fig:PV-diagram-construction}). The C$^{18}$O molecule remains the most precise tracer when the temperature is higher than $\sim$20K because below, the C$^{18}$O molecule freezes onto dust ice mantles. To determine the transition radius $R_\mathrm{trans}$ between the two tracers, we calculate the C$^{18}$O and N$_{2}$H$^{+}$ column densities along the equatorial axis from the combined integrated intensity maps (see Appendix \ref{sec:column-density} and green points in Fig. \ref{fig:PV-diagram-construction}).\\
\indent $\bullet$ At radii $r>R_\mathrm{trans}$, the N$_{2}$H$^{+}$ emission traces better the envelope dense gas. We use the combined N$_{2}$H$^{+}$ emission maps to analyze the envelope kinematics at larger intermediate scales. When the N$_{2}$H$^{+}$ column density profile reaches a minimum value due to the sensitivity of the combined dataset, this dataset is no longer the better dataset to provide a robust information on the velocity (see label "N$_{2}$H$^{+}$ combined" in Fig. \ref{fig:PV-diagram-construction}). \\
\indent $\bullet$ Finally, we use the 30m N$_{2}$H$^{+}$ emission map to populate the PV$_\mathrm{rot}$ diagram at the largest scales of the envelope (see label "N$_{2}$H$^{+}$ 30m" in Fig. \ref{fig:PV-diagram-construction}).\\

The CALYPSO datasets allow us to continuously estimate the velocity variations along the equatorial axis in the envelope over scales from 50~au up to 5000~au homogeneously for each protostar. 
Figure~\ref{fig:PV-diagrams-1} shows the PV$_\mathrm{rot}$ diagrams built for all sources of the sample. The systemic velocity used in the PV$_\mathrm{rot}$ diagrams are determined in Appendix \ref{sec:systemic-velocity}.

The method of building PV$_\mathrm{rot}$ diagrams described above and in Appendix \ref{details-diagram-PV-construction} corresponds to an ideal case with a detection of a continuous blue-red velocity gradient along the direction perpendicular to the outflow axis in the velocity maps. In practice, the direction of velocity gradients is not always continuous at all scales probed by our observations (see Table \ref{table:gradient-velocity-fit} and Figure \ref{fig:evolution-theta-plot}).
For some sources, to constrain the PV$_\mathrm{rot}$ diagrams, we did not take the kinematic information at all scales of the envelope into account. Velocity gradients can be considered as probing rotational motions if the following criteria are met:\\
$\indent \bullet$ We only consider the significant velocity gradients reported in Table \ref{table:gradient-velocity-fit} with a blue- and a red-shifted velocity components observed on each side of the protostellar embryo, itself at the systemic velocity $\mathrm{v}_0$. For example, we only take the C$^{18}$O emission from the 30m map into account for L1521F (see Fig. \ref{fig:velocity-maps-L1521F}). \\
$\indent \bullet$ We only take the velocity gradients aligned with the equatorial axis ($\Delta \Theta <$60$^{\circ}$) into account in order to minimize contamination by infall and ejection motions. For example, we do not report in the PV$_\mathrm{rot}$ diagrams the N$_{2}$H$^{+}$ velocity gradients from the combined maps for L1448-NB, IRAS2A, SVS13-B, and GF9-2 (see Table \ref{table:gradient-velocity-fit}).\\
$\indent \bullet$ We do not consider the discontinuous velocity gradients which show an inversion of the blue- and red-shifted velocity components along the equatorial axis from inner to outer envelope scales. For example, we do not report in the PV$_\mathrm{rot}$ diagrams the N$_{2}$H$^{+}$ velocity gradients from the 30m maps for IRAM04191 and L1157 (see Figs. \ref{fig:velocity-maps-IRAM04191} and \ref{fig:velocity-maps-L1157}). \\

When velocity gradients with a blue- and a red-shifted velocity components observed on each side of the protostellar embryo are continuous from inner to outer envelope scales but shifted from the equatorial axis ($\Delta \Theta \geq$60$^{\circ}$), we only report upper limits on rotational velocities in the PV$_\mathrm{rot}$ diagrams.
The sources in our sample show specific individual behaviors, therefore we followed as closely as possible the method of building the PV$_\mathrm{rot}$ diagram adapting it on a case-by-case basis.

\begin{figure*}[!ht]
\centering
\includegraphics[width=9cm]{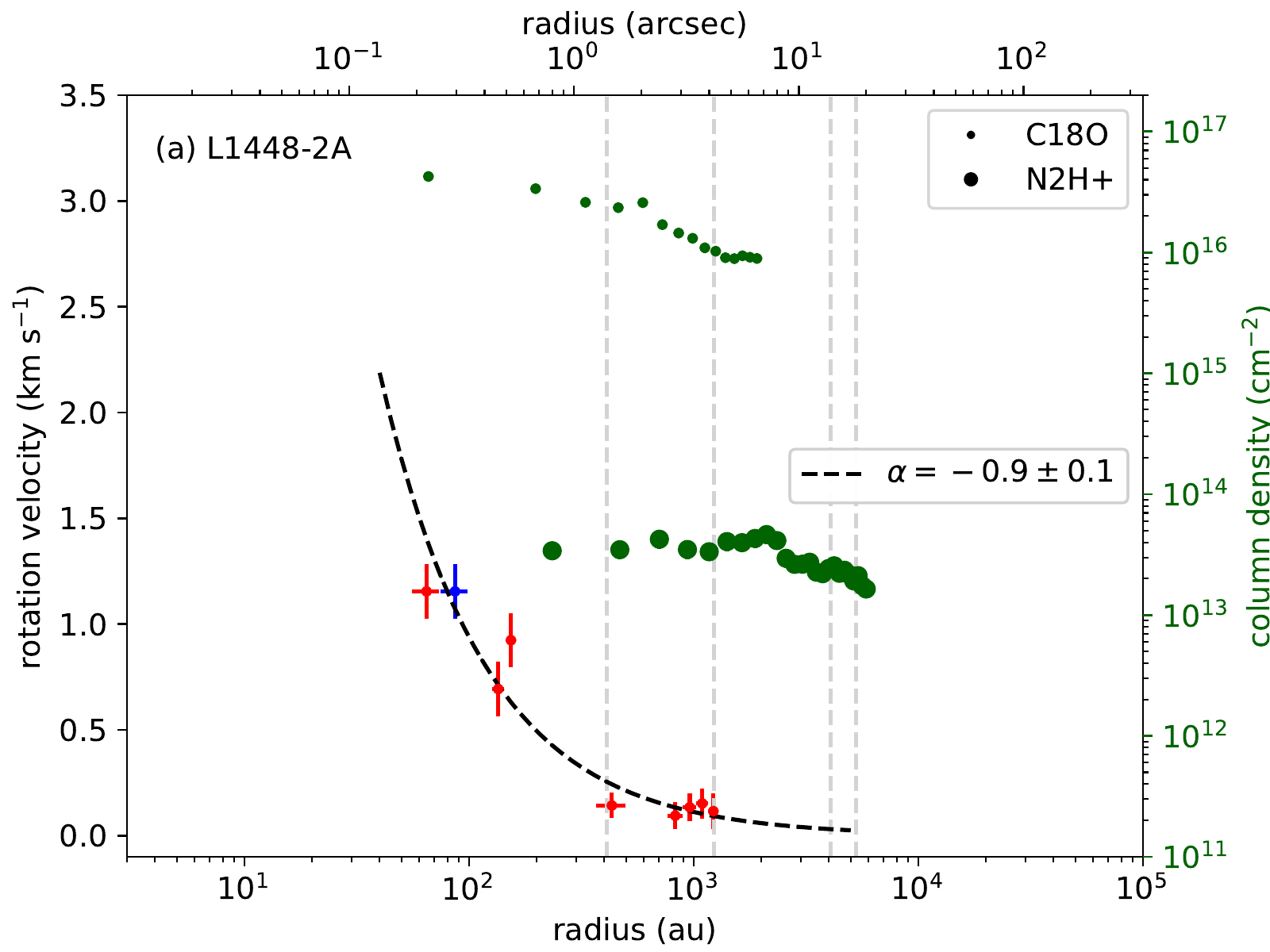}
\includegraphics[width=9cm]{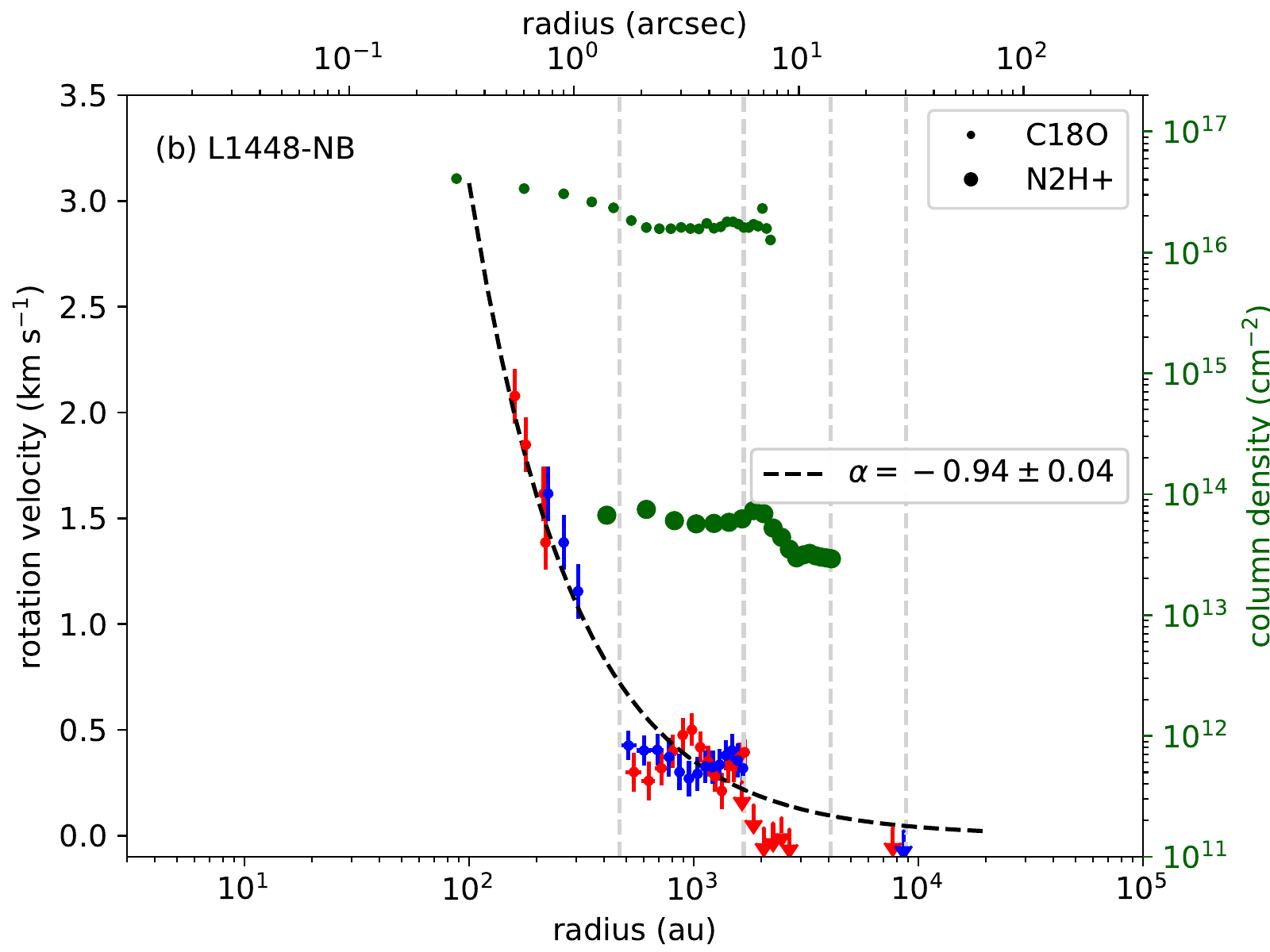}
\includegraphics[width=9cm]{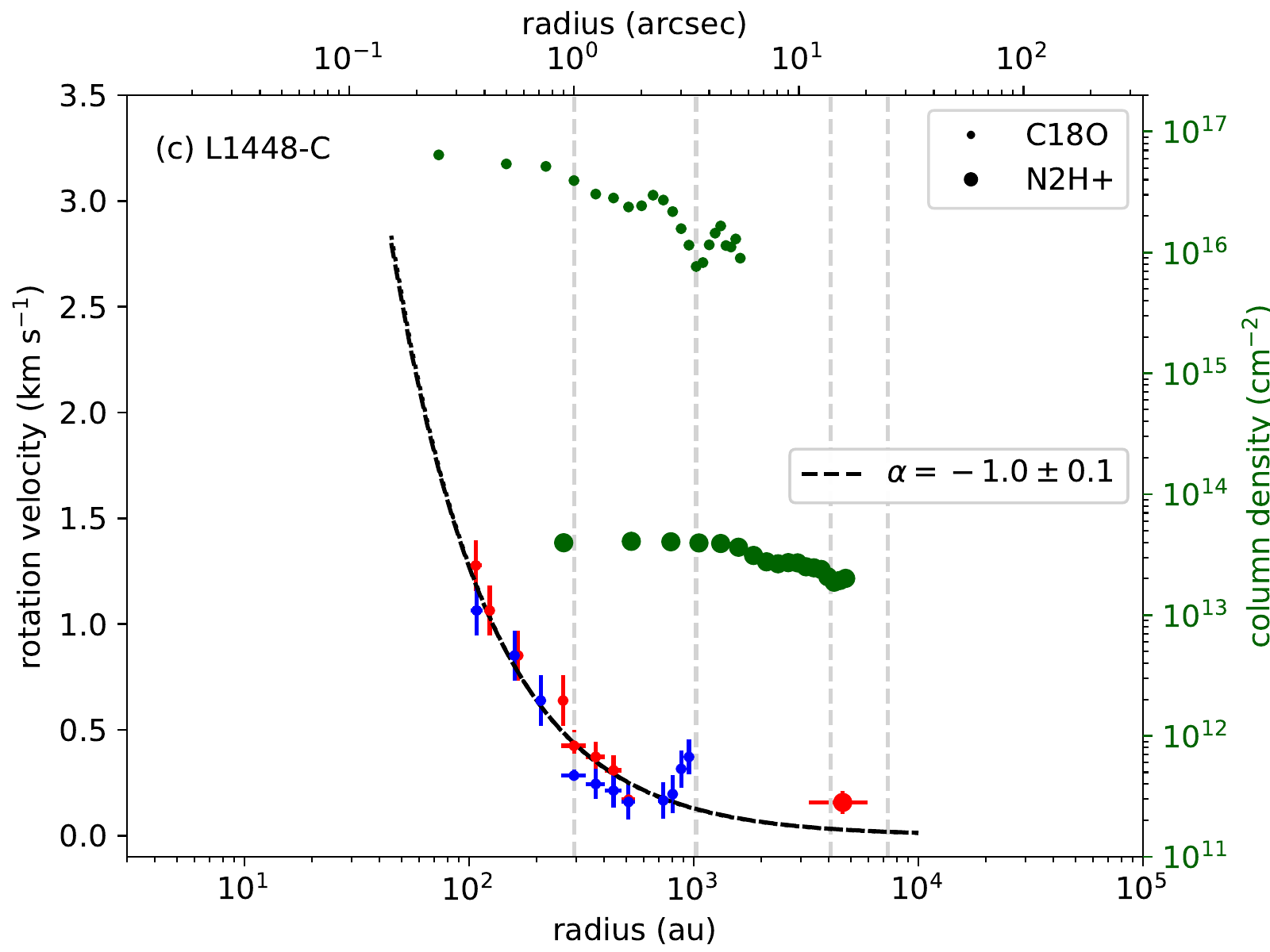}
\includegraphics[width=9cm]{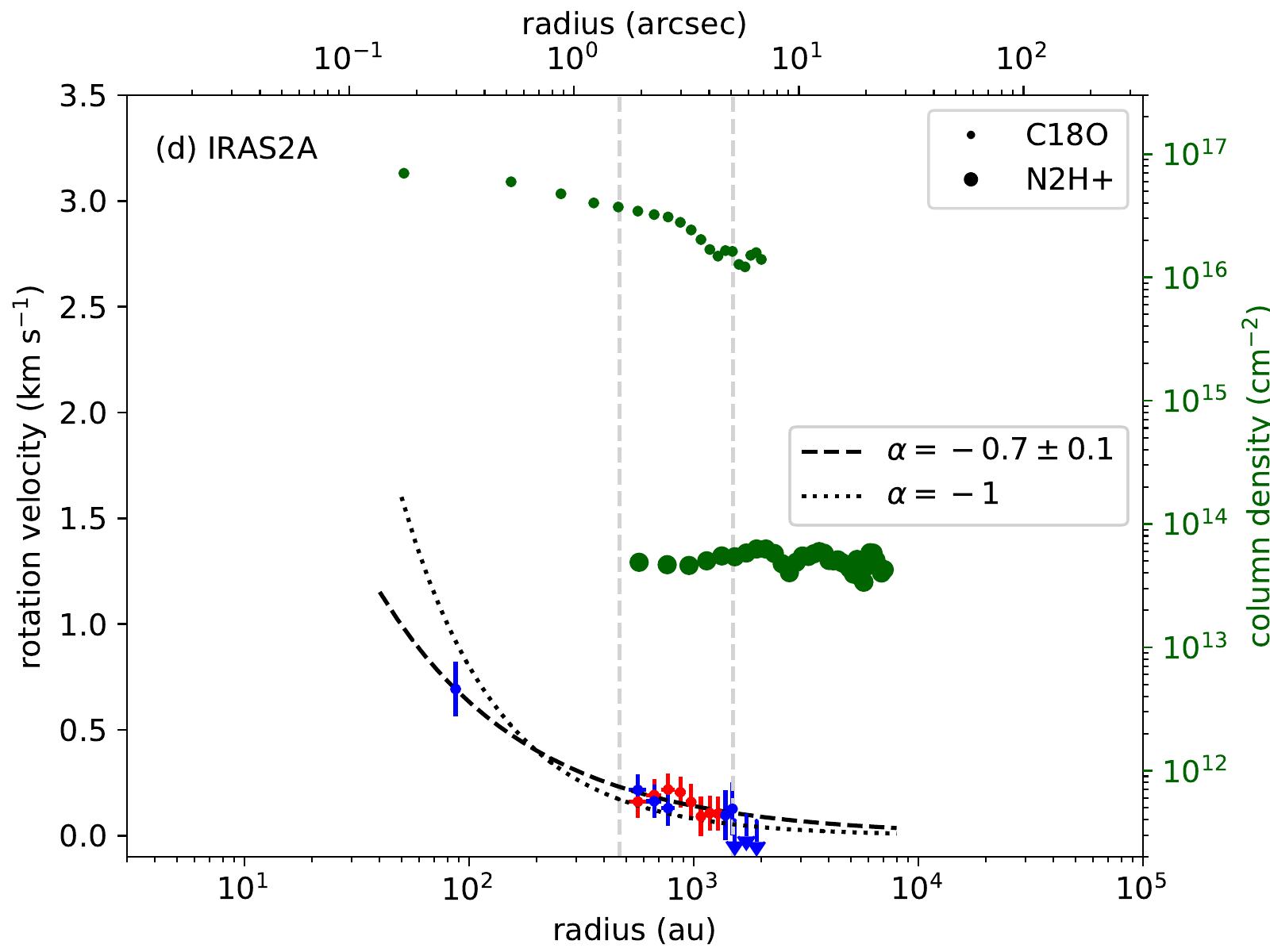}
\includegraphics[width=9cm]{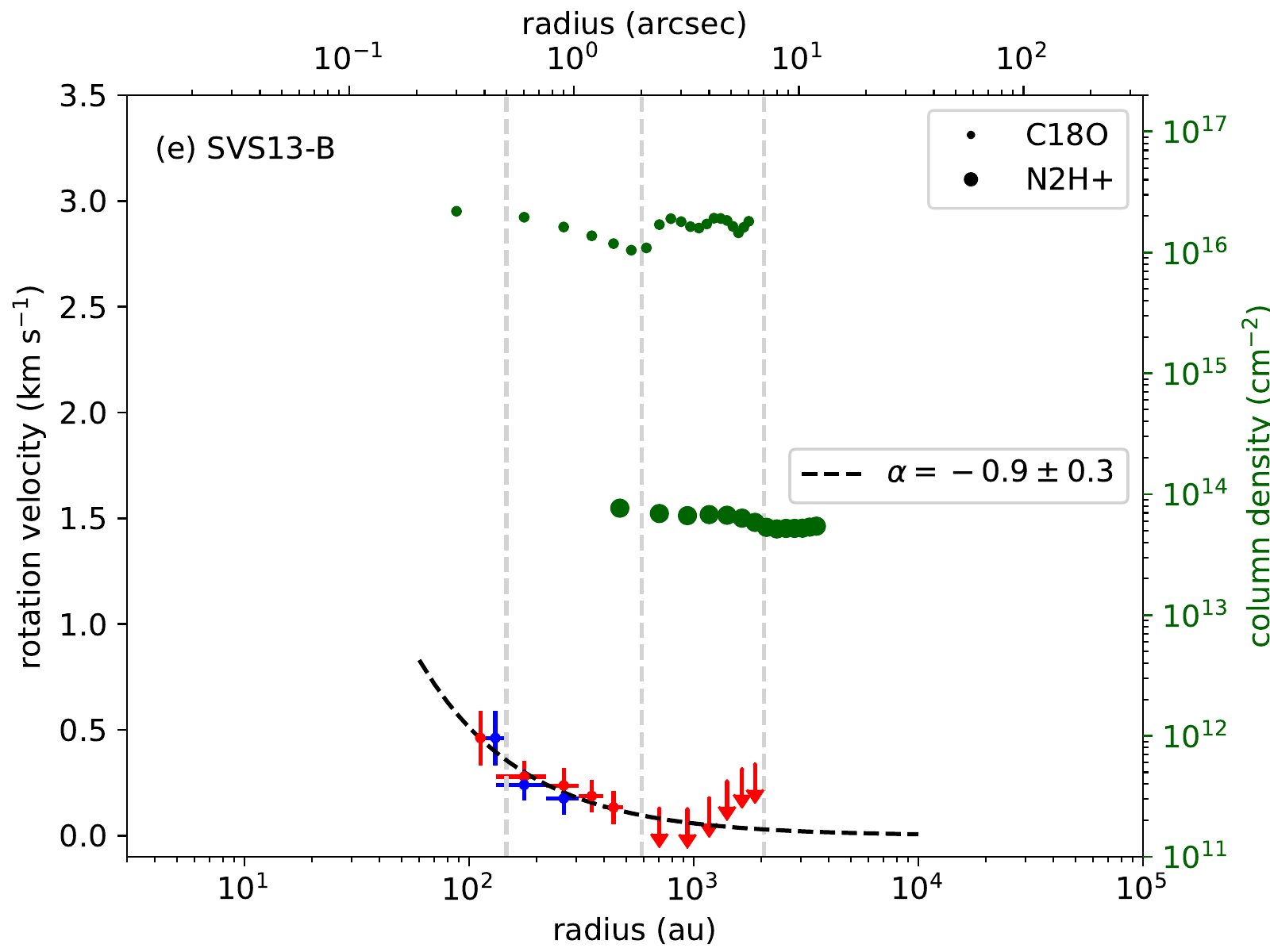}
\includegraphics[width=9cm]{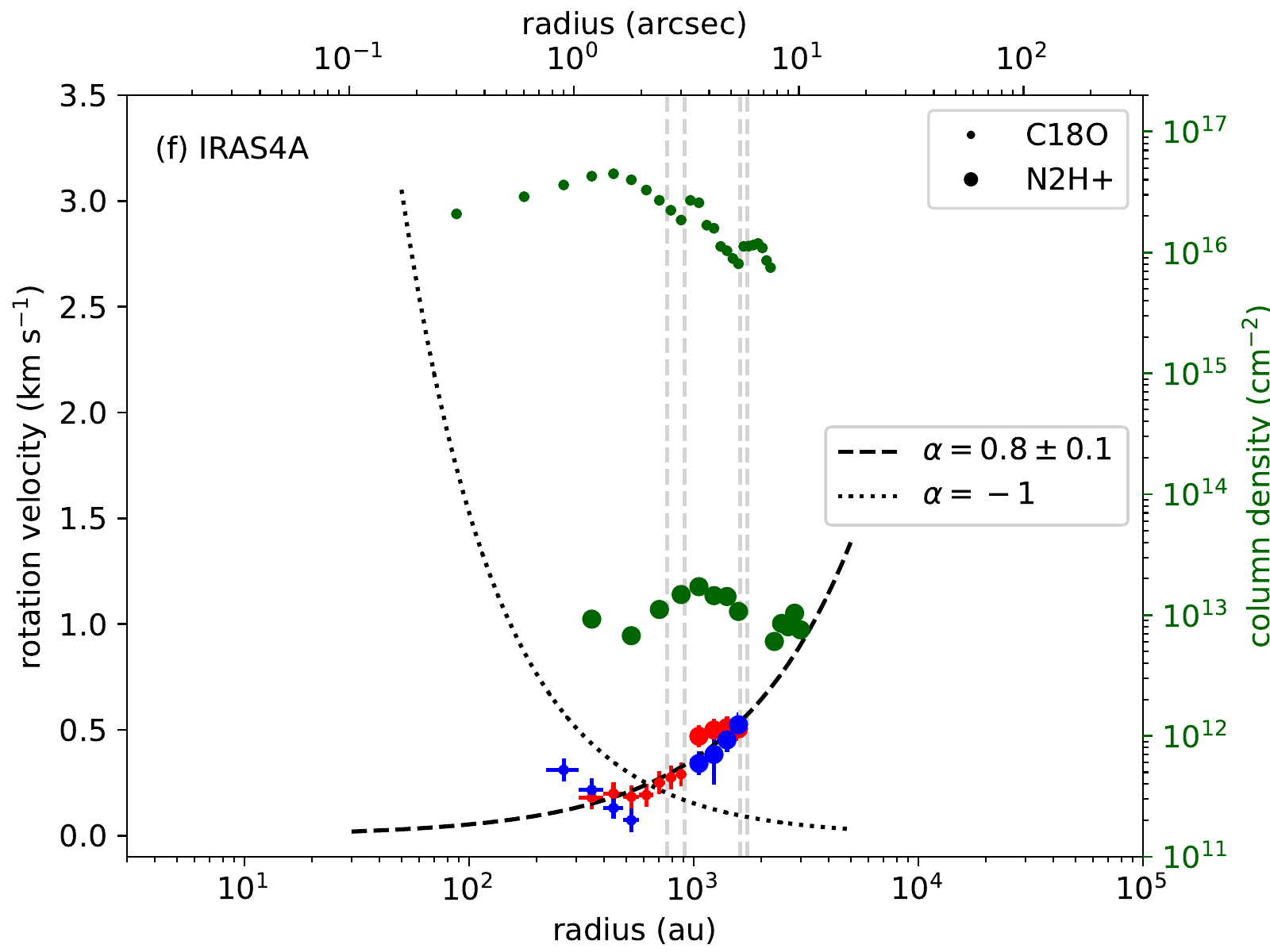}
\caption{Position-velocity diagram along the equatorial axis of the CALYPSO protostellar envelopes. Blue and red dots show the blue- and red-shifted velocities, respectively. The arrows display the upper limits of $\mathrm{v}_\mathrm{rot}$ determined from velocity maps that do not exhibit a spatial distribution of velocities as organized as one would expect from rotation motions (see Sect. \ref{subsec:diagram-PV-construction} and Appendix \ref{details-diagram-PV-construction}). Green dots show the column density profiles along the equatorial axis. Dots and large dots show the C$^{18}$O and N$_{2}$H$^{+}$ data, respectively. The dashed curve shows the best fit with a power-law model leaving the index $\alpha$ as a free parameter ($\mathrm{v}_\mathrm{rot} \propto r^{\alpha}$) whereas the dotted curve shows the best fit with a power-law model with a fixed index $\alpha$=-1. The vertical dashed lines show the transition radii between the different datasets (PdBI, combined, and 30m) and the two tracers as illustrated in Fig. \ref{fig:PV-diagram-construction} and given in Table \ref{table:radius-lines}.
}
\label{fig:PV-diagrams-1}
\end{figure*}

\begin{figure*}[!ht]
\addtocounter{figure}{-1}
\centering
\includegraphics[width=9cm]{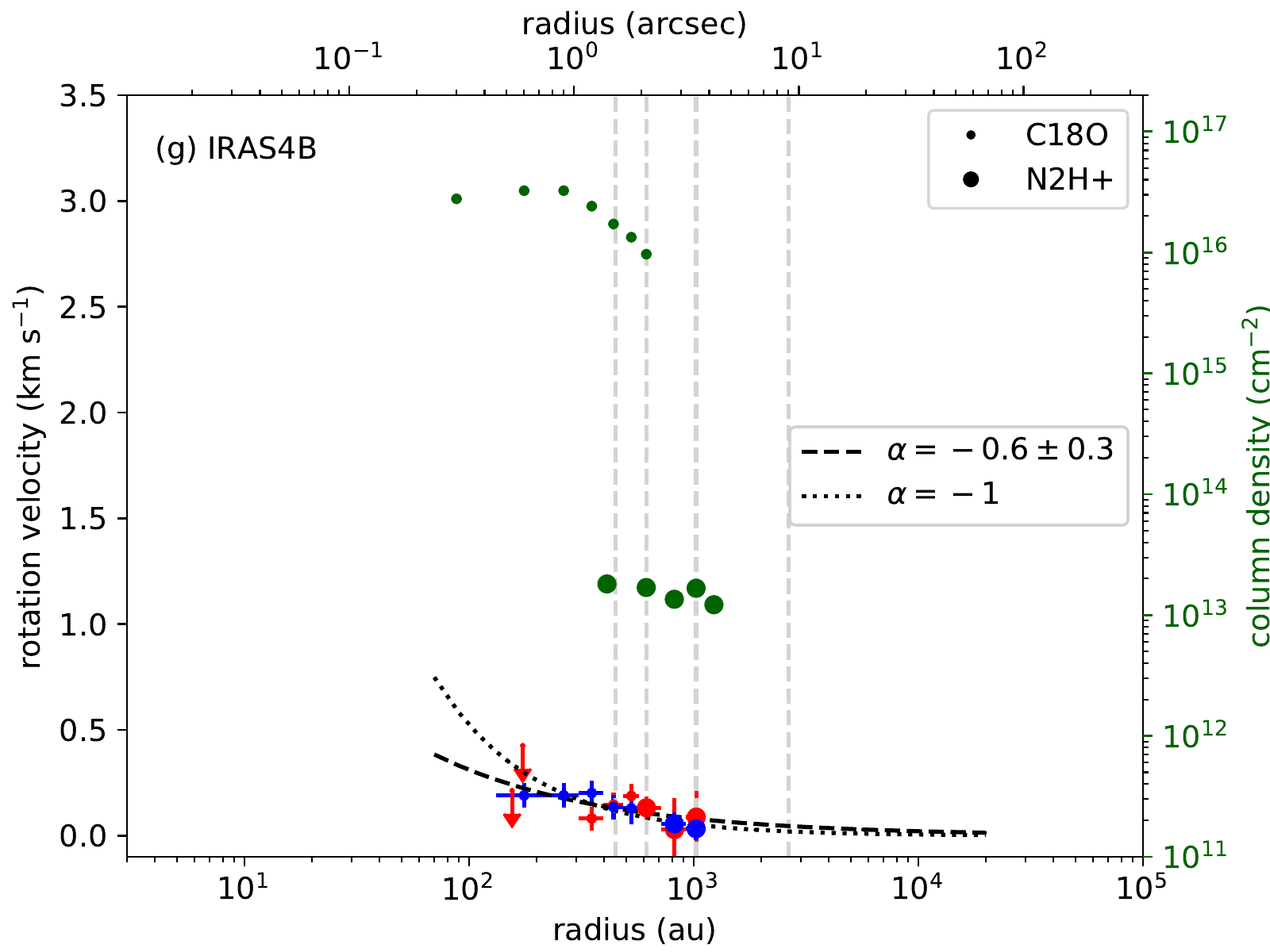}
\includegraphics[width=9cm]{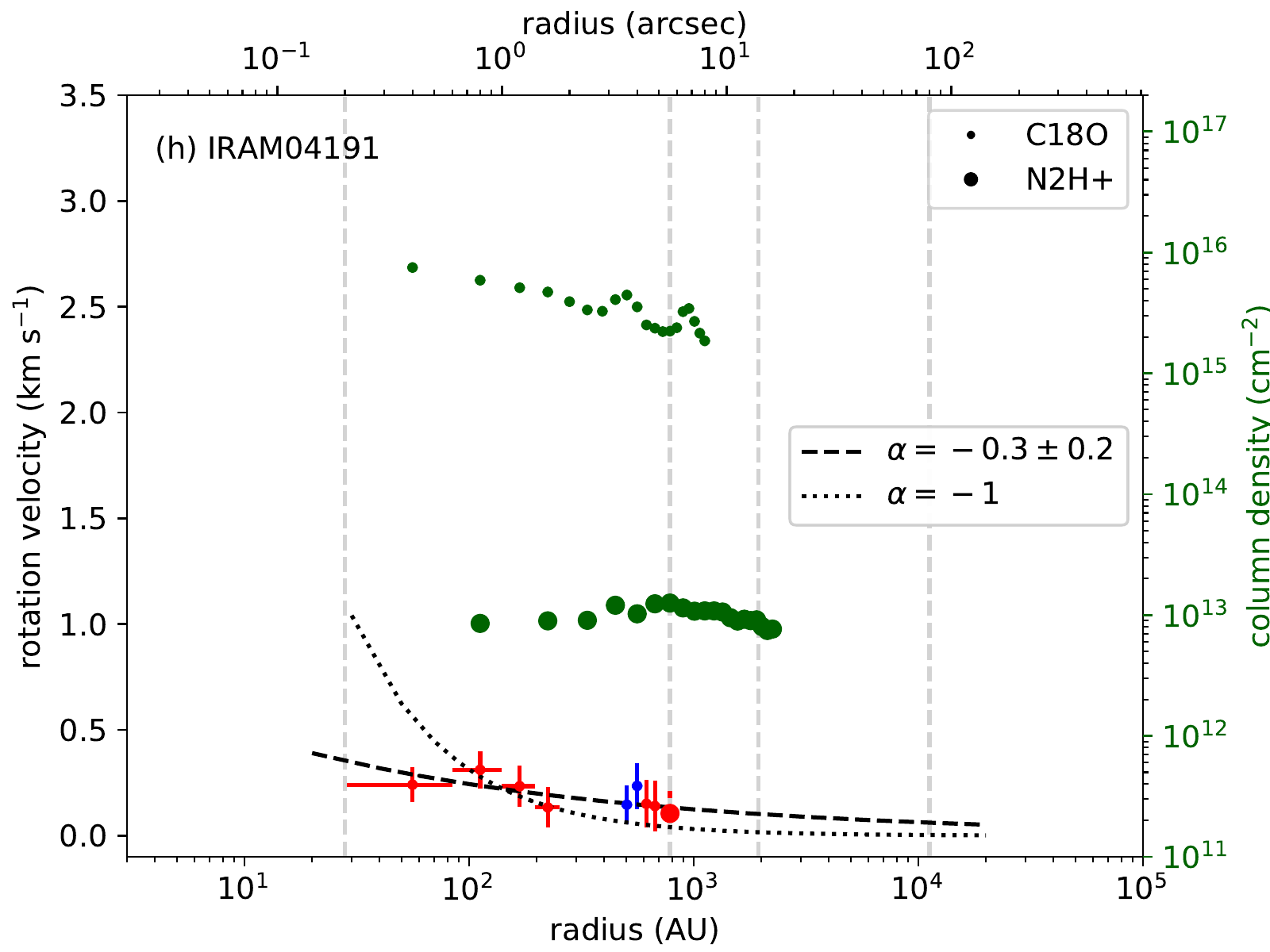}
\includegraphics[width=9cm]{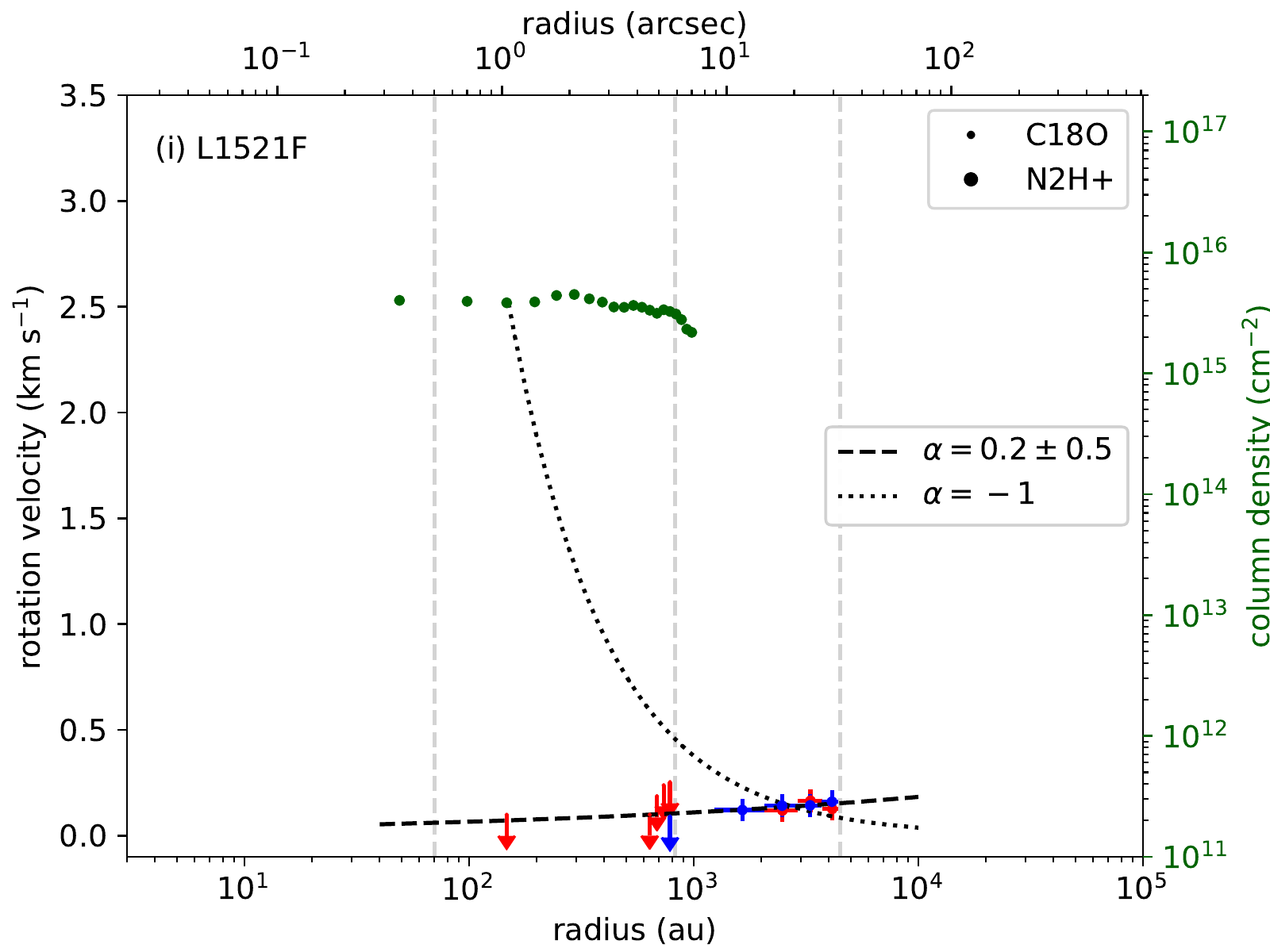}
\includegraphics[width=9cm]{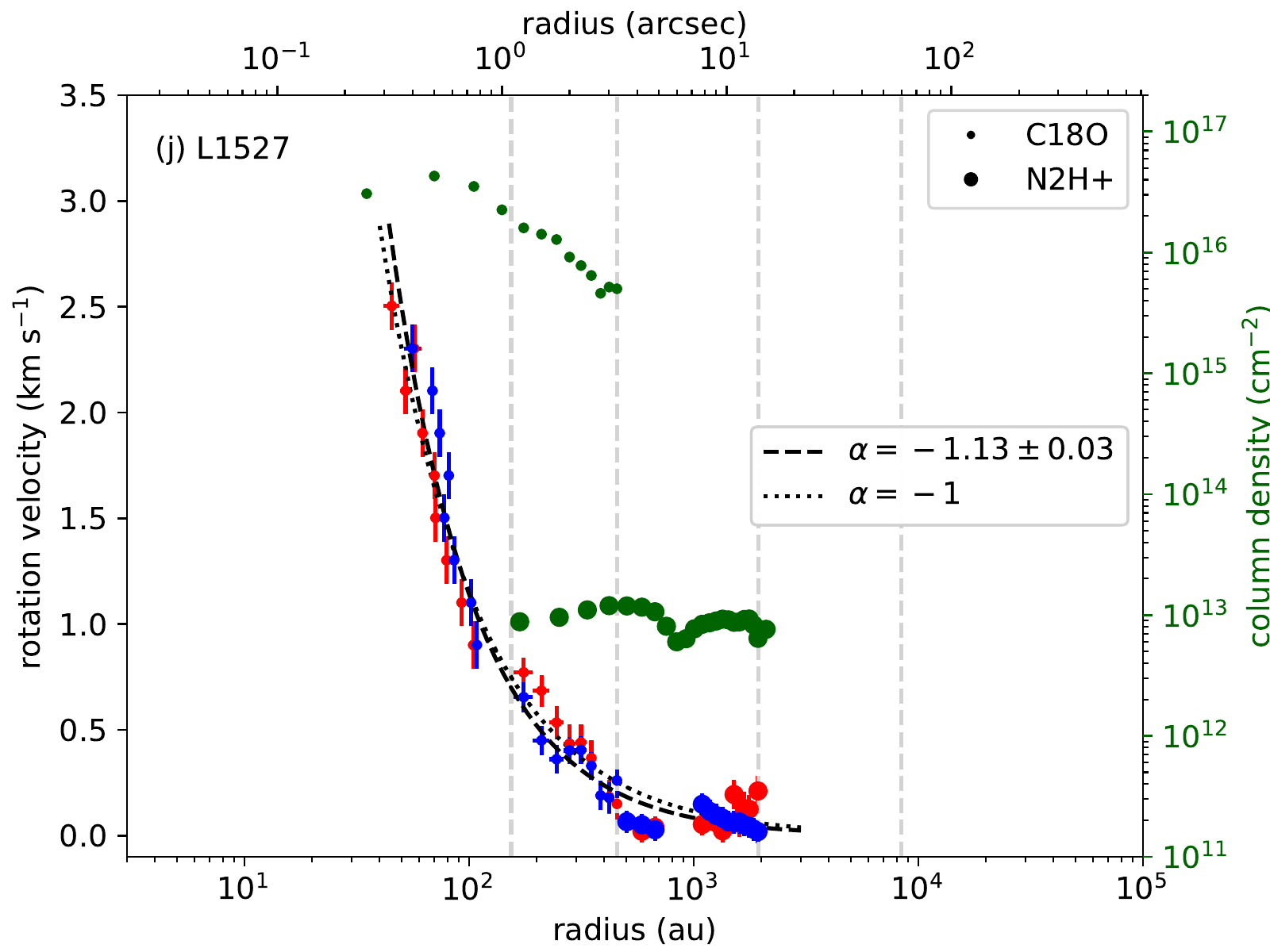}
\includegraphics[width=9cm]{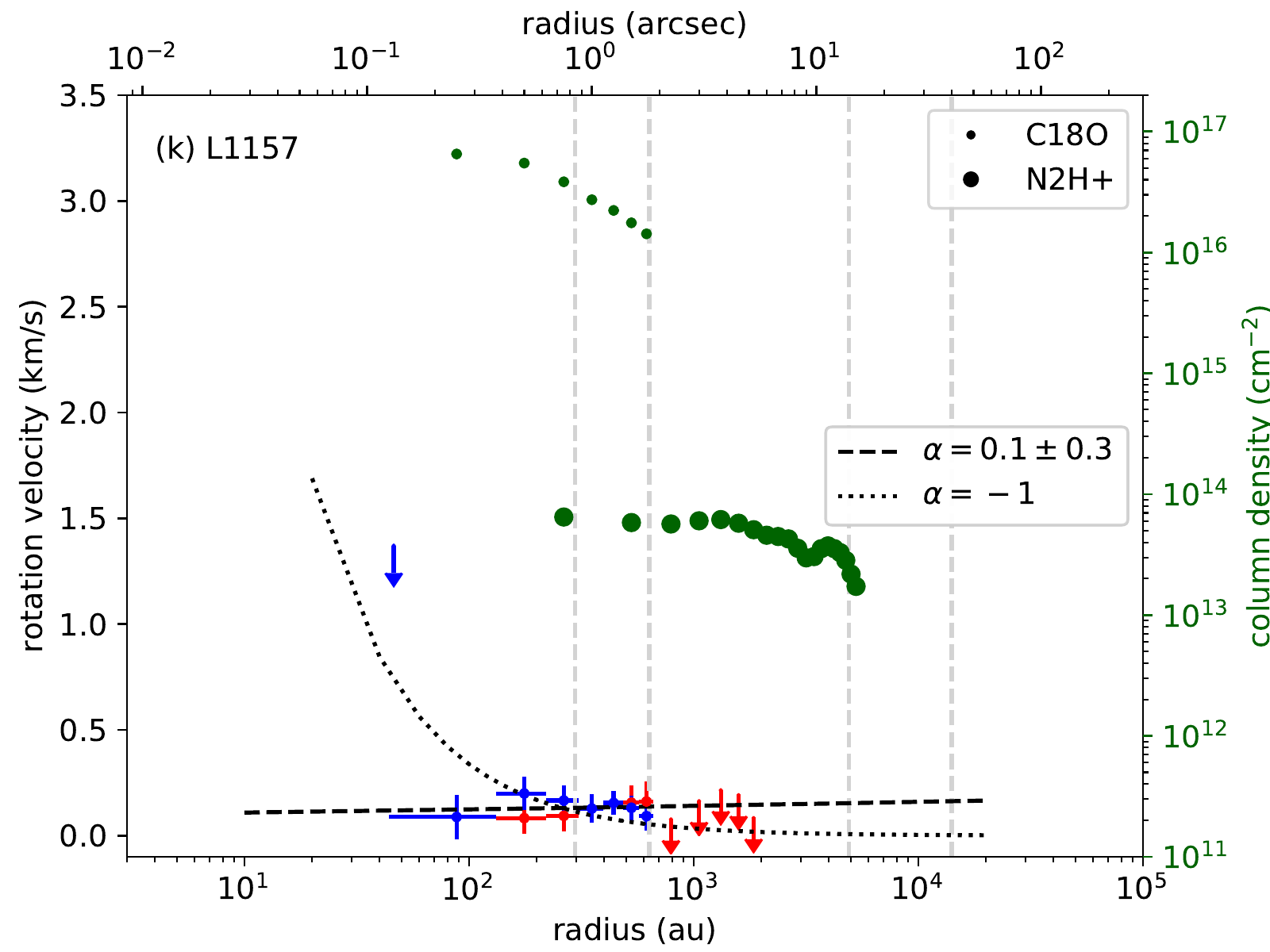}
\includegraphics[width=9cm]{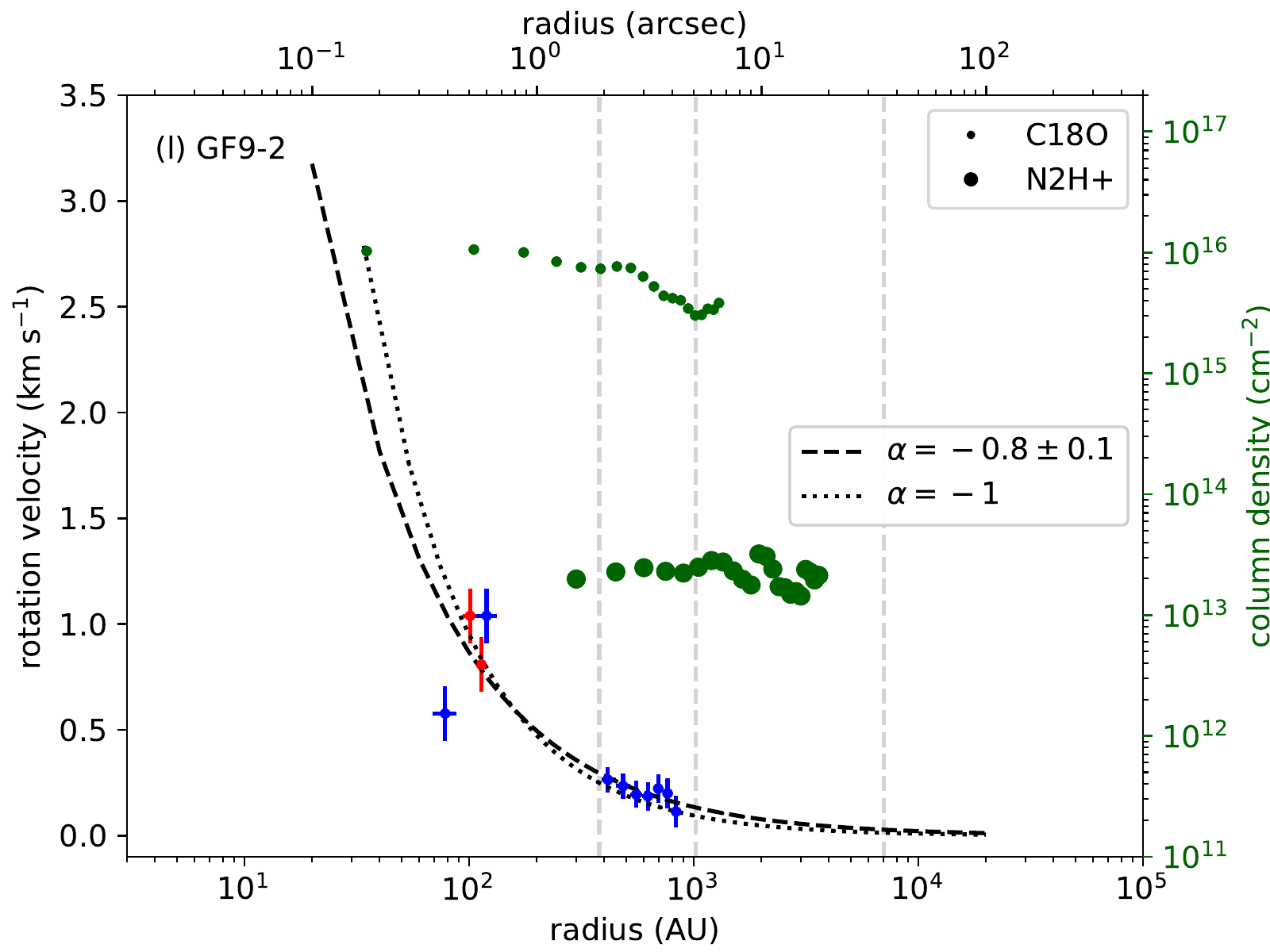}
\caption{Continued.
}
\end{figure*}

%\clearpage

%__________________________________________________________________

%\newpage
\section{Discussion}
In this section, we discuss the presence of rotation in the protostellar envelopes from the PV$_\mathrm{rot}$ diagrams (see Sect. \ref{sec:results-PV-diagram} and Fig. \ref{fig:PV-diagrams-1}). We build the distribution of specific angular momentum associated with the PV$_\mathrm{rot}$ diagrams (see Sect. \ref{subsec:distribution-j}) and explore the possible solutions to explain the $j(r)$ profiles observed in the inner ($r<$1600~au, see Sect. \ref{sec:j-inner-envelopes}) and outer ($r>$1600~au, see Sect. \ref{sec:velocity-gradient-1000au}) parts of the envelopes.

\subsection{Characterization of rotational motions} \label{sec:results-PV-diagram}
We assume that the protostellar envelopes are axisymmetric around their rotation axes, and thus, the velocity gradients observed along the equatorial axis, and reported in the PV$_\mathrm{rot}$ diagrams, are mostly due to the rotational motions of the envelopes. We model the rotational velocity variations by a simple power-law model $\mathrm{v} \propto r^{\alpha}$ without taking the upper limits into account. This method has been tested with an axisymmetric model of collapsing-rotating envelopes by \cite{Yen13}. As long as rotation dominates the velocity field on the line of sight, which depends on the inclination and flattening of the envelope, \cite{Yen13} obtain robust estimates of the rotation motions at work in the envelopes.
First, we fix the power-law index at $\alpha$=-1 to compare to what is theoretically expected for an infalling and rotating envelope from a progenitor core in solid-body rotation \citep{Ulrich76, Cassen81, Terebey84, Basu98}. The reduced $\chi^{2}$ values of fits by an orthogonal least-square model are reported in the second column of Table \ref{table:chi2-fit-profil-rotation}.
Then, we let the power-law index vary as a free parameter: the best power-law index and the reduced $\chi^{2}$ found for each protostellar envelope in our sample are reported in the third column of Table \ref{table:chi2-fit-profil-rotation}.
Figure \ref{fig:PV-diagrams-1} shows the PV$_\mathrm{rot}$ diagrams adjusted by a power-law for the sources of the CALYPSO sample.

\begin{table}[h]
\centering
\caption{Parameters of best power-law fits to the PV$_\mathrm{rot}$ diagrams.}
\label{table:chi2-fit-profil-rotation}
\resizebox{\hsize}{!}{\begin{tabular}{l|c|cc|ccc}
\hline \hline
\multicolumn{3}{c}{\hfill} & \multicolumn{2}{c}{\hfill}  \\
Source & Radial range~\tablefootmark{a} & \multicolumn{2}{c}{Power law fit $\alpha$=-1~\tablefootmark{b}} &  \multicolumn{3}{c}{Power law fit~\tablefootmark{c}}                                                  \\
    &  & DoF  & $\chi^{2}$                &  DoF & $\alpha$       & $\chi^{2}$  \\
        &         (au)&                  &                &                 \\
        
\hline
L1448-2A  & 60$-$1250 &  8 &  1.3         &  7       & -0.9 $\pm$0.1  & 1.4  \\
\hline
\multirow{2}{*}{L1448-NB} & 150$-$1700  &  34  & 2.3       &     33      & -0.94 $\pm$0.04  & 2.3   \\
 & 150$-$350 & ...   &    ...    &     5     & -0.9 $\pm$0.2 &   0.4 \\
 \hline
L1448-C & 100$-$4000& 19  & 1.6      &      18     & -1.0 $\pm$0.1  & 1.7  \\
 \hline
IRAS2A  &85$-$1500& 13   & 0.9      &     12     & -0.7 $\pm$0.1  & 0.2     \\
\hline
SVS13-B &110$-$450 & 7  &       0.2   &     6 &   -0.9 $\pm$0.3  &  0.2   \\
\hline
\multirow{2}{*}{IRAS4A} &  250$-$1600 & 18 &   20.4     &   17      &  0.8$\pm$0.1 &   1.5        \\
& 250$-$550 &  ...  &  ...      &      5    & -1.3$\pm$0.6 &  0.6  \\
\hline
IRAS4B &  175$-$1050&12 &  0.6       &   11      & -0.6 $\pm$0.3  & 0.5  \\
\hline
IRAM04191 & 55$-$800    &9  &     1.1   &     8    & -0.3 $\pm$0.2  &   0.3           \\
\hline
L1521F &1500$-$4200  & 6 &    1.0      &     5   & 0.2 $\pm$0.5  & 0.1      \\
\hline
L1527& 45$-$2000  &58  & 1.9        &    57    & -1.13 $\pm$ 0.03 & 1.6      \\
\hline
L1157 & 85$-$650  & 10  & 1.2        &    9     & 0.1 $\pm$0.3  & 0.3     \\
\hline
GF9-2 &  75$-$850  & 10  & 2.0      &      9     & -0.8 $\pm$0.1  & 1.8   \\
\hline
\end{tabular}}
\tablefoot{
\tablefoottext{a}{Range of radii over which the PV$_\mathrm{rot}$ diagrams were built and the fits were performed.}
\tablefoottext{b}{Number of degrees of freedom we used for the modeling and reduced $\chi^{2}$ value associated with the best fit with a power-law function $\mathrm{v} \propto r^{-1}$.}
\tablefoottext{c}{Number of degrees of freedom we used for the modeling, index of fit with a power-law function ($\mathrm{v} \propto r^{\alpha}$) and the reduced $\chi^{2}$ value associated with this best fit model.} 
}
\end{table}

The power-law indices of the PV$_\mathrm{rot}$ diagrams from our sample are between -1.1 and 0.8.
Five sources (L1448-2A, IRAS2A, SVS13-B, L1527, and GF9-2) show rotational velocity variations in the envelope scaling as a power law with an index close to -1. This is consistent with the expected index for collapsing and rotating protostellar envelopes. The reduced $\chi^2$ are $\sim$1.5 for these sources except for IRAS2A and SVS13-B for which it is better ($\sim$0.2).
L1521F and L1157 show a power-law index close to 0 with a very low reduced $\chi^2$ ($\leq$0.3, see Table \ref{table:chi2-fit-profil-rotation}). These flat PV$_\mathrm{rot}$ diagrams ($\mathrm{v}_\mathrm{rot} \sim$ constant) suggest differential rotation of the envelope with an angular velocity of $\Omega=\frac{v_\mathrm{rot}}{r} \propto r^{-1}$.
For two other sources (IRAS4B and IRAM04191), the best indices are compatible with -0.5, which could suggest Keplerian rotation at scales of $r<$1000~au. However, the reduced $\chi^2$ are also satisfactory ($\sim$1) when we fix the power-law index at $\alpha$=-1 (see Table \ref{table:chi2-fit-profil-rotation}). Thus, for these two sources, our CALYPSO datasets only allow us to estimate a range of the power-law indices between -1 and -0.5 (see Table \ref{table:chi2-fit-profil-rotation}).

Rotational velocity variations along the equatorial axis between 50 and 5000~au in L1448-NB cannot be reproduced satisfactorily by any single power-law model ($\chi^2 >$2, see Table~\ref{table:chi2-fit-profil-rotation}).
However, considering only the points at $r<$400~au for L1448-NB, we obtain a power-law index of -0.9$\pm$0.2 with a good reduced $\chi^2$ of 0.4, as expected for a collapsing and rotating envelope (see Table \ref{table:chi2-fit-profil-rotation}).

We found a positive index $\alpha$ for IRAS4A of 0.8 (see Table \ref{table:chi2-fit-profil-rotation}). It could be an indication of solid-body rotation of $\Omega=\frac{v_\mathrm{rot}}{r} \sim$constant. However, we observe that the velocity in the PV$_\mathrm{rot}$ diagram decreases from 2000 to $\sim$600~au and re-increases at small scales (see panel (f) of Fig. \ref{fig:PV-diagrams-1}). Thus, the velocity gradient is not uniform on the scales traced by the PV$_\mathrm{rot}$ diagram as would be expected for a solid-body rotation ($\mathrm{v}_\mathrm{rot} \propto r$). 
Moreover, points at radii $<$600~au are consistent with an infalling and rotating envelope (see panel (f) of Fig. \ref{fig:PV-diagrams-1}): considering only these points, we obtain a power-law index of -1.3$\pm$0.6 with a good reduced $\chi^2$ value of 0.6 (see Table \ref{table:chi2-fit-profil-rotation}). There is a dip in the C$^{18}$O emission at $r<$350~au that could be due to the opacity (see Figures \ref{fig:intensity-maps-IRAS4A} and \ref{fig:column-density-maps-IRAS4A}), thus, below this radius the information on velocities could be altered. 
To date, no observations have identified any solid-body rotating protostellar envelope. Numerical models also favor differential rotation of the envelope \citep{Basu98}. The interpretation of the velocity field as tracing solid-body rotation in the envelope of IRAS4A is therefore unlikely to be correct.

For the sources IRAS2A, IRAM04191, and L1157, the reduced $\chi^2$ is also good ($\sim$1) when the PV$_\mathrm{rot}$ diagrams of these sources are ajusted by a model with a fixed index of $\alpha$=-1 (see Table \ref{table:chi2-fit-profil-rotation}). We determine position and velocity from four different and independent methods and we did not consider the uncertainties of the connection between the different tracers and datasets. The uncertainty on the indices reported in Table~\ref{table:chi2-fit-profil-rotation} may thus be underestimated.
On the other hand, although we determined the systemic velocity by maximizing the overlap of the blue and red points, this method does not allow a more accurate determination than 0.05~km~s$^{-1}$. The systematic error of 0.05~km~s$^{-1}$, added to previous velocity errors of the points in the PV$_\mathrm{rot}$ diagrams to take this uncertainty on the systemic velocity into account (see Appendix \ref{sec:systemic-velocity}), can be overestimated and thus lead us to underestimate the $\chi^2$. For these three sources, the CALYPSO dataset only allow us to estimate a range of power-law indices between -1 and the $\alpha$ value reported in the fifth column of Table~\ref{table:chi2-fit-profil-rotation}.
The uncertainties on the indices reported in Table \ref{sec:systemic-velocity} are statistical errors and a systematic uncertainty of $\pm$0.1 has to be added to account for the uncertainties in the outflow directions and thus the equatorial axis directions (see Table \ref{table:sample}).
Moreover, despite the choice of the equatorial axis, the rotational velocities could be contaminated by infall at the small scales along this axis due to the envelope geometry.

To conclude, the organized motions reported in the PV$_\mathrm{rot}$ diagrams and modeled by a power-law function with an index $\alpha$ ranging from -2 to 0 are consistent with differential rotational motions ($\Omega \propto r^{\epsilon}$, with -3$< \epsilon <$-1 here). We identified rotational motions in all protostellar envelopes in our sample except in IRAS4A.

\subsection{Distribution of specific angular momentum in the CALYPSO Class~0 envelopes} \label{subsec:distribution-j}

\subsubsection{Specific angular momentum due to rotation motions}

\begin{figure*}[!ht]
\centering
\includegraphics[scale=0.7,angle=0,trim=0cm 4cm 0cm 5cm,clip=true]{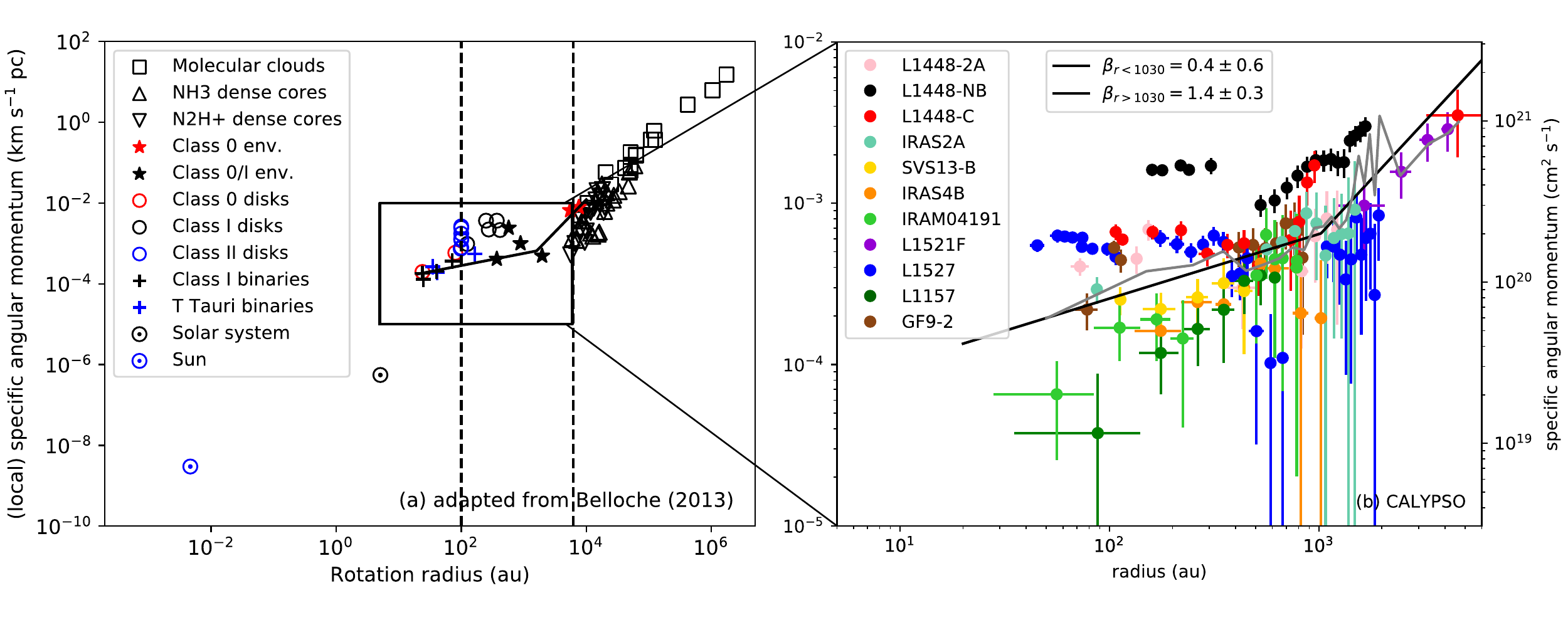}
\caption{Radial distribution of specific angular momentum. Panel (a): Figure adapted from Figure 8 of \cite{Belloche13} and from \cite{Ohashi97}. Panel (b): zoom on the region where the angular momentum profiles due to rotation of the CALYPSO sources lie. The gray curve shows the median profile $j(r)$. In the two panels, the solid black line shows the best fit with a broken power-law model.
}
\label{fig:diagramme-j-Belloche13+CALYPSO}
\end{figure*}

Assuming the motions detected along the equatorial axis are dominated by differential rotation for 11 of the 12 sources in our sample, we use the measurements reported in the PV$_\mathrm{rot}$ diagrams to derive the radial distribution of specific angular momentum in the protostellar envelopes due to rotation. In this part of the study, IRAS4A is excluded. The specific angular momentum is $j=\frac{J}{M}=\frac{I \Omega}{M}$ with the moment of inertia $I$ defined as $I \propto M r^2$ \citep{Belloche13}. Thus, the specific angular momentum is calculated from the rotational velocity: $j= \mathrm{v}_\mathrm{rot}(r) \times r$. 
We plot all the specific angular momentum profiles obtained for the CALYPSO subsample in panel (b) of Fig. \ref{fig:diagramme-j-Belloche13+CALYPSO}.
The individual distribution of specific angular momentum $j(r)$ for each source is given in Appendix \ref{sec:comments-indiv-sources}. This is the first time that the specific angular momentum distribution as a function of radius within a protostellar envelope is determined homogeneously for a large sample of 11 Class~0 protostars.
We performed a least-square fit of the $j(r)$ profiles for each source individually, using a model of a simple power-law and a broken power-law model to identify the change of regimes. 
The broken power-law model function is defined with a break radius $r_\mathrm{break}$ as follows:
$$j(r) \propto \left( \frac{r}{r_\mathrm{break}} \right)^{\beta_{1}}~~\mathrm{when}~~r<r_\mathrm{break},$$
$$j(r) \propto \left( \frac{r}{r_\mathrm{break}} \right)^{\beta_2}~~\mathrm{when}~~r>r_\mathrm{break}.$$
We report in Table \ref{table:chi2-fit-profil-moment-ang} the power-law indices fitting the best individual profiles and the associated reduced $\chi^2$. For the broken power-law fits, only results with a reduced $\chi^2$ better than the one obtained with a simple power-law model and with a break radius value $r_\mathrm{break}$ to which the $j(r)$ profile is really sensitive, have been retained.

\begin{figure*}[!ht]
\centering
\includegraphics[width=10cm]{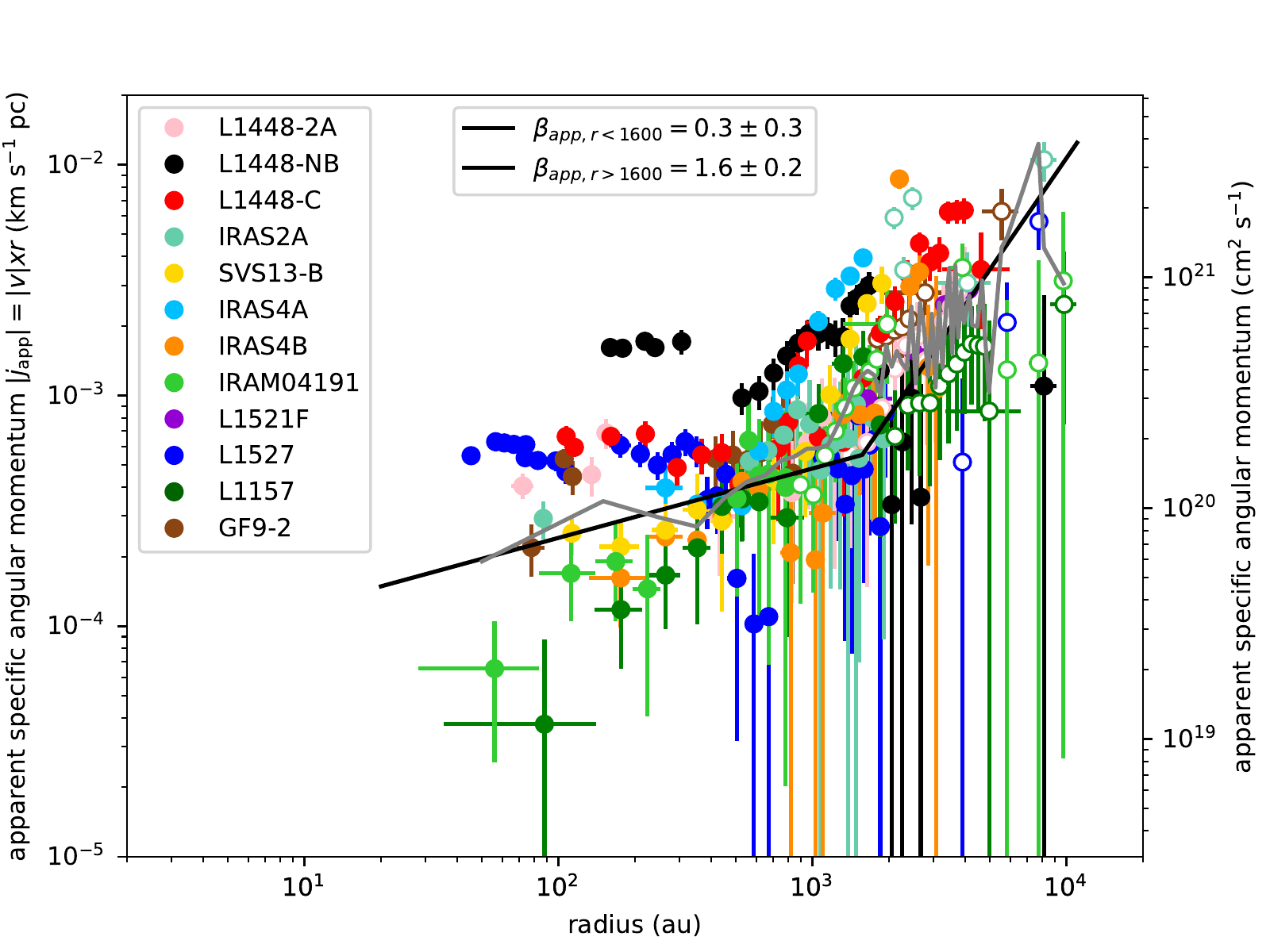}
\caption{Radial distribution of apparent specific angular momentum $|j_\mathrm{app}|= |v| \times r$ along the equatorial axis of the CALYPSO sources, considering all the velocity gradients observed at all envelope scales, including the reversed gradients and the shifted ones at scales of $r \gtrsim$1600~au (see Fig. \ref{fig:evolution-theta-plot} and Sect. \ref{sec:velocity-gradient-1000au}) which were excluded in the construction of the PV$_\mathrm{rot}$ diagrams in Fig. \ref{fig:PV-diagrams-1}, and in panel (b) of Fig. \ref{fig:diagramme-j-Belloche13+CALYPSO} for our analysis of rotational motions. The empty circles show the negative apparent specific angular momentum from the reversed gradients along the equatorial axis in the outer scales ($r >$1600~au) of the L1448-2A, IRAS2A, IRAM04191, L1527, L1157 and GF9-2 envelopes (see Sect. \ref{subsec:Counter-rotation}). The gray curve shows the median profile |$j_\mathrm{app}$| and the solid black line shows the best fit with a broken power-law model. }
\label{fig:profil-j-CALYPSO-tous-gradients}
\end{figure*}

Two sources (L1448-NB and L1448-C) are better reproduced by a broken power-law model than a simple power-law model where the $\chi^2$ are $\sim$2: this allows us to identify a change of slope from a relatively flat profile to an increasing profile at larger radius in the envelope ($\beta \sim$1), with break radii between 500 and 700~au. 
For the other sources, we also identified at scales of $r<$1300~au a flat profile of specific angular momentum with $\beta<0.5$ (L1448-2A, IRAS2A, SVS13-B, IRAS4B, L1527, and GF9-2) while the specific angular momentum profile at scales of $r>$1000~au shows a steeper slope with $\beta \sim$1 (L1521-F). 
However, two sources of the sample (IRAM04191 and L1157) stand out as the sources showing a steep increase in their specific angular momentum profile at scales of $r<$1000~au ($\beta \geq$0.7), similar to the indices found at large radii in the sources showing a break in their $j(r)$ profiles.
We note that for the flat profiles ($\beta <$0.5; L1448-2A, SVS13-B, IRAS4B, and GF9-2), and IRAM04191 and L1157, the specific angular momentum distribution is only constrained at scales $<$1300~au.
Most of the sources in our sample are better reproduced by a broken power-law model with a break radius (1000 $\pm$ 500)~au and an increasing profile at larger radius in the envelope ($\beta \sim$1.4) than a simple power-law model.

In his review, \cite{Belloche13} plotted the observed specific angular momentum as a function of rotation radius for several objects along the star-forming sequence. In this plot (panel (a) of Figure \ref{fig:diagramme-j-Belloche13+CALYPSO}), he identifies three regimes in the distribution of specific angular momentum, that can be broadly associated with different evolutionary stages: \\
\indent $\bullet$ prestellar regime: on large scales, the apparent angular momentum of molecular clouds \citep{Goldsmith85} and dense cores \citep{Goodman93, Caselli02} appears to follow the power-law relation $j \propto r^{1.6}$,\\
\indent $\bullet$ protostellar regime: between 100~au and $\sim$6000~au (0.03~pc), a few points in different protostellar envelopes suggest the specific angular momentum is relatively constant ($j \sim 10^{-3}$~km~s$^{-1}$~pc, \citealt{Ohashi97, Belloche02, Chen07}),\\
\indent $\bullet$ disk and binary regime: below 100~au, measurements in disks and Class~II binaries \citep{Chen07} show a decrease of $j$ following a trend characteristic of Keplerian rotation ($j \propto r^{0.5}$).\\

Thus, from previous observational studies on rotational motions, finding a break at $r \sim$1000~au between two trends of specific angular momentum within Class~0 protostars was unexpected. Although the velocity gradients observed in the outer part of the protostellar envelopes ($r>$1000~au) are consistent with rotational motions, the observed $j$ regime at these scales is not expected from pure rotational motions.

\subsubsection{Apparent specific angular momentum}

The radius range of $j(r)$ distribution due to rotation motions is not homogeneous between sources (see Table \ref{table:chi2-fit-profil-moment-ang}). To identify whether the radius of $\sim$1000~au is a critical radius between two trends of specific angular momentum in each source, we derive the radial distribution of the apparent specific angular momentum |$j_\mathrm{app}$| at all scales in the envelopes. To build |$j_\mathrm{app}$|$(r)$ distribution, we consider the gradients observed at all envelope scales, including also the reversed gradients and the shifted ones at scales of $r \gtrsim$1000~au (see Fig. \ref{fig:evolution-theta-plot} and Sect. \ref{sec:velocity-gradient-1000au}) which were excluded in the construction of the PV$_\mathrm{rot}$ diagrams in Fig. \ref{fig:PV-diagrams-1} because they are not consistent with rotational motions. By considering these velocity gradients, we add points in the outer envelopes but the trend observed in the inner envelopes do not change (see Tables \ref{table:chi2-fit-profil-moment-ang} and \ref{table:chi2-fit-profil-japp}). Thus, the |$j_\mathrm{app}$|$(r)$ distribution helps us to understand the origin of the trend and the velocity gradients observed at $r >$1000~au.
We plot all the apparent specific angular momentum profiles obtained for the CALYPSO subsample in Fig. \ref{fig:profil-j-CALYPSO-tous-gradients}. We also report the apparent specific angular momentum of IRAS4A which was identified as the only source that did not show any rotational motions in our sample (see Sect. \ref{sec:results-PV-diagram}). As for $j$ profiles, we performed a least-square fit of the |$j_\mathrm{app}$|$(r)$ profiles for each source individually and we report the indices of the power-law models in Table \ref{table:chi2-fit-profil-japp}.

We create the median |$j_\mathrm{app}$|$(r)$ profile of the CALYPSO subsample. We first resampled the individual profile of each source in steps of 100~au and normalized it by the value at 600~au, then we took the median value of individual profiles at each radius step. The median profile is shown in gray on Figure \ref{fig:profil-j-CALYPSO-tous-gradients}.
From a broken power-law fit, we obtain a relatively flat profile ($j_\mathrm{app} \propto r^{0.3 \pm 0.3}$) at radii smaller than 1570$\pm$300~au and an increasing profile ($j_\mathrm{app} \propto r^{1.6 \pm 0.2}$) in the outer envelope. The radius of $\sim$1600~au therefore appears to be a critical radius which delimits two regimes of angular momentum in protostellar envelopes: the specific angular momentum decreases down to $\sim$1600~au and then tends to become constant.

The change of behavior of $j_\mathrm{app}$ above the break radius could be due to a change of tracer to study the kinematics in the outer envelope. However, we do not find any systematic consistency between $r_\mathrm{app,break}$ and the transition radius $R_\mathrm{trans}$ between the two tracers C$^{18}$O and N$_{2}$H$^{+}$. Even if for SVS13-B, $R_\mathrm{trans}$ is in the error bars of $r_\mathrm{app,break}$, for three sources (L1448-NB, L1448-C, and IRAS4A) it is not consistent, and for IRAS4B, we do not observe a change of regime for $j_\mathrm{app}$ at $r \sim$1600~au (see Tables \ref{table:chi2-fit-profil-moment-ang} and \ref{table:chi2-fit-profil-japp}). Moreover, for L1521F, only the C$^{18}$O emission shows a velocity gradient allowing us to constrain the kinematics at scales of $r>$1600~au (see Fig. \ref{fig:velocity-maps-L1521F}) and we find the same trend of $j_{app}$ ($\beta_\mathrm{app} \sim$1.2) than in all other sources where we used N$_{2}$H$^{+}$ to constrain the outer part of the envelopes. 
The other sources (L1448-2A, IRAS2A, IRAM04191, L1527, L1157, and GF9-2) show a negative value of the apparent angular momentum at outer envelope scales due to a renversal of the velocity gradients (see Fig. \ref{fig:profil-j-CALYPSO-tous-gradients}, Table \ref{table:chi2-fit-profil-japp}, and Sect. \ref{subsec:Counter-rotation}). For two of these sources (L1448-2A and GF9-2) the radius where the gradient reverses along the equatorial axis, resulting in a negative $j_\mathrm{app}$ with respect to the inner envelope scales, is consistent with $R_\mathrm{trans}$ and $r_\mathrm{app,break}$. For two sources (IRAS2A and IRAM04191), $R_\mathrm{trans}$ is consistent with the radius where the gradient reverses along the equatorial axis but not with $r_\mathrm{app,break}$. For the last two sources (L1527 and L1157), the three radii are all different from each other.
The different individual behaviors in the CALYPSO sample allow us to conclude that our finding that protostellar envelopes are characterized by two regimes of angular momentum does not result from our use of two different tracers.

From the median |$j_\mathrm{app}$|$(r)$ profile without normalization of the individual profiles at 600~au, we find a mean value of specific angular momentum in the inner parts of the envelopes ($r<$1600~au) of $\sim$6 $\times$10$^{-4}$~km~s$^{-1}$~pc. This value is slightly lower but compatible with the estimates made by \cite{Ohashi97} and \cite{Chen07} in four Class~0 or I sources ($j \sim$10$^{-3}$~km~s$^{-1}$~pc at $r<$5000~au). It is also consistent with the studies by \cite{Yen15b} and \cite{Yen15} which find values between 5 $\times$ 10$^{-3}$~km~s$^{-1}$~pc and 5 $\times$ 10$^{-5}$~km~s$^{-1}$~pc in the inner envelope (r$<$1500~au). 
\cite{Yen15b} estimate a specific angular momentum of $\sim 5 \times$ 10$^{-4}$~km~s$^{-1}$~pc at $r \sim$100~au for L1448-C and L1527. Moreover, our values for L1157 are consistent with their upper limit estimate of $5 \times$ 10$^{-5}$~km~s$^{-1}$~pc in the inner envelope ($r<$100~au) of L1157.

The high angular resolution and the high dynamic range of the CALYPSO observations allow us to identify the first two regimes within individual protostellar envelopes: values at radii $\gtrsim$1600~au ($j_\mathrm{app} \propto r^{1.6}$ on average, see Table \ref{table:chi2-fit-profil-japp}) seem to correspond to the trend found in dense cores at scales $>$6000~au while the values stabilize around $\sim$6 $\times$10$^{-4}$~km~s$^{-1}$~pc on average at radii $<$1600~au. This study resolves for the first time the break radius between these two regimes deeper within the protostellar envelopes at around $\sim$1600~au instead of $\sim$6000~au. 
In a study of ammonia emission in the outer envelopes of two Class~0 objects, \cite{Pineda19} find an increasing angular momentum profile scaling as $r^{1.8}$ from 1000~au to 10000~au, with values $\sim$3 $\times$10$^{-4}$~km~s$^{-1}$~pc at radii $\sim$1000~au. They do not detect the break around $\sim$1600~au found in the CALYPSO sample.
This break radius from which the profiles are found to be flat in the inner envelope may depend on the evolutionary stage of the accretion process during the Class~0 phase as suggested by \cite{Yen15}. It could be due to the propagation of the inside-out expansion wave during the collapse \citep{Shu77}: assuming a median lifetime or half life of $\sim$5 $\times$10$^4$~yr for Class~0 protostar envelopes (\citealt{Maury11}; see also \citealt{Evans09}) at sound velocity ($\sim$ 0.2~km~s$^{-1}$), one obtains a radius $\sim$2000~au.
This radius is on the same order of magnitude as the observed break radius between the two regimes observed in the distribution of specific angular momentum of sources in our sample. In this case, the break radius could be an indication of the age of the protostars, except for four sources (L1448-NB, IRAM04191, L1521F, and L1157) in our sample where we do not observe this break radius. 
Beyond this radius, the outer envelope may not have collapsed yet, and could therefore retain the initial conditions in angular momentum of the progenitor prestellar core. 

This could be an explanation for the increase in angular momentum observed at the scales of $r>$1600~au ($j_\mathrm{app} \propto r^{1.6}$ on average, see Table \ref{table:chi2-fit-profil-japp}), consistent with the prestellar stage ($j \propto r^{1.6}$). 
We discuss the properties and physical origin of these two regimes in more details in the next sections.

\subsection{Conservation of angular momentum in Class~0 inner envelopes} \label{sec:j-inner-envelopes}

In this section, we focus on the relatively constant values of specific angular momentum observed in the inner envelopes at scales of $r \le$1600~au in the $j(r)$ profiles due to rotation motions (see Fig. \ref{fig:diagramme-j-Belloche13+CALYPSO}). From these flat profiles, we find that the matter directly involved in the formation of the stellar embryo has a specific angular momentum $\sim$3 orders of magnitude higher than the one in T-Tauri stars ($j \sim$2 $\times$10$^{-7}$~km~s$^{-1}$~pc, \citealt{Bouvier93}). We discuss constant values of specific angular momentum as conservation of angular momentum to test disk formation as a possible solution to the angular momentum problem.

It is difficult to constrain the time evolution of specific angular momentum for a given particle from angular momentum distributions which are snapshots of the angular momentum distribution of all particles at a given time during the collapse phase. During the collapse of a core initially in either solid-body rotation or differential rotation, particles conserve their specific angular momentum during the accretion on the stellar embryo \citep{Cassen81,Terebey84, Goodwin04}. 
In the case of a protostellar envelope with a density profile $\rho \propto r^{-2}$, an observed flat profile $j(r)=$constant requires, since each particle at different radii has the same specific angular momentum, an initially uniform distribution of angular momentum. This does not agree with the steep increase in specific angular momentum we observe at scales of $r>$1600~au in the $j(r)$ profiles. The break in the specific angular momentum profile could be due either to a faster collapse of the inner envelope caused by an initial inner density plateau \citep{Takahashi16} or to a change of dominant mechanisms responsible for the observed velocity gradients from inner to outer scales of the envelope.

In our sample, we distinguish eight sources with a relatively flat $j(r)$ profile in the inner envelope ($\beta<$0.5, see Table \ref{table:chi2-fit-profil-moment-ang}): L1448-2A, L1448-NB, L1448-C, IRAS2A, SVS13-B, IRAS4B, L1527, and GF9-2. 
We estimate a centrifugal radius that would be obtained when the mass currently observed at $\sim$100~au collapses and based on the mean value of specific angular momentum observed today $<j_\mathrm{100~au}>$ as follows:
\begin{equation}
R_\mathrm{cent}=\frac{<j_\mathrm{100~au}>^2}{G~M_\mathrm{100~au}}.
\label{eq:Rcent}
\end{equation}

The lower limit of the mass enclosed within 100~au, $M_\mathrm{100~au}$, is the mass of the envelope $M_\mathrm{100~au}^\mathrm{dust}$ estimated from the PdBI 1.3~mm dust continuum flux \citep{Maury18}, assuming optically thin emission, a dust temperature at 100~au computed with Eq. \eqref{Tdust} and corrected by the assumed distance (see Table \ref{table:sample}). This mass estimate does not include the mass of the central stellar object, $M_{\star}$: since the embryo mass is unknown for most sources in our sample, we consider an upper limit of $M_\mathrm{100~au}=M_{\star}+M_\mathrm{100~au}^\mathrm{dust}$ assuming $M_{\star}=$ 0.2~M$_{\odot}$ for each source in our sample. This value of 0.2~M$_{\odot}$ corresponds to the stellar mass in the Class~0/I protostar L1527 from kinematic models of the Keplerian pattern in the disk \citep{Tobin12, Ohashi14, Aso17}. The range of values for $M_\mathrm{100~au}$ are reported in the third column in Table \ref{table:rayon-disque-brot}. The calculated range of centrifugal radii associated with $M_\mathrm{100~au}$ is listed for each source in the fourth column in Table \ref{table:rayon-disque-brot}. We note that if $M_{\star}$ of a source is smaller than that of L1527, then the centrifugal radius value we calculated is underestimated.

Since the embryo mass is uncertain and $M_\mathrm{100~au}$ may be underestimated if the dust emission is not optically thin, we compute the mass enclosed within $r<$100~au, including the stellar embryo mass, needed to form a disk the size of $R_\mathrm{disk}^\mathrm{dust}$ with the $<j_\mathrm{100~au}>$ observed. The values are reported in the last column of Table \ref{table:rayon-disque-brot}.

\begin{table*}[!ht]
\centering
\caption{Centrifugal radius $R_\mathrm{cent}$ assuming angular momentum conservation.}
\label{table:rayon-disque-brot}
\begin{tabular}{lccccc}
\hline
\hline
Source  & $<j_\mathrm{100~au}>$~\tablefootmark{a}              & $M_\mathrm{100~au}$~\tablefootmark{b} & $R_\mathrm{cent}$~\tablefootmark{c} & $R_\mathrm{disk}^\mathrm{dust}$~~\tablefootmark{d} & $M_{100~au}^\mathrm{min}$~\tablefootmark{e} \\
           & (10$^{-4}$~km~s$^{-1}$~pc) & ($M_{\odot}$) & (au)          &    (au) & ($M_{\odot}$) \\
          \hline
L1448-2A  & 4.5$\pm$0.2                & 0.005 $-$ 0.2                   &  50$-$1810      & $<$50   & 0.2  \\
L1448-NB  & 16.0$\pm$0.4               &  0.042 $-$ 0.2                   & 500$-$2920    & $<$50    & 3.0    \\
L1448-C   & 6.0$\pm$0.2                & 0.025 $-$ 0.2             & 70$-$690      & 41$\pm$15   & 0.5   \\
IRAS2A   & 3.8$\pm$0.4                & 0.020 $-$  0.2         & 30$-$330    & $\lesssim$65   & 0.1     \\
SVS13-B  & 2.5$\pm$0.2                & 0.019 $-$ 0.2         & 10$-$160   & $\lesssim$75   & 0.05     \\
IRAS4B   & 2.5$\pm$0.3                & 0.003 $-$ 0.2           & 10$-$840    & 155$\pm$30   & 0.02    \\
L1527     & 5.6$\pm$0.1                & 0.001 $-$ 0.2              & 70$-$1060        & 54$\pm$10  & 0.3   \\
GF9-2    & 3.9$\pm$0.3                & 0.002 $-$ 0.2              & 40$-$3080   & 36$\pm$9  & 0.2 \\
\hline
\end{tabular}
\tablefoot{
\tablefoottext{a}{Weighted mean of specific angular momentum in the inner envelopes (50~au$< r \le$1600~au).}
\tablefoottext{b}{Range of the object mass at 100~au, the minimum and maximum values are defined in Sect. \ref{sec:j-inner-envelopes}.}
\tablefoottext{c}{Centrifugal radii estimated from $<j_\mathrm{100~au}>$ and $M_\mathrm{100~au}$ using Eq. \eqref{eq:Rcent}, assuming conservation of angular momentum.}
\tablefoottext{d}{Candidate disk radius determined from the CALYPSO study of PdBI dust continuum emission at 1.3 and 3~mm \citep{Maury18}, corrected by the assumed distance (see Table \ref{table:sample}).}
\tablefoottext{e}{Total minimum mass that needs to be enclosed at $r<$100~au to form a disk equal to $R_\mathrm{disk}^\mathrm{dust}$ if the angular momentum $<j_\mathrm{100~au}>$ was conserved. This minimum mass considers the mass of the stellar embryo and the mass of the optically thick inner envelope enclosed within 100~au.}
}
\end{table*}

For all the sources in our sample, the upper limits of the $R_\mathrm{cent}$ range are larger than 150~au and systematically larger than the continuum disk candidate radii $R_\mathrm{disk}^\mathrm{dust}$ from \cite{Maury18} reported in the fifth column of Table \ref{table:rayon-disque-brot}. 
Moreover, \cite{Maret20} only detect possible Keplerian rotation in two protostars in our sample (L1527 at radii $\sim$90~au and L1448-C at $r \sim$200~au) from the CALYPSO data. Thus, most $R_\mathrm{cent}$ values is expected to be less than 100~au. The upper $R_\mathrm{cent}$ values are probably overestimated because the contribution of the embryo mass to $M_\mathrm{100~au}$ is excluded.

Comparing the lower limits of the $R_\mathrm{cent}$ range with the candidate disk radius, we find a good agreement for most sources in our sample except for L1448-NB.
We find a larger centrifugal radius ($\sim$500~au) than the observed disk size ($<$50~au) calculated considering only the main protostar L1448-NB1 of the binary system. Since in this study, we are interested in the kinematics of the whole system, we must consider all the continuum structure and not only that of the main protostar. Considering NB1 and NB2, \cite{Maury18} resolve a circumbinary structure with a radius of (320 $\pm$ 90)~au centered on the middle of the two components. Given the uncertainties, the latter value is consistent with the lower centrifugal radius estimated here. At these scales, \cite{Tobin16} observe a spiral structure surrounding the multiple system and interpreted it as a gravitationally unstable circumbinary disk. On the other hand, \cite{Maury18} suggest that this component is due to orbital motions and tidal arms between the companions and \cite{Maret20} do not detect any Keplerian rotation at radii $<$170~au. Thus, the nature of this additional structure surrounding the multiple system is still unclear. As a consequence, the increase in specific angular momentum we measured at small scales could not only trace the rotation of {the disk or} the envelope but may be contaminated by gravitational instabilities due to orbital motions or a fragmented disk surrounding the system.
Given the large uncertainties on the dust disk radii, we found a good agreement between centrifugal radii and $R_\mathrm{disk}^\mathrm{dust}$ for L1448-C and L1527. Moreover, the dust radius (50~au in L1527, \citealt{Maury18}) does not necessarily exactly correspond to the centrifugal radius which was first detected in L1527 from observations of SO emission at 100$\pm$20~au \citep{Sakai14Nat}. For this source, our estimate of $R_\mathrm{cent}$ ($\sim$70~au) is consistent with previous kinematic studies which detect a proto-planetary disk candidate with a radius of 50$-$90~au \citep{Ohashi14, Aso17, Maret20}. Moreover, we observe a slight increase in the specific angular momentum we measured at $r<$80~au. It could be due to the transition from the envelope to the disk.

The hypothesis of collapsing material with conservation of angular momentum, resulting in disk formation, at $r<$100~au is therefore plausible for most sources in our sample. 
We notice that L1448-NB, in which \cite{Tobin16} claim the detection of a large candidate disk, shows the highest value of specific angular momentum at $r<$1600~au of the CALYPSO sample, consistent with the angular momentum observed in the proto-planetary disks surrounding the T-Tauri stars which are estimated to be 1$-$6$\times$10$^{-3}$~km~s$^{-1}$~pc \citep{Simon00,Kurtovic18,Perez18}. It could suggest an increase in the angular momentum of the disk during its evolution. In this case, the mean value of $j(r)$ in the inner envelope would be lower in the less evolved than in the more evolved Class~0 protostars, and it would increase with time until reaching the value contained in the T-Tauri disks. In this scenario, L1448-NB would be one of the most evolved objects in the sample. However, the borderline Class~0/I protostar L1527, which is the most evolved object of the CALYPSO sample, has a specific angular momentum of $\sim$6 $\times$ 10$^{-4}$~km~s$^{-1}$~pc at the inner envelope scales (see Table \ref{table:rayon-disque-brot}). In the same way, L1448-C has a specific angular momentum less than one order of magnitude lower than the values observed in the Class~II disks while \cite{Maret20} suggest the presence of a Keplerian disk in the inner envelope. As most of the CALYPSO inner protostellar envelopes have an order of magnitude less specific angular momentum than in Class II disks, we discuss below several possible explanations:  \\
(i) a part of the angular momentum inherited by the T-Tauri disks may not come from the rotating matter contained in the inner envelope accreted during the Class~0 phase. During the Class~I phase, the mass accreted could come from regions further away from the envelope ($r \gg$1600~au) with a possibly higher specific angular momentum.\\
(ii) disks may expand with time due to the transfer of angular momentum from their inner regions to their outer ones. Unfortunately, the specific angular momentum does not contain information about the mass. Large values of $j(r)$ may be carried by low masses at the outer disk radius but may remain difficult to quantify. To this day, the mechanisms at work in disk evolution remain an open question. Some studies, for example, suggest that viscous friction may be responsible for the disk expansion \citep{Najita18}. \\
(iii) the specific angular momentum of the proto-planetary disks may be biased toward high values from historical, large and massive disks. A new population of small T-Tauri disks with radii between 10 and 30~au has been observed thanks to PdBI and ALMA \citep{Pietu14, Cieza19}. Assuming a small rotationally-supported disks around a stellar object including a total mass of 0.1$-$1~$M_{\odot}$ \citep{Pietu14}, one expects a specific angular momentum between 10$^{-5}$~km~s$^{-1}$~pc and 10$^{-4}$~km~s$^{-1}$~pc, values which are similar to those we obtained in the inner Class~0 protostellar envelopes with the CALYPSO sample. However, to this day, no resolved observations of gas kinemactics of these small Class II disks allow us to estimate observationally their specific angular momentum.\\

\subsection{Origin of the velocity gradients at $r>$1600~au} \label{sec:velocity-gradient-1000au}
\begin{figure}[!ht]
\centering
\includegraphics[scale=0.45,angle=0,trim=6.8cm 0cm 6.8cm 3cm,clip=true]{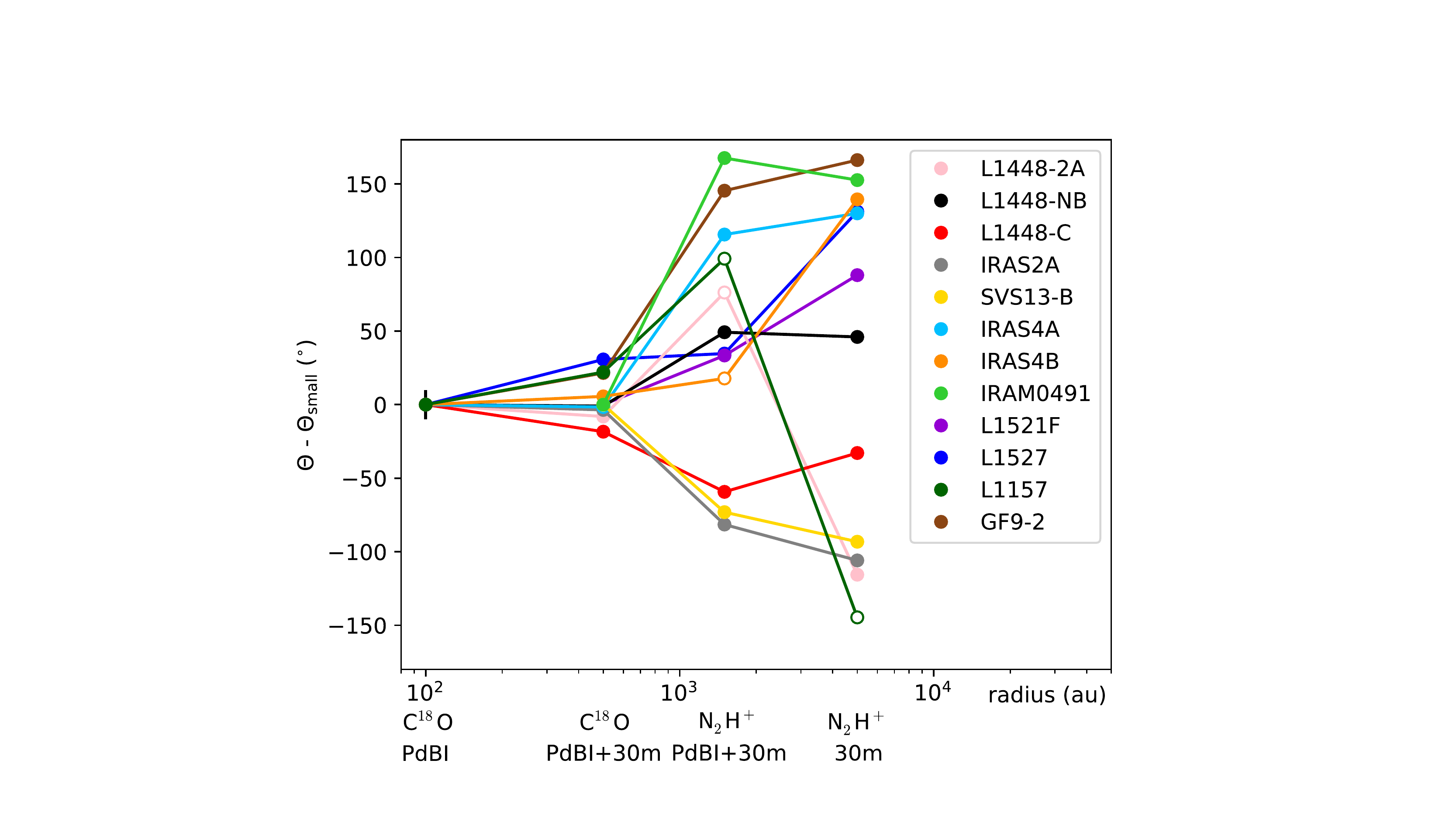}
\caption{Evolution of the orientation of the mean velocity gradient in the different datasets used to build the PV$_\mathrm{rot}$ diagrams and angular momentum distributions with respect to the PA of the velocity gradient observed at small scales $\Theta_\mathrm{small}$ (PdBI C$^{18}$O emission). The error bars of the orientation $\Theta$ are given in Table \ref{table:gradient-velocity-fit}. They are smaller than $10^{\circ}$ except for 7 of the 67 gradient measurements. For these 7 measurements, the large error bars are generally due to the absence of a clear gradient on either side of the central position of the source. Gradient measurements with large error are indicated by an empty circle. A typical error of $\pm 10^{\circ}$ is shown on the first point of the plot.}
\label{fig:evolution-theta-plot}
\end{figure}

At outer envelope scales, we detect velocity gradients ($\sim$2~km~s$^{-1}$~pc$^{-1}$ at $\sim$10000~au, see Table \ref{table:gradient-velocity-fit}) in the CALYPSO single-dish maps. They may not be directly related to rotational motions of the envelopes but rather to other mechanisms. Indeed, we observe in the CALYPSO dataset a systematic evolution of the orientation of the gradients between the inner and outer scales in the envelope (see Table \ref{table:gradient-velocity-fit}). Figure \ref{fig:evolution-theta-plot} shows the orientation of the mean velocity gradient observed at different scales of the envelope with respect to the position angle of the gradient observed at scales $\sim$100~au. The clear dispersion ($\sim$100$^{\circ}$ on average, see Fig. \ref{fig:evolution-theta-plot}) of gradient position angle across scales within individual objects may be due to a change of dominant mechanisms responsible for the observed gradients from inner to outer scales of the envelope. 
From the literature, velocity gradients are often measured in the outer protostellar envelopes along the equatorial axis and they are interpreted as due to rotational motions or infall from a filamentary structure at scales of 1500$-$10000~au \citep{Ohashi97, Belloche02, Tobin11}. In this section, we explore the possible origins of the velocity gradients found at scales of $r>1600$~au {and used to build the |$j_\mathrm{app}$|$(r)$ profiles (see Fig. \ref{fig:profil-j-CALYPSO-tous-gradients}).}

\subsubsection{Questioning the interpretation of counter-rotation}
\label{subsec:Counter-rotation}
Six sources in the sample show a clear reversal of the orientation of the mean velocity gradient ($| \Theta - \Theta_\mathrm{small} | > 130^{\circ}$) from the inner to the outer envelope scales: IRAS4A, IRAS4B, L1527, IRAM04191, L1157, and GF9-2. We note that the kinematics at scales where we observed reversed velocity gradients ($r>$1600~au) with respect to the small scales were not taken into account to build the PV$_\mathrm{rot}$ diagrams in Fig. \ref{fig:PV-diagrams-1}, or the specific angular momentum profiles shown for the full sample in panel (b) of Fig. \ref{fig:diagramme-j-Belloche13+CALYPSO}. Indeed, these profiles were aimed at characterizing the rotational motions in the envelopes and the angular momentum due to this rotation: a reversal of the rotation, if real, would require a more complex model than the power-law ($\mathrm{v} \propto r^{\alpha}$) model we adopted in Sects. \ref{subsec:diagram-PV-construction} to \ref{sec:j-inner-envelopes}. In this section, we discuss these complex patterns in more detail.

In IRAM04191, we observed velocity gradients at outer envelope scales of $r>$1600~au consistent with those observed previously by \cite{Belloche02} and \cite{Lee05} ($\Theta \sim$100$^{\circ}$, see Table \ref{table:gradient-velocity-fit}). However, in the inner envelope, we noticed a velocity gradient with a direction of $\Theta =$-83$^{\circ}$ (see bottom middle panel in Fig. \ref{fig:velocity-maps-IRAM04191} and Table \ref{table:gradient-velocity-fit}). In L1527, we found small-scale velocity gradients ($\Theta \sim$0$^{\circ}$ at $r \sim$ 1000~au) consistent with those previously observed by \cite{Tobin11} which are in the opposite direction compared to the large-scale one ($r \sim$8000~au, \citealt{Goodman93}). \cite{Tobin11} interpret this reversal of velocity gradients as counter-rotation but it also could be due to infalling motions that dominated the velocity field at the outer envelope scales \citep{Harsono14}. 

Our study suggests that reversals of velocity gradients are common in Class~0 protostellar envelopes. However, the asymmetrical velocity gradients (for IRAS4B, GF9-2), the filamentary structures traced by the integrated intensity at scales of $r>$2000~au (for IRAS4A, IRAS4B, L1527, and GF9-2), and a strong external compression of the cloud hosting IRAS4A and IRAS4B \citep{Belloche06} lead us to exclude the observed reversed gradients as counter-rotation of the envelope. 
Moreover, only MHD models with Hall effect succeed to form envelopes in counter-rotation. These models form a thin layer of counter-rotating envelopes at the outer radius of the disk ($r \sim$50-200~au; \citealt{Tsukamoto17}). This envelope layer is in counter-rotation compared to the formed disk and the protostellar envelope at $r>$200~au as a consequence of the Hall effect generated by the rotation of the disk which changes the angular momentum of the gas at the disk outer radius. Therefore, these models cannot explain the inversions of rotation in the different layers of the envelope at scales of $r>$3000~au as observed in our sample.
Historically, the gradients observed from single-dish mapping at $r>$3000~au have been used to quantify the amplitude for the angular momentum problem. However, incorrectly interpreted as pure rotational motions in the envelope, the resulting angular momentum measurements and the expected disk radii would be significantly overestimated. 

Recent studies on the angular momentum of the protostellar cores from hydrodynamical simulations of star formation are questioning the standard model of star formation from a collapsing core initially in solid-body rotation \citep{Kuznetsova19, Verliat20}. They show that the angular momentum of synthetic protostellar cores is not directly related to the initial rotation of the synthetic cloud, and Keplerian disks can be formed from a simple non-uniform perturbation in the initial density distribution.
In this scenario, the angular momentum observed in inner protostellar envelopes and disks may not been inherited from larger-scale initial conditions but generated during the collapse itself.

\subsubsection{Contribution of infalling motions and core-forming motions}

The misalignments between the gradients observed in the envelopes at inner and outer envelope scales suggest a change of dominant mechanisms at $r>$1600~au. At large scales, infalling motions of the envelope can dominate rotational motions. In the hypothesis of a flattened infalling envelope, infall motions are expected to produce a velocity gradient projected in the plane of the sky that is oriented along the minor axis of the envelope, namely at the same position angle as the outflow. 
In L1448-NB, SVS13-B and L1527, we detect velocity gradients aligned with the outflow axis at $r>$3000~au while at small scales the gradients are consistent with the equatorial axis (see Table \ref{table:gradient-velocity-fit}). These three sources could be good candidates of the transition from collapse to rotation between large and small scales.
This scenario is also suggested in the study of \cite{Ohashi97b}. They suggested that at outer envelope scales of $r \sim$2000~au, the protostellar envelope L1527 is not rotationally supported ($\mathrm{v}_\mathrm{rot} \sim$0.05~km~s$^{-1}$) but is dominated by the collapse ($\mathrm{v}_\mathrm{inf} \sim$0.3~km~s$^{-1}$). 

Currently, there are very few constraints on the infall velocities at scales of $r>$1600~au in the CALYPSO protostellar envelopes. \cite{Belloche02} estimate an infall velocity of $\mathrm{v}_\mathrm{inf} \sim$0.15~km~s$^{-1}$ at $r \sim$1000~au from radiative transfer modeling of CS and C$^{34}$S emission in IRAM04191. In the dense core L1544, \cite{Tafalla98} suggest also an infall velocity of $\sim$0.1~km~s$^{-1}$ at scales $>$3000~au. The velocity offset, with respect to the systemic velocity assumed
for each source, found along the equatorial axis at $>$1000~au with CALYPSO is reported in Table \ref{table:velocity-100-1000au}.
For most sources, we find typical velocity offsets $\lesssim$0.3~km~s$^{-1}$ at scales of 1600~au (see Table \ref{table:velocity-100-1000au}), consistent with infall velocities found in IRAM04191 and L1544, except for IRAS4A.
IRAS4A harbors a velocity of $\sim$0.5~km~s$^{-1}$ at $r \sim$1000~au. This result is consistent with those of \cite{Belloche06} at $\sim$2000~au. They suggest that a fast collapse is triggered by an external compression from the cloud in which the source is embedded.
Thus, for all sources in our sample, the velocity gradient misalignment could be due to a change of mechanism dominating the velocities projected on the line of sight. This suggests either rotational velocities much smaller than infall velocities or a non axisymmetric geometry of the kinematics at outer envelope scales.

Moreover, \textit{Herschel} observations have shown that most solar-type prestellar cores and protostars form in filaments \citep{Andre14}. Indeed, the column density maps of the Herschel Gould Belt Survey program\footnote{See \url{http://gouldbelt-herschel.cea.fr/archives}} \citep{Andre10} reveal that the CALYPSO protostars are embedded in or lie in the immediate vicinity of filamentary structures with $N_{H_2} >$10$^{21}$~cm$^{-2}$. Thus, the large-scale kinematics in protostellar envelopes could be contaminated or dominated by the kinematics of the filaments. 
\cite{Kirk13} studied the velocity field of Serpens-South in the Aquila molecular cloud and showed a complex kinematics with longitudinal collapse along the main filament, radial contractions, and accretion streams from the cloud to the main filament. The longitudinal collapse of the filament could be responsible for the large-scale gradients observed in our protostellar envelopes as observed in the Serpens-Main region by \cite{Dhabal18}. 
Several studies also highlighted transverse velocity gradients perpendicular to the main filament that suggested the material may be accreting along perpendicular striations \citep{Palmeirim13,Dhabal18,Arzoumanian18, Shimajiri19}.
\cite{Palmeirim13} estimate the velocity of the infalling material to be ~0.5$-$1~km~s$^{-1}$ at $r \sim$0.4~pc in the B211/L1495 region. In our velocity maps at 10000~au along the equatorial axis, we measure typical velocities $<$0.3~km~s$^{-1}$ in most of the sources (see Table \ref{table:velocity-100-1000au}) except in L1448-2A, IRAS4A, and IRAS4B. These three sources exhibit velocities of 0.5$-$1~km~s$^{-1}$ consistent with infall velocities estimated at filamentary scales. In these cases, host-filament motions could dominate the kinematics in outer protostellar envelopes ($r>$1600~au).

\begin{table}[!ht]
\centering
\caption{Value of velocity offset, with respect to the systemic velocity assumed for each source, at 100, 1000, and 10000~au along the equatorial axis from the velocity maps and considering all the gradients observed, even those not consistent with rotational motions.}
\label{table:velocity-100-1000au}
\begin{tabular}{lccc}
\hline
\hline
Source  & $\mathrm{v}_\mathrm{100~au}$              & $\mathrm{v}_\mathrm{1000~au}$  &  $\mathrm{v}_\mathrm{10000~au}$ \\
           & (km~s$^{-1}$) & (km~s$^{-1}$)  & (km~s$^{-1}$) \\
          \hline
L1448-2A  & 0.7                & 0.1       & 1.1    \\
L1448-NB  & 2.1               & 0.3        & 0.02     \\
L1448-C   & 1.3                & 0.1       & 0.2    \\
IRAS2A   & 0.7                & 0.1        &  0.3  \\
SVS13-B  & 0.5                & 0.1         &  0.2  \\
IRAS4A   & ...                & 0.4     &  0.9       \\
IRAS4B   & ...                & 0.1    &  0.5        \\
IRAM04191    & 0.3                & 0.1    &   0.1     \\
L1521F    & 0.1                & 0.2 &    0.2       \\
L1527     & 1.1               & 0.1        &  0.2 \\
L1157    & 0.1                & 0.2     &  0.1     \\
GF9-2    & 1.0               & 0.2         &  0.2 \\
\hline
\end{tabular}
\end{table}

\subsubsection{Contamination by turbulent motions from cloud scales}

All sources of the CALYPSO subsample (except L1448-NB) reveal a steep increase in apparent specific angular momentum with the radius at $\sim$1600$-$10000~au scales, with an average trend of $j_\mathrm{app} \propto r^{1.6 \pm 0.2}$ (see Table \ref{table:chi2-fit-profil-japp} and Fig. \ref{fig:profil-j-CALYPSO-tous-gradients}). This trend is similar to that observed in prestellar cores and clumps at scales $>$10000~au (see Figs. \ref{fig:diagramme-j-Belloche13+CALYPSO} and \ref{fig:profil-j-CALYPSO-tous-gradients}). Indeed, \cite{Goodman93}, \cite{Caselli02} and \cite{Tatematsu16} show a trend between the size of prestellar cores and their observed mean angular momenta at scales on the order of 0.1~pc: $j(r)$ distribution scaling as $r^{1.2-1.7}$. From this dependency of $j$ with core radius and the linewidth-size relation, \cite{Tatematsu16} suggest that non-thermal motions (turbulence) are related to the origin of angular momentum observed in these 0.1~pc cores.

\cite{Burkert00} studied numerical models of turbulent molecular clouds with a symmetric density profile and a Gaussian or random velocity field. They showed that 0.1~pc cores with random motions exhibit most of the time velocity gradients that, interpreted as rotation, would have specific angular momentum values of $j \sim 3 \times 10^{-3}$~km~s$^{-1}$~pc (10$^{21}$~cm$^2$~s$^{-1}$) and would scale as $j \propto r^{1.5}$. This is in good agreement with observed values at 0.1~pc from the literature \citep{Goodman93,Caselli02,Tatematsu16}. Our observations showing a trend of $j \propto r^{1.6}$ at scales $\sim$1600-5000~au could be either a signature of the turbulent cascade from the large-scale ISM propagating with subsonic properties to 1600~au envelope scales, or gravitationally-driven turbulence due to large-scale collapse motions at the interface between filaments and cores \citep{Kirk13}.

From the analysis of the gas velocity dispersion in molecular line observations, \cite{Goodman98} and \cite{Pineda10} identify the dense cores at a typical scale of 0.1~pc as the first velocity-coherent structures decoupled from the turbulent cloud. In this case, we would expect a quiescent structure with subsonic motions at radii $<$0.1~pc and the interpretation of ISM turbulent cascade with supersonic motions as a consequence of the steep increase of $j_\mathrm{app}$ at scales $<10000$~au would no longer be valid.
Except L1448-2A, IRAS4A, and IRAS4B which exhibit velocities 0.5$-$1~km~s$^{-1}$ consistent with supersonic turbulent motions, all sources in our sample show typical velocities $\lesssim$0.3~km~s$^{-1}$ consistent with subsonic-transonic turbulent motions.
This could suggest that the power-law behavior of $j_\mathrm{app} \propto r^{1.6}$ observed in the outer envelopes ($r>$1600~au) could be a scaling law due to the tail of a low velocity subsonic-transonic turbulent cascade.

At scales of $r>$1600~au, we observe typical velocity linewidths $\lesssim$1~km~s$^{-1}$ (see Fig. \ref{fig:line-width-diagrams}). We note that the linewidths tend to decrease from $\sim$1600~au to larger scales in the outer envelopes and they do not show scaling laws with the radius as expected from turbulent motions in the ISM (\citealt{Larson81}; see Fig. \ref{fig:line-width-diagrams} and Table \ref{table:chi2-fit-profil-dV}), but the velocity structure at these scales seems to show multiple components in velocity for several sources (L1527, L1448-C, IRAS2A, SVS13-B, IRAS4A, IRAS4B). As we can not disentangle them and identify exactly which component comes from the outer envelope or the host cloud for example, we need either a more elaborate model than a Gaussian or a HFS model to analyze the spectra or a more suitable tracer to determine more robustly the linewidths of the outer envelopes.

\begin{figure*}[!ht]
\centering
\includegraphics[width=9cm]{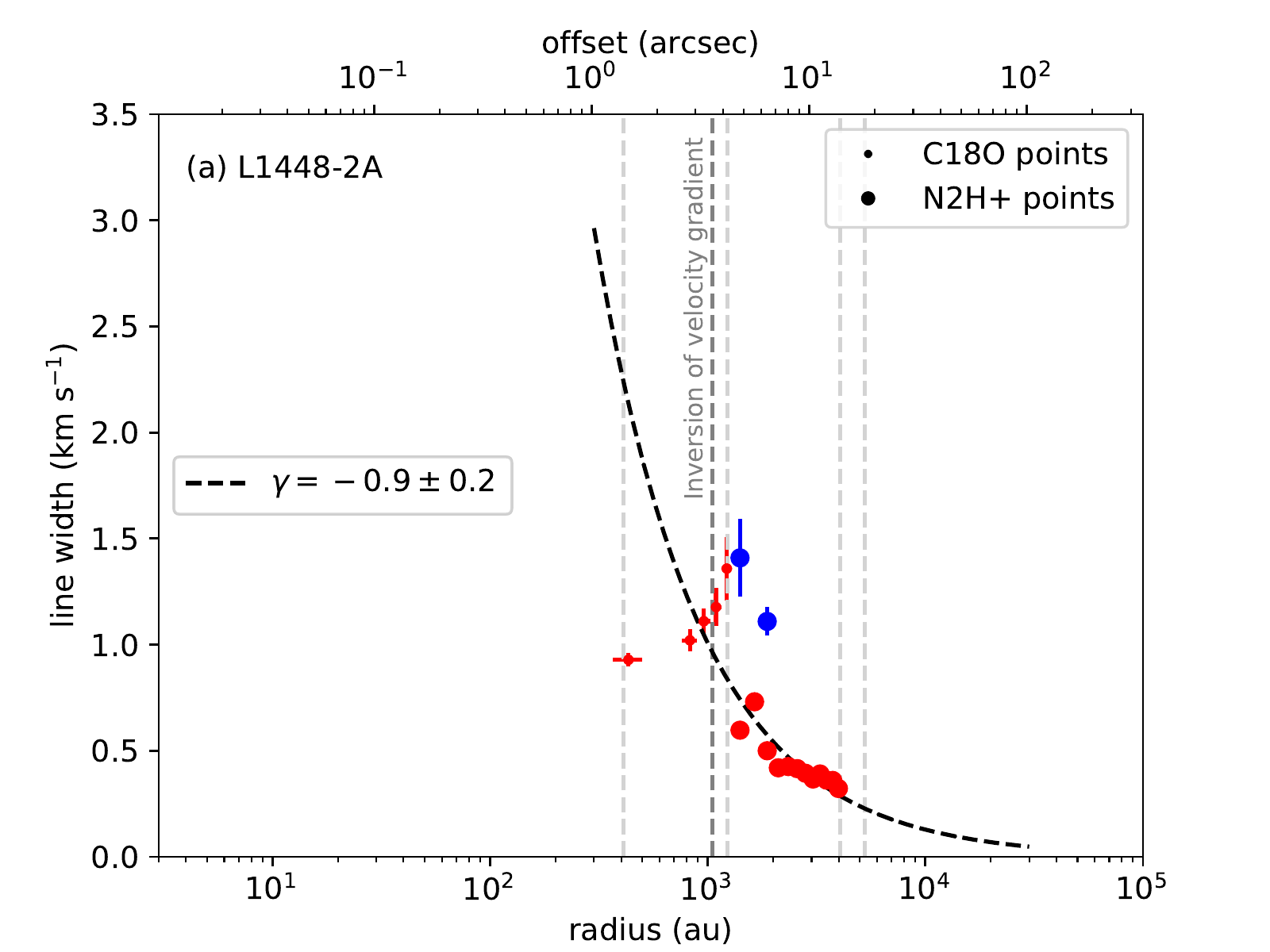}
\includegraphics[width=9cm]{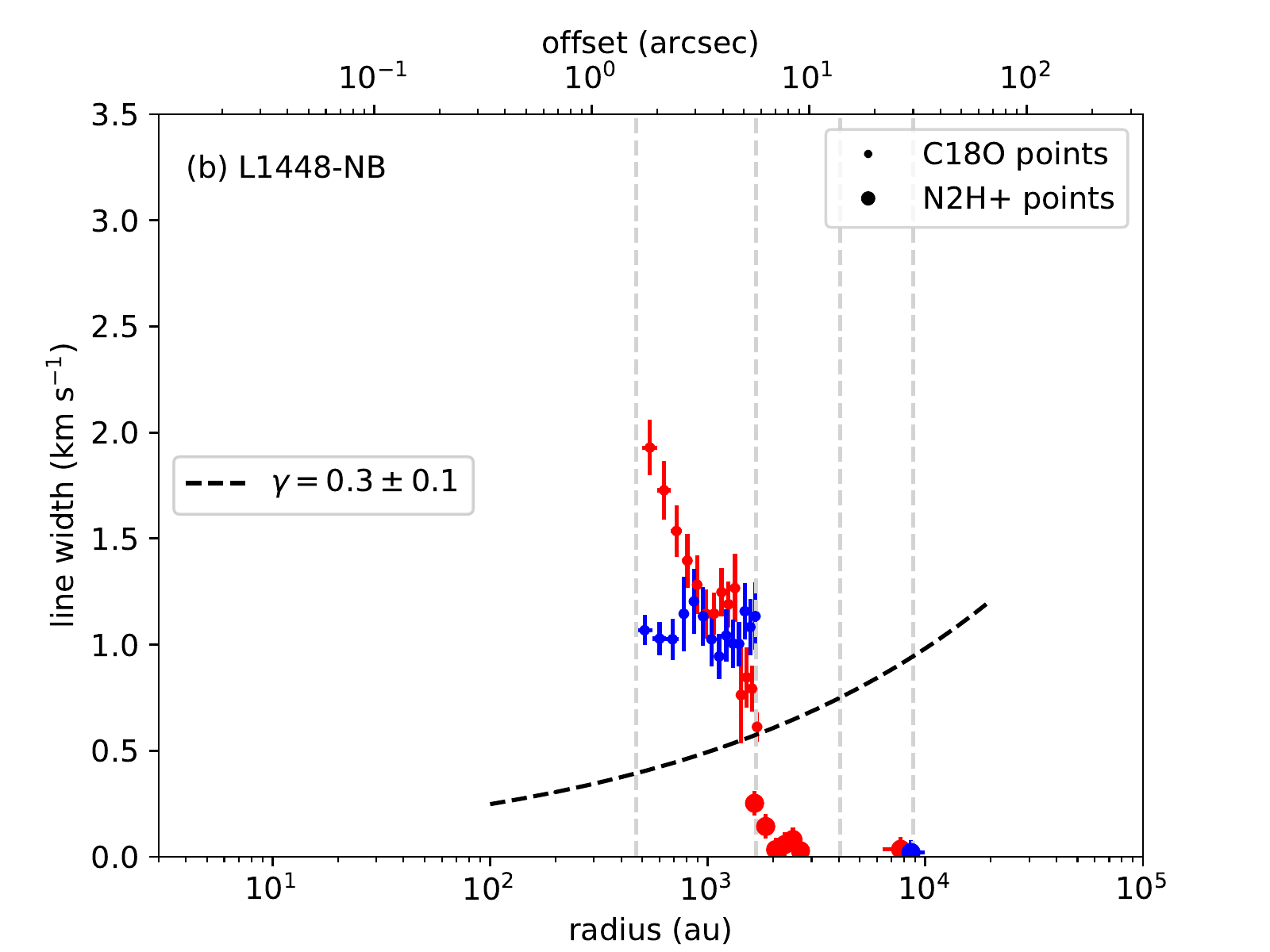}
\includegraphics[width=9cm]{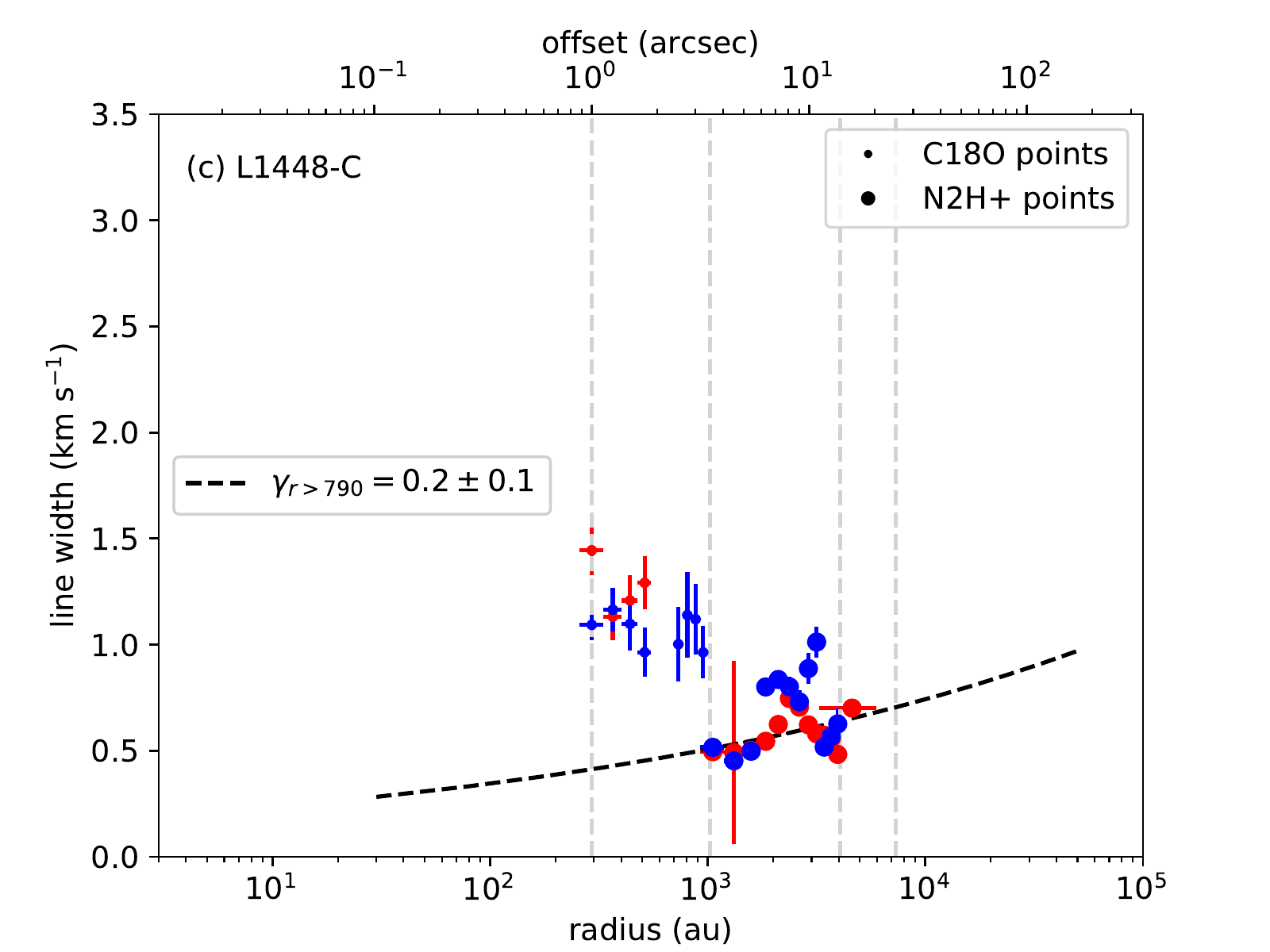}
\includegraphics[width=9cm]{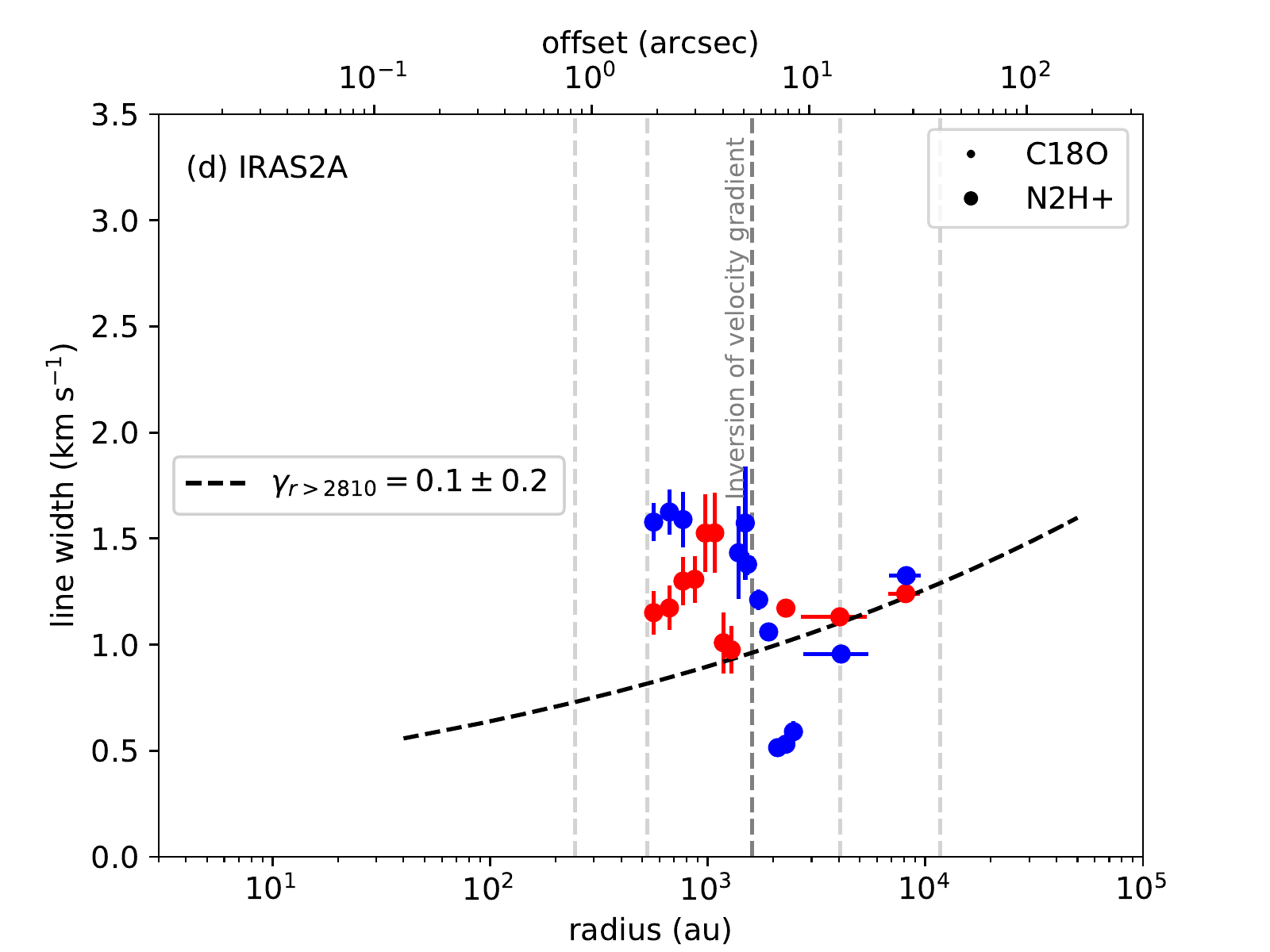}
\includegraphics[width=9cm]{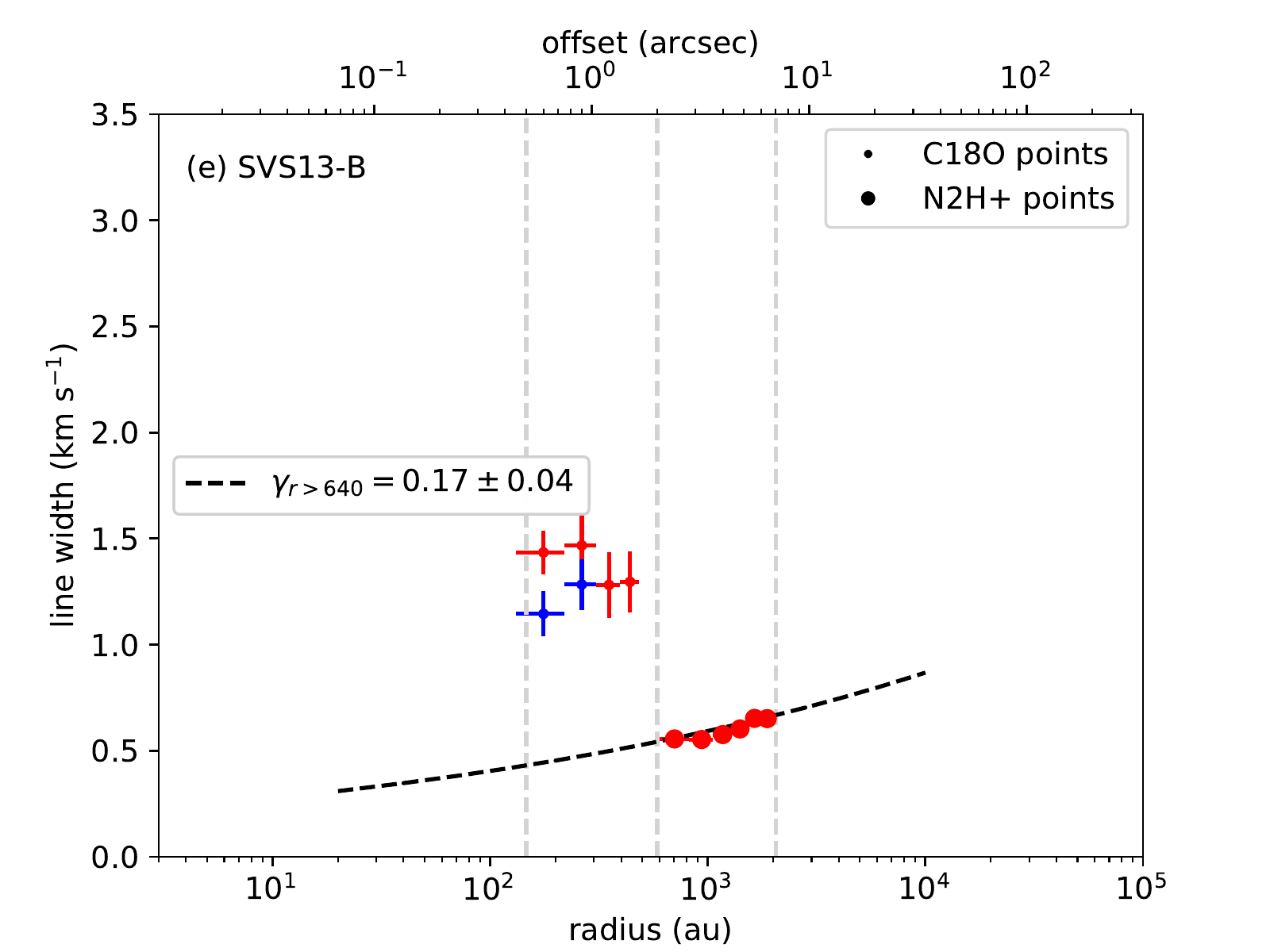}
\includegraphics[width=9cm]{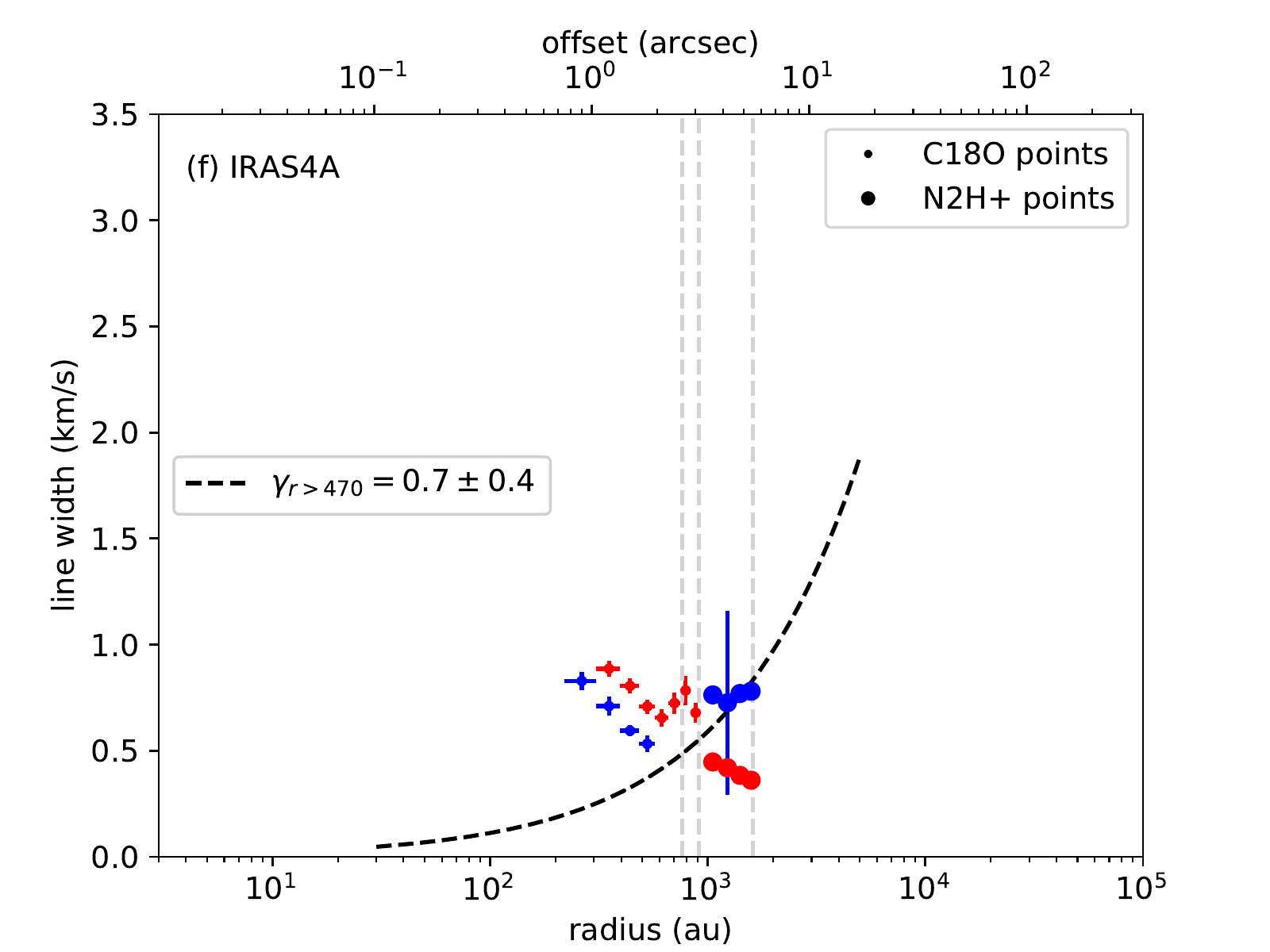}
\caption{Linewidth along the equatorial axis of the CALYPSO protostellar envelopes. Blue and red dots show the linewidths at positions that have blue- and red-shifted velocities, respectively. Dots and large dots show the C$^{18}$O and N$_{2}$H$^{+}$ data, respectively. The dashed curve shows the best fit with a power-law model leaving the index $\gamma$ as a free parameter ($D \mathrm{v} \propto r^{\gamma}$) in the outer envelope (see Appendix \ref{sec:details-Dv-distribution}). The radius of the outer envelope is given by the break radius of the $j(r)$ or $j_\mathrm{app}(r)$ profiles (see Tables \ref{table:chi2-fit-profil-moment-ang} and \ref{table:chi2-fit-profil-japp}) or the radius where we observe a reversal of the velocity gradients with respect to the inner envelope. The vertical dashed lines show the transition radii between the different datasets (PdBI, combined, and 30m) and the two tracers as given in Table \ref{table:radius-lines}.
}
\label{fig:line-width-diagrams}
\end{figure*}

\begin{figure*}[!ht]
\addtocounter{figure}{-1}
\centering
\includegraphics[width=9cm]{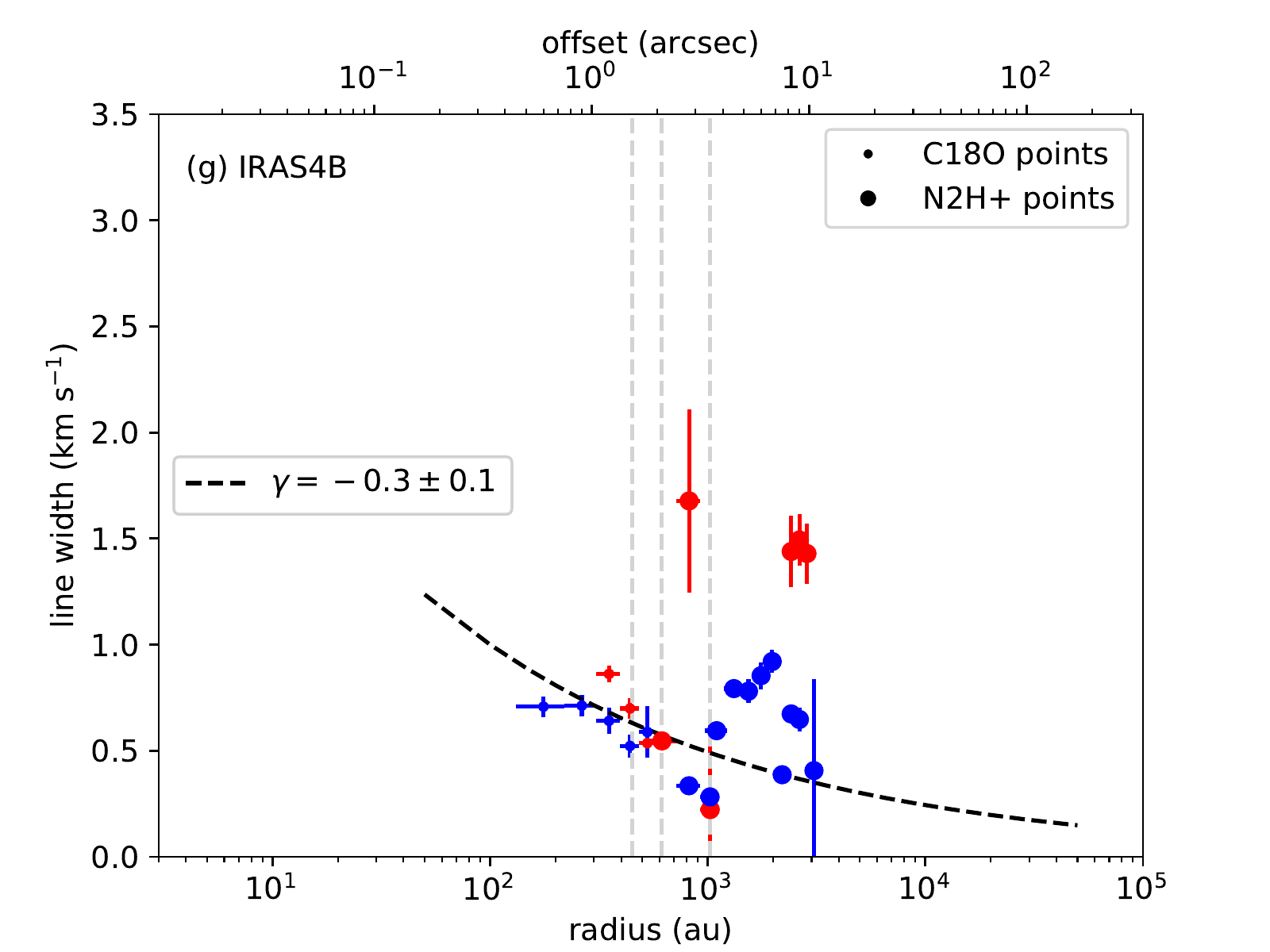}
\includegraphics[width=9cm]{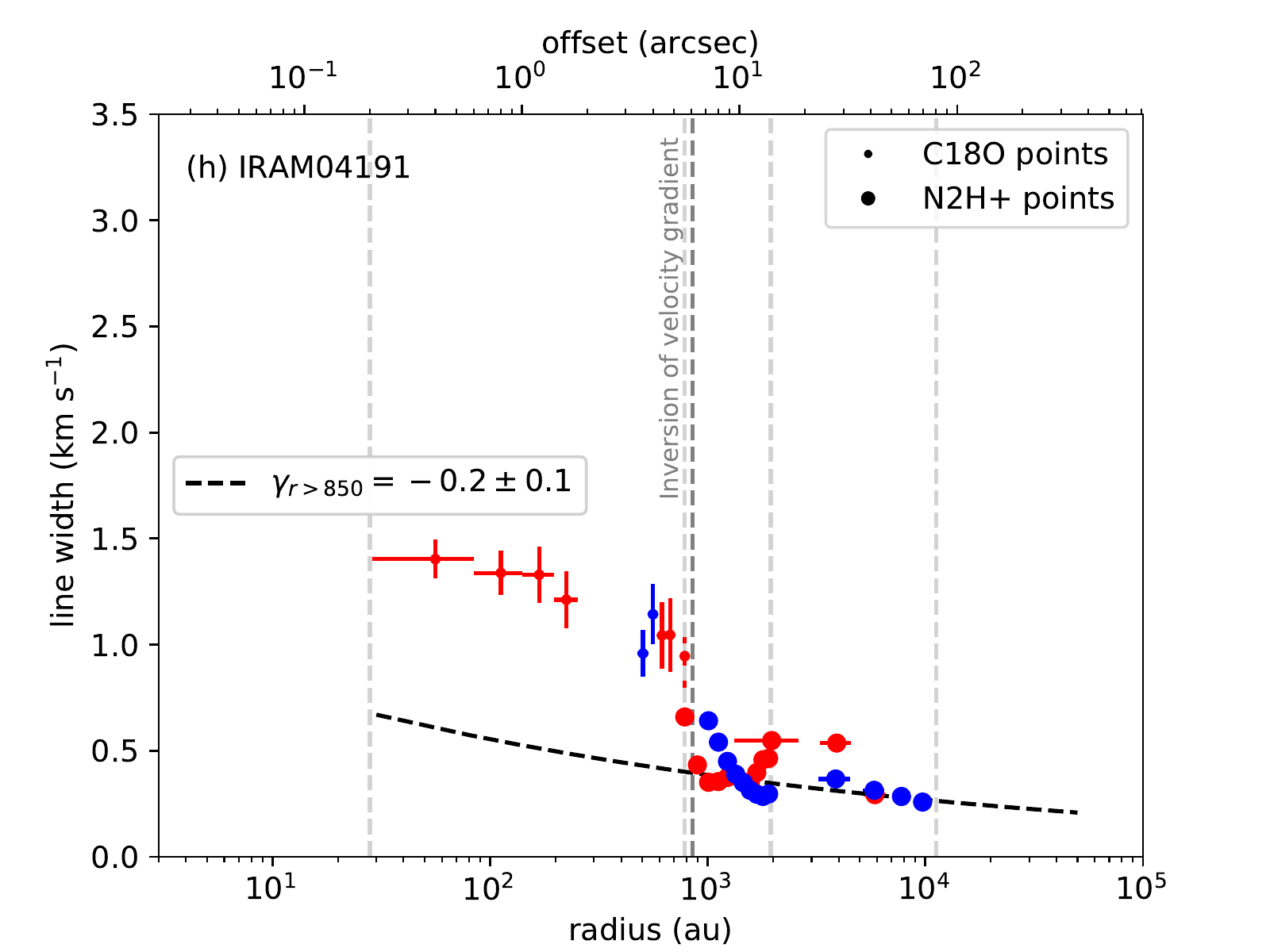}
\includegraphics[width=9cm]{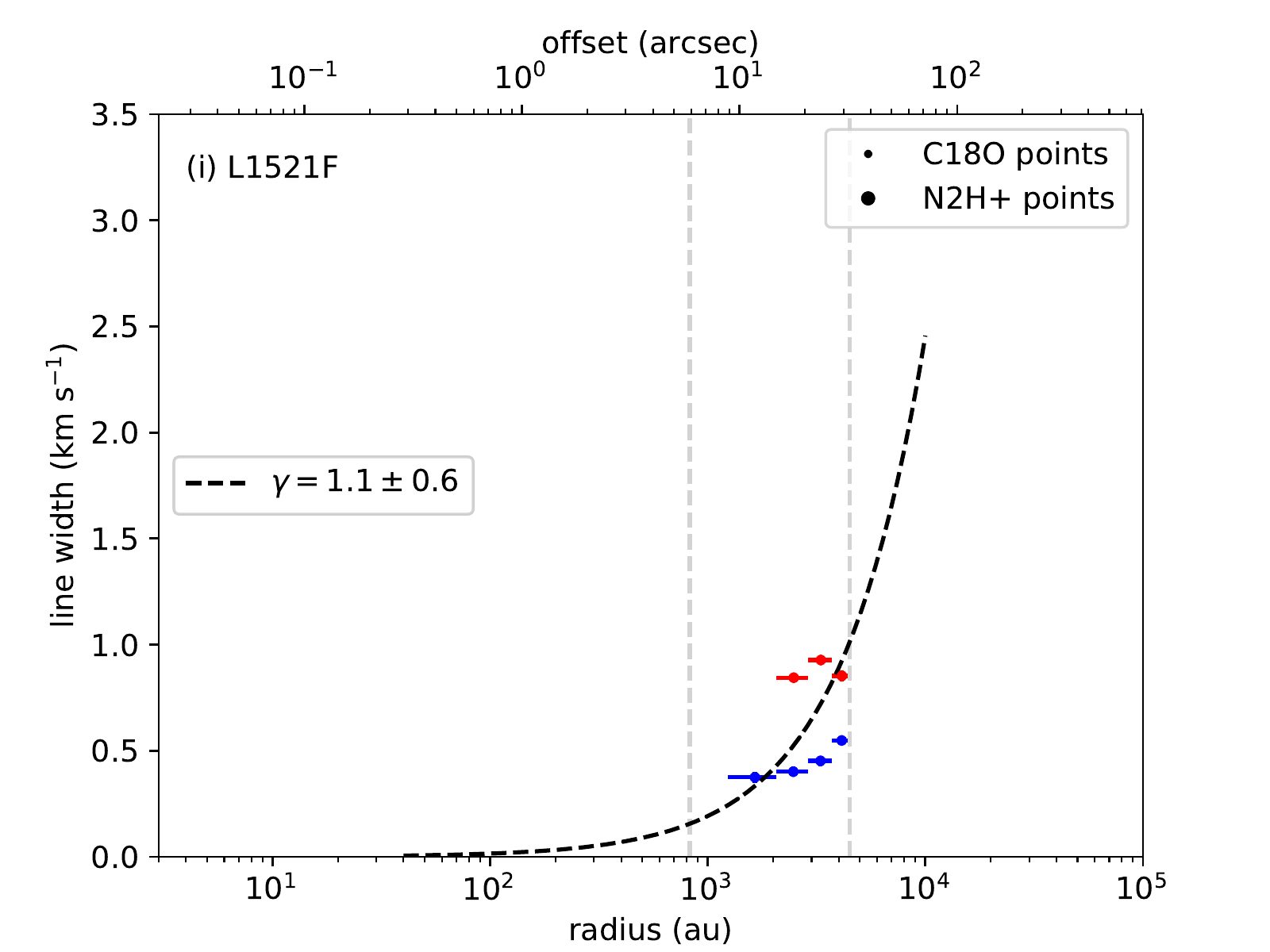}
\includegraphics[width=9cm]{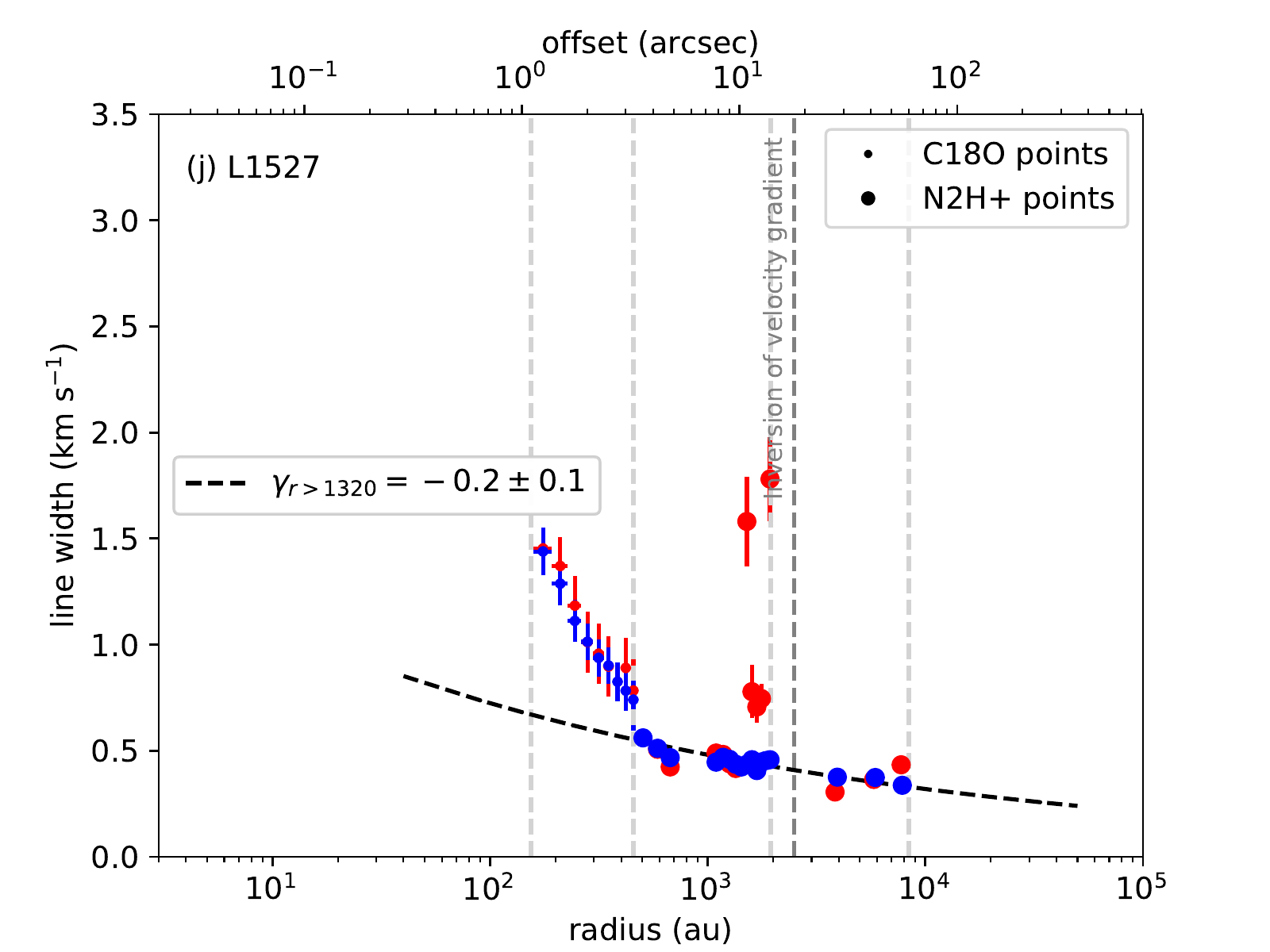}
\includegraphics[width=9cm]{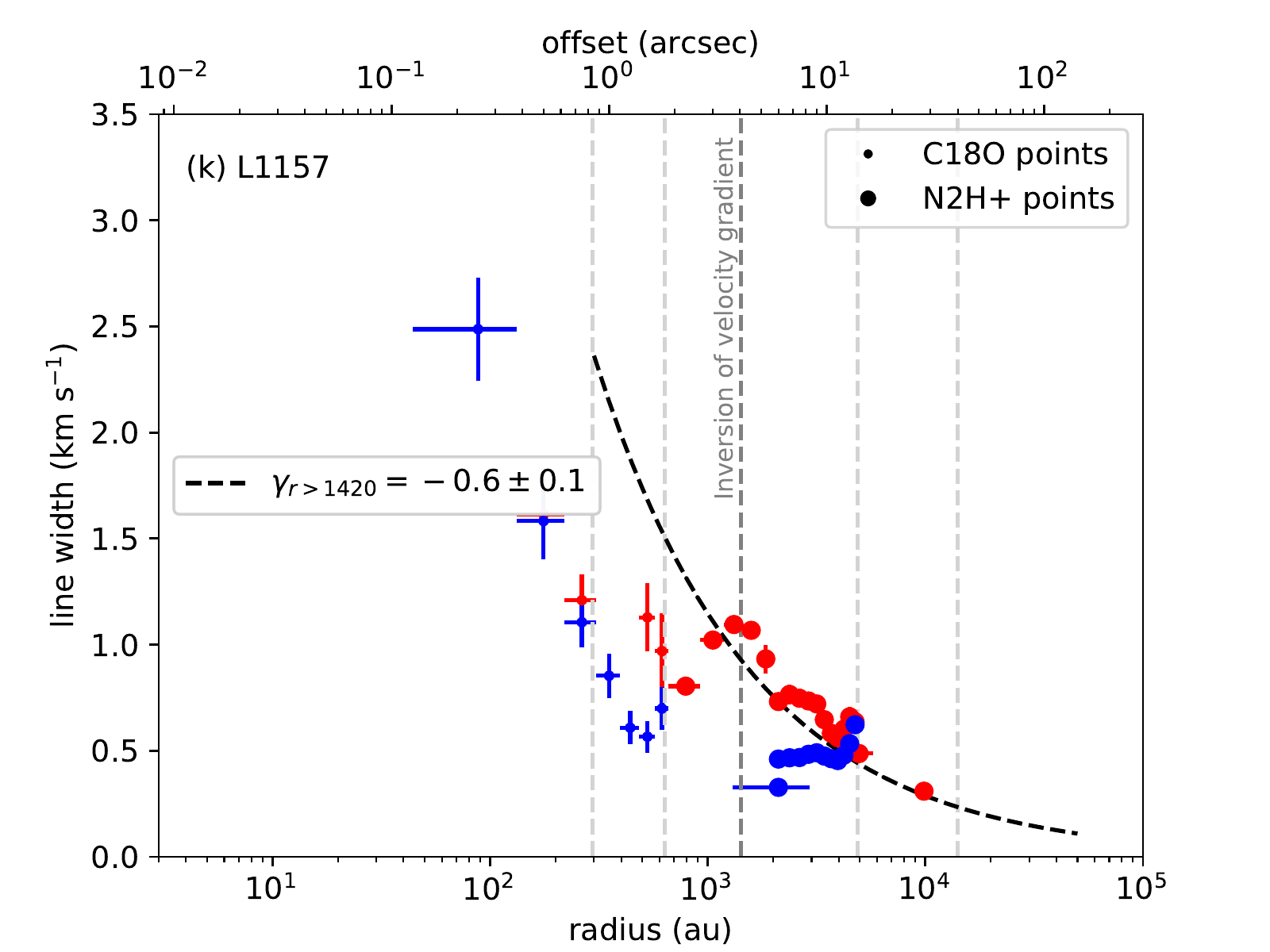}
\includegraphics[width=9cm]{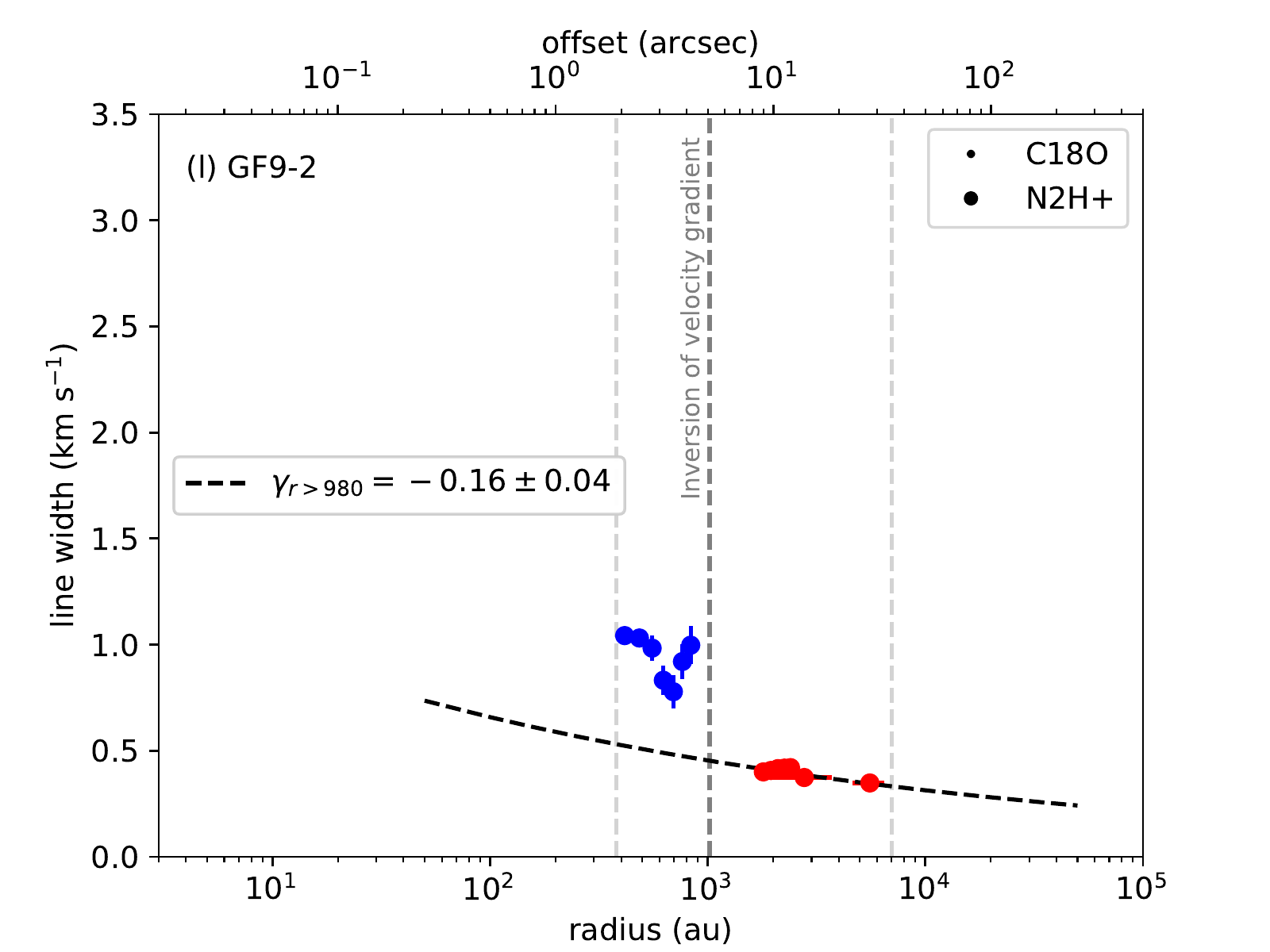}
\caption{Continued.
}
\end{figure*}

%__________________________________________________________________
\section{Conclusions}
In the framework of the CALYPSO survey, we analyzed the kinematics of Class~0 protostellar envelopes. The main results of our study are listed below.
\begin{enumerate}
\item We identify differential rotation motions in 11 sources in a sample of 12 Class~0 protostellar envelopes. The only exception is IRAS4A : the motions reported in the PV$_\mathrm{rot}$ and modeled by a power-law function are consistent with a solid-body rotation, but the velocity gradient is not uniform in the inner envelope at $r<$2000~au as would be expected.

\item This is the first time that the specific angular momentum distribution as a function of envelope radius is determined homogeneously for a large sample of 11 Class~0 protostars. The high angular resolution and the high dynamic range of the CALYPSO observations allow us to identify two distinct regimes: the apparent specific angular momentum decreases as $j_\mathrm{app} \propto r^{1.6 \pm 0.2}$ down to $\sim$1600~au and then tends to become relatively constant around $\sim$6 $\times$ 10$^{-4}$~km~s$^{-1}$~pc down to $\sim$50~au. 

\item The values of specific angular momentum measured in the inner Class~0 envelopes suggest that material directly involved in the star formation process ($<$1600~au) typically encloses the same order of magnitude in specific angular momentum as what is inferred in small T-Tauri disks ($r \sim$10~au). The constant values of $j$ at 50$-$1600~au allow us to determine good estimates of the centrifugal radius in the Class~0 protostars of the CALYPSO sample, which compare well with the disk radii estimated from the dust continuum \citep{Maury18}. This suggests that the specific angular momentum is conserved during the accretion on the stellar embryo, resulting in disk formation.

\item At scales of $r>$1600~au, we conclude that the velocity gradients observed in the outer envelope with respect to small scales are not due to pure rotational motions or counter-rotation motions but related to other mechanisms. Historically, the gradients observed from single-dish mapping at $r>$3000~au have been interpreted as rotation and used to quantify the amplitude of the angular momentum problem for star formation. Thus, if the gradients are incorrectly interpreted as rotation, the angular momentum problem for star formation and the expected disk radii may have been significantly over-estimated. Moreover, we find no robust hints that envelopes are rotating with typical velocities higher than the sound speed at scales of $r>$1600~au. This suggests that the origin of angular momentum in the outer protostellar envelopes could be the gravitationally-driven turbulence due to large-scale collapse motions at the interface between filaments and cores, or the dissipation of the large-scale ISM turbulent cascade propagating with subsonic velocities to 1600~au envelope scales.
\end{enumerate}

\begin{acknowledgements}
We thank the IRAM staff for their support carrying out the CALYPSO observations. 
This work has benefited from the support of the European Research Council under the European Union’s Seventh Framework Programme (Advanced Grant ORISTARS with grant agreement no. 291294 and Starting Grant MagneticYSOs with grant agreement no. 679937). M.G. thanks the Max-Planck Institute for Radio Astronomy for its support toward the end of this work. We would like to thank Cecilia Cecarelli for comments and suggestions on the estimation of column density, and Nagayoshi Ohashi and Jaime E. Pineda for valuable discussions on the interpretation. We also thank the referee for the useful comments which helps to improve this paper.
\end{acknowledgements}

%\newpage

\bibliographystyle{aa}
\bibliography{bibliographie}

\begin{thebibliography}{189}
\expandafter\ifx\csname natexlab\endcsname\relax\def\natexlab#1{#1}\fi

\bibitem[{{Anderl} {et~al.}(2016){Anderl}, {Maret}, {Cabrit}, {Belloche},
  {Maury}, {Andr{\'e}}, {Codella}, {Bacmann}, {Bontemps}, {Podio}, {Gueth}, \&
  {Bergin}}]{Anderl16}
{Anderl}, S., {Maret}, S., {Cabrit}, S., {et~al.} 2016, \aap, 591, A3

\bibitem[{{Andr{\'e}} {et~al.}(2014){Andr{\'e}}, {Di Francesco},
  {Ward-Thompson}, {Inutsuka}, {Pudritz}, \& {Pineda}}]{Andre14}
{Andr{\'e}}, P., {Di Francesco}, J., {Ward-Thompson}, D., {et~al.} 2014, in
  Protostars and Planets VI, ed. H.~{Beuther}, R.~S. {Klessen}, C.~P.
  {Dullemond}, \& T.~{Henning}, 27

\bibitem[{{Andr{\'e}} {et~al.}(2010){Andr{\'e}}, {Men'shchikov}, {Bontemps},
  {K{\"o}nyves}, {Motte}, {Schneider}, {Didelon}, {Minier}, {Saraceno}, {Ward-
  Thompson}, {di Francesco}, {White}, {Molinari}, {Testi}, {Abergel},
  {Griffin}, {Henning}, {Royer}, {Mer{\'\i}n}, {Vavrek}, {Attard},
  {Arzoumanian}, {Wilson}, {Ade}, {Aussel}, {Baluteau}, {Benedettini},
  {Bernard}, {Blommaert}, {Cambr{\'e}sy}, {Cox}, {di Giorgio}, {Hargrave},
  {Hennemann}, {Huang}, {Kirk}, {Krause}, {Launhardt}, {Leeks}, {Le Pennec},
  {Li}, {Martin}, {Maury}, {Olofsson}, {Omont}, {Peretto}, {Pezzuto}, {Prusti},
  {Roussel}, {Russeil}, {Sauvage}, {Sibthorpe}, {Sicilia-Aguilar}, {Spinoglio},
  {Waelkens}, {Woodcraft}, \& {Zavagno}}]{Andre10}
{Andr{\'e}}, P., {Men'shchikov}, A., {Bontemps}, S., {et~al.} 2010, \aap, 518,
  L102

\bibitem[{{Andr{\'e}} {et~al.}(1999){Andr{\'e}}, {Motte}, \&
  {Bacmann}}]{Andre99}
{Andr{\'e}}, P., {Motte}, F., \& {Bacmann}, A. 1999, \apj, 513, L57

\bibitem[{{Andr\'e} {et~al.}(1993){Andr\'e}, {Ward-Thompson}, \&
  {Barsony}}]{Andre93}
{Andr\'e}, P., {Ward-Thompson}, D., \& {Barsony}, M. 1993, \apj, 406, 122

\bibitem[{{Andr\'e} {et~al.}(2000){Andr\'e}, {Ward-Thompson}, \&
  {Barsony}}]{Andre00}
{Andr\'e}, P., {Ward-Thompson}, D., \& {Barsony}, M. 2000, Protostars and
  Planets IV, 59

\bibitem[{{Andrews} {et~al.}(2009){Andrews}, {Wilner}, {Hughes}, {Qi}, \&
  {Dullemond}}]{Andrews09}
{Andrews}, S.~M., {Wilner}, D.~J., {Hughes}, A.~M., {Qi}, C., \& {Dullemond},
  C.~P. 2009, \apj, 700, 1502

\bibitem[{{Anglada} {et~al.}(1989){Anglada}, {Rodriguez}, {Torrelles},
  {Estalella}, {Ho}, {Canto}, {Lopez}, \& {Verdes-Montenegro}}]{Anglada89}
{Anglada}, G., {Rodriguez}, L.~F., {Torrelles}, J.~M., {et~al.} 1989, \apj,
  341, 208

\bibitem[{{Arzoumanian} {et~al.}(2018){Arzoumanian}, {Shimajiri}, {Inutsuka},
  {Inoue}, \& {Tachihara}}]{Arzoumanian18}
{Arzoumanian}, D., {Shimajiri}, Y., {Inutsuka}, S.-i., {Inoue}, T., \&
  {Tachihara}, K. 2018, Publications of the Astronomical Society of Japan, 70,
  96

\bibitem[{{Aso} {et~al.}(2017){Aso}, {Ohashi}, {Aikawa}, {Machida}, {Saigo},
  {Saito}, {Takakuwa}, {Tomida}, {Tomisaka}, \& {Yen}}]{Aso17}
{Aso}, Y., {Ohashi}, N., {Aikawa}, Y., {et~al.} 2017, \apj, 849, 56

\bibitem[{{Bachiller} {et~al.}(1991){Bachiller}, {Andre}, \&
  {Cabrit}}]{Bachiller91}
{Bachiller}, R., {Andre}, P., \& {Cabrit}, S. 1991, \aap, 241, L43

\bibitem[{{Bachiller} {et~al.}(1990){Bachiller}, {Cernicharo},
  {Martin-Pintado}, {Tafalla}, \& {Lazareff}}]{Bachiller90}
{Bachiller}, R., {Cernicharo}, J., {Martin-Pintado}, J., {Tafalla}, M., \&
  {Lazareff}, B. 1990, \aap, 231, 174

\bibitem[{{Bachiller} {et~al.}(2000){Bachiller}, {Gueth}, {Guilloteau},
  {Tafalla}, \& {Dutrey}}]{Bachiller00}
{Bachiller}, R., {Gueth}, F., {Guilloteau}, S., {Tafalla}, M., \& {Dutrey}, A.
  2000, \aap, 362, L33

\bibitem[{{Bachiller} {et~al.}(1995){Bachiller}, {Guilloteau}, {Dutrey},
  {Planesas}, \& {Martin-Pintado}}]{Bachiller95}
{Bachiller}, R., {Guilloteau}, S., {Dutrey}, A., {Planesas}, P., \&
  {Martin-Pintado}, J. 1995, \aap, 299, 857

\bibitem[{{Bachiller} {et~al.}(1998){Bachiller}, {Guilloteau}, {Gueth},
  {Tafalla}, {Dutrey}, {Codella}, \& {Castets}}]{Bachiller98}
{Bachiller}, R., {Guilloteau}, S., {Gueth}, F., {et~al.} 1998, \aap, 339, L49

\bibitem[{{Bachiller} \& {P{\'e}rez Guti{\'e}rrez}(1997)}]{Bachiller97}
{Bachiller}, R. \& {P{\'e}rez Guti{\'e}rrez}, M. 1997, \apj, 487, L93

\bibitem[{{Bachiller} {et~al.}(2001){Bachiller}, {P{\'e}rez Guti{\'e}rrez},
  {Kumar}, \& {Tafalla}}]{Bachiller01}
{Bachiller}, R., {P{\'e}rez Guti{\'e}rrez}, M., {Kumar}, M.~S.~N., \&
  {Tafalla}, M. 2001, \aap, 372, 899

\bibitem[{{Barsony} {et~al.}(1998){Barsony}, {Ward-Thompson}, {Andr{\'e}}, \&
  {O'Linger}}]{Barsony98}
{Barsony}, M., {Ward-Thompson}, D., {Andr{\'e}}, P., \& {O'Linger}, J. 1998,
  \apj, 509, 733

\bibitem[{{Basu}(1998)}]{Basu98}
{Basu}, S. 1998, \apj, 509, 229

\bibitem[{{Belloche}(2013)}]{Belloche13}
{Belloche}, A. 2013, in EAS Publications Series, Vol.~62, EAS Publications
  Series, ed. P.~{Hennebelle} \& C.~{Charbonnel}, 25--66

\bibitem[{{Belloche} \& {Andr{\'e}}(2004)}]{Belloche04}
{Belloche}, A. \& {Andr{\'e}}, P. 2004, \aap, 419, L35

\bibitem[{{Belloche} {et~al.}(2002){Belloche}, {Andr{\'e}}, {Despois}, \&
  {Blinder}}]{Belloche02}
{Belloche}, A., {Andr{\'e}}, P., {Despois}, D., \& {Blinder}, S. 2002, \aap,
  393, 927

\bibitem[{{Belloche} {et~al.}(2006){Belloche}, {Hennebelle}, \&
  {Andr{\'e}}}]{Belloche06}
{Belloche}, A., {Hennebelle}, P., \& {Andr{\'e}}, P. 2006, \aap, 453, 145

\bibitem[{{Belloche} {et~al.}(2020){Belloche}, {Maury}, {Maret}, {Anderl},
  {Bacmann}, {Andr{\'e}}, {Bontemps}, {Cabrit}, {Codella}, {Gaudel}, {Gueth},
  {Lef{\`e}vre}, {Lefloch}, {Podio}, \& {Testi}}]{Belloche20}
{Belloche}, A., {Maury}, A.~J., {Maret}, S., {et~al.} 2020, \aap, 635, A198

\bibitem[{{Beltr{\'a}n} {et~al.}(2004){Beltr{\'a}n}, {Gueth}, {Guilloteau}, \&
  {Dutrey}}]{Beltran04}
{Beltr{\'a}n}, M.~T., {Gueth}, F., {Guilloteau}, S., \& {Dutrey}, A. 2004,
  \aap, 416, 631

\bibitem[{{Benson} \& {Myers}(1989)}]{Benson89}
{Benson}, P.~J. \& {Myers}, P.~C. 1989, The Astrophysical Journal Supplement
  Series, 71, 89

\bibitem[{{Bergin} {et~al.}(2002){Bergin}, {Alves}, {Huard}, \&
  {Lada}}]{Bergin02}
{Bergin}, E.~A., {Alves}, J., {Huard}, T., \& {Lada}, C.~J. 2002, \apjl, 570,
  L101

\bibitem[{{Blandford} \& {Payne}(1982)}]{Blandford82}
{Blandford}, R.~D. \& {Payne}, D.~G. 1982, \mnras, 199, 883

\bibitem[{{Bodenheimer}(1995)}]{Bodenheimer95}
{Bodenheimer}, P. 1995, \araa, 33, 199

\bibitem[{{Bourke} {et~al.}(2006){Bourke}, {Myers}, {Evans}, {Dunham},
  {Kauffmann}, {Shirley}, {Crapsi}, {Young}, {Huard}, {Brooke}, {Chapman},
  {Cieza}, {Lee}, {Teuben}, \& {Wahhaj}}]{Bourke06}
{Bourke}, T.~L., {Myers}, P.~C., {Evans}, Neal~J., I., {et~al.} 2006, \apj,
  649, L37

\bibitem[{{Bouvier} {et~al.}(1993){Bouvier}, {Cabrit}, {Fernandez}, {Martin},
  \& {Matthews}}]{Bouvier93}
{Bouvier}, J., {Cabrit}, S., {Fernandez}, M., {Martin}, E.~L., \& {Matthews},
  J.~M. 1993, \aap, 272, 176

\bibitem[{{Brinch} {et~al.}(2009){Brinch}, {J{\o}rgensen}, \&
  {Hogerheijde}}]{Brinch09}
{Brinch}, C., {J{\o}rgensen}, J.~K., \& {Hogerheijde}, M.~R. 2009, \aap, 502,
  199

\bibitem[{{Burkert} \& {Bodenheimer}(2000)}]{Burkert00}
{Burkert}, A. \& {Bodenheimer}, P. 2000, \apj, 543, 822

\bibitem[{{Carter} {et~al.}(2012){Carter}, {Lazareff}, {Maier}, {Chenu},
  {Fontana}, {Bortolotti}, {Boucher}, {Navarrini}, {Blanchet}, {Greve}, {John},
  {Kramer}, {Morel}, {Navarro}, {Pe{\~n}alver}, {Schuster}, \&
  {Thum}}]{Carter12}
{Carter}, M., {Lazareff}, B., {Maier}, D., {et~al.} 2012, \aap, 538, A89

\bibitem[{{Caselli} {et~al.}(2002){Caselli}, {Benson}, {Myers}, \&
  {Tafalla}}]{Caselli02}
{Caselli}, P., {Benson}, P.~J., {Myers}, P.~C., \& {Tafalla}, M. 2002, \apj,
  572, 238

\bibitem[{{Caselli} {et~al.}(1995){Caselli}, {Myers}, \&
  {Thaddeus}}]{Caselli95}
{Caselli}, P., {Myers}, P.~C., \& {Thaddeus}, P. 1995, \apj, 455, L77

\bibitem[{{Cassen} \& {Moosman}(1981)}]{Cassen81}
{Cassen}, P. \& {Moosman}, A. 1981, \icarus, 48, 353

\bibitem[{{Ceccarelli} {et~al.}(1996){Ceccarelli}, {Hollenbach}, \&
  {Tielens}}]{Ceccarelli96}
{Ceccarelli}, C., {Hollenbach}, D.~J., \& {Tielens}, A. G.~G.~M. 1996, \apj,
  471, 400

\bibitem[{{Chandler} \& {Richer}(2000)}]{Chandler00}
{Chandler}, C.~J. \& {Richer}, J.~S. 2000, \apj, 530, 851

\bibitem[{{Chen} {et~al.}(2007){Chen}, {Launhardt}, \& {Henning}}]{Chen07}
{Chen}, X., {Launhardt}, R., \& {Henning}, T. 2007, \apj, 669, 1058

\bibitem[{{Chen} {et~al.}(2009){Chen}, {Launhardt}, \& {Henning}}]{Chen09}
{Chen}, X., {Launhardt}, R., \& {Henning}, T. 2009, \apj, 691, 1729

\bibitem[{{Chiang} {et~al.}(2010){Chiang}, {Looney}, {Tobin}, \&
  {Hartmann}}]{Chiang10}
{Chiang}, H.-F., {Looney}, L.~W., {Tobin}, J.~J., \& {Hartmann}, L. 2010, \apj,
  709, 470

\bibitem[{{Ching} {et~al.}(2016){Ching}, {Lai}, {Zhang}, {Yang}, {Girart}, \&
  {Rao}}]{Ching16}
{Ching}, T.-C., {Lai}, S.-P., {Zhang}, Q., {et~al.} 2016, \apj, 819, 159

\bibitem[{{Chini} {et~al.}(1997){Chini}, {Reipurth}, {Sievers},
  {Ward-Thompson}, {Haslam}, {Kreysa}, \& {Lemke}}]{Chini97}
{Chini}, R., {Reipurth}, B., {Sievers}, A., {et~al.} 1997, \aap, 325, 542

\bibitem[{{Choi}(2001)}]{Choi01}
{Choi}, M. 2001, \apj, 553, 219

\bibitem[{{Ciardi} {et~al.}(2000){Ciardi}, {Woodward}, {Clemens}, {Harker}, \&
  {Rudy}}]{Ciardi00}
{Ciardi}, D.~R., {Woodward}, C.~E., {Clemens}, D.~P., {Harker}, D.~E., \&
  {Rudy}, R.~J. 2000, \aj, 120, 393

\bibitem[{{Cieza} {et~al.}(2019){Cieza}, {Ru{\'\i}z-Rodr{\'\i}guez}, {Hales},
  {Casassus}, {P{\'e}rez}, {Gonzalez-Ruilova}, {C{\'a}novas}, {Williams},
  {Zurlo}, {Ansdell}, {Avenhaus}, {Bayo}, {Bertrang}, {Christiaens}, {Dent},
  {Ferrero}, {Gamen}, {Olofsson}, {Orcajo}, {Pe{\~n}a Ram{\'\i}rez},
  {Principe}, {Schreiber}, \& {van der Plas}}]{Cieza19}
{Cieza}, L.~A., {Ru{\'\i}z-Rodr{\'\i}guez}, D., {Hales}, A., {et~al.} 2019,
  \mnras, 482, 698

\bibitem[{{Codella} {et~al.}(2014{\natexlab{a}}){Codella}, {Cabrit}, {Gueth},
  {Podio}, {Leurini}, {Bachiller}, {Gusdorf}, {Lefloch}, {Nisini}, {Tafalla},
  \& {Yvart}}]{Codella14}
{Codella}, C., {Cabrit}, S., {Gueth}, F., {et~al.} 2014{\natexlab{a}}, \aap,
  568, L5

\bibitem[{{Codella} {et~al.}(2004){Codella}, {Lorenzani}, {Gallego},
  {Cesaroni}, \& {Moscadelli}}]{Codella04}
{Codella}, C., {Lorenzani}, A., {Gallego}, A.~T., {Cesaroni}, R., \&
  {Moscadelli}, L. 2004, \aap, 417, 615

\bibitem[{{Codella} {et~al.}(2014{\natexlab{b}}){Codella}, {Maury}, {Gueth},
  {Maret}, {Belloche}, {Cabrit}, \& {Andr{\'e}}}]{Codella14-bis}
{Codella}, C., {Maury}, A.~J., {Gueth}, F., {et~al.} 2014{\natexlab{b}}, \aap,
  563, L3

\bibitem[{{Codella} {et~al.}(1997){Codella}, {Welser}, {Henkel}, {Benson}, \&
  {Myers}}]{Codella97}
{Codella}, C., {Welser}, R., {Henkel}, C., {Benson}, P.~J., \& {Myers}, P.~C.
  1997, \aap, 324, 203

\bibitem[{{Crapsi} {et~al.}(2004){Crapsi}, {Caselli}, {Walmsley}, {Tafalla},
  {Lee}, {Bourke}, \& {Myers}}]{Crapsi04}
{Crapsi}, A., {Caselli}, P., {Walmsley}, C.~M., {et~al.} 2004, \aap, 420, 957

\bibitem[{{Curiel} {et~al.}(1990){Curiel}, {Raymond}, {Rodriguez}, {Canto}, \&
  {Moran}}]{Curiel90}
{Curiel}, S., {Raymond}, J.~C., {Rodriguez}, L.~F., {Canto}, J., \& {Moran},
  J.~M. 1990, \apj, 365, L85

\bibitem[{{Curiel} {et~al.}(1999){Curiel}, {Torrelles}, {Rodr{\'\i}guez},
  {G{\'o}mez}, \& {Anglada}}]{Curiel99}
{Curiel}, S., {Torrelles}, J.~M., {Rodr{\'\i}guez}, L.~F., {G{\'o}mez}, J.~F.,
  \& {Anglada}, G. 1999, \apj, 527, 310

\bibitem[{{Daniel} {et~al.}(2005){Daniel}, {Dubernet}, {Meuwly}, {Cernicharo},
  \& {Pagani}}]{Daniel05}
{Daniel}, F., {Dubernet}, M.~L., {Meuwly}, M., {Cernicharo}, J., \& {Pagani},
  L. 2005, \mnras, 363, 1083

\bibitem[{{De Simone} {et~al.}(2017){De Simone}, {Codella}, {Testi},
  {Belloche}, {Maury}, {Anderl}, {Andr{\'e}}, {Maret}, \& {Podio}}]{Simone17}
{De Simone}, M., {Codella}, C., {Testi}, L., {et~al.} 2017, \aap, 599, A121

\bibitem[{{Desmurs} {et~al.}(2009){Desmurs}, {Codella},
  {Santiago-Garc{\'{\i}}a}, {Tafalla}, \& {Bachiller}}]{Desmurs09}
{Desmurs}, J.-F., {Codella}, C., {Santiago-Garc{\'{\i}}a}, J., {Tafalla}, M.,
  \& {Bachiller}, R. 2009, \aap, 498, 753

\bibitem[{{Dhabal} {et~al.}(2018){Dhabal}, {Mundy}, {Rizzo}, {Storm}, \&
  {Teuben}}]{Dhabal18}
{Dhabal}, A., {Mundy}, L.~G., {Rizzo}, M.~J., {Storm}, S., \& {Teuben}, P.
  2018, \apj, 853, 169

\bibitem[{{di Francesco} {et~al.}(2007){di Francesco}, {Evans}, {Caselli},
  {Myers}, {Shirley}, {Aikawa}, \& {Tafalla}}]{DiFrancesco07}
{di Francesco}, J., {Evans}, N.~J., I., {Caselli}, P., {et~al.} 2007, in
  Protostars and Planets V, ed. B.~{Reipurth}, D.~{Jewitt}, \& K.~{Keil}, 17

\bibitem[{{Dunham} {et~al.}(2006){Dunham}, {Evans}, {Bourke}, {Dullemond},
  {Young}, {Brooke}, {Chapman}, {Myers}, {Porras}, {Spiesman}, {Teuben}, \&
  {Wahhaj}}]{Dunham06}
{Dunham}, M.~M., {Evans}, Neal~J., I., {Bourke}, T.~L., {et~al.} 2006, \apj,
  651, 945

\bibitem[{{Endres} {et~al.}(2016){Endres}, {Schlemmer}, {Schilke}, {Stutzki},
  \& {M{\"u}ller}}]{Endres16}
{Endres}, C.~P., {Schlemmer}, S., {Schilke}, P., {Stutzki}, J., \&
  {M{\"u}ller}, H. S.~P. 2016, Journal of Molecular Spectroscopy, 327, 95

\bibitem[{{Enoch} {et~al.}(2009){Enoch}, {Evans}, {Sargent}, \&
  {Glenn}}]{Enoch09}
{Enoch}, M.~L., {Evans}, Neal~J., I., {Sargent}, A.~I., \& {Glenn}, J. 2009,
  \apj, 692, 973

\bibitem[{{Evans} {et~al.}(2009){Evans}, {Dunham}, {J{\o}rgensen}, {Enoch},
  {Mer{\'\i}n}, {van Dishoeck}, {Alcal{\'a}}, {Myers}, {Stapelfeldt}, {Huard},
  {Allen}, {Harvey}, {van Kempen}, {Blake}, {Koerner}, {Mundy}, {Padgett}, \&
  {Sargent}}]{Evans09}
{Evans}, Neal~J., I., {Dunham}, M.~M., {J{\o}rgensen}, J.~K., {et~al.} 2009,
  The Astrophysical Journal Supplement Series, 181, 321

\bibitem[{{Flower}(2001)}]{Flower01}
{Flower}, D.~R. 2001, \mnras, 328, 147

\bibitem[{{Furuya} {et~al.}(2006){Furuya}, {Kitamura}, \&
  {Shinnaga}}]{Furuya06}
{Furuya}, R.~S., {Kitamura}, Y., \& {Shinnaga}, H. 2006, \apj, 653, 1369

\bibitem[{{Girart} \& {Acord}(2001)}]{Girart01}
{Girart}, J.~M. \& {Acord}, J.~M.~P. 2001, \apjl, 552, L63

\bibitem[{{Girart} {et~al.}(2006){Girart}, {Rao}, \& {Marrone}}]{Girart06}
{Girart}, J.~M., {Rao}, R., \& {Marrone}, D.~P. 2006, Science, 313, 812

\bibitem[{{Goldsmith} \& {Arquilla}(1985)}]{Goldsmith85}
{Goldsmith}, P.~F. \& {Arquilla}, R. 1985, in Protostars and Planets II, ed.
  D.~C. {Black} \& M.~S. {Matthews}, 137--149

\bibitem[{{Goldsmith} \& {Langer}(1999)}]{Goldsmith99}
{Goldsmith}, P.~F. \& {Langer}, W.~D. 1999, \apj, 517, 209

\bibitem[{{Goodman} {et~al.}(1998){Goodman}, {Barranco}, {Wilner}, \&
  {Heyer}}]{Goodman98}
{Goodman}, A.~A., {Barranco}, J.~A., {Wilner}, D.~J., \& {Heyer}, M.~H. 1998,
  \apj, 504, 223

\bibitem[{{Goodman} {et~al.}(1993){Goodman}, {Benson}, {Fuller}, \&
  {Myers}}]{Goodman93}
{Goodman}, A.~A., {Benson}, P.~J., {Fuller}, G.~A., \& {Myers}, P.~C. 1993,
  \apj, 406, 528

\bibitem[{{Goodwin} {et~al.}(2004){Goodwin}, {Whitworth}, \&
  {Ward-Thompson}}]{Goodwin04}
{Goodwin}, S.~P., {Whitworth}, A.~P., \& {Ward-Thompson}, D. 2004, \aap, 414,
  633

\bibitem[{{Grossman} {et~al.}(1987){Grossman}, {Masson}, {Sargent}, {Scoville},
  {Scott}, \& {Woody}}]{Grossman87}
{Grossman}, E.~N., {Masson}, C.~R., {Sargent}, A.~I., {et~al.} 1987, \apj, 320,
  356

\bibitem[{{Gueth} {et~al.}(2003){Gueth}, {Bachiller}, \& {Tafalla}}]{Gueth03}
{Gueth}, F., {Bachiller}, R., \& {Tafalla}, M. 2003, \aap, 401, L5

\bibitem[{{Gueth} {et~al.}(1996){Gueth}, {Guilloteau}, \&
  {Bachiller}}]{Gueth96}
{Gueth}, F., {Guilloteau}, S., \& {Bachiller}, R. 1996, \aap, 307, 891

\bibitem[{{Guilloteau} {et~al.}(1992){Guilloteau}, {Bachiller}, {Fuente}, \&
  {Lucas}}]{Guilloteau92}
{Guilloteau}, S., {Bachiller}, R., {Fuente}, A., \& {Lucas}, R. 1992, \aap,
  265, L49

\bibitem[{{Harsono} {et~al.}(2014){Harsono}, {J{\o}rgensen}, {van Dishoeck},
  {Hogerheijde}, {Bruderer}, {Persson}, \& {Mottram}}]{Harsono14}
{Harsono}, D., {J{\o}rgensen}, J.~K., {van Dishoeck}, E.~F., {et~al.} 2014,
  \aap, 562, A77

\bibitem[{{Hartmann} {et~al.}(1998){Hartmann}, {Calvet}, {Gullbring}, \&
  {D'Alessio}}]{Hartmann98}
{Hartmann}, L., {Calvet}, N., {Gullbring}, E., \& {D'Alessio}, P. 1998, \apj,
  495, 385

\bibitem[{{Hirano} {et~al.}(2010){Hirano}, {Ho}, {Liu}, {Shang}, {Lee}, \&
  {Bourke}}]{Hirano10}
{Hirano}, N., {Ho}, P. P.~T., {Liu}, S.-Y., {et~al.} 2010, \apj, 717, 58

\bibitem[{{Hirota} {et~al.}(2011){Hirota}, {Honma}, {Imai}, {Sunada}, {Ueno},
  {Kobayashi}, \& {Kawaguchi}}]{Hirota11}
{Hirota}, T., {Honma}, M., {Imai}, H., {et~al.} 2011, \pasj, 63, 1

\bibitem[{{Hull} {et~al.}(2014){Hull}, {Plambeck}, {Kwon}, {Bower},
  {Carpenter}, {Crutcher}, {Fiege}, {Franzmann}, {Hakobian}, {Heiles}, {Houde},
  {Hughes}, {Lamb}, {Looney}, {Marrone}, {Matthews}, {Pillai}, {Pound},
  {Rahman}, {Sandell}, {Stephens}, {Tobin}, {Vaillancourt}, {Volgenau}, \&
  {Wright}}]{Hull14}
{Hull}, C. L.~H., {Plambeck}, R.~L., {Kwon}, W., {et~al.} 2014, The
  Astrophysical Journal Supplement Series, 213, 13

\bibitem[{{Isella} {et~al.}(2009){Isella}, {Carpenter}, \&
  {Sargent}}]{Isella09}
{Isella}, A., {Carpenter}, J.~M., \& {Sargent}, A.~I. 2009, \apj, 701, 260

\bibitem[{{Jennings} {et~al.}(1987){Jennings}, {Cameron}, {Cudlip}, \&
  {Hirst}}]{Jennings87}
{Jennings}, R.~E., {Cameron}, D.~H.~M., {Cudlip}, W., \& {Hirst}, C.~J. 1987,
  \mnras, 226, 461

\bibitem[{{J{\o}rgensen} {et~al.}(2007){J{\o}rgensen}, {Bourke}, {Myers}, {Di
  Francesco}, {van Dishoeck}, {Lee}, {Ohashi}, {Sch{\"o}ier}, {Takakuwa},
  {Wilner}, \& {Zhang}}]{Jorgensen07}
{J{\o}rgensen}, J.~K., {Bourke}, T.~L., {Myers}, P.~C., {et~al.} 2007, \apj,
  659, 479

\bibitem[{{J{\o}rgensen} {et~al.}(2004){J{\o}rgensen}, {Hogerheijde}, {van
  Dishoeck}, {Blake}, \& {Sch{\"o}ier}}]{Jorgensen04-n2hp}
{J{\o}rgensen}, J.~K., {Hogerheijde}, M.~R., {van Dishoeck}, E.~F., {Blake},
  G.~A., \& {Sch{\"o}ier}, F.~L. 2004, \aap, 413, 993

\bibitem[{{J{\o}rgensen} \& {van Dishoeck}(2010)}]{Jorgensen10}
{J{\o}rgensen}, J.~K. \& {van Dishoeck}, E.~F. 2010, \apj, 710, L72

\bibitem[{{Karska} {et~al.}(2013){Karska}, {Herczeg}, {van Dishoeck},
  {Wampfler}, {Kristensen}, {Goicoechea}, {Visser}, {Nisini}, {San
  Jos{\'e}-Garc{\'\i}a}, {Bruderer}, {{\'S}niady}, {Doty}, {Fedele},
  {Y{\i}ld{\i}z}, {Benz}, {Bergin}, {Caselli}, {Herpin}, {Hogerheijde},
  {Johnstone}, {J{\o}rgensen}, {Liseau}, {Tafalla}, {van der Tak}, \&
  {Wyrowski}}]{Karska13}
{Karska}, A., {Herczeg}, G.~J., {van Dishoeck}, E.~F., {et~al.} 2013, \aap,
  552, A141

\bibitem[{{Kirk} {et~al.}(2013){Kirk}, {Myers}, {Bourke}, {Gutermuth},
  {Hedden}, \& {Wilson}}]{Kirk13}
{Kirk}, H., {Myers}, P.~C., {Bourke}, T.~L., {et~al.} 2013, \apj, 766, 115

\bibitem[{{Knee} \& {Sandell}(2000)}]{Knee00}
{Knee}, L.~B.~G. \& {Sandell}, G. 2000, \aap, 361, 671

\bibitem[{{Kurono} {et~al.}(2013){Kurono}, {Saito}, {Kamazaki}, {Morita}, \&
  {Kawabe}}]{Kurono13}
{Kurono}, Y., {Saito}, M., {Kamazaki}, T., {Morita}, K.-I., \& {Kawabe}, R.
  2013, \apj, 765, 85

\bibitem[{{Kurtovic} {et~al.}(2018){Kurtovic}, {P{\'e}rez}, {Benisty}, {Zhu},
  {Zhang}, {Huang}, {Andrews}, {Dullemond}, {Isella}, {Bai}, {Carpenter},
  {Guzm{\'a}n}, {Ricci}, \& {Wilner}}]{Kurtovic18}
{Kurtovic}, N.~T., {P{\'e}rez}, L.~M., {Benisty}, M., {et~al.} 2018, \apj, 869,
  L44

\bibitem[{{Kuznetsova} {et~al.}(2019){Kuznetsova}, {Hartmann}, \&
  {Heitsch}}]{Kuznetsova19}
{Kuznetsova}, A., {Hartmann}, L., \& {Heitsch}, F. 2019, \apj, 876, 33

\bibitem[{{Kwon} {et~al.}(2006){Kwon}, {Looney}, {Crutcher}, \&
  {Kirk}}]{Kwon06}
{Kwon}, W., {Looney}, L.~W., {Crutcher}, R.~M., \& {Kirk}, J.~M. 2006, \apj,
  653, 1358

\bibitem[{{Ladd} {et~al.}(1991){Ladd}, {Adams}, {Casey}, {Davidson}, {Fuller},
  {Harper}, {Myers}, \& {Padman}}]{Ladd91}
{Ladd}, E.~F., {Adams}, F.~C., {Casey}, S., {et~al.} 1991, \apj, 382, 555

\bibitem[{{Larson}(1981)}]{Larson81}
{Larson}, R.~B. 1981, \mnras, 194, 809

\bibitem[{{Lay} {et~al.}(1995){Lay}, {Carlstrom}, \& {Hills}}]{Lay95}
{Lay}, O.~P., {Carlstrom}, J.~E., \& {Hills}, R.~E. 1995, \apjl, 452, L73

\bibitem[{{Lee} {et~al.}(2005){Lee}, {Ho}, \& {White}}]{Lee05}
{Lee}, C.-F., {Ho}, P. T.~P., \& {White}, S.~M. 2005, \apj, 619, 948

\bibitem[{{Lee} {et~al.}(2015){Lee}, {Dunham}, {Myers}, {Tobin}, {Kristensen},
  {Pineda}, {Vorobyov}, {Offner}, {Arce}, {Li}, {Bourke}, {J{\o}rgensen},
  {Goodman}, {Sadavoy}, {Chandler}, {Harris}, {Kratter}, {Looney}, {Melis},
  {Perez}, \& {Segura-Cox}}]{Lee15}
{Lee}, K.~I., {Dunham}, M.~M., {Myers}, P.~C., {et~al.} 2015, \apj, 814, 114

\bibitem[{{Lef{\`e}vre} {et~al.}(2017){Lef{\`e}vre}, {Cabrit}, {Maury},
  {Gueth}, {Tabone}, {Podio}, {Belloche}, {Codella}, {Maret}, {Anderl},
  {Andr{\'e}}, \& {Hennebelle}}]{Lefevre17}
{Lef{\`e}vre}, C., {Cabrit}, S., {Maury}, A.~J., {et~al.} 2017, \aap, 604

\bibitem[{{Looney} {et~al.}(2000){Looney}, {Mundy}, \& {Welch}}]{Looney00}
{Looney}, L.~W., {Mundy}, L.~G., \& {Welch}, W.~J. 2000, \apj, 529, 477

\bibitem[{{Looney} {et~al.}(2003){Looney}, {Mundy}, \& {Welch}}]{Looney03}
{Looney}, L.~W., {Mundy}, L.~G., \& {Welch}, W.~J. 2003, \apj, 592, 255

\bibitem[{{Looney} {et~al.}(2007){Looney}, {Tobin}, \& {Kwon}}]{Looney07}
{Looney}, L.~W., {Tobin}, J.~J., \& {Kwon}, W. 2007, \apjl, 670, L131

\bibitem[{{L{\'o}pez-Sepulcre} {et~al.}(2017){L{\'o}pez-Sepulcre}, {Sakai},
  {Neri}, {Imai}, {Oya}, {Ceccarelli}, {Higuchi}, {Aikawa}, {Bottinelli},
  {Caux}, {Hirota}, {Kahane}, {Lefloch}, {Vastel}, {Watanabe}, \&
  {Yamamoto}}]{Lopez-Sepulcre17}
{L{\'o}pez-Sepulcre}, A., {Sakai}, N., {Neri}, R., {et~al.} 2017, \aap, 606,
  A121

\bibitem[{{Lynden-Bell} \& {Pringle}(1974)}]{Lynden-Bell74}
{Lynden-Bell}, D. \& {Pringle}, J.~E. 1974, \mnras, 168, 603

\bibitem[{{Machida} {et~al.}(2014){Machida}, {Inutsuka}, \&
  {Matsumoto}}]{Machida14}
{Machida}, M.~N., {Inutsuka}, S.-i., \& {Matsumoto}, T. 2014, \mnras, 438, 2278

\bibitem[{{Mangum} \& {Shirley}(2015)}]{Mangum15}
{Mangum}, J.~G. \& {Shirley}, Y.~L. 2015, \pasp, 127, 266

\bibitem[{{Maret} {et~al.}(2014){Maret}, {Belloche}, {Maury}, {Gueth},
  {Andr{\'e}}, {Cabrit}, {Codella}, \& {Bontemps}}]{Maret14}
{Maret}, S., {Belloche}, A., {Maury}, A.~J., {et~al.} 2014, \aap, 563, L1

\bibitem[{{Maret} {et~al.}(2007){Maret}, {Bergin}, \& {Lada}}]{Maret07}
{Maret}, S., {Bergin}, E.~A., \& {Lada}, C.~J. 2007, \apjl, 670, L25

\bibitem[{{Maret} {et~al.}(2002){Maret}, {Ceccarelli}, {Caux}, {Tielens}, \&
  {Castets}}]{Maret02}
{Maret}, S., {Ceccarelli}, C., {Caux}, E., {Tielens}, A.~G.~G.~M., \&
  {Castets}, A. 2002, \aap, 395, 573

\bibitem[{{Maret} {et~al.}(2020){Maret}, {Maury}, {Belloche}, {Gaudel},
  {Andr{\'e}}, {Cabrit}, {Codella}, {Lef{\'e}vre}, {Podio}, {Anderl}, {Gueth},
  \& {Hennebelle}}]{Maret20}
{Maret}, S., {Maury}, A.~J., {Belloche}, A., {et~al.} 2020, \aap, 635, A15

\bibitem[{{Masson} {et~al.}(2016){Masson}, {Chabrier}, {Hennebelle}, {Vaytet},
  \& {Commer{\c{c}}on}}]{Masson16}
{Masson}, J., {Chabrier}, G., {Hennebelle}, P., {Vaytet}, N., \&
  {Commer{\c{c}}on}, B. 2016, \aap, 587, A32

\bibitem[{{Mather} {et~al.}(1994){Mather}, {Cheng}, {Cottingham}, {Eplee},
  {Fixsen}, {Hewagama}, {Isaacman}, {Jensen}, {Meyer}, {Noerdlinger}, {Read},
  {Rosen}, {Shafer}, {Wright}, {Bennett}, {Boggess}, {Hauser}, {Kelsall},
  {Moseley}, {Silverberg}, {Smoot}, {Weiss}, \& {Wilkinson}}]{Mather94}
{Mather}, J.~C., {Cheng}, E.~S., {Cottingham}, D.~A., {et~al.} 1994, \apj, 420,
  439

\bibitem[{{Maury} {et~al.}(2010){Maury}, {Andr{\'e}}, {Hennebelle}, {Motte},
  {Stamatellos}, {Bate}, {Belloche}, {Duch{\^e}ne}, \& {Whitworth}}]{Maury10}
{Maury}, A.~J., {Andr{\'e}}, P., {Hennebelle}, P., {et~al.} 2010, \aap, 512,
  A40+

\bibitem[{{Maury} {et~al.}(2011){Maury}, {Andr{\'e}}, {Men'shchikov},
  {K{\"o}nyves}, \& {Bontemps}}]{Maury11}
{Maury}, A.~J., {Andr{\'e}}, P., {Men'shchikov}, A., {K{\"o}nyves}, V., \&
  {Bontemps}, S. 2011, \aap, 535

\bibitem[{{Maury} {et~al.}(2019){Maury}, {Andr{\'e}}, {Testi}, {Maret},
  {Belloche}, {Hennebelle}, {Cabrit}, {Codella}, {Gueth}, {Podio}, {Anderl},
  {Bacmann}, {Bontemps}, {Gaudel}, {Ladjelate}, {Lef{\`e}vre}, {Tabone}, \&
  {Lefloch}}]{Maury18}
{Maury}, A.~J., {Andr{\'e}}, P., {Testi}, L., {et~al.} 2019, \aap, 621, A76

\bibitem[{{Maury} {et~al.}(2014){Maury}, {Belloche}, {Andr{\'e}}, {Maret},
  {Gueth}, {Codella}, {Cabrit}, {Testi}, \& {Bontemps}}]{Maury14}
{Maury}, A.~J., {Belloche}, A., {Andr{\'e}}, P., {et~al.} 2014, \aap, 563, L2

\bibitem[{{Mizuno} {et~al.}(1994){Mizuno}, {Onishi}, {Hayashi}, {Ohashi},
  {Sunada}, {Hasegawa}, \& {Fukui}}]{Mizuno94}
{Mizuno}, A., {Onishi}, T., {Hayashi}, M., {et~al.} 1994, \nat, 368, 719

\bibitem[{{Motte} \& {Andr{\'e}}(2001)}]{Motte01}
{Motte}, F. \& {Andr{\'e}}, P. 2001, \aap, 365, 440

\bibitem[{{Najita} \& {Bergin}(2018)}]{Najita18}
{Najita}, J.~R. \& {Bergin}, E.~A. 2018, \apj, 864, 168

\bibitem[{{Ohashi} {et~al.}(1997{\natexlab{a}}){Ohashi}, {Hayashi}, {Ho}, \&
  {Momose}}]{Ohashi97b}
{Ohashi}, N., {Hayashi}, M., {Ho}, P. T.~P., \& {Momose}, M.
  1997{\natexlab{a}}, \apj, 475, 211

\bibitem[{{Ohashi} {et~al.}(1997{\natexlab{b}}){Ohashi}, {Hayashi}, {Ho},
  {Momose}, {Tamura}, {Hirano}, \& {Sargent}}]{Ohashi97}
{Ohashi}, N., {Hayashi}, M., {Ho}, P.~T.~P., {et~al.} 1997{\natexlab{b}}, \apj,
  488, 317

\bibitem[{{Ohashi} {et~al.}(1999){Ohashi}, {Lee}, {Wilner}, \&
  {Hayashi}}]{Ohashi99}
{Ohashi}, N., {Lee}, S., {Wilner}, D., \& {Hayashi}, M. 1999, \apjl, 518, L41

\bibitem[{{Ohashi} {et~al.}(2014){Ohashi}, {Saigo}, {Aso}, {Aikawa},
  {Koyamatsu}, {Machida}, {Saito}, {Takahashi}, {Takakuwa}, {Tomida},
  {Tomisaka}, \& {Yen}}]{Ohashi14}
{Ohashi}, N., {Saigo}, K., {Aso}, Y., {et~al.} 2014, \apj, 796, 131

\bibitem[{{O'Linger} {et~al.}(1999){O'Linger}, {Wolf-Chase}, {Barsony}, \&
  {Ward-Thompson}}]{Olinger99}
{O'Linger}, J., {Wolf-Chase}, G., {Barsony}, M., \& {Ward-Thompson}, D. 1999,
  \apj, 515, 696

\bibitem[{{Onishi} {et~al.}(1999){Onishi}, {Mizuno}, \& {Fukui}}]{Onishi99}
{Onishi}, T., {Mizuno}, A., \& {Fukui}, Y. 1999, Publications of the
  Astronomical Society of Japan, 51, 257

\bibitem[{{Ortiz-Le{\'o}n} {et~al.}(2018){Ortiz-Le{\'o}n}, {Loinard}, {Dzib},
  {Kounkel}, {Galli}, {Tobin}, {Evans}, {Hartmann}, {Rodr{\'\i}guez},
  {Brice{\~n}o}, {Torres}, \& {Mioduszewski}}]{OrtizLeon18}
{Ortiz-Le{\'o}n}, G.~N., {Loinard}, L., {Dzib}, S.~A., {et~al.} 2018, \apj,
  869, L33

\bibitem[{{Palmeirim} {et~al.}(2013){Palmeirim}, {Andr{\'e}}, {Kirk},
  {Ward-Thompson}, {Arzoumanian}, {K{\"o}nyves}, {Didelon}, {Schneider},
  {Benedettini}, {Bontemps}, {Di Francesco}, {Elia}, {Griffin}, {Hennemann},
  {Hill}, {Martin}, {Men'shchikov}, {Molinari}, {Motte}, {Nguyen Luong},
  {Nutter}, {Peretto}, {Pezzuto}, {Roy}, {Rygl}, {Spinoglio}, \&
  {White}}]{Palmeirim13}
{Palmeirim}, P., {Andr{\'e}}, P., {Kirk}, J., {et~al.} 2013, \aap, 550, A38

\bibitem[{{Pelletier} \& {Pudritz}(1992)}]{Pelletier92}
{Pelletier}, G. \& {Pudritz}, R.~E. 1992, \apj, 394, 117

\bibitem[{{P{\'e}rez} {et~al.}(2018){P{\'e}rez}, {Benisty}, {Andrews},
  {Isella}, {Dullemond}, {Huang}, {Kurtovic}, {Guzm{\'a}n}, {Zhu}, {Birnstiel},
  {Zhang}, {Carpenter}, {Wilner}, {Ricci}, {Bai}, {Weaver}, \&
  {{\"O}berg}}]{Perez18}
{P{\'e}rez}, L.~M., {Benisty}, M., {Andrews}, S.~M., {et~al.} 2018, \apj, 869,
  L50

\bibitem[{{Pety}(2005)}]{Pety05}
{Pety}, J. 2005, in SF2A-2005: Semaine de l'Astrophysique Francaise, ed.
  F.~{Casoli}, T.~{Contini}, J.~M. {Hameury}, \& L.~{Pagani}, 721

\bibitem[{{Pi{\'e}tu} {et~al.}(2014){Pi{\'e}tu}, {Guilloteau}, {Di Folco},
  {Dutrey}, \& {Boehler}}]{Pietu14}
{Pi{\'e}tu}, V., {Guilloteau}, S., {Di Folco}, E., {Dutrey}, A., \& {Boehler},
  Y. 2014, \aap, 564, A95

\bibitem[{{Pineda} {et~al.}(2010){Pineda}, {Goodman}, {Arce}, {Caselli},
  {Foster}, {Myers}, \& {Rosolowsky}}]{Pineda10}
{Pineda}, J.~E., {Goodman}, A.~A., {Arce}, H.~G., {et~al.} 2010, \apj, 712,
  L116

\bibitem[{{Pineda} {et~al.}(2019){Pineda}, {Zhao}, {Schmiedeke}, {Segura-Cox},
  {Caselli}, {Myers}, {Tobin}, \& {Dunham}}]{Pineda19}
{Pineda}, J.~E., {Zhao}, B., {Schmiedeke}, A., {et~al.} 2019, \apj, 882, 103

\bibitem[{{Plunkett} {et~al.}(2013){Plunkett}, {Arce}, {Corder}, {Mardones},
  {Sargent}, \& {Schnee}}]{Plunkett13}
{Plunkett}, A.~L., {Arce}, H.~G., {Corder}, S.~A., {et~al.} 2013, \apj, 774, 22

\bibitem[{{Podio} {et~al.}(2016){Podio}, {Codella}, {Gueth}, {Cabrit}, {Maury},
  {Tabone}, {Lef{\`e}vre}, {Anderl}, {Andr{\'e}}, {Belloche}, {Bontemps},
  {Hennebelle}, {Lefloch}, {Maret}, \& {Testi}}]{Podio16}
{Podio}, L., {Codella}, C., {Gueth}, F., {et~al.} 2016, \aap, 593, L4

\bibitem[{{Pudritz} {et~al.}(2007){Pudritz}, {Ouyed}, {Fendt}, \&
  {Brandenburg}}]{Pudritz07}
{Pudritz}, R.~E., {Ouyed}, R., {Fendt}, C., \& {Brandenburg}, A. 2007, in
  Protostars and Planets V, ed. B.~{Reipurth}, D.~{Jewitt}, \& K.~{Keil}, 277

\bibitem[{{Redman} {et~al.}(2004){Redman}, {Keto}, {Rawlings}, \&
  {Williams}}]{Redman04}
{Redman}, M.~P., {Keto}, E., {Rawlings}, J.~M.~C., \& {Williams}, D.~A. 2004,
  \mnras, 352, 1365

\bibitem[{{Reid} {et~al.}(2016){Reid}, {Dame}, {Menten}, \&
  {Brunthaler}}]{Reid16}
{Reid}, M.~J., {Dame}, T.~M., {Menten}, K.~M., \& {Brunthaler}, A. 2016, \apj,
  823, 77

\bibitem[{{Ricci} {et~al.}(2010){Ricci}, {Testi}, {Natta}, {Neri}, {Cabrit}, \&
  {Herczeg}}]{Ricci10}
{Ricci}, L., {Testi}, L., {Natta}, A., {et~al.} 2010, \aap, 512, A15

\bibitem[{{Sadavoy} {et~al.}(2014){Sadavoy}, {Di Francesco}, {Andr{\'e}},
  {Pezzuto}, {Bernard}, {Maury}, {Men'shchikov}, {Motte}, {Nguyen-Lu'o'ng},
  {Schneider}, {Arzoumanian}, {Benedettini}, {Bontemps}, {Elia}, {Hennemann},
  {Hill}, {K{\"o}nyves}, {Louvet}, {Peretto}, {Roy}, \& {White}}]{Sadavoy14}
{Sadavoy}, S.~I., {Di Francesco}, J., {Andr{\'e}}, P., {et~al.} 2014, \apj,
  787, L18

\bibitem[{{Sakai} {et~al.}(2014){Sakai}, {Sakai}, {Hirota}, {Watanabe},
  {Ceccarelli}, {Kahane}, {Bottinelli}, {Caux}, {Demyk}, {Vastel}, {Coutens},
  {Taquet}, {Ohashi}, {Takakuwa}, {Yen}, {Aikawa}, \& {Yamamoto}}]{Sakai14Nat}
{Sakai}, N., {Sakai}, T., {Hirota}, T., {et~al.} 2014, \nat, 507, 78

\bibitem[{{Sandell} \& {Knee}(2001)}]{Sandell01}
{Sandell}, G. \& {Knee}, L. 2001, in Science with the Atacama Large Millimeter
  Array, ed. A.~{Wootten}, Vol. 235, 154

\bibitem[{{Sandell} {et~al.}(1994){Sandell}, {Knee}, {Aspin}, {Robson}, \&
  {Russell}}]{Sandell94}
{Sandell}, G., {Knee}, L.~B.~G., {Aspin}, C., {Robson}, I.~E., \& {Russell},
  A.~P.~G. 1994, \aap, 285, L1

\bibitem[{{Santangelo} {et~al.}(2015){Santangelo}, {Codella}, {Cabrit},
  {Maury}, {Gueth}, {Maret}, {Lefloch}, {Belloche}, {Andr{\'e}}, {Hennebelle},
  {Anderl}, {Podio}, \& {Testi}}]{Santangelo15}
{Santangelo}, G., {Codella}, C., {Cabrit}, S., {et~al.} 2015, \aap, 584, A126

\bibitem[{{Schneider} \& {Elmegreen}(1979)}]{Schneider79}
{Schneider}, S. \& {Elmegreen}, B.~G. 1979, The Astrophysical Journal
  Supplement Series, 41, 87

\bibitem[{{Segura-Cox} {et~al.}(2016){Segura-Cox}, {Harris}, {Tobin}, {Looney},
  {Li}, {Chandler}, {Kratter}, {Dunham}, {Sadavoy}, {Perez}, \&
  {Melis}}]{Segura-Cox16}
{Segura-Cox}, D.~M., {Harris}, R.~J., {Tobin}, J.~J., {et~al.} 2016, \apj, 817,
  L14

\bibitem[{{Shimajiri} {et~al.}(2019){Shimajiri}, {Andr{\'e}}, {Palmeirim},
  {Arzoumanian}, {Bracco}, {K{\"o}nyves}, {Ntormousi}, \&
  {Ladjelate}}]{Shimajiri19}
{Shimajiri}, Y., {Andr{\'e}}, P., {Palmeirim}, P., {et~al.} 2019, \aap, 623,
  A16

\bibitem[{{Shu}(1977)}]{Shu77}
{Shu}, F.~H. 1977, \apj, 214, 488

\bibitem[{{Simon} {et~al.}(2000){Simon}, {Dutrey}, \& {Guilloteau}}]{Simon00}
{Simon}, M., {Dutrey}, A., \& {Guilloteau}, S. 2000, \apj, 545, 1034

\bibitem[{{Spezzi} {et~al.}(2013){Spezzi}, {Cox}, {Prusti}, {Mer{\'\i}n},
  {Ribas}, {Alves de Oliveira}, {Winston}, {K{\'o}sp{\'a}l}, {Royer}, {Vavrek},
  {Andr{\'e}}, {Pilbratt}, {Testi}, {Bressert}, {Ricci}, {Men'shchikov}, \&
  {K{\"o}nyves}}]{Spezzi13}
{Spezzi}, L., {Cox}, N.~L.~J., {Prusti}, T., {et~al.} 2013, \aap, 555, A71

\bibitem[{{Tafalla} \& {Bachiller}(1995)}]{Tafalla95}
{Tafalla}, M. \& {Bachiller}, R. 1995, \apj, 443, L37

\bibitem[{{Tafalla} {et~al.}(1998){Tafalla}, {Mardones}, {Myers}, {Caselli},
  {Bachiller}, \& {Benson}}]{Tafalla98}
{Tafalla}, M., {Mardones}, D., {Myers}, P.~C., {et~al.} 1998, \apj, 504, 900

\bibitem[{{Takahashi} {et~al.}(2016){Takahashi}, {Tomida}, {Machida}, \&
  {Inutsuka}}]{Takahashi16}
{Takahashi}, S.~Z., {Tomida}, K., {Machida}, M.~N., \& {Inutsuka}, S.-i. 2016,
  \mnras, 463, 1390

\bibitem[{{Takakuwa} {et~al.}(2003){Takakuwa}, {Ohashi}, \&
  {Hirano}}]{Takakuwa03}
{Takakuwa}, S., {Ohashi}, N., \& {Hirano}, N. 2003, \apj, 590, 932

\bibitem[{{Tatematsu} {et~al.}(2016){Tatematsu}, {Ohashi}, {Sanhueza}, {Nguyen
  Luong}, {Umemoto}, \& {Mizuno}}]{Tatematsu16}
{Tatematsu}, K., {Ohashi}, S., {Sanhueza}, P., {et~al.} 2016, Publications of
  the Astronomical Society of Japan, 68, 24

\bibitem[{{Terebey} {et~al.}(1993){Terebey}, {Chandler}, \&
  {Andre}}]{Terebey93}
{Terebey}, S., {Chandler}, C.~J., \& {Andre}, P. 1993, \apj, 414, 759

\bibitem[{{Terebey} {et~al.}(2009){Terebey}, {Fich}, {Noriega-Crespo},
  {Padgett}, {Fukagawa}, {Audard}, {Brooke}, {Carey}, {Evans}, {Guedel},
  {Hines}, {Huard}, {Knapp}, {McCabe}, {Menard}, {Monin}, \&
  {Rebull}}]{Terebey09}
{Terebey}, S., {Fich}, M., {Noriega-Crespo}, A., {et~al.} 2009, \apj, 696, 1918

\bibitem[{{Terebey} {et~al.}(1984){Terebey}, {Shu}, \& {Cassen}}]{Terebey84}
{Terebey}, S., {Shu}, F.~H., \& {Cassen}, P. 1984, \apj, 286, 529

\bibitem[{{Tobin} {et~al.}(2015){Tobin}, {Dunham}, {Looney}, {Li}, {Chandler},
  {Segura-Cox}, {Sadavoy}, {Melis}, {Harris}, {Perez}, {Kratter},
  {J{\o}rgensen}, {Plunkett}, \& {Hull}}]{Tobin15}
{Tobin}, J.~J., {Dunham}, M.~M., {Looney}, L.~W., {et~al.} 2015, \apj, 798, 61

\bibitem[{{Tobin} {et~al.}(2011){Tobin}, {Hartmann}, {Chiang}, {Looney},
  {Bergin}, {Chandler}, {Masqu{\'e}}, {Maret}, \& {Heitsch}}]{Tobin11}
{Tobin}, J.~J., {Hartmann}, L., {Chiang}, H.-F., {et~al.} 2011, \apj, 740, 45

\bibitem[{{Tobin} {et~al.}(2012){Tobin}, {Hartmann}, {Chiang}, {Wilner},
  {Looney}, {Loinard}, {Calvet}, \& {D'Alessio}}]{Tobin12}
{Tobin}, J.~J., {Hartmann}, L., {Chiang}, H.-F., {et~al.} 2012, \nat, 492, 83

\bibitem[{{Tobin} {et~al.}(2010){Tobin}, {Hartmann}, {Looney}, \&
  {Chiang}}]{Tobin10}
{Tobin}, J.~J., {Hartmann}, L., {Looney}, L.~W., \& {Chiang}, H.-F. 2010, \apj,
  712, 1010

\bibitem[{{Tobin} {et~al.}(2016{\natexlab{a}}){Tobin}, {Kratter}, {Persson},
  {Looney}, {Dunham}, {Segura-Cox}, {Li}, {Chandler}, {Sadavoy}, {Harris},
  {Melis}, \& {P{\'e}rez}}]{Tobin16}
{Tobin}, J.~J., {Kratter}, K.~M., {Persson}, M.~V., {et~al.}
  2016{\natexlab{a}}, \nat, 538, 483

\bibitem[{{Tobin} {et~al.}(2016{\natexlab{b}}){Tobin}, {Looney}, {Li},
  {Chandler}, {Dunham}, {Segura-Cox}, {Sadavoy}, {Melis}, {Harris}, {Kratter},
  \& {Perez}}]{Tobin16b}
{Tobin}, J.~J., {Looney}, L.~W., {Li}, Z.-Y., {et~al.} 2016{\natexlab{b}},
  \apj, 818, 73

\bibitem[{{Tobin} {et~al.}(2007){Tobin}, {Looney}, {Mundy}, {Kwon}, \&
  {Hamidouche}}]{Tobin07}
{Tobin}, J.~J., {Looney}, L.~W., {Mundy}, L.~G., {Kwon}, W., \& {Hamidouche},
  M. 2007, \apj, 659, 1404

\bibitem[{{Tokuda} {et~al.}(2016){Tokuda}, {Onishi}, {Matsumoto}, {Saigo},
  {Kawamura}, {Fukui}, {Inutsuka}, {Machida}, {Tomida}, {Tachihara}, \&
  {Andr{\'e}}}]{Tokuda16}
{Tokuda}, K., {Onishi}, T., {Matsumoto}, T., {et~al.} 2016, \apj, 826, 26

\bibitem[{{Tokuda} {et~al.}(2017){Tokuda}, {Onishi}, {Saigo}, {Hosokawa},
  {Matsumoto}, {Inutsuka}, {Machida}, {Tomida}, {Kunitomo}, {Kawamura},
  {Fukui}, \& {Tachihara}}]{Tokuda17}
{Tokuda}, K., {Onishi}, T., {Saigo}, K., {et~al.} 2017, \apj, 849, 101

\bibitem[{{Tokuda} {et~al.}(2014){Tokuda}, {Onishi}, {Saigo}, {Kawamura},
  {Fukui}, {Matsumoto}, {Inutsuka}, {Machida}, {Tomida}, \&
  {Tachihara}}]{Tokuda14}
{Tokuda}, K., {Onishi}, T., {Saigo}, K., {et~al.} 2014, \apjl, 789, L4

\bibitem[{{Torres} {et~al.}(2009){Torres}, {Loinard}, {Mioduszewski}, \&
  {Rodr{\'\i}guez}}]{Torres09}
{Torres}, R.~M., {Loinard}, L., {Mioduszewski}, A.~J., \& {Rodr{\'\i}guez},
  L.~F. 2009, \apj, 698, 242

\bibitem[{{Tsukamoto} {et~al.}(2015){Tsukamoto}, {Iwasaki}, {Okuzumi},
  {Machida}, \& {Inutsuka}}]{Tsukamoto15}
{Tsukamoto}, Y., {Iwasaki}, K., {Okuzumi}, S., {Machida}, M.~N., \& {Inutsuka},
  S. 2015, \mnras, 452, 278

\bibitem[{{Tsukamoto} {et~al.}(2017){Tsukamoto}, {Okuzumi}, {Iwasaki},
  {Machida}, \& {Inutsuka}}]{Tsukamoto17}
{Tsukamoto}, Y., {Okuzumi}, S., {Iwasaki}, K., {Machida}, M.~N., \& {Inutsuka},
  S.-i. 2017, Publications of the Astronomical Society of Japan, 69, 95

\bibitem[{{Ulrich}(1976)}]{Ulrich76}
{Ulrich}, R.~K. 1976, \apj, 210, 377

\bibitem[{{Umemoto} {et~al.}(1992){Umemoto}, {Iwata}, {Fukui}, {Mikami},
  {Yamamoto}, {Kameya}, \& {Hirano}}]{Umemoto92}
{Umemoto}, T., {Iwata}, T., {Fukui}, Y., {et~al.} 1992, \apj, 392, L83

\bibitem[{{Verliat} {et~al.}(2020){Verliat}, {Hennebelle}, {Maury}, \&
  {Gaudel}}]{Verliat20}
{Verliat}, A., {Hennebelle}, P., {Maury}, A.~J., \& {Gaudel}, M. 2020, \aap,
  635, A130

\bibitem[{{Viotti}(1969)}]{Viotti69}
{Viotti}, R. 1969, \memsai, 40, 75

\bibitem[{{Ward-Thompson} {et~al.}(2007){Ward-Thompson}, {Andr{\'e}},
  {Crutcher}, {Johnstone}, {Onishi}, \& {Wilson}}]{Ward-Thompson07}
{Ward-Thompson}, D., {Andr{\'e}}, P., {Crutcher}, R., {et~al.} 2007, in
  Protostars and Planets V, ed. B.~{Reipurth}, D.~{Jewitt}, \& K.~{Keil}, 33

\bibitem[{{Wiesemeyer}(1997)}]{Wiesemeyer97}
{Wiesemeyer}, H. 1997, PhD thesis, Institut de Radio Astronomie
  Millim{\'e}trique, 300, Rue de la Piscine, Domaine Universitaire de Grenoble,
  F-38406 Saint- Martin-d'H{\`e}res CEDEX, France

\bibitem[{{Wiesemeyer} {et~al.}(1999){Wiesemeyer}, {Cox}, {Gusten}, \&
  {Zylka}}]{Wiesemeyer99}
{Wiesemeyer}, H., {Cox}, P., {Gusten}, R., \& {Zylka}, R. 1999, in The Universe
  as Seen by ISO, ed. P.~{Cox} \& M.~{Kessler}, Vol. 427, 533

\bibitem[{{Wiesemeyer} {et~al.}(1998){Wiesemeyer}, {Gusten}, {Cox}, {Zylka}, \&
  {Wright}}]{Wiesemeyer98}
{Wiesemeyer}, H., {Gusten}, R., {Cox}, P., {Zylka}, R., \& {Wright}, M.~C.~H.
  1998, in Astronomical Society of the Pacific Conference Series, Vol. 132,
  Star Formation with the Infrared Space Observatory, ed. J.~{Yun} \&
  L.~{Liseau}, 189

\bibitem[{{Williams} {et~al.}(2006){Williams}, {Lee}, \& {Myers}}]{Williams06}
{Williams}, J.~P., {Lee}, C.~W., \& {Myers}, P.~C. 2006, \apj, 636, 952

\bibitem[{{Wolf-Chase} {et~al.}(2000){Wolf-Chase}, {Barsony}, \&
  {O'Linger}}]{Wolf-Chase00}
{Wolf-Chase}, G.~A., {Barsony}, M., \& {O'Linger}, J. 2000, \aj, 120, 1467

\bibitem[{{Yen} {et~al.}(2015{\natexlab{a}}){Yen}, {Koch}, {Takakuwa}, {Ho},
  {Ohashi}, \& {Tang}}]{Yen15b}
{Yen}, H.-W., {Koch}, P.~M., {Takakuwa}, S., {et~al.} 2015{\natexlab{a}}, \apj,
  799, 193

\bibitem[{{Yen} {et~al.}(2017){Yen}, {Koch}, {Takakuwa}, {Krasnopolsky},
  {Ohashi}, \& {Aso}}]{Yen17}
{Yen}, H.-W., {Koch}, P.~M., {Takakuwa}, S., {et~al.} 2017, \apj, 834, 178

\bibitem[{{Yen} {et~al.}(2015{\natexlab{b}}){Yen}, {Takakuwa}, {Koch}, {Aso},
  {Koyamatsu}, {Krasnopolsky}, \& {Ohashi}}]{Yen15}
{Yen}, H.-W., {Takakuwa}, S., {Koch}, P.~M., {et~al.} 2015{\natexlab{b}}, \apj,
  812, 129

\bibitem[{{Yen} {et~al.}(2011){Yen}, {Takakuwa}, \& {Ohashi}}]{Yen11}
{Yen}, H.-W., {Takakuwa}, S., \& {Ohashi}, N. 2011, \apj, 742, 57

\bibitem[{{Yen} {et~al.}(2013){Yen}, {Takakuwa}, {Ohashi}, \& {Ho}}]{Yen13}
{Yen}, H.-W., {Takakuwa}, S., {Ohashi}, N., \& {Ho}, P.~T.~P. 2013, \apj, 772,
  22

\bibitem[{{Yorke} \& {Bodenheimer}(1999)}]{Yorke99}
{Yorke}, H.~W. \& {Bodenheimer}, P. 1999, \apj, 525, 330

\bibitem[{{Zhang} {et~al.}(1995){Zhang}, {Ho}, {Wright}, \& {Wilner}}]{Zhang95}
{Zhang}, Q., {Ho}, P. T.~P., {Wright}, M.~C.~H., \& {Wilner}, D.~J. 1995, \apj,
  451, L71

\bibitem[{{Zucker} {et~al.}(2019){Zucker}, {Speagle}, {Schlafly}, {Green},
  {Finkbeiner}, {Goodman}, \& {Alves}}]{Zucker19}
{Zucker}, C., {Speagle}, J.~S., {Schlafly}, E.~F., {et~al.} 2019, arXiv
  e-prints, arXiv:1902.01425

\end{thebibliography}

%\Online

%%%%%%%%%%%%%%%%%%%%%%%%%%%%%%%%%%%%%%%%%%%%%%%%
%%%%%%%%%%%%%%%%%%%%%%%%%%%%%%%%%%%%%%%%%%%%%%%%
%                 ANNEXES
%%%%%%%%%%%%%%%%%%%%%%%%%%%%%%%%%%%%%%%%%%%%%%%%
%%%%%%%%%%%%%%%%%%%%%%%%%%%%%%%%%%%%%%%%%%%%%%%%
\onecolumn
\clearpage
\begin{appendix}

%%%%%%%%%%%% DETAILS OBS 30m %%%%%%%%%%%%
\section{Details of IRAM 30m observations} 
\label{details-obs}

\begin{table*}[h]
\centering
\caption{Details of observations carried out with the IRAM 30m telescope.}
\label{table:temps-observations-30m}
\begin{tabular}{lc|cc|l}
\hline \hline
\multicolumn{5}{c}{\hfill}  \\
Source                  & Line               & $\tau_{225}$~\tablefootmark{a} & Time~\tablefootmark{b} & \thead{Dates of observation}   \\
                  &               &  & (h) & \thead{(dd-month-yy)} \\
\hline
\multirow{2}{*}{L1448-2A} & C$^{18}$O (2$-$1)      &   0.29       &  1.9    &    04 June 2012; 16 October 2012   \\
                         & N$_{2}$H$^{+}$ (1$-$0) &   0.69      &  2.3    &   17 November 2011; 30 May 2012    \\
\hline
\multirow{2}{*}{L1448-NB} & C$^{18}$O (2$-$1)      &   0.28       &  2.3    &  16 November 2011; 17 November 2011; 16 October 2014     \\
                         & N$_{2}$H$^{+}$ (1$-$0) &    0.43      &  3.1    &   17 November 2011; 18 November 2011; 03 June 2012    \\
\hline
\multirow{2}{*}{L1448-C} & C$^{18}$O (2$-$1)      & 0.18         &  2.3    &  16 November 2011 ; 17 October 2014     \\
                         & N$_{2}$H$^{+}$ (1$-$0) & 0.41         &  3.1    &  17 November 2011; 18 November 2011; 03 June 2012    \\
\hline
\multirow{2}{*}{IRAS2A}  & C$^{18}$O (2$-$1)      & 0.14         &  3.1    &    15 November 2011; 16 November 2011; 06 November 2014    \\
                         & N$_{2}$H$^{+}$ (1$-$0) & 0.31         &  1.5    &  15 November 2011; 17 November 2011    \\
\hline
\multirow{2}{*}{SVS13-B}   & C$^{18}$O (2$-$1)      &   0.18       &  1.6    &  16 November 2012;  06 November 2014     \\
                         & N$_{2}$H$^{+}$ (1$-$0) &    0.64      &  5.4    &   31 May 2012; 02 June 2012; 03 June 2012; 04 June 2012    \\
\hline
\multirow{2}{*}{IRAS4A}  & C$^{18}$O (2$-$1)      &    0.13      &  1.6    &   16 November 2012    \\
                         & N$_{2}$H$^{+}$ (1$-$0) &   0.73       &  3.1   &   01 June 2012; 04 June 2012    \\
\hline
\multirow{2}{*}{IRAS4B}  & C$^{18}$O (2$-$1)      & 0.15         &   1.6   &  16 November 2012     \\
                         & N$_{2}$H$^{+}$ (1$-$0) & 0.62         &   2.7   &  02 June 2012     \\
\hline                         
\multirow{2}{*}{IRAM04191}   & C$^{18}$O (2$-$1)      &    0.17      &  1.9    &  16 November 2011    \\
                         & N$_{2}$H$^{+}$ (1$-$0) &   0.59       &  3.5   &  17 November 2011; 18 November 2011; 02 June 2012     \\
\hline
\multirow{2}{*}{L1521F}   & C$^{18}$O (2$-$1)      &   0.23      &  2.3    &  21 October 2014     \\
                         & N$_{2}$H$^{+}$ (1$-$0) &   0.58       &  5.8    &  18 November 2011; 30 May 2012; 03 June 2012; 04 June 2012     \\
\hline
\multirow{2}{*}{L1527}   & C$^{18}$O (2$-$1)      & 0.18         &  2.3    &  21 October 2014; 05 November 2014; 06 November 2014; 10 November 2014     \\
                         & N$_{2}$H$^{+}$ (1$-$0) & 0.64         &  5.2    &  01 June 2012; 02 June 2012; 04 June 2012; 31 May 2015    \\
\hline
\multirow{2}{*}{L1157}   & C$^{18}$O (2$-$1)      & 0.10         &   2.8   &   04 June 2012; 16 November 2012    \\
                         & N$_{2}$H$^{+}$ (1$-$0) & 0.60         &   6.1   &   30 May 2012; 31 May 2012; 01 June 2012    \\
                         \hline
\multirow{2}{*}{GF9-2}   & C$^{18}$O (2$-$1)      & 0.09         &  3.1    &  04 June 2012; 16 November 2012; 05 November 2014     \\
                         & N$_{2}$H$^{+}$ (1$-$0) & 0.55         &  4.2    &   30 May 2012; 03 June 2012    \\
\hline
\end{tabular}
\tablefoot{\tablefoottext{a}{$\tau_{225}$ is the zenith optical depth $\tau$ at a frequency of 225~GHz from the tipping taumeter.}
\tablefoottext{b}{Time is the total observed time spent on each molecular line for a given source.}}
\end{table*}

%%%%%%%%%%%% PROPERTIES MAPS %%%%%%%%%%%%
\clearpage
\section{Properties of molecular line emission maps} 
\label{properties-maps}

% Beam sizes C18O %
\begin{table*}[h]
\centering
\caption{Beam sizes of the C$^{18}$O (2$-$1) emission maps.}
\label{table:beams-c18o}
\begin{tabular}{l|ccc}
\hline \hline
 \multicolumn{4}{c}{\hfill}  \\
  Source      & PdBI         & PdBI+30m           & 30m             \\
 \hfill & \hfill & \hfill & \hfill \\
        & Major ($\arcsec$) $\times$ Minor ($\arcsec$), Pos. Ang. ($^{\circ}$)  & Major ($\arcsec$) $\times$ Minor ($\arcsec$), Pos. Ang. ($^{\circ}$)  & Major ($\arcsec$) $\times$ Minor ($\arcsec$), Pos. Ang. ($^{\circ}$)  \\

        \hline
L1448-2A &   0.94 $\times$ 0.69, 24.7            &   0.96  $\times$ 0.70, 25.1 &     11.8  $\times$ 11.8, 0   \\
L1448-NB &   0.76  $\times$ 0.71, 65.5             &   0.77  $\times$ 0.71, 48.3         &    11.8  $\times$ 11.8,0    \\
L1448-C &   0.63  $\times$ 0.40, 32.0      &   0.63  $\times$ 0.40, 32.0      &    11.8  $\times$ 11.8, 0    \\
IRAS2A &   0.78  $\times$ 0.72, 49.5          &   0.79  $\times$ 0.72, 46.8        &    11.8  $\times$ 11.8, 0    \\
SVS13-B &   0.71  $\times$ 0.58, 36.5      &   0.73  $\times$ 0.61, 33.0          &    11.8  $\times$ 11.8, 0     \\
IRAS4A &   0.77  $\times$ 0.63, 31.0         &   0.81  $\times$ 0.67, 26.3          &    11.8  $\times$ 11.8, 0    \\
IRAS4B &   0.77  $\times$ 0.65, 28.9         &   0.82  $\times$ 0.69, 22.6           &    11.8  $\times$ 11.8, 0     \\
IRAM04191   &  0.88  $\times$ 0.72, -166.7     	    &  0.91  $\times$ 0.75, -170.5    	     &    11.8  $\times$ 11.8, 0    \\
L1521F   &  1.0  $\times$ 0.72, 26.0     &  1.04  $\times$ 0.75, 26.8         &    11.8  $\times$ 11.8, 0     \\
L1527   &  0.68  $\times$ 0.58, 50.7   &   0.69  $\times$ 0.60, 48.9    &    11.8  $\times$ 11.8, 0     \\
L1157   &  0.60  $\times$ 0.46, -1.3    		&  0.61  $\times$ 0.48, -1.7              &    11.8  $\times$ 11.8, 0     \\
GF9-2   &  0.85  $\times$ 0.75, -161.8  	&  0.88  $\times$ 0.78, 22.0            &    11.8  $\times$ 11.8, 0   \\
\hline
\end{tabular}
\end{table*}

% Beam sizes N2H+ %
\begin{table*}[h]
\centering
\caption{Beam sizes of the N$_{2}$H$^{+}$ (1$-$0) emission maps.}
\label{table:beams-n2hp}
\begin{tabular}{l|ccc}
\hline \hline
 \multicolumn{4}{c}{\hfill}  \\
 Source       & PdBI                    & PdBI+30m                 & 30m                  \\
  \hfill & \hfill &  \hfill & \hfill \\
        & Major ($\arcsec$) $\times$ Minor ($\arcsec$), Pos. Ang. ($^{\circ}$)  & Major ($\arcsec$) $\times$ Minor ($\arcsec$), Pos. Ang. ($^{\circ}$)  & Major ($\arcsec$) $\times$ Minor ($\arcsec$), Pos. Ang. ($^{\circ}$) \\
\hline
L1448-2A &   1.73  $\times$ 1.36, 64.5     &   1.84  $\times$ 1.46, 68.5    & 27.8  $\times$ 27.8, 0      \\
L1448-NB &   1.73  $\times$ 1.33, 52.4      &   1.88  $\times$ 1.48, 53.6    & 27.8  $\times$ 27.8, 0      \\
L1448-C &   1.77  $\times$ 1.40, 63.4     &   1.93  $\times$ 1.56, 67.0    & 27.8  $\times$ 27.8, 0     \\
IRAS2A &   1.72  $\times$ 1.33, 55.2   &   1.86  $\times$ 1.47, 55.3    & 27.8  $\times$ 27.8, 0     \\
SVS13-B &   1.82  $\times$ 1.29, 38.1     &   1.91  $\times$ 1.40, 40.4   & 27.8  $\times$ 27.8, 0      \\
IRAS4A &   1.83  $\times$ 1.33, 38.4     &   1.92  $\times$ 1.41, 42.3    & 27.8  $\times$ 27.8, 0      \\
IRAS4B &   1.77  $\times$ 1.27, 38.0      &   1.86  $\times$ 1.37, 39.7  & 27.8  $\times$ 27.8, 0      \\
IRAM04191    &  1.84  $\times$ 1.55, 53.2       &   2.06  $\times$ 1.72, 52.2     &     27.8  $\times$ 27.8, 0       \\
L1521F    &  1.83  $\times$ 1.50, 80.1    &   1.98  $\times$ 1.62, 81.5    &      27.8  $\times$ 27.8, 0       \\
L1527    &  1.74  $\times$ 1.33, 45.8      &   1.84  $\times$ 1.45, 50.2    &        27.8  $\times$ 27.8, 0        \\
L1157  &   1.49  $\times$ 1.10, 60.3    &   1.55  $\times$ 1.18, 59.6     &           27.8  $\times$ 27.8, 0     \\
GF9-2  &   1.50  $\times$ 1.11, 68.5     &   1.56  $\times$ 1.16, 67.0       &           27.8  $\times$ 27.8, 0      \\
\hline
\end{tabular}
\end{table*}

% Emission sizes and rms C18O %
\begin{table*}[h]
\centering
\caption{Emission size, rms noise, and integrated fluxes in the C$^{18}$O (2$-$1) emission maps.}
\label{table:details-obs-c18o}
\begin{tabular}{l|ccc|ccc|ccc}
\hline \hline
 \multicolumn{10}{c}{\hfill}  \\
  Source      & \multicolumn{3}{c|}{PdBI}                                   & \multicolumn{3}{c|}{PdBI+30m}                                & \multicolumn{3}{c}{30m}                                     \\
 \hfill & \multicolumn{3}{c|}{\hfill}&  \multicolumn{3}{c|}{\hfill} & \multicolumn{3}{c}{\hfill} \\
        &  r$_{max}$~\tablefootmark{a} & rms~\tablefootmark{b}              & Flux~\tablefootmark{c} & r$_{max}$~\tablefootmark{a}  & rms~\tablefootmark{b}              & Flux~\tablefootmark{c} &  r$_{max}$~\tablefootmark{a} & rms~\tablefootmark{b}              & Flux~\tablefootmark{c} \\
        & ($\arcsec$)  & (mJy~beam$^{-1}$) & (Jy) & ($\arcsec$)  & (mJy~beam$^{-1}$) & (Jy) & ($\arcsec$)  & (Jy~beam$^{-1}$) & (Jy)  \\
\hline
L1448-2A &  1.4  &  20.4 &   4.2             &  3.5  &  20.7 &  71.1           &  35   &  0.7  &   91.1 \\
L1448-NB & 1.6  &  32.9 &   0.7             &   6.9 &  32.6 &  32.6           &   25   &  0.4  &   191.7 \\
L1448-C & 1.0 &  23.5 &   1.1             &  3.0  &  23.4 &  10.4           &   25   &  0.6  &   95.5 \\
IRAS2A &  1.7  &  33.8 &    1.6         &  4.5  &  33.1 &  27.2          &   37  & 0.3  &   214.9 \\
SVS13-B &  0.5  &  20.1 &    5.8         &  7.0  &  19.8 &  35.8          &   42    & 0.9  &   252.3 \\
IRAS4A & 2.6  &  26.7 &    11.2         &  4.3  &  26.2 &  24.2          &   22   & 0.4  &   33.9 \\
IRAS4B &  1.5  &  25.8 &    4.8         &  2.2  &  25.2 &  7.8         &   35    & 0.4  &   137.6 \\
IRAM04191   &  ...  &  16.7 &  0.02    	    &  2.5  &  16.5 &   7.0   	     &   42    &  0.3  &  34.2  \\
L1521F   & 0.5  &  11.6 &  0.06    	    & 9.6 &  12.2 &   21.1   	     &   40    &  0.1  &  49.0  \\
L1527   &  1.1  &  20.3 &  2.1    	    & 1.5  &  19.9 &   3.4   	     &   29    &  0.4  &  53.6  \\
L1157   & 0.8   & 27.7  &   1.3   		&  1.2  &  27.1  &   3.8           &   21    &  0.3  &   40.1 \\
GF9-2   & 1.9  & 13.5  &   0.9   		& 4.3  &  13.3  &   11.9           &  35    &  0.3  &   34.0 \\
\hline
\end{tabular}
\tablefoot{
\tablefoottext{a}{r$_{max}$ value defined as the mean radius of the area where integrated molecular line emission is detected above a 5$\sigma$ contour.}
\tablefoottext{b}{rms noise levels per channel computed with a spectral resolution of 0.27~km~s$^{-1}$ for all maps.}
\tablefoottext{c}{Integrated fluxes computed inside the PdBI primary beam at 1.3~mm ($\sim$20$\arcsec$) for the PdBI and combined PdBI+30m datasets and inside the primary beam (2 $\times$ HPBW $\sim$25$\arcsec$ at 1.3~mm) for the 30m datasets.}
}
\end{table*}

% Emission sizes and rms N2H+ %
\begin{table*}[h]
\centering
\caption{Emission size, rms noise, and integrated fluxes in the N$_{2}$H$^{+}$ (1$-$0) emission maps.}
\label{table:details-obs-n2hp}
\begin{tabular}{l|ccc|ccc|ccc}
\hline \hline
 \multicolumn{10}{c}{\hfill}  \\
 Source       & \multicolumn{3}{c|}{PdBI}                                    & \multicolumn{3}{c|}{PdBI+30m}                                & \multicolumn{3}{c}{30m}                                      \\
  \hfill & \multicolumn{3}{c|}{\hfill}&  \multicolumn{3}{c|}{\hfill} & \multicolumn{3}{c}{\hfill} \\
        &  r$_{max}$~\tablefootmark{a} & rms~\tablefootmark{b}              & Flux~\tablefootmark{c} &  r$_{max}$~\tablefootmark{a} & rms~\tablefootmark{b}              & Flux~\tablefootmark{c} &  r$_{max}$~\tablefootmark{a} & rms~\tablefootmark{b}              & Flux~\tablefootmark{c}  \\
        &  ($\arcsec$) & (mJy~beam$^{-1}$) & (Jy) &  ($\arcsec$) & (mJy~beam$^{-1}$) & (Jy) & ($\arcsec$)  & (Jy~beam$^{-1}$) & (Jy)  \\
\hline
L1448-2A &  13.1   &  9.0  &  2.8   &  21.4  &   9.5 &  153.4 &  35  &  0.8   &  136.7   \\
L1448-NB &  12.9   &  9.9  &  17.0   &  21.5   &   10.4 &  355.0 & 39 &  1.1   &  397.4   \\
L1448-C &   9.4  &  7.9  &  2.8   &  15.7  &   8.5 &  160.6 &  53  &  0.9   &  202.4   \\
IRAS2A &    14.0  &  9.9  &  10.2   &   23.0   &   10.0 &  174.6 &  40  &  0.7   &  217.0  \\
SVS13-B & 11   &  6.5  &  13.3   &  14.9   &   8.1 &   275.8 & 53  &  0.5   &  311.9   \\
IRAS4A &  10.0   &  9.2  &  3.6   &  23.1   &   9.3 &  160.0 & 50  &  0.6   &  247.4  \\
IRAS4B &  6.2   &  7.9  &  2.7   &  22.4  &   8.3 &  191.2 & 45  &  0.7   &  203.2   \\
IRAM04191    &  4.7  &   5.6   &  2.7    &  19.6   &   5.5  &  46.7 &    32    &   0.3  & 52.5   \\
L1521F    &  5.8 &   5.6   &  0.05    &  16.0  &   5.8  &  78.1 &  46      &   0.4  & 115.8   \\
L1527    & 13.8  &   6.6   &  0.2    &  22.2   &   6.6  &  48.1    &   55       &   0.3  & 45.5    \\
L1157  &  6.5   &  9.1   &   3.0  &  11.3   & 9.0 &  70.9    &    38          &   0.6  &  66.0 \\
GF9-2  &  6.4   &  5.8   &   1.3  &   18.7   & 5.8 &  51.7    &   32          &   0.5  &  66.0 \\
\hline
\end{tabular}
\tablefoot{
\tablefoottext{a}{r$_{max}$ value defined as the mean radius of the area where integrated molecular line emission is detected above a 5$\sigma$ contour.}
\tablefoottext{b}{rms noise levels per channel computed with a spectral resolution of 0.13~km~s$^{-1}$ for all maps.}
\tablefoottext{c}{Integrated fluxes computed inside the PdBI primary beam at 3~mm ($\sim$50$\arcsec$) for the PdBI and combined PdBI+30m datasets and inside the primary beam (2 $\times$ HPBW $\sim$60$\arcsec$ at 3~mm) for the 30m datasets.}
}
\end{table*}

%%%%%%%%%%%% VELOCITY GRADIENTS %%%%%%%%%%%%
\clearpage 
\section{Details of construction method of PV$_\mathrm{rot}$ diagrams} \label{details-diagram-PV-construction}

In this appendix, we present in detail our method of quantifying centroid velocity variations. We have built large dynamic range PV$_\mathrm{rot}$ diagrams (see Sect. \ref{subsec:diagram-PV-construction}) by determining the position and the velocity in the three datasets. We analyzed the PdBI datasets in the (u,v) plane while the combined and 30m datasets are analyzed in the image plane.

\subsection{Analysis of PdBI datasets in the (u,v) plane}
The PdBI datasets provide the best spatial resolution available in the CALYPSO sample, allowing us to constrain the velocity variations at the smallest scales of the protostellar envelopes.
Because the (u,v)-coverage of our PdBI dataset is limited, inside the C$^{18}$O PdBI emitting size of the source (see R$_{i}$ in Tables \ref{table:radius-lines} and \ref{table:details-obs-c18o}), we work directly on the visibilities to avoid the complex imaging and deconvolution processes involved in image plane analysis and to characterize the peak of emission at a given velocity with a higher astrometric precision.
We fit the PdBI visibilities of each channel with an elliptical Gaussian source in order to determine the centroid position of the emission in each velocity bin. In the cases where the emission shows irregular intensity distributions with multiple peaks, this analysis recovers the most intense peak at a given velocity. Our results are consistent with those of \cite{Maret20} who are doing a similar analysis in the (u,v) plane. We note that they are not strictly identical because \cite{Maret20} consider only pixels at scales $<$2$\arcsec$ from the central position to study the disk kinematics whereas we are interested in the kinematics of the envelope scales between 50 and 5000~au. \cite{Maret20} also did some tests from synthetic visibilities with an interferometer and find that working in the uv-plane provides a good estimate of the rotation profile at small scales.

We report in the PV$_\mathrm{rot}$ diagrams the fitting results of sources showing a velocity gradient in at least two channels in the PdBI channel maps (see Appendix \ref{sec:comments-indiv-sources}). 
We identify a gradient as a variation in the central position of the C$^{18}$O emission from one side of the continuum peak in the redshifted velocities to the other side in the blueshifted velocities with respect to the systemic velocity. The fitting results in channel maps of sources which do not exhibit such an organized spatial distribution of velocities are considered as upper limits.
We only consider the fitting results in channels where the C$^{18}$O emission is detected with a signal-to-noise ratio higher than 5 and with a fit central position consistent with the equatorial axis, namely at a position angle $<$|45$^{\circ}|$ with respect to the equatorial axis. These criteria on the position angle prevent the selection of PdBI C$^{18}$O points from being affected by the typical error of $\pm$10$^{\circ}$ on the direction of the equatorial axis.
When these three criteria are satisfied, we project the fit centroid position onto the equatorial axis to constrain the PV$_\mathrm{rot}$ diagram at small scales. The position errors are derived from our modeled elliptical Gaussian emitting source. The errors on the velocity are related to the channel width (0.2~km~s$^{-1}$ in C$^{18}$O emission from the PdBI datasets). We note that most of the CALYPSO sources show an optically thin emission at small scales (except IRAS4A, see Appendix \ref{fig:column-density-maps-IRAS4A}), thus, the intensity peak may be located in the equatorial plane where the density is typically higher. In the case that the centroid position in a channel map projected onto the plane of the sky is not along the equatorial axis while the centroid position in 3D actually belongs to the equatorial plane, the apparent distance in the channel map would underestimate the true distance of the centroid emission to the rotation axis because of its location in the third dimension. We did not apply any further correction on the velocities due to the position angle with respect to the equatorial axis of the fit central position. The associated rotational velocity of a fit central position close to 45$^{\circ}$ with respect to the equatorial axis could be over-estimated because more contaminated by the ejection or collapse motions at small scales than a fit central position closer to the equatorial axis. 
Restraining the analysis to the fit central positions with a position angle $<$|20$^{\circ}|$ with respect to the equatorial axis results in an additional uncertainty of $\pm$0.1 on the indices reported in Table \ref{sec:systemic-velocity}. This uncertainty is consistent with the systematic uncertainty of $\pm$0.1 which has to be added to account for the uncertainties in the equatorial axis directions (see Sect. \ref{sec:results-PV-diagram}).

\begin{figure*}[!ht]
\centering
\includegraphics[scale=0.5,angle=0,trim=0cm 1.5cm 0cm 1.5cm,clip=true]{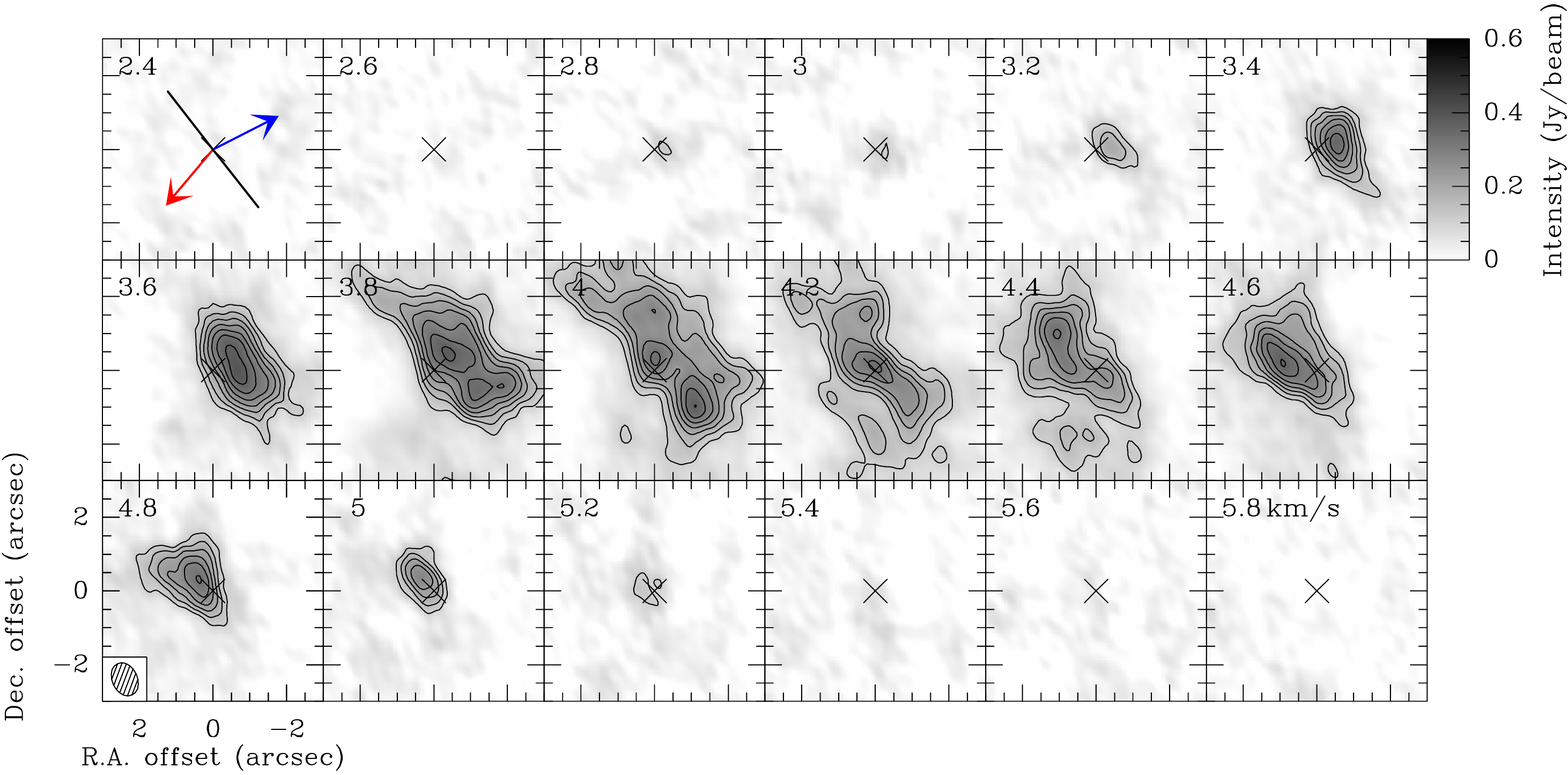}
\caption{Channel maps of the C$^{18}$O (2$-$1) emission from PdBI in L1448-2A. In the first panel, the outflow directions are shown by blue and red solid arrows and the equatorial axis is represented by a solid black line. The clean beam is shown by an ellipse on the bottom left. Contours are drawn in black solid lines at 5$\sigma$, 7$\sigma$ and so on. The systemic velocity is estimated to be $\mathrm{v}_\mathrm{sys}=$4.0~km~s$^{-1}$ (see Table \ref{table:chi2-fit-lines}).
}
\label{fig:channel-maps-L1448-2A}
\end{figure*}

\begin{figure*}[!ht]
\centering
\includegraphics[scale=0.5,angle=0,trim=0cm 1.5cm 0cm 1.5cm,clip=true]{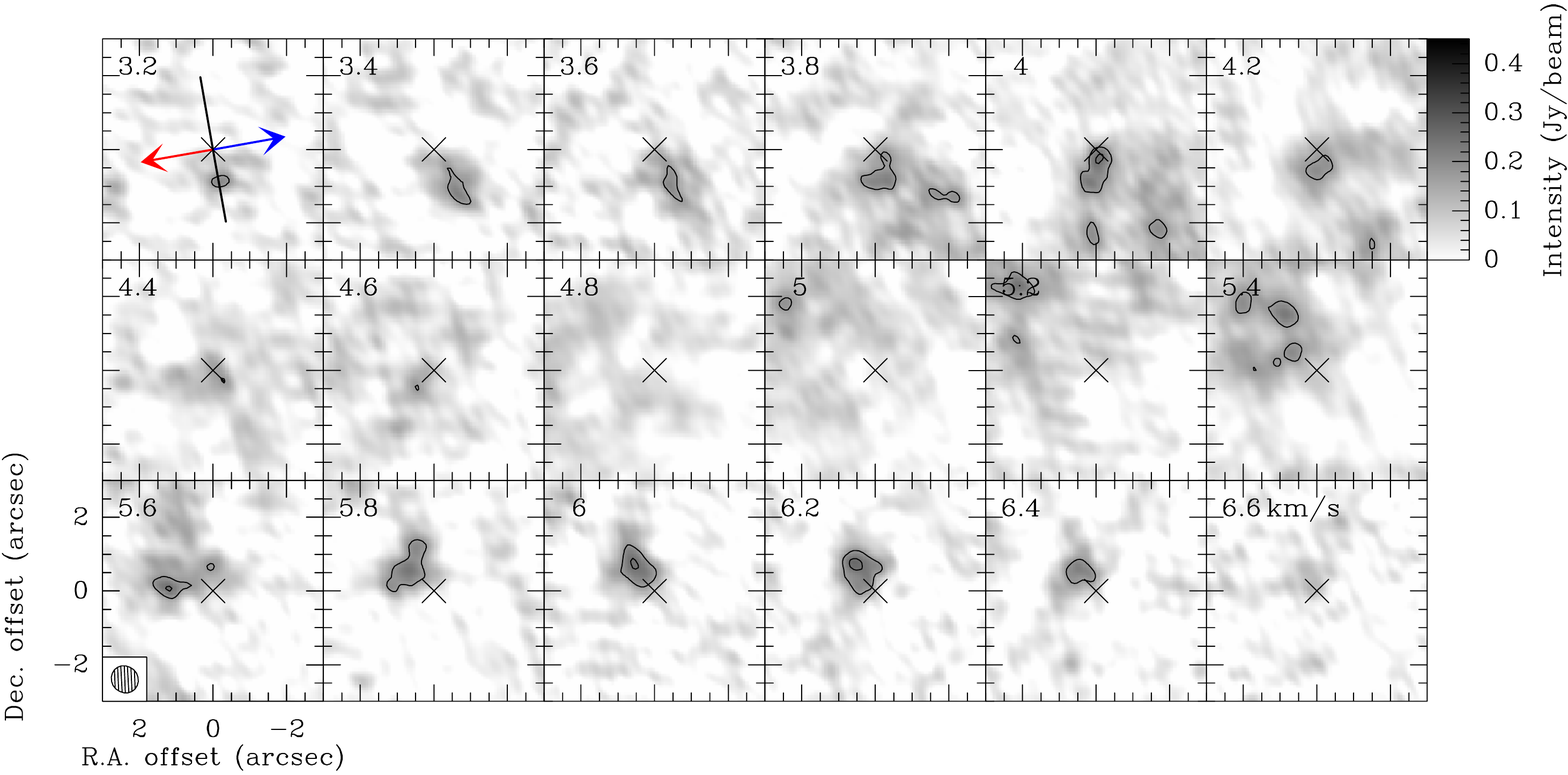}
\caption{Same as Figure \ref{fig:channel-maps-L1448-2A}, but for L1448-NB. The systemic velocity is estimated to be $\mathrm{v}_\mathrm{sys}=$4.6~km~s$^{-1}$ (see Table \ref{table:chi2-fit-lines}).
}
\label{fig:channel-maps-L1448NB}
\end{figure*}

\begin{figure*}[!ht]
\centering
\includegraphics[scale=0.5,angle=0,trim=0cm 1.5cm 0cm 1.5cm,clip=true]{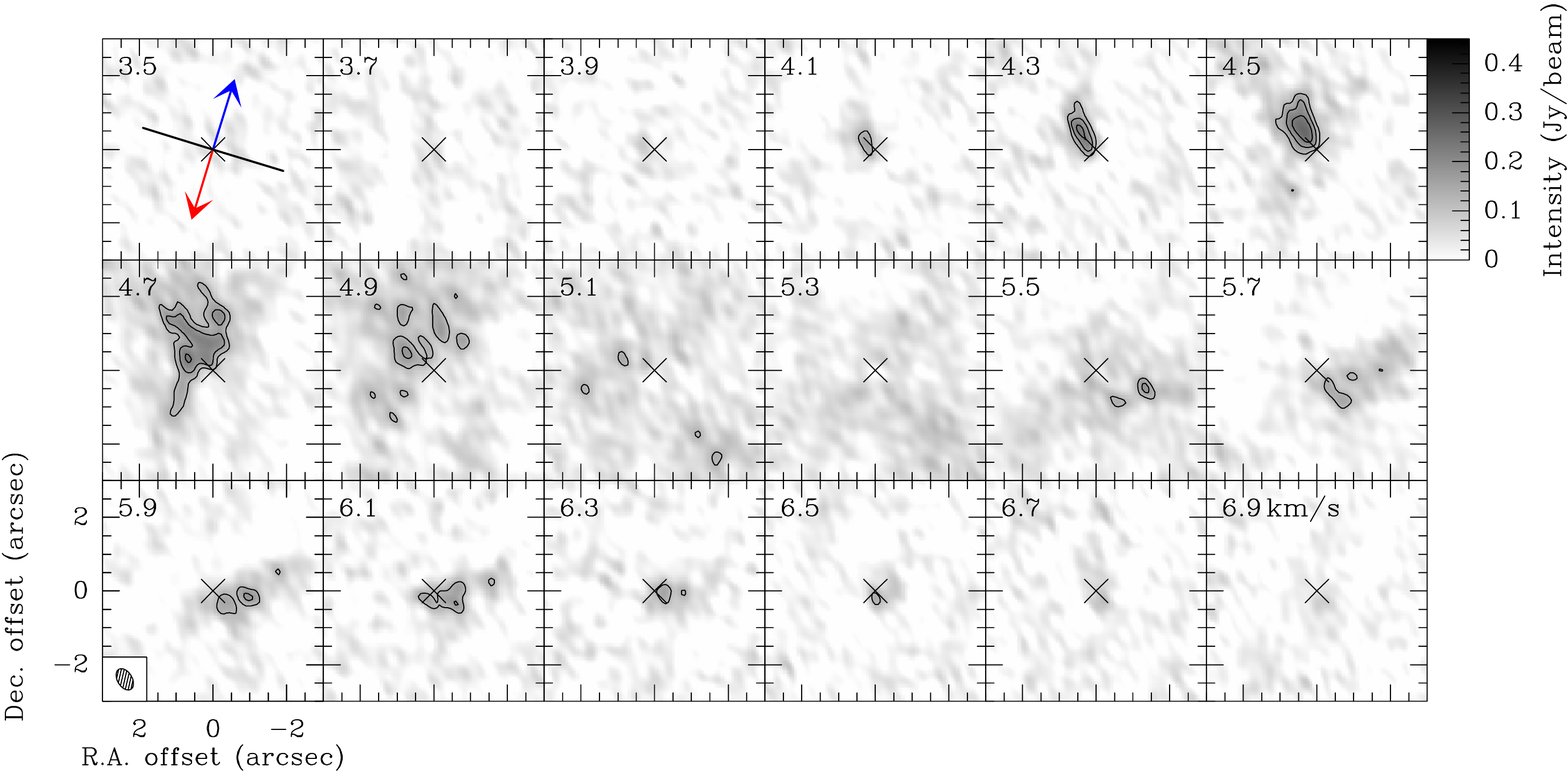}
\caption{Same as Figure \ref{fig:channel-maps-L1448-2A}, but for L1448-C. The systemic velocity is estimated to be $\mathrm{v}_\mathrm{sys}=$5.1~km~s$^{-1}$ (see Table \ref{table:chi2-fit-lines}).
}
\label{fig:channel-maps-L1448C}
\end{figure*}

\begin{figure*}[!ht]
\centering
\includegraphics[scale=0.5,angle=0,trim=0cm 1cm 0cm 1.5cm,clip=true]{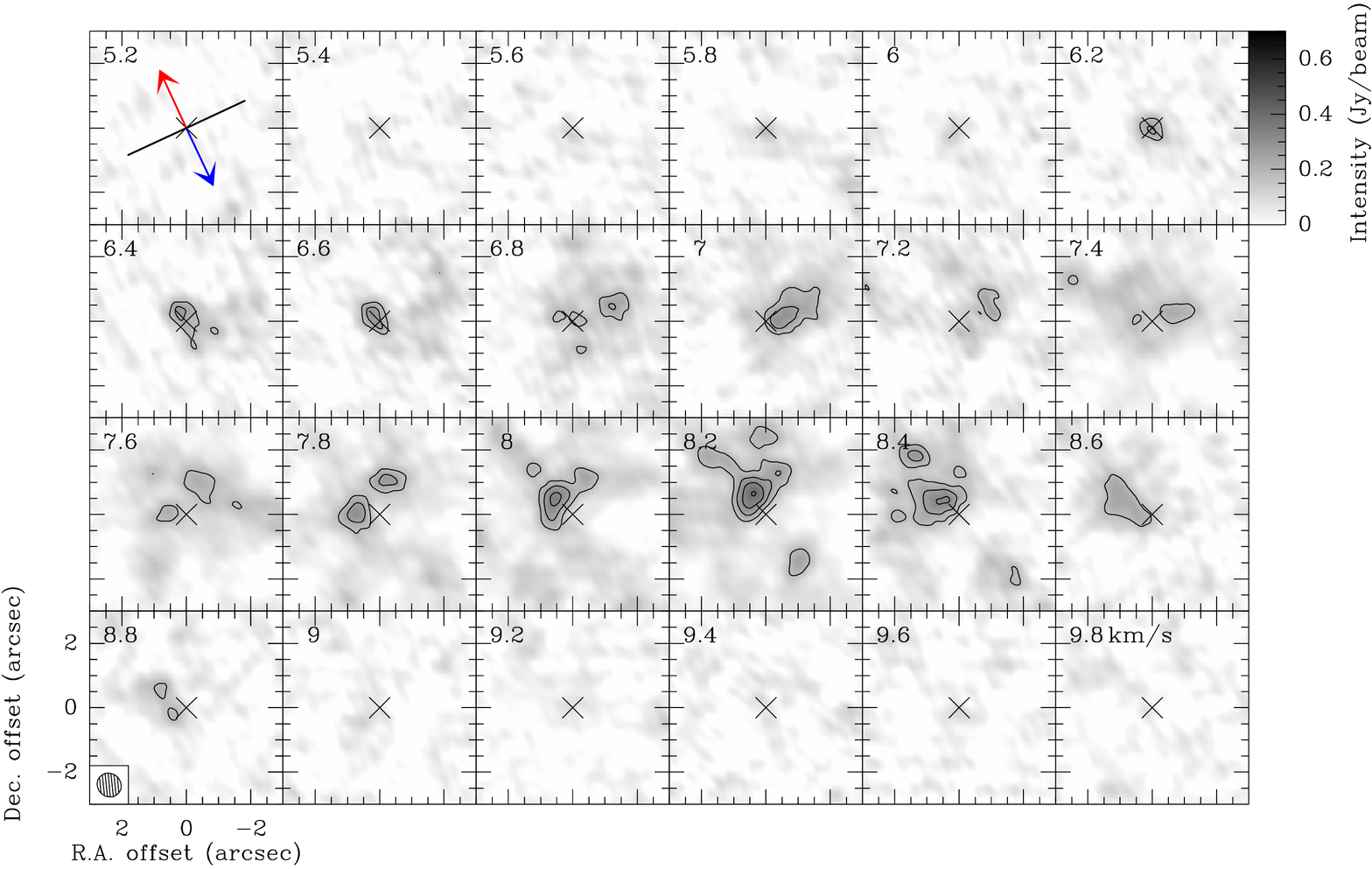}
\caption{Same as Figure \ref{fig:channel-maps-L1448-2A}, but for IRAS2A. The systemic velocity is estimated to be $\mathrm{v}_\mathrm{sys}=$7.6~km~s$^{-1}$ (see Table \ref{table:chi2-fit-lines}).
}
\label{fig:channel-maps-IRAS2A}
\end{figure*}

\begin{figure*}[!ht]
\centering
\includegraphics[scale=0.5,angle=0,trim=0cm 1.5cm 0cm 1.5cm,clip=true]{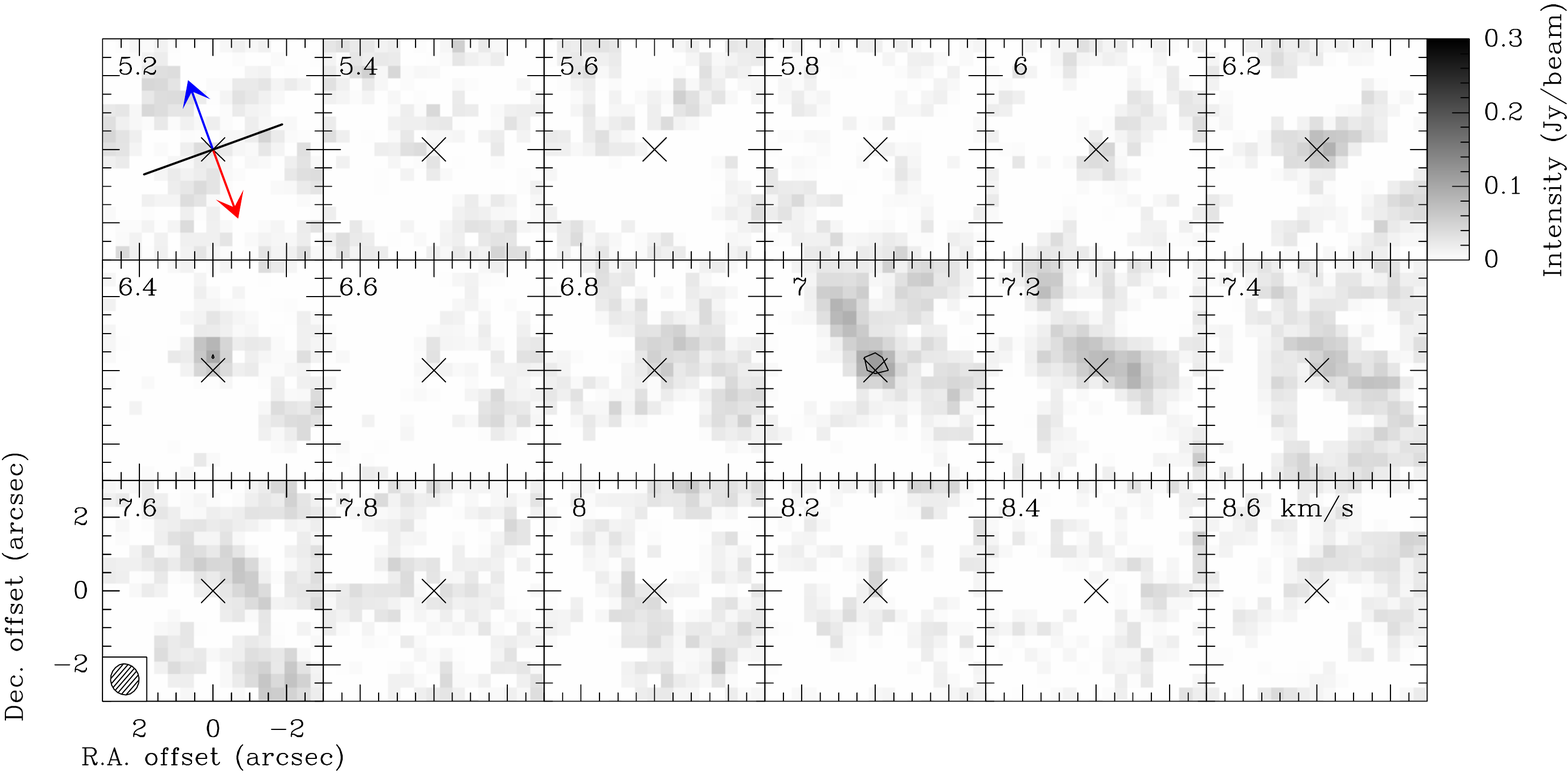}
\caption{Same as Figure \ref{fig:channel-maps-L1448-2A}, but for IRAM04191. The systemic velocity is estimated to be $\mathrm{v}_\mathrm{sys}=$6.6~km~s$^{-1}$ (see Table \ref{table:chi2-fit-lines}).
}
\label{fig:channel-maps-IRAM04191}
\end{figure*}

\begin{figure*}[!ht]
\centering
\includegraphics[scale=0.5,angle=0,trim=0cm 1.5cm 0cm 1.5cm,clip=true]{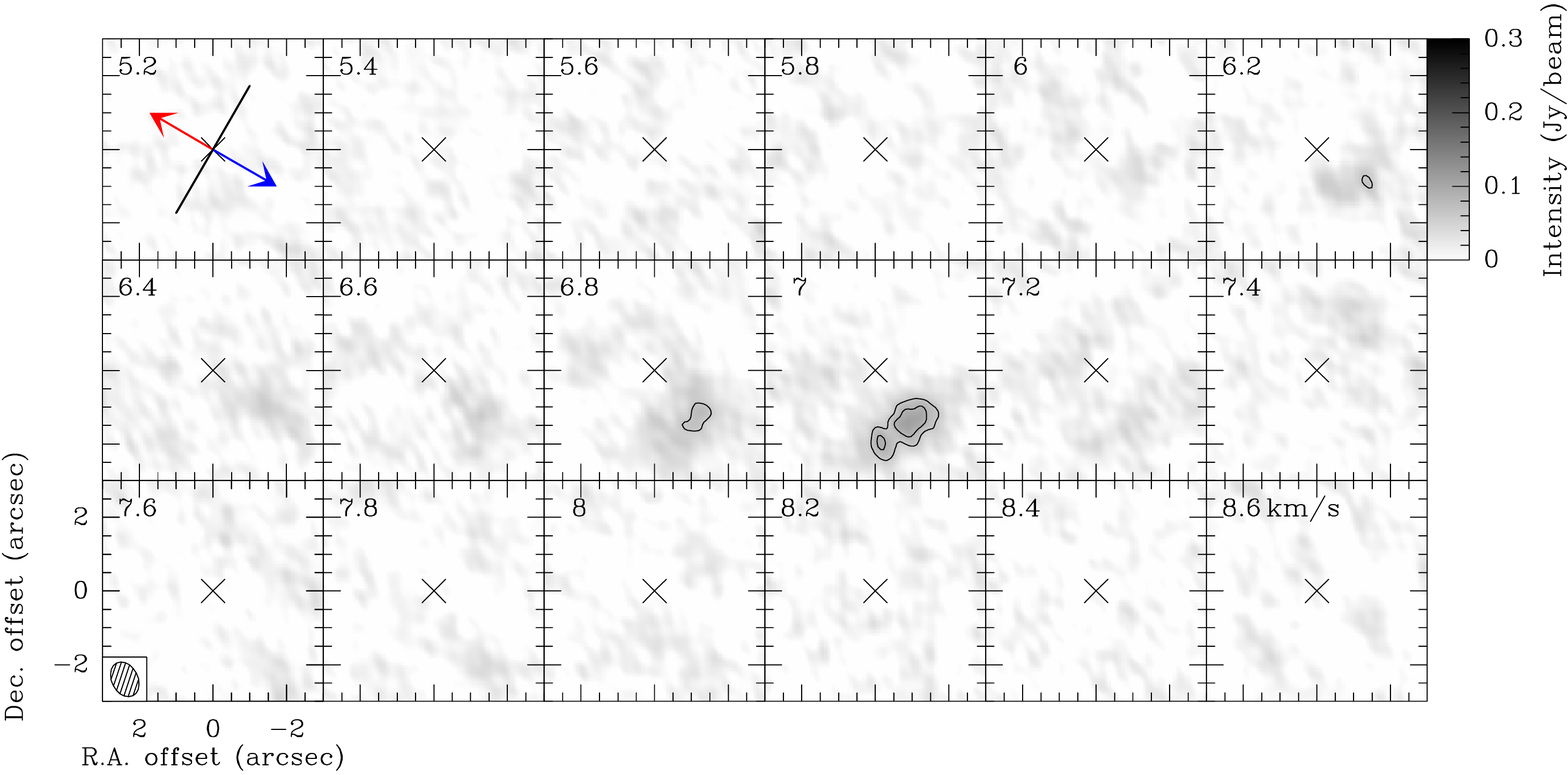}
\caption{Same as Figure \ref{fig:channel-maps-L1448-2A}, but for L1521F. The systemic velocity is estimated to be $\mathrm{v}_\mathrm{sys}=$6.55~km~s$^{-1}$ (see Table \ref{table:chi2-fit-lines}).
}
\label{fig:channel-maps-L1521F}
\end{figure*}

\begin{figure*}[!ht]
\centering
\includegraphics[scale=0.5,angle=0,trim=0cm 1cm 0cm 1.5cm,clip=true]{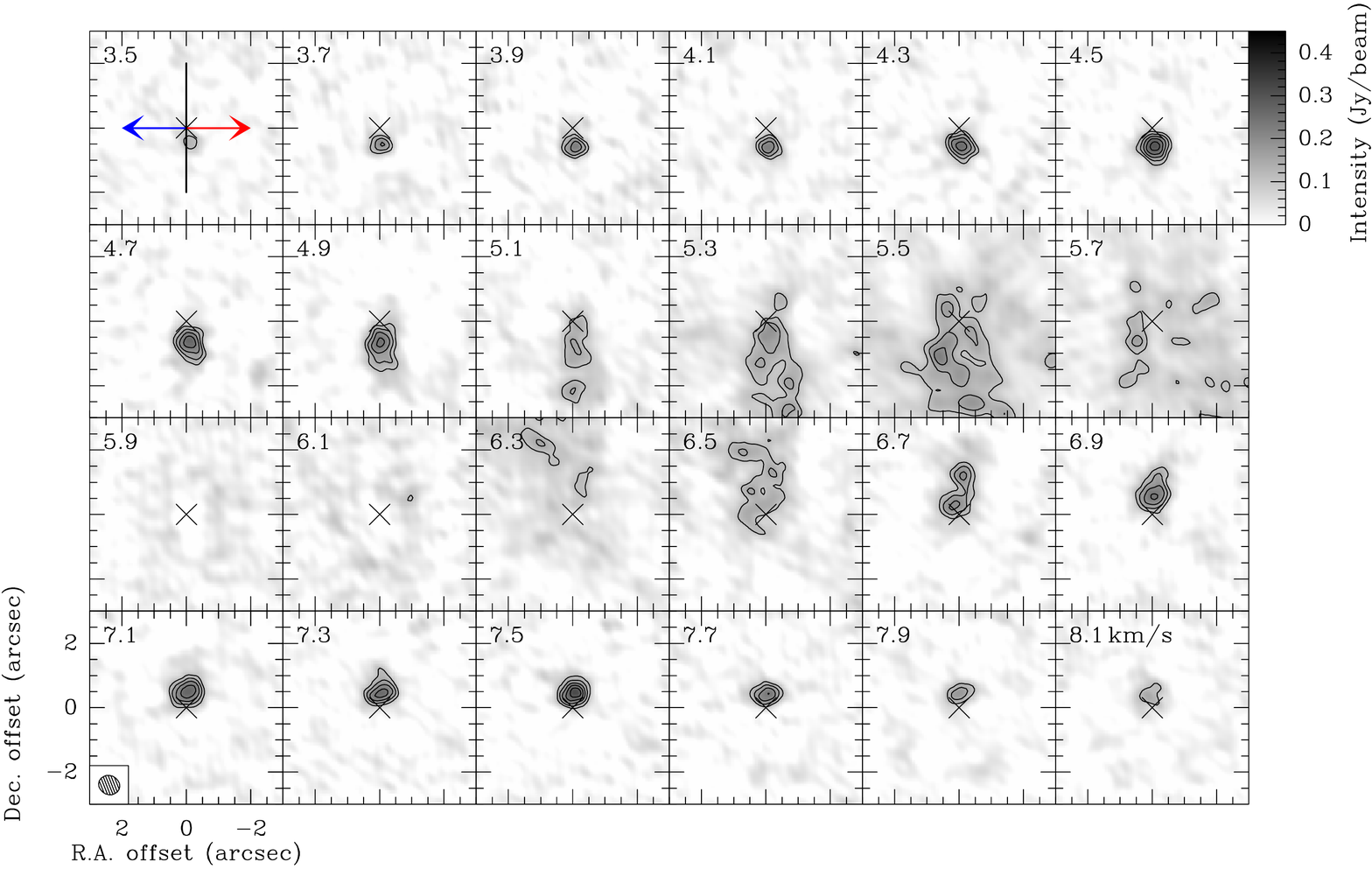}
\caption{Same as Figure \ref{fig:channel-maps-L1448-2A}, but for L1527. The systemic velocity is estimated to be $\mathrm{v}_\mathrm{sys}=$5.8~km~s$^{-1}$ (see Table \ref{table:chi2-fit-lines}).
}
\label{fig:channel-maps-L1527}
\end{figure*}

\begin{figure*}[!ht]
\centering
\includegraphics[scale=0.5,angle=0,trim=0cm 1.5cm 0cm 1.5cm,clip=true]{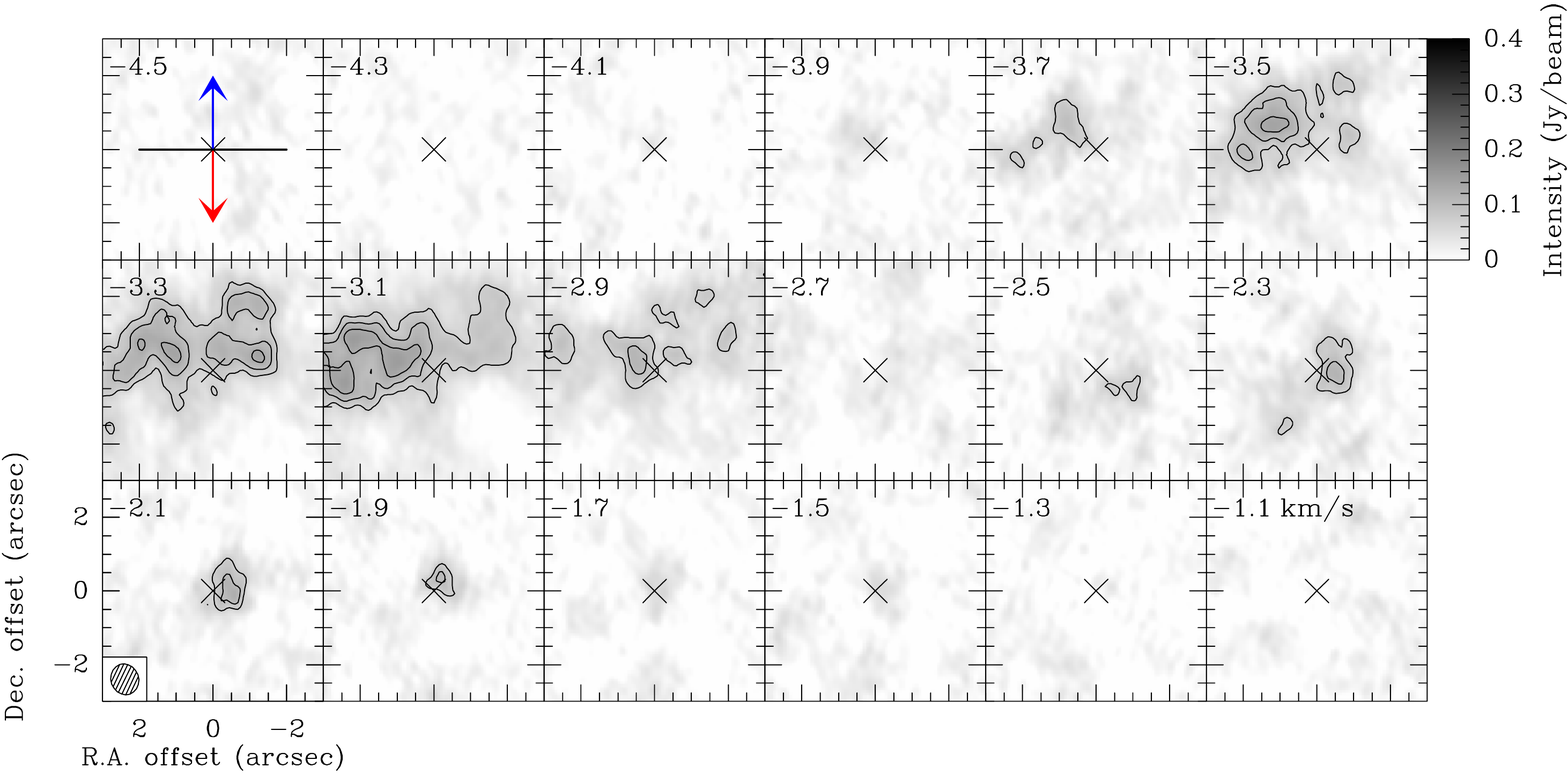}
\caption{Same as Figure \ref{fig:channel-maps-L1448-2A}, but for GF9-2. The systemic velocity is estimated to be $\mathrm{v}_\mathrm{sys}=$-2.8~km~s$^{-1}$ (see Table \ref{table:chi2-fit-lines}).
}
\label{fig:channel-maps-GF92}
\end{figure*}

Seven sources in our sample show a small-scale velocity gradient aligned along the equatorial axis in C$^{18}$O emission: L1448-2A, L1448-NB, L1448-C, IRAS2A, SVS13-B, L1527, and GF9-2 (see Figs. \ref{fig:channel-maps-L1448-2A}, \ref{fig:channel-maps-L1448NB}, \ref{fig:channel-maps-L1448C}, \ref{fig:channel-maps-IRAS2A}, \ref{fig:channel-maps-SVS13B}, \ref{fig:channel-maps-L1527}, and \ref{fig:channel-maps-GF92}, respectively).
For IRAM04191 and L1521F (see Figures \ref{fig:channel-maps-IRAM04191} and \ref{fig:channel-maps-L1521F}, respectively), the weak detection in C$^{18}$O emission from the PdBI datasets does not allow the diagram to be constrained to the smallest scales of the envelope ($r<$60~au).
For IRAS4A, IRAS4B, and L1157 (see Figures \ref{fig:channel-maps-IRAS4A}, \ref{fig:channel-maps-IRAS4B}, and \ref{fig:channel-maps-L1157}, respectively), the channel maps do not exhibit an organized spatial distribution of velocities along the equatorial axis at $r\lesssim$350~au: the central emission fit show a position angle $>$|45$^{\circ}|$ with respect to the equatorial axis, suggesting a contamination by the outflow kinematics. However, PdBI velocity maps exhibit weak gradients along the equatorial axis but working with the visibilities does not allow us to disentangle it from the emission from the outflows. Thus, for these three sources, we analyzed the PdBI observations in the image plane as described in the next section to constrain the PV$_\mathrm{rot}$ diagrams at scales of $r<$600~au.

\subsection{Analysis of combined and 30m datasets in the image plane}

We analyze the combined and 30m datasets in the image plane to probe the intermediate and outer scales of the envelope. For the pixel (i.e., a position) along the equatorial axis, we report the centroid velocity determined in the velocity maps in Sect. \ref{sec:velocity-maps}.
The errors on the velocity given by the Gaussian or HFS functions are reported on the PV$_\mathrm{rot}$ diagrams (see Fig. \ref{fig:PV-diagrams-1} and Appendix \ref{sec:comments-indiv-sources}). The position errors are related to the pixel size, which is itself related to the spatial resolution of the datasets.

To build PV$_\mathrm{rot}$ diagrams we only consider the velocity gradients with blue and red-shifted components with respect to the systemic velocity, centered on the central protostar and along the equatorial axis axis as expected from rotational motions. The values from velocity maps that do not exhibit such an organized spatial distribution of velocities are considered as upper limits on the rotational velocity.
Another requirement is the robustness of the centroid velocity offset with respect to the systemic velocity.
We only considered spectra with a signal-to-noise ratio higher than 5. From a Gaussian measurement, we assume the centroid velocity cannot be robustly determined with an accuracy better than one third of the spectral resolution: we only consider the C$^{18}$O points where the relative centroid velocity, $\mathrm{v}-\mathrm{v}_\mathrm{lsr}$, is $\geq \frac{\Delta \mathrm{v}}{3}$, with $\Delta \mathrm{v}$ the spectral resolution.
From a HFS measurement, the centroid velocity is determined much more precisely thanks to the larger number of components: we only consider the N$_{2}$H$^{+}$ points where $\mathrm{v}-\mathrm{v}_\mathrm{lsr}\geq \frac{\Delta \mathrm{v}}{3 \times \sqrt{6}}$. We note that the weakest component which is 7 times weaker than the strongest one is never detected with a signal-to-noise ratio higher than 2 in the spectra.

\subsection{Construction method of PV$_\mathrm{rot}$ diagrams}  \label{sec:method-construction-PV-diagrams}
The analysis described above allows us to determine the centroid position of the emission at a given velocity and the centroid velocity at a given position, respectively. By putting the results end-to-end, we build a PV$_\mathrm{rot}$ diagram with a high dynamic range from 50~au up to 5000~au for each source as follows (see the example of L1448-C in Fig. \ref{fig:PV-diagram-construction}):\\
\indent $\bullet$ At the smallest scales resolved by our dataset ($\sim$0.5$\arcsec$) and up to the PdBI C$^{18}$O emission size radius of the sources $R_i$ (see Tables \ref{table:details-obs-c18o} and \ref{table:radius-lines}, and Fig. \ref{fig:PV-diagram-construction}), we use the PdBI C$^{18}$O datasets to constrain the PV$_\mathrm{rot}$ diagram (see label "C$^{18}$O PdBI" in Fig. \ref{fig:PV-diagram-construction}). \\
\indent $\bullet$ Since the C$^{18}$O extended emission is filtered out by the interferometer, we used the combined C$^{18}$O emission to populate the PV$_\mathrm{rot}$ diagram at angular radii between $R_i$ and the 30m half-power beam width (HPBW $\sim$6$\arcsec$ at the C$^{18}$O frequency, see Table \ref{table:beams-c18o}). However, because the C$^{18}$O molecule freezes onto dust ice mantles at temperatures below $\sim$20K, the maximum radius up to which it remains the best tracer of the inner envelope can be smaller than the 30m HPBW. To determine the maximal radius $R_\mathrm{trans}$ where the C$^{18}$O molecule remains the best tracer, we calculate the C$^{18}$O and N$_{2}$H$^{+}$ column densities along the equatorial axis from the combined integrated intensity maps (see Appendix \ref{sec:column-density} and green points in Fig. \ref{fig:PV-diagram-construction}). $R_\mathrm{trans}$ is defined one of the following criteria, depending on the source (see Table \ref{table:radius-lines}):\\
(i) the radius from which the C$^{18}$O column density changes from a smooth decreasing profile to a noisy dispersion,\\
(ii) the maximum radius where the C$^{18}$O emission is detected with a signal-to-noise ratio higher than 5 along the equatorial axis,\\
(iii) the radius below which the N$_{2}$H$^{+}$ column density no longer traces the inner region and the profile flattens,\\
(iv) the radius where the N$_{2}$H$^{+}$ profile reaches its maximum in column density.\\
For most sources, $R_\mathrm{trans}$ is smaller than the 30m HPBW. We used the combined C$^{18}$O emission map to constrain the gas kinematics in the PV$_\mathrm{rot}$ diagram up to R$_\mathrm{trans}$ (see label "C$^{18}$O combined" in Fig. \ref{fig:PV-diagram-construction}).\\
\indent $\bullet$ At radii $r>R_\mathrm{trans}$, the N$_{2}$H$^{+}$ emission traces better the envelope dense gas. We use the combined N$_{2}$H$^{+}$ emission maps to analyze the envelope kinematics at intermediate scales, from $R_\mathrm{trans}$ to the 30m HPBW ($\sim$14$\arcsec$ at the N$_{2}$H$^{+}$ frequency, see Table \ref{table:beams-n2hp}). However, the N$_{2}$H$^{+}$ column density profile can reach a minimum value at radii $R_\mathrm{trans}< r<$ 30m HPBW due to the sensitivity of the combined datasets. In this case, the combined dataset is no longer the better dataset to provide a robust information on the velocity. We defined $R_\mathrm{int}$ the maximum radius up to which we use N$_{2}$H$^{+}$ emission from the combined dataset (see Table \ref{table:radius-lines} and see label "N$_{2}$H$^{+}$ combined" in Fig. \ref{fig:PV-diagram-construction}). \\
\indent $\bullet$ Beyond $R_\mathrm{int}$, we use the 30m N$_{2}$H$^{+}$ emission map to populate the PV$_\mathrm{rot}$ diagram up to the largest scales of the envelope (see label "N$_{2}$H$^{+}$ 30m" in Fig. \ref{fig:PV-diagram-construction}). We note $R_\mathrm{out}$ the maximum radius of the protostellar envelopes (see Table \ref{table:radius-lines}).\\

The sources in our sample show specific individual behaviors, therefore we adapted the method of building the PV$_\mathrm{rot}$ diagram described above on a case-by-case basis. Nevertheless we respected the transition radii as closely as possible (see Table \ref{table:radius-lines}). 

\begin{table*}[ht]
\centering
\caption{Transition radii between the different datasets (PdBI, combined, and 30m) and the two C$^{18}$O and N$_{2}$H$^{+}$ tracers used to build a high dynamic range position-velocity diagram for sources from the CALYPSO survey.}
\label{table:radius-lines}
\begin{tabular}{l|cc|ccl|cc|cc}
\hline \hline
\multicolumn{9}{c}{\hfill}  \\
 Source       & \multicolumn{5}{c|}{C$^{18}$O (2$-$1)} & \multicolumn{4}{c}{N$_{2}$H$^{+}$ (1$-$0)} \\
\hfill &     \multicolumn{5}{c|}{\hfill} &     \multicolumn{4}{c}{\hfill}  \\
\hfill & \multicolumn{2}{c|}{$R_\mathrm{i}$~\tablefootmark{a}}      & \multicolumn{3}{c|}{$R_\mathrm{trans}$~\tablefootmark{b}}     & \multicolumn{2}{c|}{$R_\mathrm{int}$~\tablefootmark{c}}   & \multicolumn{2}{c}{$R_\mathrm{out}$~\tablefootmark{d}}  \\
        & ($\arcsec$)  & (au)  & ($\arcsec$)  &  (au) &   & ($\arcsec$) & (au) & ($\arcsec$)   &    (au)       \\
\hline
L1448-2A &   1.4 &  410 &  4.2 & 1230 & (iii)  &  13.9  & 4070 &    18$\pm$5 &  5270$\pm$1470  \\
L1448-NB &   1.6   & 470  &  5.9 & 1730 &  (iii) &   13.9    & 4070   & 33$\pm$5  &  9670$\pm$1470   \\
L1448-C & 1.0  &  290  &  3.5  & 1230 &  (iii)  &  13.9 & 4070    &  25$\pm$2  &  7320$\pm$590       \\
IRAS2A  &  1.6 &  470  &   5.1 & 1490 & (i)  &  9  &    2640  & 34$\pm$6 &  9960$\pm$1760  \\
SVS13-B  &  0.5    &  145  &  2.0 & 590  & (i)  &  7  &  2050  &       7$\pm$3  &  2050$\pm$880    \\
IRAS4A~$^{\star}$ &   2.6 & 760  & 3.1 & 910 &  (i) & 5.5  &  1610    &  5.9$\pm$1.0 &  1730$\pm$ 290  \\
IRAS4B~$^{\star}$  & 1.5 & 440  & 2.1 & 615 & (ii)   &  3.5   &  1030  &  11$\pm$2 &  3220$\pm$590   \\
IRAM04191   &  ...   &  ...  &    5.6 & 780 &  (iv)  &  13.9 &  1950   &   100$\pm$11  & 14000$\pm$1540   \\
L1521F~$^{\star \star}$  &   0.5   &  70   &      5.9  & 820 &   ...     &    ...   & ...    &   32$\pm$3   &   4480$\pm$420      \\
L1527   & 1.1  &  150   &  3.3 & 460  &   (ii)   &  13.9 &  1950  &  $>$121  &   $>$16940      \\
L1157   & 0.8  &  280    &  1.8  & 630 &   (ii)  &  13.9 &   4890  &  45$\pm$5 &    15840$\pm$1760     \\
GF9-2  &   1.9 &  380   &     5.1 & 1020 &   (i)      &   ...      &   ... &   35$\pm$5 &      7000$\pm$1000    \\
\hline
\end{tabular}
\tablefoot{ 
\tablefoottext{a}{Outer radius of the C$^{18}$O emission from the PdBI datasets.}
\tablefoottext{b}{Transition radius between C$^{18}$O and N$_{2}$H$^{+}$ determined from column density profiles along the equatorial axis (see Appendix \ref{sec:column-density}). $R_\mathrm{trans}$ is defined following criteria (i), (ii), (iii), and (iv) which are detailed in Sect. \ref{sec:method-construction-PV-diagrams}.}
\tablefoottext{c}{Intermediate radius in N$_{2}$H$^{+}$ emission between the combined and 30m datasets.}
\tablefoottext{d}{Envelope radius from PdBI or 30m dust continuum emission. See Table \ref{table:sample} for the references.}
\tablefoottext{$\star$}{For IRAS4A and IRAS4B, we used the N$_{2}$H$^{+}$ emission from the PdBI datasets instead of the combined datasets to build the PV$_\mathrm{rot}$ diagram. In these cases, $R_\mathrm{int}$ is the maximum radius below which we used the PdBI dataset.}
\tablefoottext{$\star$ $\star$}{For L1521F, we only used the C$^{18}$O emission from the combined and 30m datasets to build the PV$_\mathrm{rot}$ diagram (see Appendix \ref{sec:comments-indiv-sources}). For this case, we used the combined dataset up to $R_\mathrm{trans}$ and we used the 30m dataset up to $R_\mathrm{out}$.}
}
\end{table*}

%%%%%%%%%%%% OPACITY VALUES %%%%%%%%%%%%
\clearpage
\section{Details of column density estimate} \label{sec:column-density}
To interpret the gas kinematics using both C$^{18}$O and N$_{2}$H$^{+}$ emission in a robust way, it is important to constrain from which scales in the envelope their emission comes from. This is especially important because we merged single-dish and interferometric datasets which are sensitive to different scales of the envelope (see Tables \ref{table:details-obs-c18o} and \ref{table:details-obs-n2hp}). 
Therefore, we built column density maps to quantify the scales probed by C$^{18}$O and N$_{2}$H$^{+}$ molecular lines. 
We assumed that the observed molecular transitions are optically thin ($\tau_{\nu} <<$1) and we used Eq. \eqref{column-density-thin} defined in \citet{Mangum15}:
\begin{equation}
N_{thin}~= ~\frac{3~k}{8~\pi^{3} ~\mu^{2}~ B_\mathrm{0}~ R_\mathrm{i}} ~ (T_\mathrm{ex}+\frac{h B}{3 k})~\frac{\exp(\frac{E_\mathrm{u}}{T_\mathrm{ex}})}{\exp(\frac{h \nu}{k T_\mathrm{ex}})-1} ~~ \frac{\int T_\mathrm{R} \, \mathrm{d}v}{\mathrm{J}_{\nu}(T_\mathrm{ex})-\mathrm{J}_{\nu}(T_\mathrm{bg})},
\label{column-density-thin}
\end{equation}
where $E_\mathrm{u}$ is the energy of the upper level, $\mu$ is the dipole moment of the molecule, $B_\mathrm{0}$ the rotational constant, $h$ Planck's constant, $k$ Boltzmann's constant, $R_\mathrm{i}$ the relative intensity of the component if the transition has a hyperfine structure, $T_\mathrm{ex}$ the excitation temperature, and $T_\mathrm{bg}$ the cosmic background temperature \citep{Mather94}. $\mathrm{J}_{\nu}$ is the 
effective radiation temperature defined by $\mathrm{J}_{\nu}(T)=\frac{h \nu}{k}~ \frac{1}{\exp(\frac{h \nu}{k T})-1}$. We report the values of each parameter for the two emission lines in Table \ref{table:parameters-column-density}.

\begin{table}[!ht]
\centering
\caption{Parameters used to determine the column density of the C$^{18}$O and N$_{2}$H$^{+}$ molecules from Eq. \eqref{column-density-thin}.}
\label{table:parameters-column-density}
\begin{tabular}{c|cc}
\hline \hline
\multicolumn{1}{c}{\hfill}  & \multicolumn{2}{c}{\hfill}   \\
Parameter     & N$_{2}$H$^{+}$ & C$^{18}$O   \\
\hfill  & \hfill  & \hfill    \\
\hline
$\nu$ (MHz)   & 93176.2595~\tablefootmark{$\star$}     & 219560.3541 \\
$E_{u}$ (K)   & 4.47172        & 15.8058      \\
$\mu$ (Debye) & 3.40           & 0.11049     \\
$B_\mathrm{0}$ (MHz) & 46586.87       & 54891.42    \\
$R_\mathrm{i}$       &    $\frac{3}{27}$~\tablefootmark{$\star$}            & 1           \\
\hline
\end{tabular}
\tablefoot{All values were obtained from the \href{http://www.astro.uni-koeln.de/cdms/catalog}{CDMS} \citep{Endres16}. \tablefoottext{$\star$}{The N$_{2}$H$^{+}$ column density is determined from the isolated hyperfine component ($1_{01}-0_{12}$).}
} 
\end{table}

Assuming local thermodynamical equilibrium (LTE), we can determine the excitation temperature from the gas temperature within the emission size of each tracer observed in the combined dataset (see Table \ref{table:radius-lines}). This hypothesis is valid when the density is higher than the critical density of the transition of C$^{18}$O (2$-$1) and N$_{2}$H$^{+}$ ($1_{01}-0_{12}$), estimated at $\sim$ 8.4 $\times$ 10$^3$~cm$^{-3}$ \citep{Flower01} and $\sim$2.6 $\times$ 10$^5$~cm$^{-3}$ \citep{Daniel05}, respectively at $T \sim$10~K. \cite{Belloche02} estimate for IRAM04191, one of the lowest envelope masses of our sample (see Table \ref{table:sample}), a density higher than 10$^5$~cm$^{-3}$ up to $r \sim$4000~au. Thus, LTE is a good assumption for both transitions in our sample.
As the dust temperature in the envelope is a good approximation for the gas kinetic temperature \citep{Ceccarelli96}, we used dust temperature profiles from CALYPSO PdBI observations assuming that the temperature distribution depends on the radius $r$ from the central stellar object and the internal luminosity of the source $L_\mathrm{int}$ (see Table \ref{table:sample}) as follows \citep{Terebey93}:
\begin{equation}
T_\mathrm{dust}(r)= 60~\times ~\left( \frac{r}{13400~au} \right)^{-q} ~ \left( \frac{L_\mathrm{int}}{10^5 L_{\odot}} \right)^{q/2},
\label{Tdust}
\end{equation}
with $q=0.4$ for the CALYPSO sample \citep{Terebey93, Maury18}.
We weighted the radial temperature distribution with the dust column density as a function of $r$, that is the amount of material at each radius, to estimate a single robust mean value of gas temperature (see Table \ref{table:tau-mean}). For C$^{18}$O, we used the PdBI flux at 1.3~mm calculated at different radii by \cite{Maury18}. For N$_{2}$H$^{+}$, we assumed a density profile of $\rho \propto r^{-p}$, with the index $p$ equal to the values determined at 3~mm by \cite{Maury18}.
The excitation temperature values adopted for each source to derive the column density are given in Table \ref{table:tau-mean}.
There are two exceptions in the CALYPSO sample: IRAM04191 and L1521F, which have the lowest luminosities of the CALYPSO sample (see Table \ref{table:sample}). The dust temperature profile of these sources from the PdBI dataset cannot be determined because the temperature profile is dominated by external heating. 
\cite{Belloche02} and \cite{Tokuda14,Tokuda16} estimate the dust temperature at $\sim$10~K at scales $\sim$2000~au. From Eq. \eqref{Tdust} and from values of $L_\mathrm{int}$ estimated for these two sources (0.05~L$_{\odot}$ for IRAM04191, \citealt{Andre00}, and 0.035~L$_{\odot}$ for L1521F, \citealt{Tokuda16}) we determined mean excitation temperatures of 20~K for the C$^{18}$O molecule and 10~K for the N$_{2}$H$^{+}$ molecule (see Table \ref{table:tau-mean}).

We note that we did not use the excitation temperature determined by fitting the HFS line profile for the N$_{2}$H$^{+}$ molecule. 
For most sources in the CALYPSO sample, the average spectrum from the combined dataset within the central 14$\arcsec \times$14$\arcsec$ does not follow the relative intensities of the hyperfine components expected under LTE conditions (3,3,7,5,3,5,1 from the isolated hyperfine component $1_{01}-0_{12}$;  \citealt{Endres16}). Thus, either the emission does not satisfy the LTE conditions or this is an effect of the partial opacity of emission. Moreover, some adjacent hyperfine components are too close to be spectroscopically separated despite the good spectral resolution of the combined dataset (0.13~km~s$^{-1}$ at 3~mm). To minimize the error propagation, we only considered the isolated hyperfine component ($1_{01}-0_{12}$) to determine the N$_{2}$H$^{+}$ column density \citep{Caselli95,Caselli02}.

\begin{figure}[!ht]
\centering
\includegraphics[scale=0.3,angle=0,trim=0cm 3cm 0cm 3.5cm,clip=true]{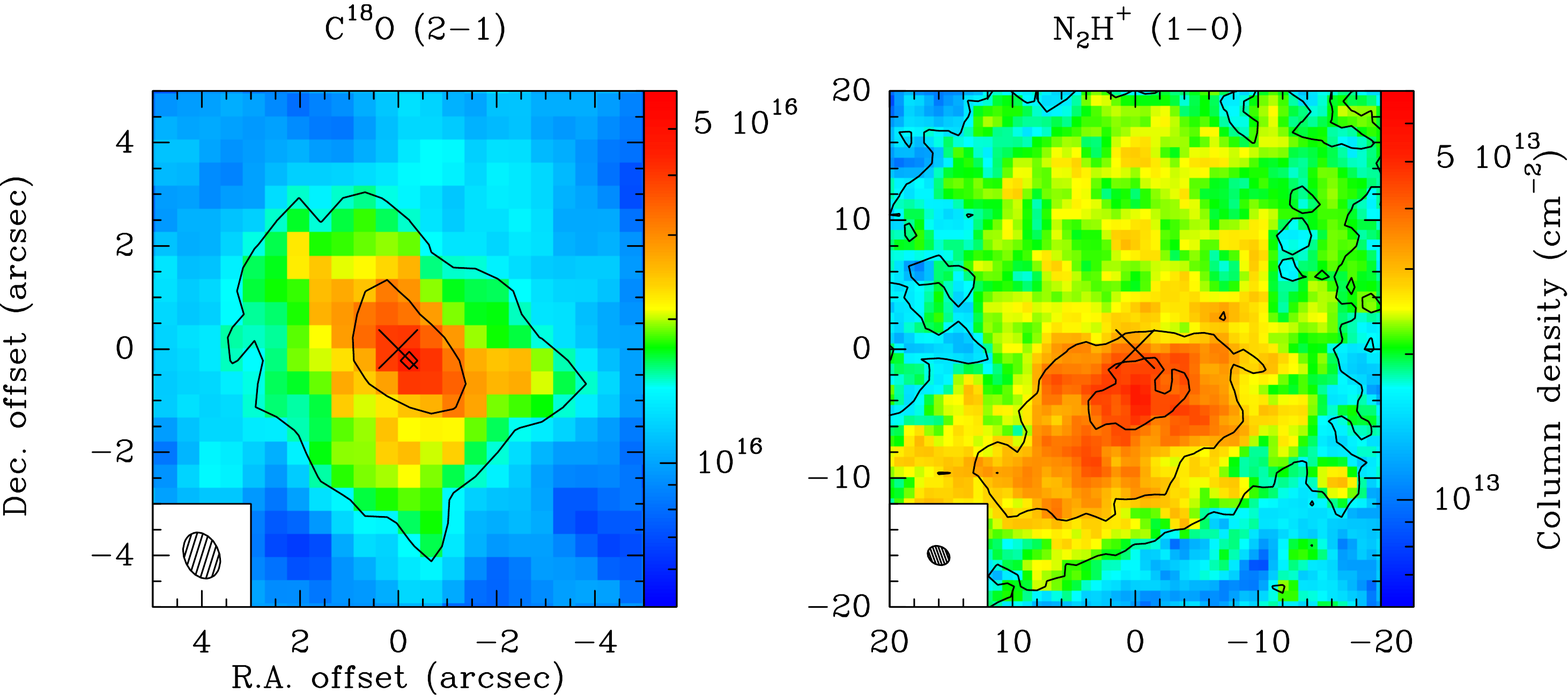}
\caption{Column density maps of C$^{18}$O (2$-$1) (left) and N$_{2}$H$^{+}$ (1$-$0) (right) emission from the combined dataset for L1448-2A. The black cross represents the middle position between the binary protostars determined with the 1.3~mm dust continuum emission. The N$_{2}$H$^{+}$ column density map was determined from the isolated hyperfine component ($1_{01}-0_{12}$) but the contours represent the integrated intensity of the 7 hyperfine components. The integrated intensity contours in black are the same as in Fig. \ref{fig:intensity-maps-L1448-2A}.}
\label{fig:column-density-maps-L1448-2A}
\end{figure}

\begin{figure*}[!ht]
\centering
\includegraphics[scale=0.3,angle=0,trim=0cm 3cm 0cm 3.5cm,clip=true]{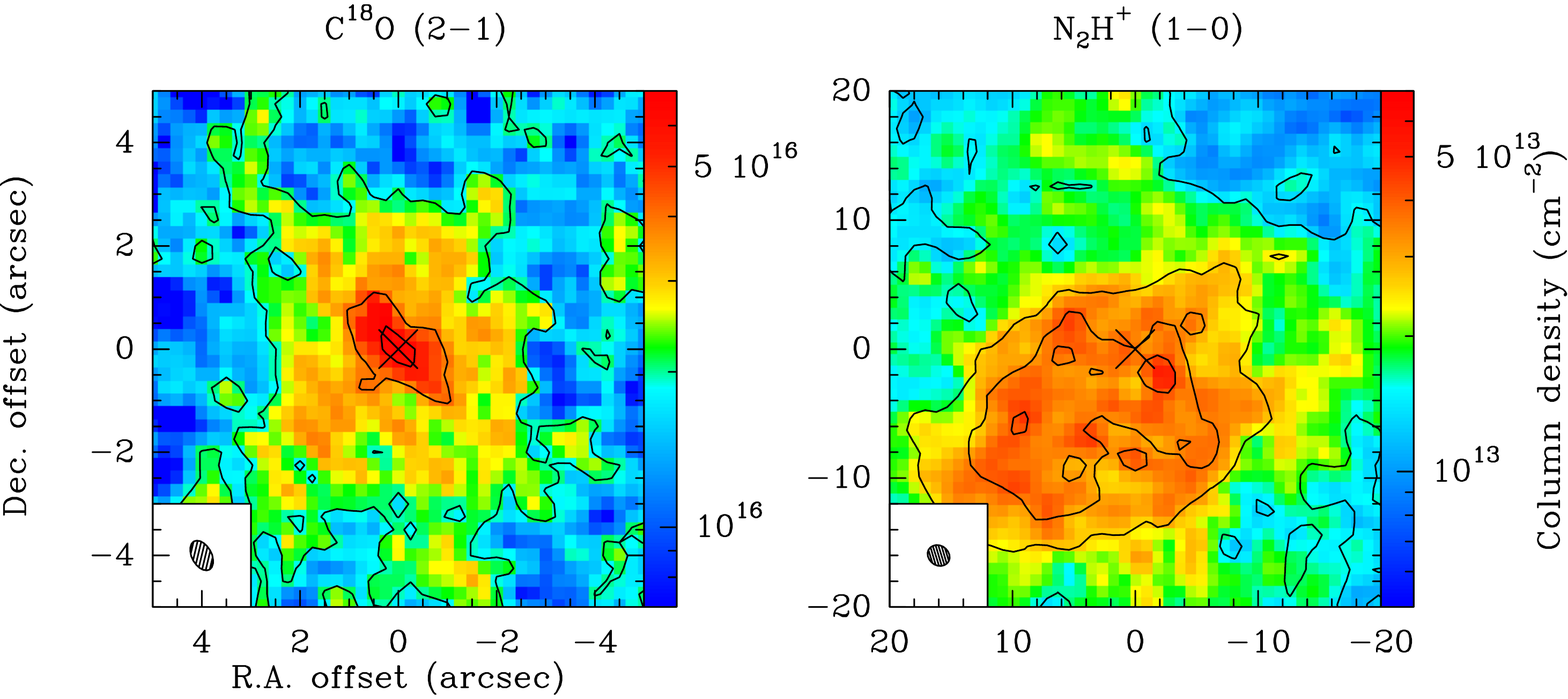}
\caption{Same Figure as \ref{fig:column-density-maps-L1448-2A} for L1448-C. 
}
\label{fig:column-density-maps-L1448-C}
\end{figure*}

\begin{figure*}[!ht]
\centering
\includegraphics[scale=0.3,angle=0,trim=0cm 3cm 0cm 3.5cm,clip=true]{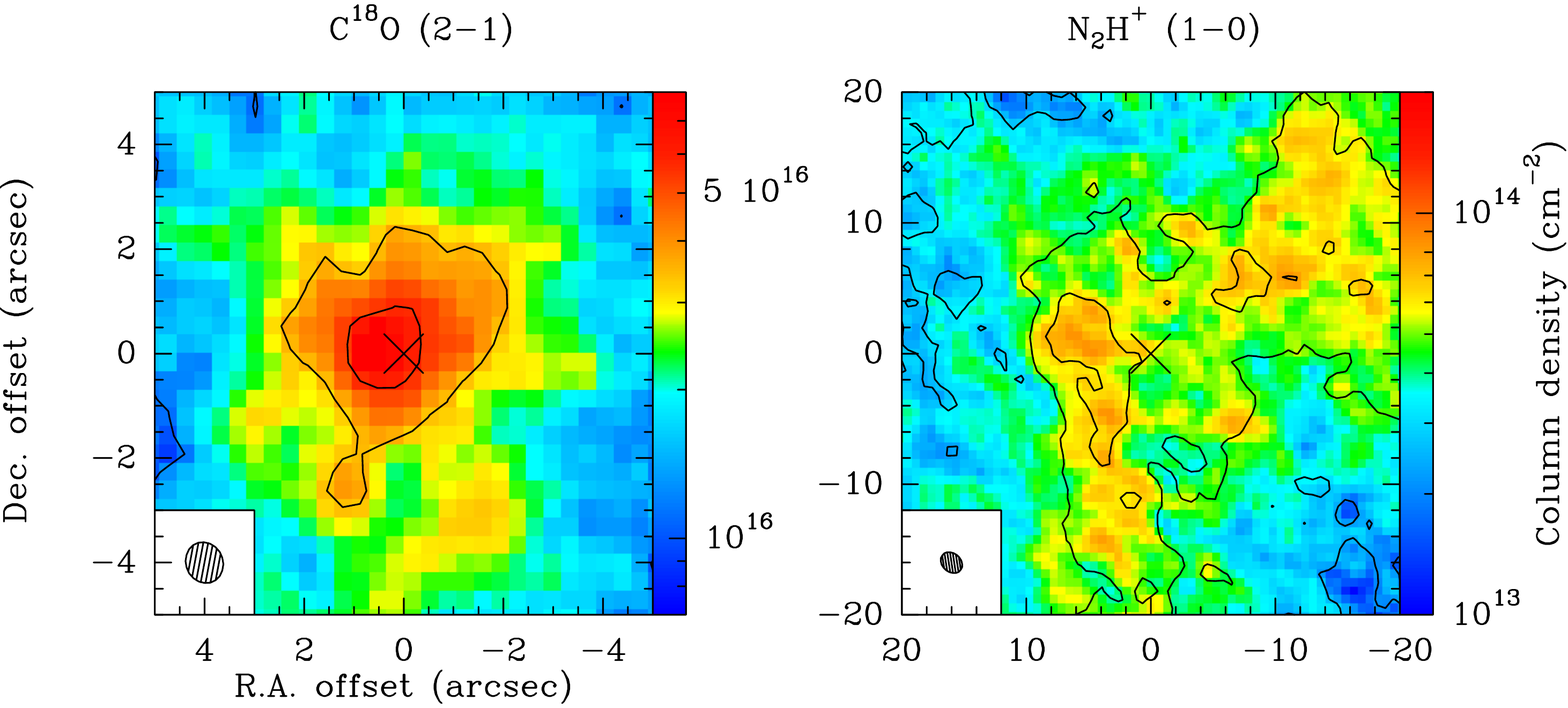}
\caption{Same Figure as \ref{fig:column-density-maps-L1448-2A} for IRAS2A. 
}
\label{fig:column-density-maps-IRAS2A}
\end{figure*}

\begin{figure*}[!ht]
\centering
\includegraphics[scale=0.3,angle=0,trim=0cm 3cm 0cm 3.5cm,clip=true]{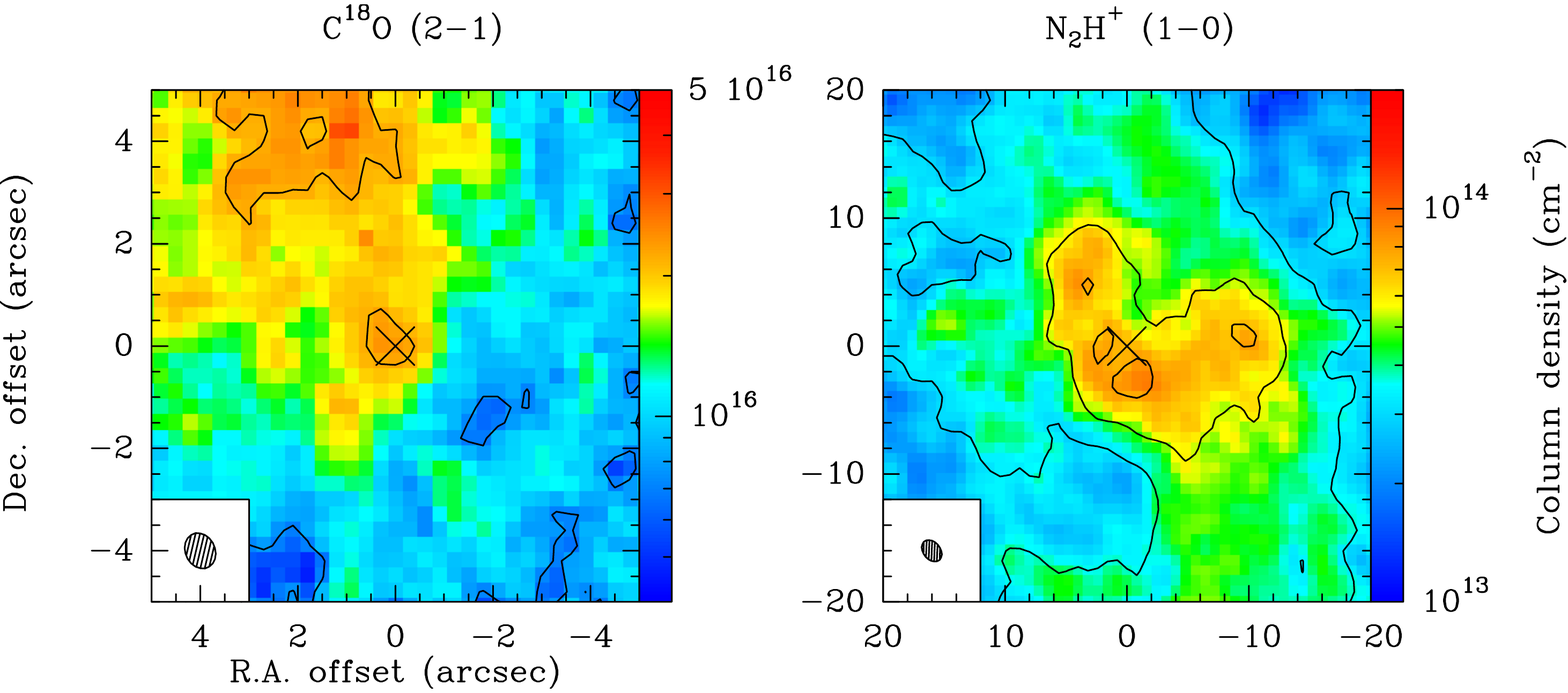}
\caption{Same Figure as \ref{fig:column-density-maps-L1448-2A} for SVS13-B
}
\label{fig:column-density-maps-SVS13B}
\end{figure*}

\begin{figure*}[!ht]
\centering
\includegraphics[scale=0.3,angle=0,trim=0cm 3cm 0cm 3.5cm,clip=true]{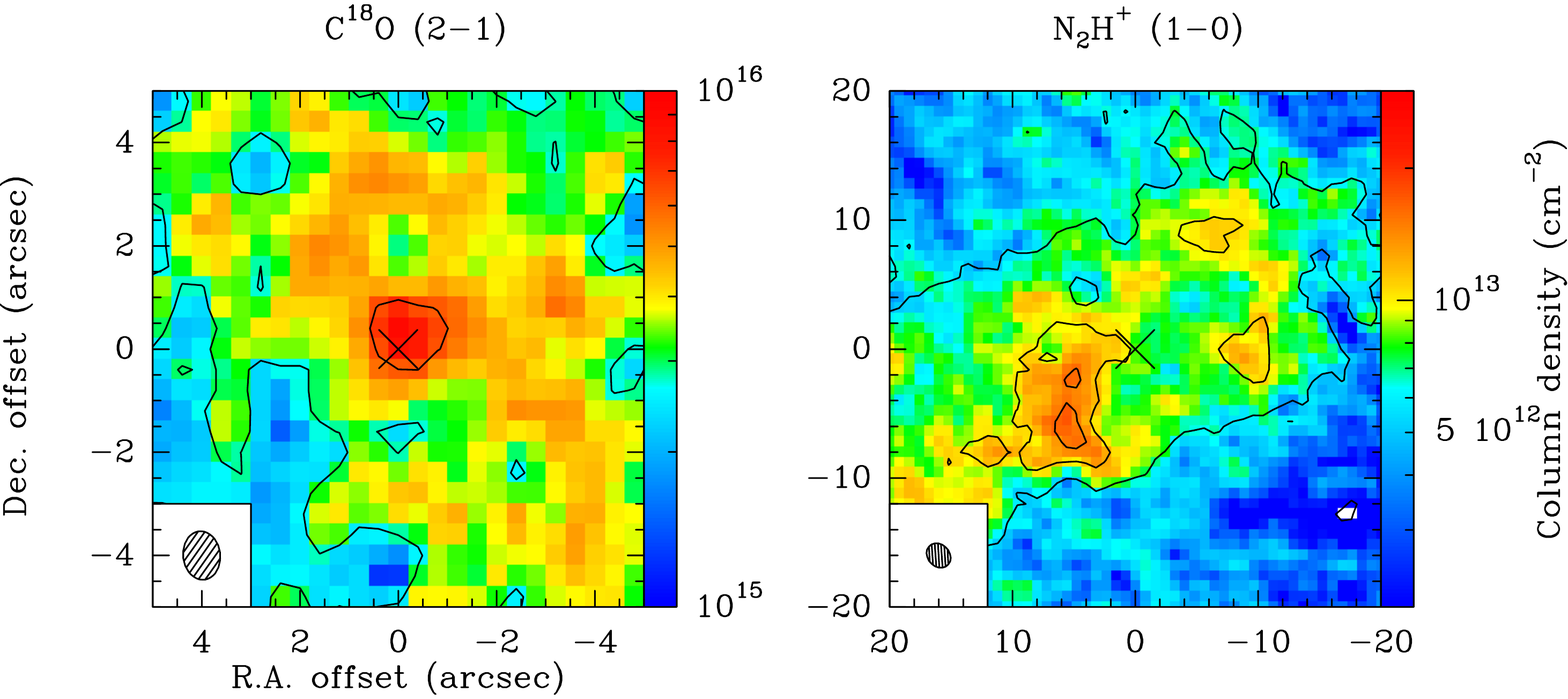}
\caption{Same Figure as \ref{fig:column-density-maps-L1448-2A} for IRAM04191. 
}
\label{fig:column-density-maps-IRAM04191}
\end{figure*}

\begin{figure*}[!ht]
\centering
\includegraphics[scale=0.3,angle=0,trim=0cm 3cm 0cm 3.5cm,clip=true]{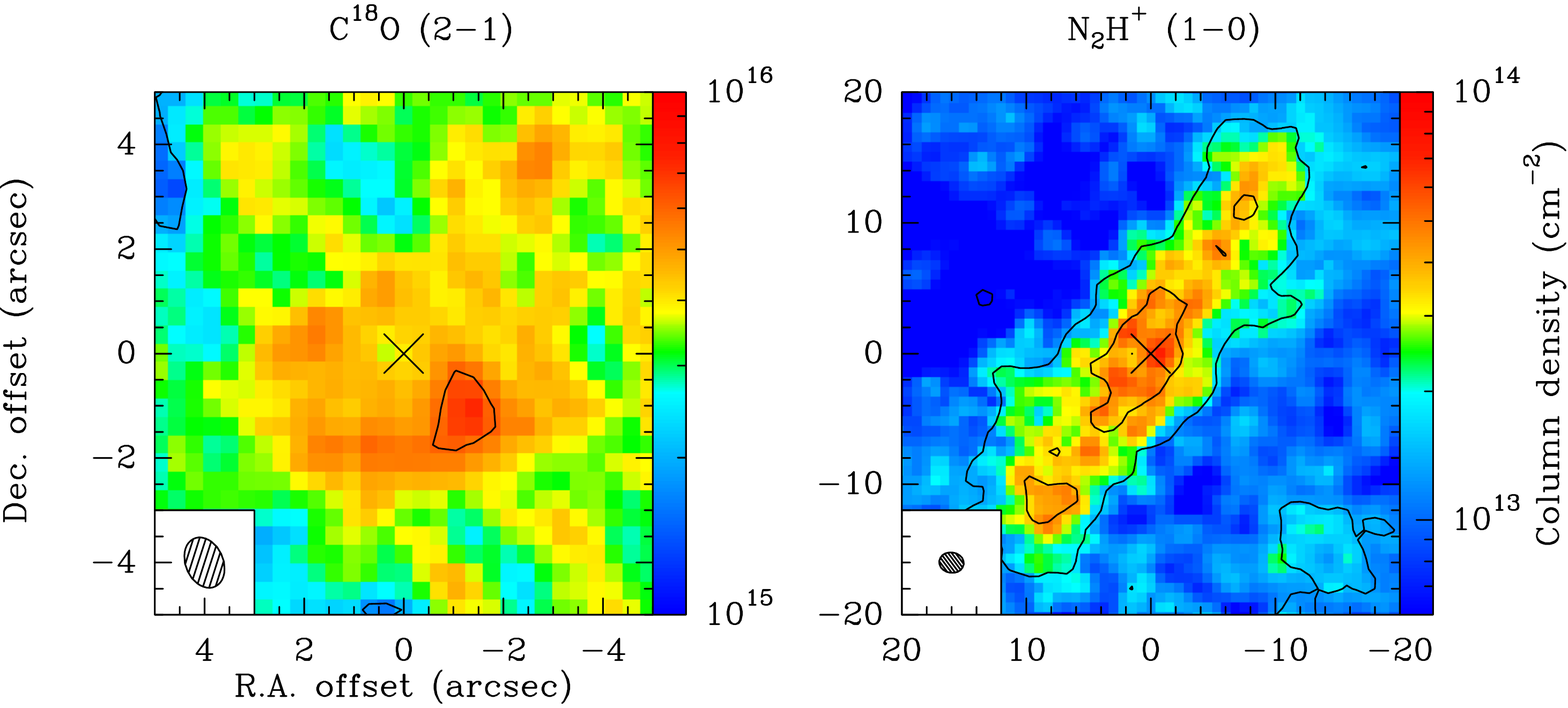}
\caption{Same Figure as \ref{fig:column-density-maps-L1448-2A} for L1521F. 
}
\label{fig:column-density-maps-L1521F}
\end{figure*}

\begin{figure*}[!ht]
\centering
\includegraphics[scale=0.3,angle=0,trim=0cm 3cm 0cm 3.5cm,clip=true]{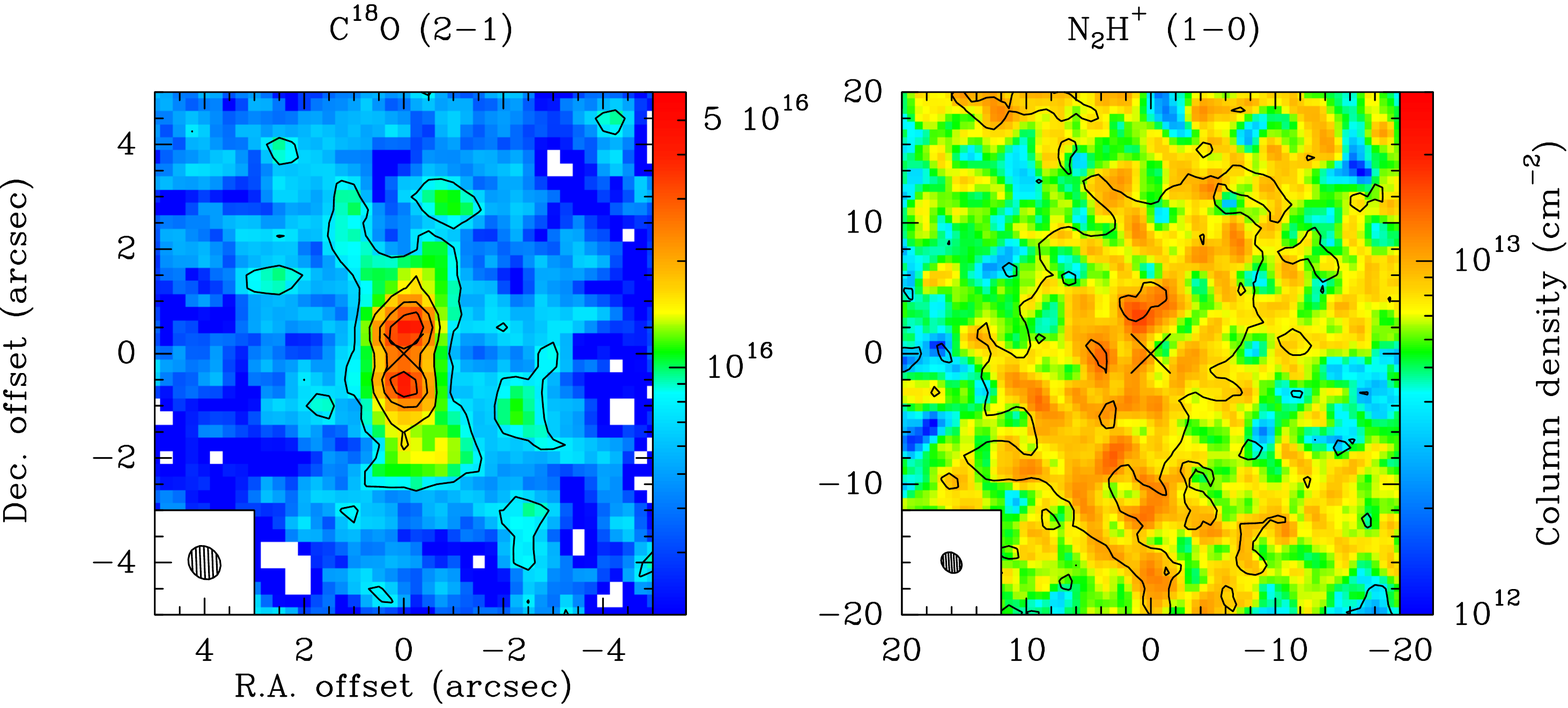}
\caption{Same Figure as \ref{fig:column-density-maps-L1448-2A} for L1527. 
}
\label{fig:column-density-maps-L1527}
\end{figure*}

\begin{figure*}[!ht]
\centering
\includegraphics[scale=0.3,angle=0,trim=0cm 3cm 0cm 3.5cm,clip=true]{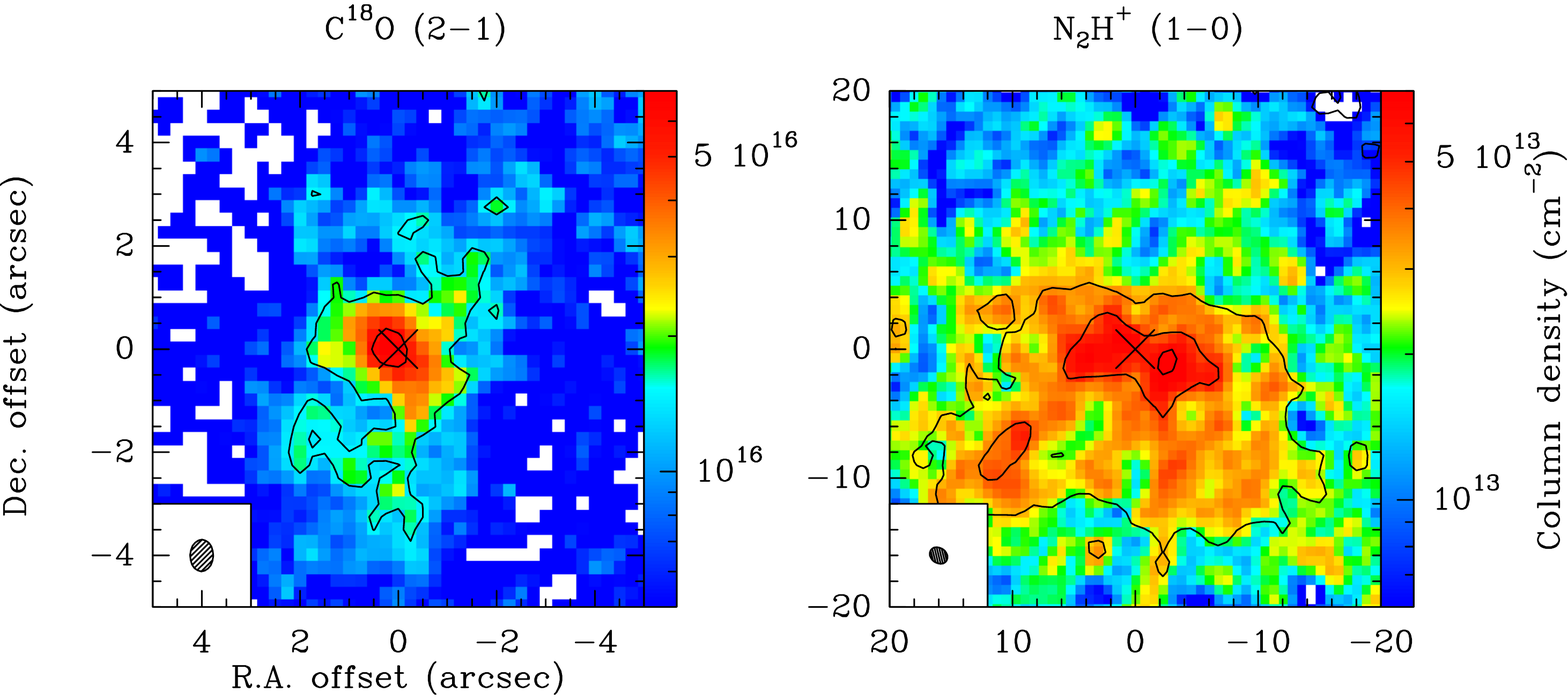}
\caption{Same Figure as \ref{fig:column-density-maps-L1448-2A}, but for L1157. 
}
\label{fig:column-density-maps-L1157}
\end{figure*}

\begin{figure*}[!ht]
\centering
\includegraphics[scale=0.3,angle=0,trim=0cm 3cm 0cm 3.5cm,clip=true]{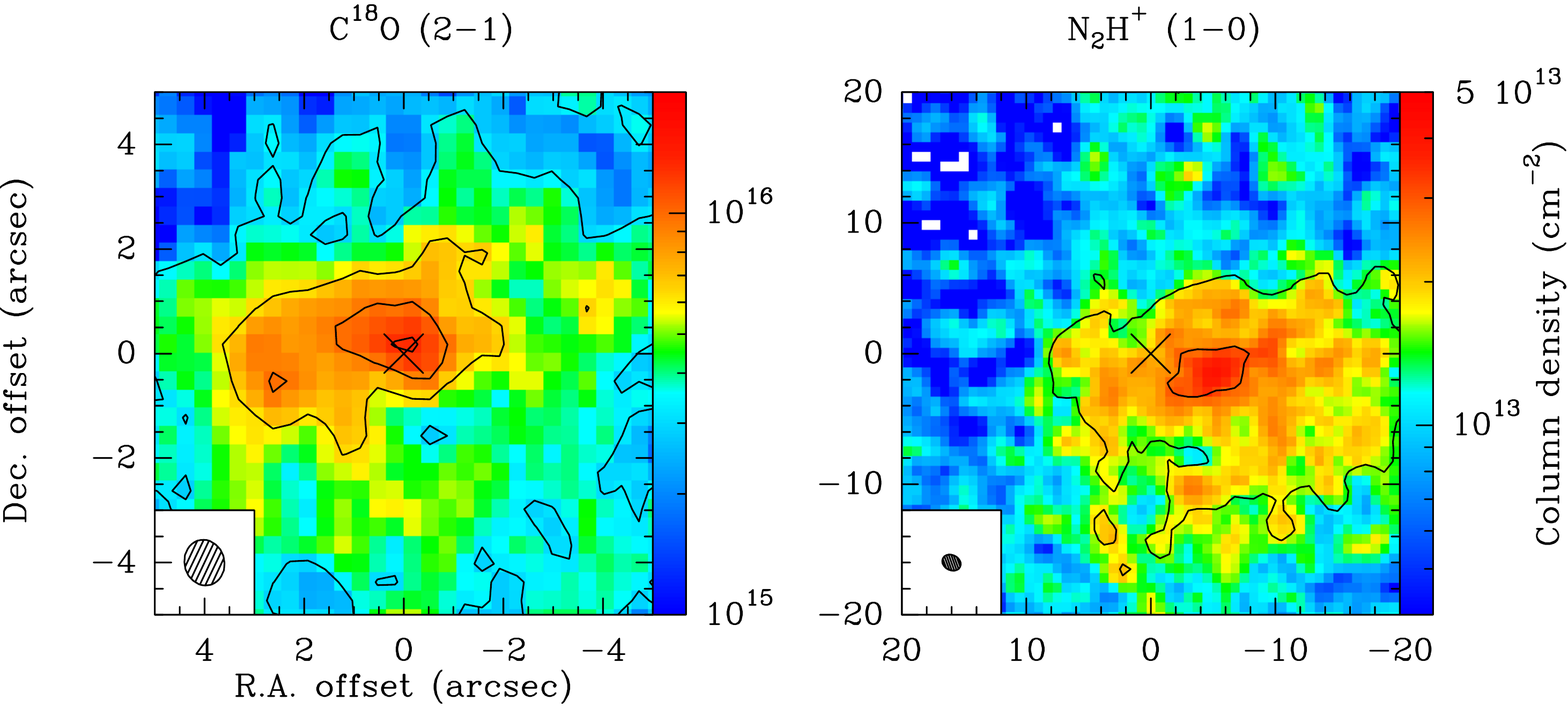}
\caption{Same Figure as \ref{fig:column-density-maps-L1448-2A} for GF9-2. 
}
\label{fig:column-density-maps-GF92}
\end{figure*}

We determined the C$^{18}$O (2$-$1) and N$_{2}$H$^{+}$ ($1_{01}-0_{12}$) opacities for our source sample assuming optically thin emission and using the maximum temperature of the emission spectrum $T_\mathrm{peak}$:
\begin{equation}
\tau_{\nu} = \frac{T_\mathrm{peak}}{\mathrm{J}_{\nu}(T_\mathrm{ex}) - \mathrm{J}_{\nu}(T_\mathrm{bg})}  .
\label{opacity-thin}
\end{equation}
We calculated the opacity on the spectra of each pixel from the combined dataset for both molecular transition. We used the excitation temperature values mentioned above and a standard value of the cosmic background temperature $T_\mathrm{bg}$=2.7~K \citep{Mather94}. We report the average $\tau_{\nu}$ on the size of the combined maps of C$^{18}$O and N$_2$H$^+$ emissions (5$\arcsec$ $\times$ 5$\arcsec$ and 20$\arcsec$ $\times$ 20$\arcsec$, respectively) in Table \ref{table:tau-mean}. We noticed that the sources have average $\tau_{\nu} <$0.4 but some pixels have opacity values $>$0.5. 
We determined the column density from each pixel using Eq. \eqref{column-density-thin} and the parameter values listed in Table \ref{table:parameters-column-density} for the C$^{18}$O and N$_{2}$H$^{+}$ molecules. For the calculation of column density, we distinguished two opacity regimes according to the value of $\tau_{\nu}$ of each spectrum:\\
$\indent \bullet$ an optically thin regime where $\tau_{\nu}<$0.4 : in this case, the density column is determined by Eq. \eqref{column-density-thin},\\
$\indent \bullet$ an intermediate regime where $\tau_{\nu} \geq$0.4 : in this case, the optically thin hypothesis is no longer a reasonable hypothesis and a correction factor $\frac{\tau_{\nu}}{1-e^{-\tau_{\nu}}}$ \citep{Goldsmith99} must be applied to the Eq. \eqref{column-density-thin} to calculate the column density. Indeed, the column density determined in the optically thin hypothesis by Eq. \eqref{column-density-thin} is underestimated by more than 20\% for a value of $\tau_{\nu}=$0.4.\\
Figure~\ref{fig:column-density-maps-L1448-2A} shows as an example the column density maps from the combined dataset obtained for L1448-2A. The column density maps of the other sources from our sample are provided in Appendix~\ref{sec:comments-indiv-sources} (see Figs. \ref{fig:column-density-maps-L1448NB}, \ref{fig:column-density-maps-L1448-C}, \ref{fig:column-density-maps-IRAS2A}, \ref{fig:column-density-maps-SVS13B}, \ref{fig:column-density-maps-IRAS4A}, \ref{fig:column-density-maps-IRAS4B}, \ref{fig:column-density-maps-IRAM04191}, \ref{fig:column-density-maps-L1521F}, \ref{fig:column-density-maps-L1527},   \ref{fig:column-density-maps-L1157}, and \ref{fig:column-density-maps-GF92}).

\begin{table*}[h]
\centering
\caption{Mean column density of C$^{18}$O and N$_{2}$H$^{+}$ and opacity of the transitions (2$-$1) and (1$-$0), respectively determined from the combined dataset for the CALYPSO sample sources.}
\label{table:tau-mean}
\begin{tabular}{l|ccc|ccc}
\hline \hline
\multicolumn{7}{c}{} \\
   \multicolumn{1}{l|}{Source}       & \multicolumn{3}{c|}{C$^{18}$O (2$-$1)} & \multicolumn{3}{c}{N$_2$H$^+$ (1$-$0)} \\
   &  $T_\mathrm{ex}$ & $N_\mathrm{mean}$           & $\tau_\mathrm{mean}$      & $T_\mathrm{ex}$     & $N_\mathrm{mean}$  & $\tau_\mathrm{mean}$  \\
          & (K) & (10$^{15}$ cm$^{-2}$)           &                 & (K)        & (10$^{13}$ cm$^{-2}$)  &                 \\
\hline
L1448-2A  &  39     &  13.4              &                 0.22    &  20   &       2.2       &      0.09           \\
L1448-NB  &  33    &    17.2             &           0.37        &   19    &       3.3       &       0.16          \\
L1448-C   &    61    &      20.3         &         0.18           &   26   &    2.1          &            0.06      \\
IRAS2A    &   58    &    23.5            &        0.17            &   33   &      4.3        &         0.06        \\
SVS13-B   &   40   &  13.4               &         0.17          &   24    &     3.9         &   0.10              \\
IRAS4A    &    33   &      16.5          &           0.30         &  24    &      0.6~\tablefootmark{$\star$}        &     0.03~\tablefootmark{$\star$}            \\
IRAS4B    &   39    &      9.5          &         0.14            &   25  &        0.4~\tablefootmark{$\star$}      &          0.03~\tablefootmark{$\star$}      \\
IRAM04191 &   20   &  3.5               &           0.17        &   10    &      0.6        &     0.10            \\
L1521F    &   20     & 3.6              &          0.22            &  10  &     1.5         &      0.33           \\
L1527     &    37     & 5.4             &              0.13         &  19 &      0.7        &        0.05         \\
L1157     &   38     &     8.1          &               0.19        & 25  &    2.2          &        0.05         \\                GF9-2     &   43   &      4.2           &             0.09           & 21 &      1.2        &           0.05      \\
\hline
\end{tabular}
\tablefoot{
The mean column density of C$^{18}$O and N$_{2}$H$^{+}$ and the opacity of the transitions (2$-$1) and (1$-$0), respectively, are calculated within the central 5$\arcsec \times$5$\arcsec$ and 20$\arcsec \times$20$\arcsec$ respectively, from the combined datasets. The N$_{2}$H$^{+}$ column density was determined from the isolated hyperfine component ($1_{01}-0_{12}$). \tablefoottext{$\star$}{For IRAS4A and IRAS4B, the calculations were made from the PdBI datasets for the N$_{2}$H$^{+}$ emission.}} 
\end{table*}

%%%%%%%%%%%%%%%%%%%%%%%%%%%%%%%%%%%%%%%%%%%%%
\clearpage
\section{Systemic velocity estimate} \label{sec:systemic-velocity} 

The systemic velocities of the sources are the reference velocities to quantify centroid velocity variations in the PV$_\mathrm{rot}$ diagrams. The systemic velocity is defined as the average velocity of an object.
The linear velocity gradient modeling allows us to determine a systemic velocity, $\mathrm{v}_0$, in the different datasets and tracers. For a given source, we noticed that $\mathrm{v}_0$ is different depending on the region of the envelope considered (see Table \ref{table:gradient-velocity-fit}).

In the first place, we determine these values by fitting the HFS line profile of the optically thin transition N$_{2}$H$^{+}$ (1$-$0) from the 30m datasets. 
Table \ref{table:vitesse-systemique-n2hp-30m} reports these first estimates and compares them to the values found in the literature.
The differences of a few tenths of a km~s$^{-1}$ between our values and the literature can be explained by the different tracers used in our study compared to the literature. Indeed, different tracers do not trace exactly the same material in the protostellar envelopes, as we observed with the C$^{18}$O and N$_{2}$H$^{+}$ in our study (see Sect. \ref{sec:velocity-maps} and Table \ref{table:gradient-velocity-fit}). There could also be systematic uncertainty on the rest frequencies or on the calibration of the spectral axis for instance.
This first estimate in the quieter parts of the envelopes, could be more robust than a value determined in the inner parts which can be dominated by rotation, collapse, or ejection. On the other hand, the turbulence at large scales could affect the systemic velocity if the outer envelope is turbulent or if there is a contamination of the turbulence on the line of sight from cloud scales. In the case of an axisymmetric envelope with a symmetric kinematics (infall and rotation), the systemic velocity could be more accurate at small scales.

\begin{table}[!ht]
\centering
\caption{Systemic velocity values determined from HFS fit on the average spectrum N$_{2}$H$^{+}$ from CALYPSO 30m dataset compared to the literature.}
\label{table:vitesse-systemique-n2hp-30m}
\begin{tabular}{lcc}
\hline \hline
\multicolumn{1}{c}{\hfill}  & \multicolumn{2}{c}{\hfill}   \\
Source   & $\mathrm{v}_\mathrm{sys}^\mathrm{30m}$           &     $\mathrm{v}_\mathrm{sys}^\mathrm{literature}$       \\
          & (km~s$^{-1}$)           &             (km~s$^{-1}$)                      \\
\hline
L1448-2A  &      4.10 $\pm$ 0.04                 &            4.20~\tablefootmark{1}                  \\
L1448-NB  &    4.47 $\pm$ 0.02                   &      4.70~\tablefootmark{2}               \\
L1448-C   &      4.73 $\pm$ 0.03                 &          5.00~\tablefootmark{1}    \\
IRAS2A    &        7.60 $\pm$ 0.04               &               7.30~\tablefootmark{3,4}   \\
SVS13-B   &       8.02 $\pm$ 0.02                &                8.40~\tablefootmark{5}    \\
IRAS4A    &       7.34 $\pm$ 0.03~\tablefootmark{$\star$}               &       7.24~\tablefootmark{6,7}             \\
IRAS4B    &          7.20 $\pm$ 0.06~\tablefootmark{$\star$}        &          7.00~\tablefootmark{8}             \\
IRAM04191 &     6.63 $\pm$ 0.06                  &           6.66~\tablefootmark{9}                    \\
L1521F    &        6.47 $\pm$ 0.01                &              6.45~\tablefootmark{10}               \\
L1527     &      5.89 $\pm$ 0.01                 &                 5.90~\tablefootmark{10}               \\
L1157     &         2.65 $\pm$ 0.03              &       2.65~\tablefootmark{10,11}              \\ 
GF9-2     &       -2.56 $\pm$ 0.02                &          -2.60~\tablefootmark{12}             \\
\hline
\end{tabular}
\tablefoot{
\tablefoottext{$\star$}{Average spectrum ajusted by fixing a second velocity component corresponding to the external compression at 7.75~km~s$^{-1}$ \citep{Belloche06}.}
}
\tablebib{(1) \cite{Tobin07}; (2) \cite{Kwon06}; (3) \cite{Jorgensen07}; (4) \cite{Codella14-bis}; (5) \cite{Chen09}; (6) \cite{Belloche06}; (7) \cite{Anderl16}; (8) \cite{Jorgensen10}; (9) \cite{Belloche02}; (10) \cite{Tobin11}; (11) \cite{Bachiller97}; (12) \cite{Goodman93}.}
\end{table}

If the reference velocity value is well chosen, blue- and red-shifted velocity points should overlap if the envelope is axisymmetric.
In practice, the gradients may not be symmetric with respect to the central position of the source. This asymmetry can cause shifts in position between the blue-shifted and the red-shifted emission. We built the PV$_\mathrm{rot}$ diagrams by maximizing the overlap between blue- and red-shifted velocity points to produce the best symmetric pattern. 
We applied this method independently for the PV$_\mathrm{rot}$ diagram points from the C$^{18}$O and N$_{2}$H$^{+}$ emission to determine the best couple ($\mathrm{v}_\mathrm{sys}$, $r_\mathrm{orig}$) adapted to each tracer, with $\mathrm{v}_\mathrm{sys}$ the systemic velocity and $r_\mathrm{orig}$ the central position of the gradient along the equatorial axis.
We fit a power-law function (see Sect. \ref{sec:results-PV-diagram}) exploring a range of $\pm$0.7~km~s$^{-1}$ around the first estimate of the systemic velocity determined at outer envelope scales and a range of $\pm$0.2$\arcsec$ along the equatorial axis around the dust continuum peak (see Table \ref{table:sample}).
Table \ref{table:chi2-fit-lines} gives the reduced $\chi^{2}$ associated with the best couple of parameters corresponding to the best overlap of the blue- and red-shifted points.
This method does not allow for a more accurate determination of the systemic velocity value than 0.05~km~s$^{-1}$ given the errors on the velocity of the points populating the PV$_\mathrm{rot}$ diagrams. We therefore added a systematic error of 0.05~km~s$^{-1}$ to previous velocity errors determined above for both tracers.
The central positions and systemic velocities obtained with this method are on average within $<$0.2~km~s$^{-1}$ and $<$0.1\arcsec of the first estimates of these values using the average N$_{2}$H$^{+}$ spectrum (see Sect. \ref{sec:systemic-velocity} and Table \ref{table:vitesse-systemique-n2hp-30m}) and the dust continuum peak (see Table \ref{table:sample}).

\begin{table*}[!ht]
\centering
\caption{Values of systemic velocity and central position giving the best overlap and $\chi^{2}$ by independent fits of C$^{18}$O and N$_{2}$H$^{+}$ points.}
\label{table:chi2-fit-lines}
\begin{tabular}{ll|ccc|cc}
\hline \hline
\multicolumn{2}{c}{\hfill} & \multicolumn{3}{c}{\hfill} & \multicolumn{2}{c}{\hfill} \\
Source                  & Line           & $v_\mathrm{sys}$~\tablefootmark{a}     & \multicolumn{2}{c|}{ $r_\mathrm{orig}$ coordinates~\tablefootmark{b}} & \multicolumn{2}{c}{Power law fit~\tablefootmark{c}}       \\
                         &                &               & R.A.              & DEC              & $\eta$        & $\chi^{2}$    \\
                         &                & (km~s$^{-1}$) & (h:m:s) (J2000)           & ($^{\circ}$:$\arcmin$:$\arcsec$) (J2000)          &                 &                                 \\
\hline
\multirow{2}{*}{L1448-2A} & C$^{18}$O      & 4.0 (-0.1)           &     03:25:22.380 (+0.06$\arcsec$)              & 30:45:13.28 (+0.08$\arcsec$)           & -0.9 $\pm$0.1   & 1.4      \\
                         & N$_{2}$H$^{+}$ &      $u$      &           $u$        &       $u$         &  ... & ...  \\
\hline
\multirow{2}{*}{L1448-NB} & C$^{18}$O      & 4.6 (+0.1)           &      03:25:36.316 (+0.01$\arcsec$)             &  30:45:15.10 (-0.05$\arcsec$)        & -0.94 $\pm$0.04   & 2.3   \\
                         & N$_{2}$H$^{+}$ & ...          &        ...           & ...                & ...   & ...    \\
\hline
\multirow{2}{*}{L1448-C} & C$^{18}$O      & 5.1 (+0.4)           &      $u$             & $u$  & -1.0 $\pm$0.1   & 1.4   \\
                         & N$_{2}$H$^{+}$ & 4.9 (+0.2)          &      $u$            &  $u$                & 0.8 $\pm$0.1   & 3.0    \\
\hline                         
\multirow{2}{*}{IRAS2A}  & C$^{18}$O      & $u$           &      $u$          & $u$                &  -0.7 $\pm$ 0.1 &   0.2          \\
                         & N$_{2}$H$^{+}$ &      ...      &       ...            &      ...           &  ...  &     ...      \\
\hline 
\multirow{2}{*}{SVS13-B}   & C$^{18}$O      &    8.3 (+0.3)       &      $u$             & $u$                &  -0.9$\pm$ 0.3  &  0.2   \\
                         & N$_{2}$H$^{+}$ &   ...        &     ...              & ...                &   ...  &   ...    \\
\hline 
\multirow{2}{*}{IRAS4A}  & C$^{18}$O      &  6.6 (-0.7)          &        $u$           &    $u$            & 0.2$\pm$0.2  &   1.8        \\
                         & N$_{2}$H$^{+}$ &    6.8 (-0.5)      &       $u$            &      $u$        & 0.6$\pm$0.3  &    0.7      \\
\hline                         
\multirow{2}{*}{IRAS4B}  & C$^{18}$O      & 6.9 (-0.3)            &      $u$            & $u$              & -0.3 $\pm$0.4  &   0.5         \\
                         & N$_{2}$H$^{+}$ & 7.0 (-0.2)            &        $u$           & $u$                & -2.0 $\pm$2.2  & 0.1          \\
\hline
\multirow{2}{*}{IRAM04191}   & C$^{18}$O      &  $u$        &   $u$           &     $u$       & -0.2 $\pm$ 0.2  & 0.3   \\
                         & N$_{2}$H$^{+}$ &    ...     &    ...              &  ...             &  ...  &  ...  \\
\hline
\multirow{2}{*}{L1521F}   & C$^{18}$O      &  6.55 (+0.1)       &     $u$            &   $u$             & 0.2 $\pm$ 0.5  &  0.1    \\
                         & N$_{2}$H$^{+}$ &     ...      &          ...         &        ...         &  ... &  ...  \\
\hline 
\multirow{2}{*}{L1527}   & C$^{18}$O      & 5.8 (-0.1)           &        $u$           & $u$          & -1.07 $\pm$ 0.04  & 1.5        \\
                         & N$_{2}$H$^{+}$ & $u$           &      $u$             &  $u$              & 0.4 $\pm$0.4   & 0.7   \\
\hline                                                 
\multirow{2}{*}{L1157}   & C$^{18}$O      & $u$      &   $u$ & $u$         & 0.1 $\pm$0.3   & 0.3    \\
                         & N$_{2}$H$^{+}$ & ...          &     ... & ...          & ... & ...   \\
\hline 
 \multirow{2}{*}{GF9-2}   & C$^{18}$O      & -2.8 (-0.2)          &   20:51:29.836   (+0.2 $\arcsec$)             &  60:18:38.44 (+0$\arcsec$)                & -0.8 $\pm$0.1   & 1.8    \\
                         & N$_{2}$H$^{+}$ &     ...      &           ...        &       ...          &  ... & ...  \\
\hline 
\end{tabular}
\tablefoot{
\tablefoottext{a,b}{Values of $\mathrm{v}_\mathrm{sys}$ and the $r_\mathrm{orig}$ coordinates when they are different from $\mathrm{v}_\mathrm{sys}^\mathrm{30m}$ (see Table \ref{table:vitesse-systemique-n2hp-30m}) and from coordinates of the continuum peak at 1.3~mm (see Table \ref{table:sample}), respectively. 
The values in parentheses represent the offset in km~s$^{-1}$ with respect to the value $\mathrm{v}_\mathrm{sys}^\mathrm{30m}$ and the offset in arcsec with respect to the continuum peak, respectively.
The $u$ symbol means "unchanged": $\mathrm{v}_\mathrm{sys}=\mathrm{v}_\mathrm{sys}^\mathrm{30m}$ (see Table \ref{table:vitesse-systemique-n2hp-30m}) and the $r_\mathrm{orig}$ coordinates are still equal to coordinates of the continuum peak at 1.3~mm (see Table \ref{table:sample}).}
\tablefoottext{c}{Index of fits by an orthogonal least-square power-law ($\mathrm{v} \propto r^{\eta}$) on the red and blue points and the reduced $\chi^{2}$ value associated with this best-fit model.
The dashes mean that no fit has been done because the tracer is not used to constrain the PV$_\mathrm{rot}$ diagram of this source.}
 }
\end{table*}

%%%%%%%%%%%% DETAILS ON J DISTRIBUTIONS %%%%%%%%%%%%
\clearpage
\section{Details on $j$ distributions} \label{sec:details-j-distribution}

\begin{table*}[!ht]
\centering
\caption{Parameters from best fits by power-law or broken power-law functions of specific angular momentum distribution $j$.}
\label{table:chi2-fit-profil-moment-ang}
\begin{tabular}{l|c|cc|cccc}
\hline \hline
\multicolumn{3}{c}{\hfill} & \multicolumn{2}{c}{\hfill}  \\
Source  &  Radial range~\tablefootmark{a} & \multicolumn{2}{c}{Power law fit~\tablefootmark{b}} & \multicolumn{4}{c}{Broken power law fit~\tablefootmark{c}}                                       \\
    &     & $\beta$           & $\chi^{2}$   & $\beta_{r<r_\mathrm{break}}$ & $r_\mathrm{break}$  & $\beta_{r>r_\mathrm{break}}$ & $\chi^{2}$ \\
  & (au) $-$ (au) 	&	&	&	& (au) &	& \\ 
	\hline
L1448-2A &  60$-$1250  & 0.1 $\pm$ 0.1    & 1.4        &        ...                &     ...       &    ...                    &     ...      \\
L1448-NB &  150$-$1700  & 0.07 $\pm$ 0.04   & 2.4       &     -0.3 $\pm$ 0.1                    &       630 $\pm$  80     &           0.9 $\pm$ 0.2              &   0.8         \\
L1448-C &  100$-$4000 &  -0.01 $\pm$ 0.08    &      1.6     &      -0.1 $\pm$ 0.1                  &    509 $\pm$ 105            &        1.0 $\pm$ 0.3                  &       1.0      \\
IRAS2A &  85$-$1500  & 0.3 $\pm$ 0.1     & 0.2        & ...          & ...  & ...           & ...      \\
SVS13-B &  110$-$450   &   0.1 $\pm$ 0.3    &    0.2   &     ...       &  ...&        ...    &   ...    \\
IRAS4B &  175$-$1050   & 0.4 $\pm$ 0.3        & 0.5            &       ...                  &           ...    &          ...               &      ...      \\
IRAM04191 &  55$-$800   &    0.7 $\pm$ 0.2   &  0.2      &   ...         &...  &       ...     &    ...   \\
L1521F  &  1500$-$4200 & 1.2 $\pm$ 0.6      & 0.1            &            ...             &        ...       &           ...              &   ...         \\
L1527 &  45$-$2000    & -0.13 $\pm$ 0.03   & 1.6         &     ...                   &     ...         &      ...                   &   ...         \\
L1157 &   85$-$650   &  1.1 $\pm$ 0.4     &   0.2    &          ...               &      ...         &       ...                  &    ...       \\
GF9-2 &   75$-$850   & 0.3 $\pm$ 0.1    & 1.6          &         ...                &       ...        &        ...                 &    ...        \\
\hline
Median profile &  45$-$4200 & 1.0 $\pm$ 0.2 &  &    0.4$\pm$0.6 &  1030$\pm$500 & 1.4$\pm$0.3 &   \\
\hline
\end{tabular}
\tablefoot{
\tablefoottext{a}{Range of radii over which the $j(r)$ profiles were built and the fits were performed.}
\tablefoottext{b}{Index of fits by a power-law function ($\mathrm{v} \propto r^{\beta}$) on the red and blue points and the reduced $\chi^{2}$ value associated with this best-fit model.}
\tablefoottext{c}{Parameters of fits by a broken power-law function and the reduced $\chi^{2}$ value associated with this best-fit model.}
The last line reports the index of the best fits of the median profile of specific angular momentum for all sources. Each individual $j$ distribution has been resampled in steps of 100~au and normalized using the value of $j$ at 600~au. }
\end{table*}

\begin{table*}[!ht]
\centering
\caption{Same as Table \ref{table:chi2-fit-profil-moment-ang} for the distribution of apparent specific angular momentum |$j_\mathrm{app}$| along the equatorial axis.}
\label{table:chi2-fit-profil-japp}
\begin{tabular}{l|c|cc|cccc}
\hline \hline
\multicolumn{3}{c}{\hfill} & \multicolumn{2}{c}{\hfill}  \\
Source &  Radial range & \multicolumn{2}{c}{Power law fit} & \multicolumn{4}{c}{Broken power law fit}                                       \\
      &   & $\beta_\mathrm{app}$           & $\chi^{2}$   & $\beta_{\mathrm{app},r<r_\mathrm{break}}$ & $r_\mathrm{\mathrm{app},break}$  & $\beta_{\mathrm{app},r>r_\mathrm{break}}$ & $\chi^{2}$ \\
 & (au) $-$ (au) 	&	&	&	& (au) &	& \\ 
	\hline
L1448-2A~\tablefootmark{$\star$ $\star$} &  60$-$4000  & 0.3 $\pm$ 0.1    & 1.3        &        0.1 $\pm$ 0.1                &     1380 $\pm$ 400       &    1.6 $\pm$ 0.5                   &     0.6     \\
L1448-NB &  150$-$8200   & 0.02 $\pm$ 0.04   & 2.3       &     ...                    &      ...     &           ...             &   ...         \\
L1448-C & 100$-$4000  & 0.69 $\pm$ 0.03    &     5.5     &     -0.1 $\pm$ 0.1                   &   790 $\pm$ 100            &       1.6 $\pm$ 0.1                  &      2.0      \\
IRAS2A~\tablefootmark{$\star$ $\star$} &   85$-$8200 & 1.4 $\pm$ 0.1     & 6.2        &  0.3 $\pm$ 0.1        &  2180$\pm$110  &   28.4 $\pm$ 22.2           & 2.5      \\
SVS13-B &  110$-$1900   &   1.0 $\pm$ 0.1    &    1.5   &     0.1 $\pm$ 0.3      &  640$\pm$190 &        2.2 $\pm$ 0.5    &   0.2    \\
IRAS4A &   250$-$1600   & 1.9 $\pm$ 0.1        & 0.9            &       -0.3 $\pm$ 0.7                  &           470$\pm$60    &          2.1$\pm$0.2               &      0.4      \\
IRAS4B &    175$-$3100  &  1.5 $\pm$ 0.1       &   4.0          &      ...               &       ...       &       ...              &   ...         \\
IRAM04191~\tablefootmark{$\star$ $\star$} &  55$-$9700   &    0.8 $\pm$ 0.1   &  0.5      &   ...         &...  &       ...     &    ...   \\
L1521F &  1500$-$4200    & 1.2 $\pm$ 0.6      & 0.1            &            ...             &        ...       &           ...              &   ...         \\
L1527~\tablefootmark{$\star$ $\star$} &   45$-$7800    & -0.12 $\pm$ 0.03   & 1.8         &   -0.14 $\pm$ 0.03                     &    1320 $\pm$ 260         &      1.3 $\pm$ 0.2                   &   1.6         \\
L1157~\tablefootmark{$\star$ $\star$} &  85$-$9800     &  0.8 $\pm$ 0.1     &   0.4    &          ...               &      ...         &       ...                  &    ...       \\
GF9-2~\tablefootmark{$\star$ $\star$} &   75$-$5600   &  0.5 $\pm$ 0.1   &    2.3       &      0.3 $\pm$ 0.1                  &    980 $\pm$ 370        &        1.2 $\pm$ 0.3                &     1.0     \\
\hline
Median profile &  45$-$9800   & 0.9 $\pm$ 0.1 &  &    0.3$\pm$0.3 &  1570$\pm$300 & 1.6$\pm$0.2 &   \\
\hline
\end{tabular}
\tablefoot{ Here, we consider all the velocity gradients observed at all envelope scales, including the reversed gradients and the shifted ones at scales of $r \gtrsim$1000~au (see Fig. \ref{fig:evolution-theta-plot} and Sect. \ref{sec:velocity-gradient-1000au}) which were excluded in the construction of the PV$_\mathrm{rot}$ diagrams in Fig. \ref{fig:PV-diagrams-1}, and in Table \ref{table:chi2-fit-profil-moment-ang} for our analysis of rotational motions.
\tablefoottext{$\star$ $\star$}{Sources with a negative apparent specific angular momentum along the equatorial axis in the outer envelope scales of $r>$1600~au (see Sect. \ref{subsec:Counter-rotation}).}
}
\end{table*}

%%%%%%%%%%%% DETAILS ON linewidth DISTRIBUTIONS %%%%%%%%%%%%
\clearpage
\section{Details on linewidth distributions} \label{sec:details-Dv-distribution}

\begin{table*}[!ht]
\centering
\caption{Parameters from best fits by power-law functions of linewidth distribution $D \mathrm{v}$ in the outer envelope.}
\label{table:chi2-fit-profil-dV}
\begin{tabular}{lccc}
\hline \hline
\multicolumn{4}{c}{\hfill}   \\
Source  & Outer envelope~\tablefootmark{a} &\multicolumn{2}{c}{Power law fit~\tablefootmark{b}}                                       \\
     &  (au)  & $\gamma$           & $\chi^{2}$  \\
 	&	&	& \\ 
	\hline
L1448-2A & $r>$1380 & -0.9 $\pm$ 0.2    & 8.0           \\
L1448-NB & $r>$630 & 0.3 $\pm$ 0.1   & 30.2        \\
L1448-C & $r>$790 & 0.2 $\pm$ 0.1    &     16.7 \\
IRAS2A  & $r>$2180 & 0.1 $\pm$ 0.2     & 65.3    \\
SVS13-B  & $r>$640 & 0.17 $\pm$ 0.04    &    2.5    \\
IRAS4A   &  $r>$470 & 0.7 $\pm$ 0.4        & 34.1     \\
IRAS4B  & $r>$0  & -0.3 $\pm$ 0.1        & 33.1     \\
IRAM04191  & $r>$850 &   -0.2 $\pm$ 0.1   &  16.3      \\
L1521F & $r>$1000  & 1.1 $\pm$ 0.6      & 7.3    \\

L1527  &$r>$1320  & -0.2 $\pm$ 0.1   & 14.1   \\

L1157 & $r>$1420  &  0.6 $\pm$ 0.1     &   13.0     \\
GF9-2  &$r>980$  & 0.16 $\pm$ 0.04    & 0.3        \\
%\hline
%Median profile & 1.0 $\pm$ 0.2 &  &    0.4$\pm$0.3 &  1300$\pm$280 & 2.0$\pm$0.3 &   \\
\hline
\end{tabular}
\tablefoot{
\tablefoottext{a}{Radius of the outer envelope defined by the break radius of the $j(r)$ or $j_\mathrm{app}(r)$ profiles (see Tables \ref{table:chi2-fit-profil-moment-ang} and \ref{table:chi2-fit-profil-japp}) or the radius where we observe a reversal of the velocity gradients with respect to the inner envelope.}
\tablefoottext{b}{Index of fits by a power-law function ($D \mathrm{v} \propto r^{\gamma}$) on the red and blue points in the outer envelope and the reduced $\chi^{2}$ value associated with this best-fit model.}
}
\end{table*}

%%%%%%%%%%%% COMMENTS ON INDIVIDUAL SOURCES %%%%%%%%%%%%
\clearpage
\section{Comments on individual sources} \label{sec:comments-indiv-sources}

%%%%%%%%%%%
\subsection{L1448-2A} \label{sec:comments-L1448-2A}
L1448-2A (also known as L1448-IRS2 or Per-emb-22) is a Class~0 protostar in the L1448N complex in the Perseus cloud at a distance previously estimated to be 235~pc \citep{Hirota11} but determined at (293 $\pm$ 20)~pc by recent Gaia parallax measurements \citep{OrtizLeon18}. 
CO emission maps showed the presence of a bipolar flow in the northwest-southeast direction \citep{Olinger99}. Another study by \cite{Wolf-Chase00} suggests that the source shows distinct bipolar outflows, namely a signature of the presence of a binary system. Observations with VLA confirmed that L1448-2A is a binary system separated by $\sim$170~au \citep{Tobin16b}. 1.3~mm dust continuum emission also resolves the binary system with a separation of $\sim$180~au (\citealt{Maury18}, see Table \ref{table:sample}). Podio \& CALYPSO (in prep.) report two angle values (+63$^{\circ}$ (blue), +140$^{\circ}$ (red), see Table \ref{table:sample}) because the observed red and blue flow cavities are not aligned.

Figures \ref{fig:intensity-maps-L1448-2A} and \ref{fig:velocity-maps-L1448-2A} show the integrated intensity and centroid velocity maps obtained for this source from the PdBI, combined, and 30m CALYPSO datasets for the C$^{18}$O and N$_2$H$^+$ emission respectively. The C$^{18}$O emission from the PdBI datasets does not peak on the main protostar L1448-2A1 but in the middle of the binary 
system (see Table \ref{table:sample} and Figure \ref{fig:intensity-maps-L1448-2A}). Thus, the origin of the coordinate offsets is chosen to be the middle of the binary system (RA: 03$^h$25$^m$22$^s$.380, Dec.: 30$^{\circ}$45$\arcmin$13$\arcsec$.21; see Table \ref{table:sample}) to study the kinematics in the protostellar envelope of L1448-2A.

The gradients observed in the PdBI and combined velocity maps of the C$^{18}$O emission (see bottom left and middle panels on Figure \ref{fig:velocity-maps-L1448-2A}) have an orientation of 99$^{\circ} \leq \Theta \leq $107$^{\circ}$, they show a difference $>$60$^{\circ}$ with respect to the equatorial axis (see Table \ref{table:gradient-velocity-fit}). Therefore, these gradients at $r <$700~au are a merging of the kinematics between several mechanisms: ejection by outflows, orbital motions of the binary system, and rotational motions of the envelope along the equatorial axis.
The N$_2$H$^+$ emission from the PdBI datasets (see top left panel on Figure \ref{fig:velocity-maps-L1448-2A}) shows reversed gradients at scales of $r \sim$2000~au with respect to those observed at smaller scales ($\Theta \sim$-87$^{\circ}$, see Table \ref{table:gradient-velocity-fit}). In the outer envelope (see top middle and right panels on Figure \ref{fig:velocity-maps-L1448-2A}), the velocity gradients are not continuous in the same direction and not source-centered. Therefore, these gradients seem to be due to an external contamination.
We notice that the gradients from the 30m datasets at scales of $r >$4500~au (see right top and bottom panels on Figure \ref{fig:velocity-maps-L1448-2A}) have a PA of -8$-$14$^{\circ}$ and are in the opposite to the direction of the bipolar outflows (see Table \ref{table:gradient-velocity-fit}). Moreover, the integrated intensity at these scales seems to trace an elongated structure from east to west that could be compatible with part of the filament.

\begin{figure*}[!ht]
\centering
\includegraphics[width=10cm]{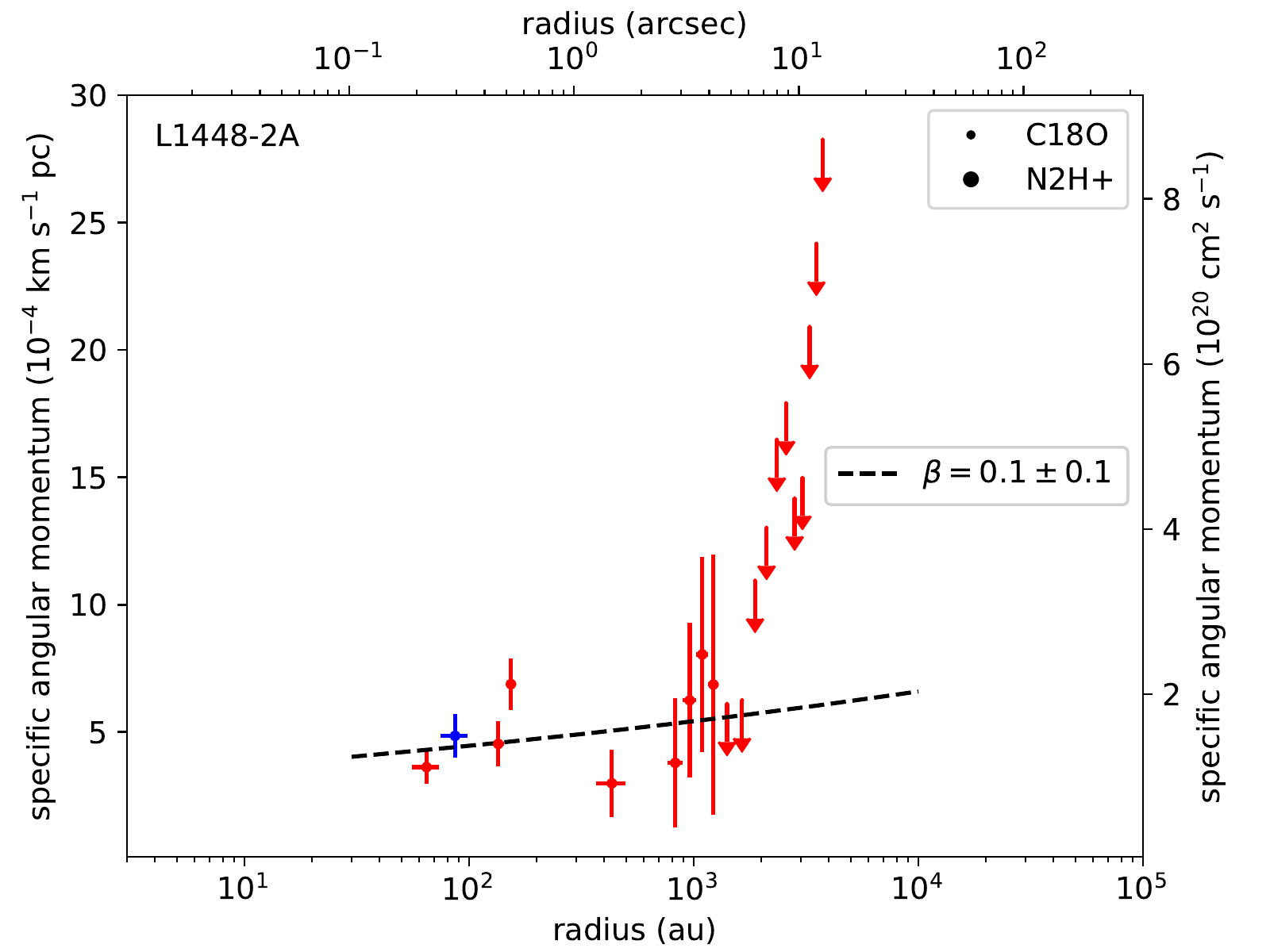}
\caption{
Radial distribution of specific angular momentum along the equatorial axis for L1448-2A. Blue and red dots show the blue- and red-shifted rotational velocities, respectively. Dots and large dots show points from C$^{18}$O and N$_{2}$H$^{+}$ emission, respectively. The arrows display upper limits of $j$ determined from velocity maps that do not exhibit a spatial distribution of velocities as organized as one would expect from rotation motions (see Sect. \ref{subsec:diagram-PV-construction} and Appendix \ref{details-diagram-PV-construction}). The dashed line shows the best least-square fitting with a power-law model leaving the index $\beta$ as a free parameter ($j \propto r^{\beta}$).}
\label{fig:angular-momentum-profil-L1448-2A}
\end{figure*}

The panel (a) of Fig. \ref{fig:PV-diagrams-1} shows the PV$_\mathrm{rot}$ diagram of L1448-2A built from the velocity gradients observed at scales of $r <$1300~au. The index of the fitting by a power-law ($\alpha \sim$-0.9, see Table \ref{table:chi2-fit-profil-rotation}) is consistent with an infalling and rotating protostellar envelope. Therefore, we constrain the radial distribution of the specific angular momentum of L1448-2A at radii of 60$-$1300~au (see Figure \ref{fig:angular-momentum-profil-L1448-2A}).

%%%%%%%%%%%
\clearpage
\subsection{L1448-NB} \label{sec:comments-L14482-NB}
L1448-NB (or L1448-IRS3) is a Class~0 protostar in the L1448N complex in the Perseus cloud at (293 $\pm$ 20)~pc \citep{OrtizLeon18}. This source is part of a multiple system: L1448-NA, L1448-NB, and L1448-NC. NA and NB are separated by 1700~au, and NB and NC are separated by 4900~au \citep{Maury18}. L1448-NB is itself also a multiple system: \cite{Maury18} find a binary system with a separation of $\sim$140~au. ALMA 1.3~mm observations with an angular resolution of $\sim$0.2$\arcsec$ showed L1448-NB as a triple protostar system with binaries separated by $\sim$60~au and a third core separated by $\sim$180~au from others \citep{Tobin16}.

\begin{figure*}[!ht]
\centering
\includegraphics[scale=0.5,angle=0,trim=0cm 1.5cm 0cm 1.5cm,clip=true]{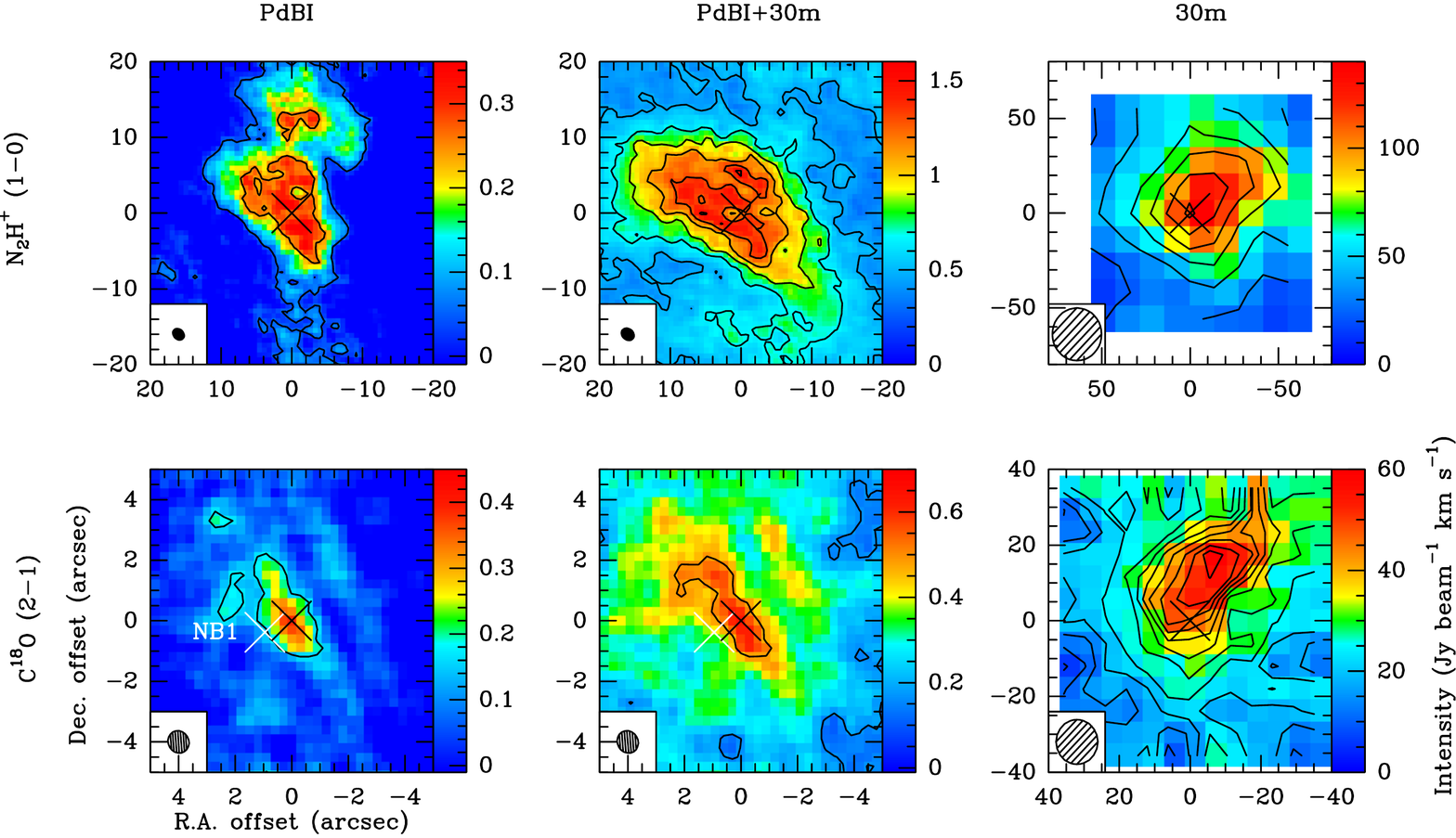}
\caption{Same as Figure \ref{fig:intensity-maps-L1448-2A}, but for L1448-NB. The white cross represents the position of main protostar L1448-NB1 determined from the 1.3~mm dust continuum emission (see Table \ref{table:sample}). The black cross represents the position of the secondary protostar L1448-NB2 of the multiple system. The black lines represent the integrated intensity contours of each tracer starting at 5$\sigma$ and increasing in steps of 20$\sigma$ for N$_{2}$H$^{+}$ and 10$\sigma$ for C$^{18}$O (see Tables \ref{table:details-obs-c18o} and \ref{table:details-obs-n2hp}).
}
\label{fig:intensity-maps-L1448NB}
\end{figure*}

Figures \ref{fig:intensity-maps-L1448NB} and \ref{fig:velocity-maps-L1448NB} show the integrated intensity and centroid velocity maps obtained for L1448-NB from the PdBI, combined, and 30m CALYPSO datasets for the C$^{18}$O and N$_2$H$^+$ emission respectively.
The C$^{18}$O emission from the PdBI dataset does not peak on the main protostar L1448-NB1 but on the secondary NB2 resolved by 1.3~mm continuum emission of the binary system (see Table \ref{table:sample} and Figure \ref{fig:intensity-maps-L1448NB}). Thus, the origin of the coordinate offsets is chosen to be the secondary protostar NB2 of the binary system
(RA: 03$^h$25$^m$36$^s$.315, Dec.: 30$^{\circ}$45$\arcmin$15$\arcsec$.15; see Table \ref{table:sample}) to study the kinematics in the protostellar envelope of L1448-NB.

The gradients observed in the PdBI and combined velocity maps of the C$^{18}$O emission (see bottom left and middle panels on Figure \ref{fig:velocity-maps-L1448NB}) are along a northeast-southwest axis and have a PA of $\Theta \sim$50$^{\circ}$, namely a difference $>$50$^{\circ}$ with respect to the equatorial axis (see Table \ref{table:gradient-velocity-fit}). Therefore, these gradients at $r <$1000~au are a merging of the kinematics between several mechanisms: ejection by outflows, orbital motions of the binary system NB1 and NB2, and rotational motions of the envelope along the equatorial axis. They are consistent with those observed from SMA observations of C$^{18}$O emission by \cite{Lee15} and \cite{Yen15b}. 
In the outer envelope at $r >$1000~au, velocity gradients are in the direction of the bipolar outflows. They have an angle of 64$^{\circ} \leq \Delta \Theta \leq $90$^{\circ}$ with respect to the equatorial axis (see Table \ref{table:gradient-velocity-fit}). Therefore, these gradients are not dominated by rotational motions along the equatorial axis but by ejection motions from bipolar outflows or due to an external contamination.

\begin{figure*}[!ht]
\centering
\includegraphics[width=10cm]{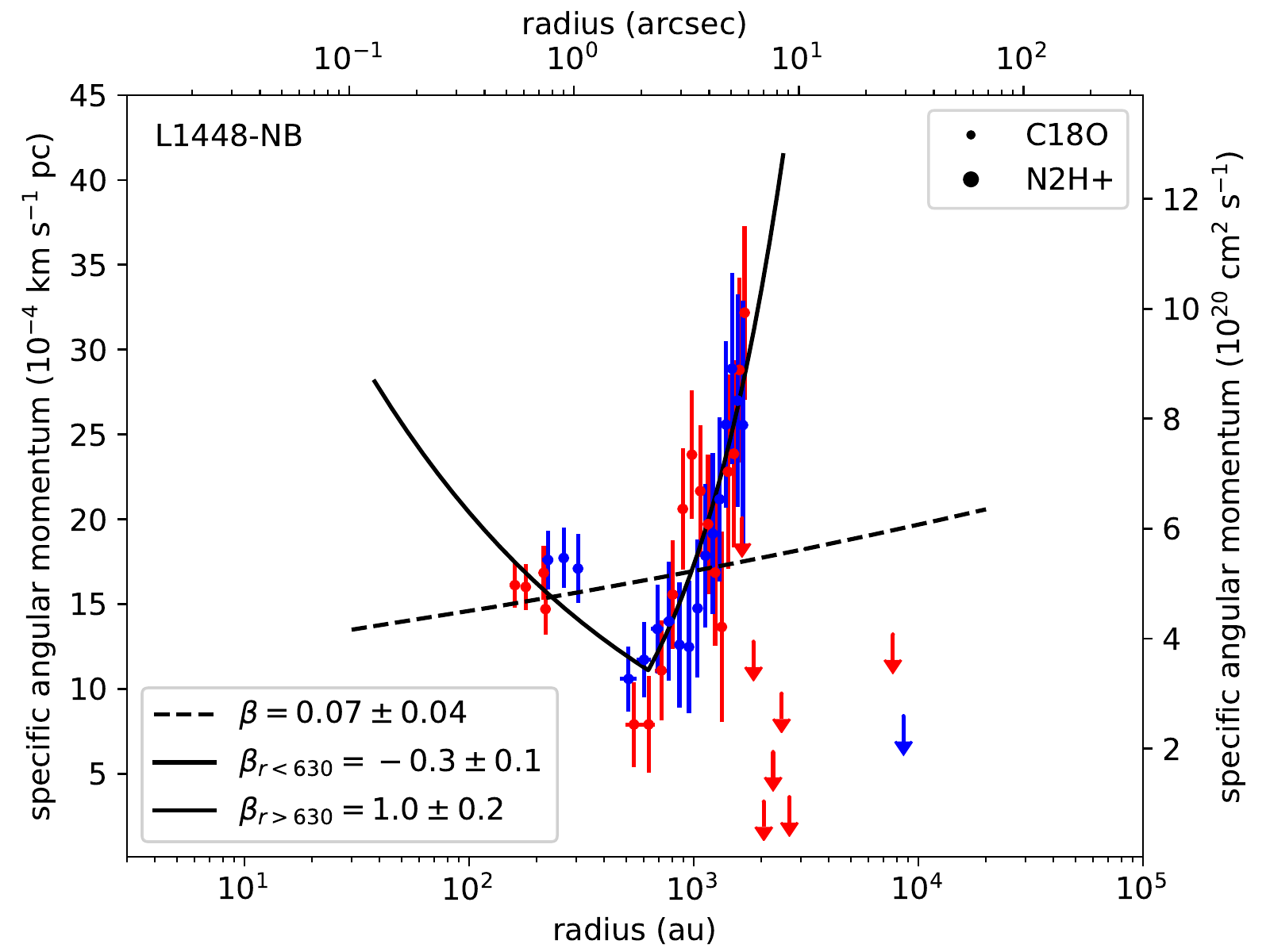}
\caption{Same as Figure \ref{fig:angular-momentum-profil-L1448-2A}, but for L1448-NB. The dashed line shows the best least-square fitting with a power-law model ($j \propto r^{\beta}$) whereas the solid line shows the best least-square fitting with broken power-law. }
\label{fig:angular-momentum-profil-L1448NB}
\end{figure*}

The panel (b) of Fig. \ref{fig:PV-diagrams-1} shows the PV$_\mathrm{rot}$ diagram of L1448-NB built from the velocity gradients observed at scales of $r <$1700~au. The index of the fitting by a power-law ($\alpha \sim$-0.9, see Table \ref{table:chi2-fit-profil-rotation}) is consistent with an infalling and rotating protostellar envelope. Therefore, we constrain the radial distribution of the specific angular momentum of L1448-NB at radii of 150$-$1700~au (see Figure \ref{fig:angular-momentum-profil-L1448NB}).

\begin{figure*}[!ht]
\centering
\includegraphics[scale=0.3,angle=0,trim=0cm 3cm 0cm 3.5cm,clip=true]{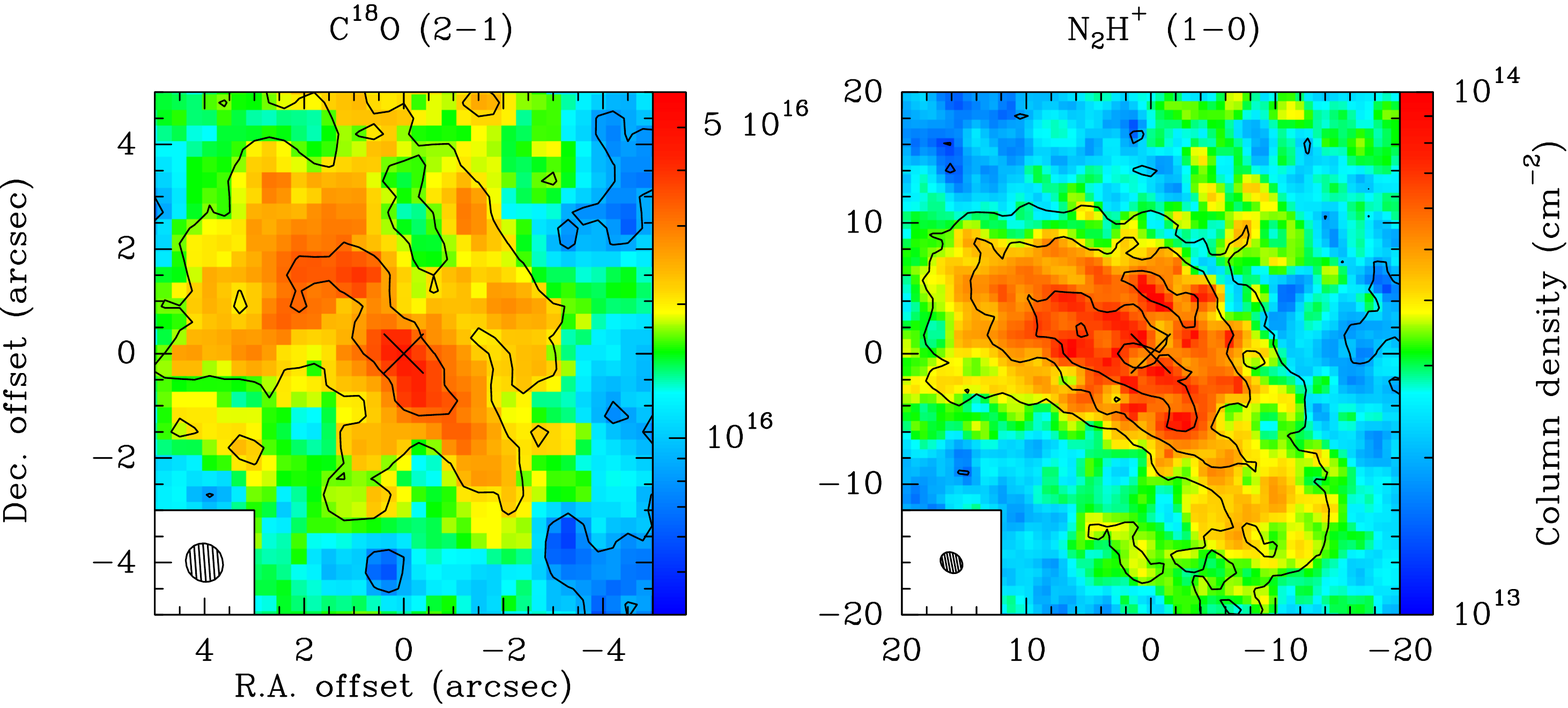}
\caption{Same Figure as \ref{fig:column-density-maps-L1448-2A} for L1448-NB. 
}
\label{fig:column-density-maps-L1448NB}
\end{figure*}

%%%%%%%%%%%
\clearpage
\subsection{L1448-C} \label{sec:comments-L1448C}
L1448-C (or L1448-mm) is located in the Perseus molecular cloud at a distance of (293 $\pm$ 20)~pc \citep{OrtizLeon18}. This source was firstly detected as a radio source at 2~cm \citep{Curiel90} associated with a strong millimetric continuum emission \citep{Bachiller91}. This Class~0 protostar \citep{Barsony98} has a powerful and very collimated outflow ($\sim$70~km~s$^{-1}$; \citealt{Bachiller90,Guilloteau92,Bachiller95,Hirano10}). Podio \& CALYPSO (in prep.) estimate the PA of this outflow at -17$^{\circ}$ (see Table \ref{table:sample}). \cite{Maret20} detect for the first time hints of Keplerian rotation at scales of $r \sim$200~au.

\begin{figure*}[!ht]
\centering
\includegraphics[scale=0.5,angle=0,trim=0cm 1.5cm 0cm 1.5cm,clip=true]{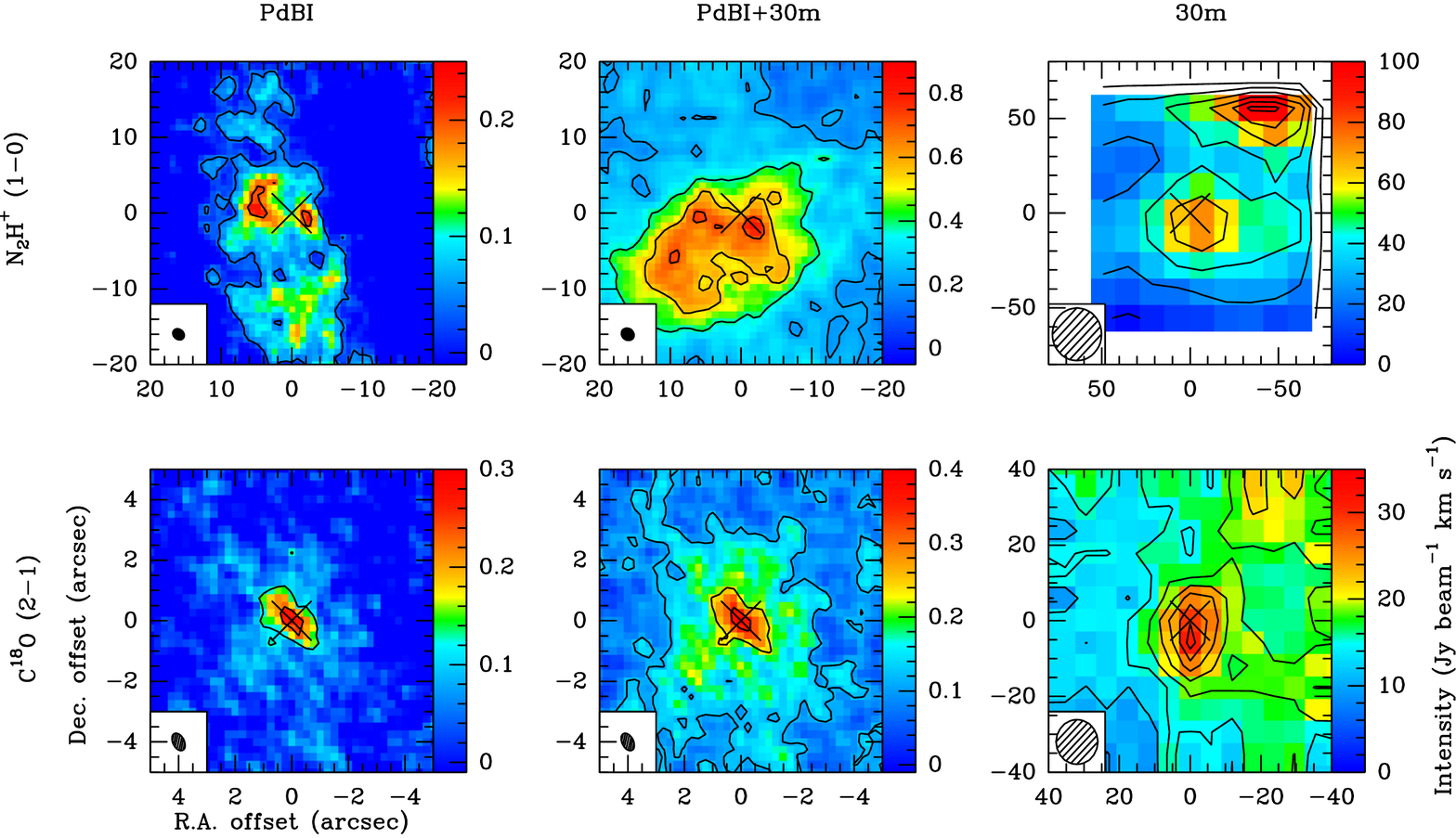}
\caption{Same as Figure \ref{fig:intensity-maps-L1448-2A}, but for L1448-C. The black lines represent the integrated intensity contours of each tracer starting at 5$\sigma$ and increasing in steps of 20$\sigma$ for N$_{2}$H$^{+}$ and 5$\sigma$ for C$^{18}$O (see Tables \ref{table:details-obs-c18o} and \ref{table:details-obs-n2hp}
}
\label{fig:intensity-maps-L1448C}
\end{figure*}

Figures \ref{fig:intensity-maps-L1448C} and \ref{fig:velocity-maps-L1448C} show the integrated intensity and centroid velocity maps obtained for L1448-C from the PdBI, combined, and 30m CALYPSO datasets for the C$^{18}$O and N$_2$H$^+$ emission respectively.
The gradients observed in the PdBI and combined velocity maps of the C$^{18}$O emission (see bottom left and middle panels on Figure \ref{fig:velocity-maps-L1448C}) have a PA of -140$^{\circ}$ < $\Theta$ <-121$^{\circ}$, namely an angle difference $<$30$^{\circ}$ with respect to the equatorial axis (see Table \ref{table:gradient-velocity-fit}). This gradient is consistent with the one observed in C$^{18}$O (2$-$1) emission with SMA \citep{Yen13,Yen15}. The 30m datasets show no velocity gradient (see bottom right panel on Figure \ref{fig:velocity-maps-L1448C}).
From the N$_{2}$H$^{+}$ emission (see top panels on Figure \ref{fig:velocity-maps-L1448C}), we observed a clear velocity gradient with an orientation consistent with the outflow axis ($\Theta$ $\geq$152$^{\circ}$, see Table \ref{table:gradient-velocity-fit}). However, we detect a weak velocity gradient along the equatorial axis. The kinematics seems to be dominated by bipolar jets and outflows or by external contamination from scales $>$1500~au. 
\cite{Curiel99} interpreted the gradient along the equatorial axis and the gradient along the outflows at larger scales as a suggestion that the envelope is rotating and contracting at similar velocities.

\begin{figure*}[!ht]
\centering
\includegraphics[width=10cm]{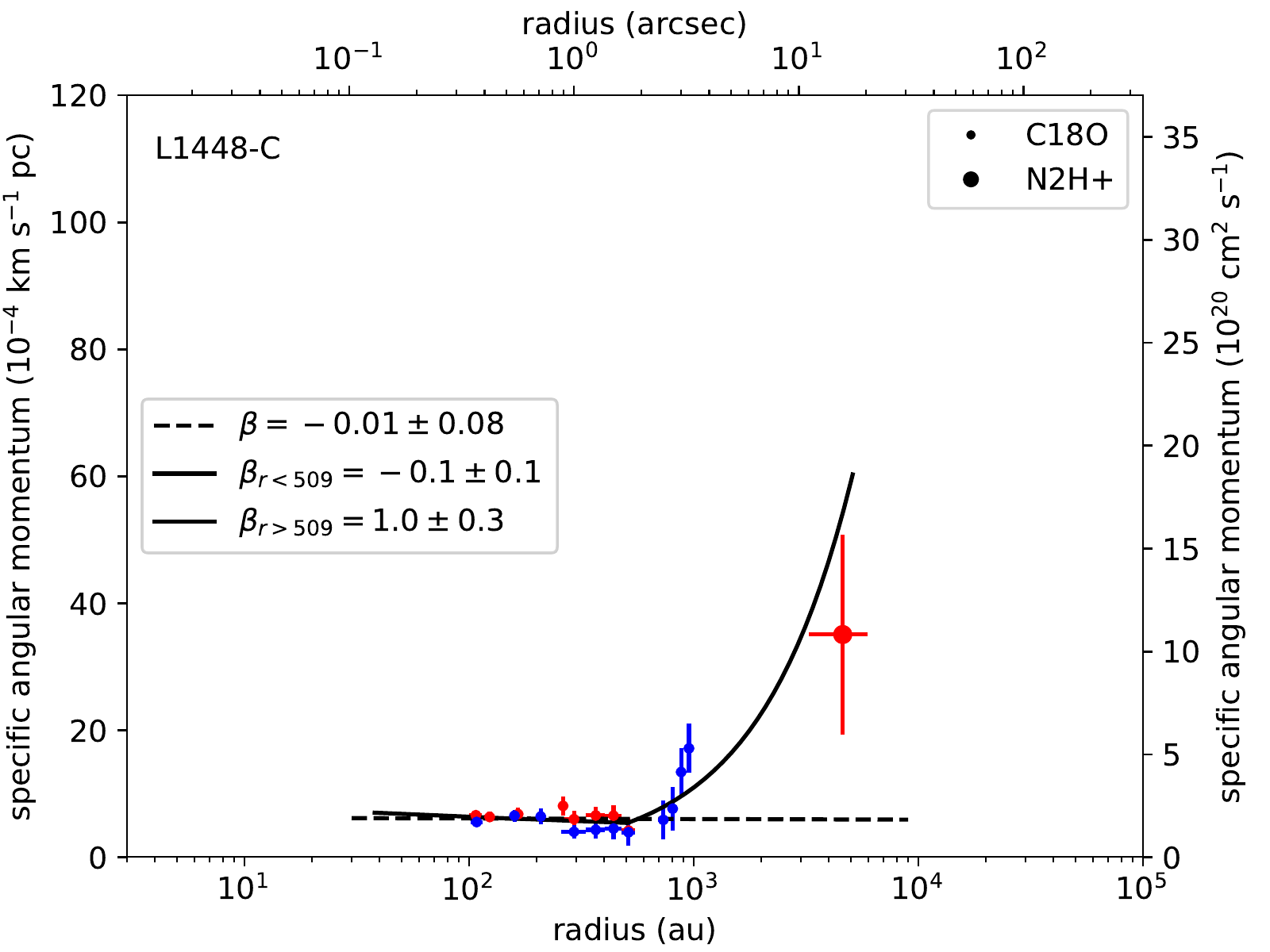}
\caption{Same as Figure \ref{fig:angular-momentum-profil-L1448-2A}, but for L1448-C. The dashed line shows the best least-square fitting with a power-law model ($j \propto r^{\beta}$) whereas the solid line shows the best least-square fitting with broken power-law. }
\label{fig:angular-momentum-profil-L1448C}
\end{figure*}

The panel (c) of Fig. \ref{fig:PV-diagrams-1} shows the PV$_\mathrm{rot}$ diagram of L1448-C built from the velocity gradients observed at scales of $r <$4000~au. The index of the fitting by a power-law ($\alpha \sim$-1, see Table \ref{table:chi2-fit-profil-rotation}) is consistent with an infalling and rotating protostellar envelope. From SMA observations, \cite{Yen13} obtain an index $\alpha \sim$-1 on scales of 100$-$3000~au, consistent with our results. Figure \ref{fig:angular-momentum-profil-L1448C} shows the radial distribution of the specific angular momentum of L1448-C at radii of 100$-$4000~au.


%%%%%%%%%%%%%
\clearpage
\subsection{IRAS2A}
IRAS2A (also known as NGC1333-IRAS2A) is located in the molecular cloud NGC1333 in the Perseus complex at (293 $\pm$ 20)~pc \citep{OrtizLeon18}. This Class~0 protostar was firstly detected via observations of dust continuum emission at 450~$\mu$m, 850~$\mu$m, and at 3~mm \citep{Sandell94,Sandell01,Jorgensen04-n2hp}. The CO emission showed the presence of two bipolar outflows in the north-south east-west direction \citep{Knee00}, suggesting the presence of a binary system. Dust continuum observations with VLA resolve IRAS2A into a protobinary system separated by $\sim$140~au \citep{Tobin15}, which is not resolved by the CALYPSO observations \citep{Maury18}.
\cite{Codella14} and Podio \& CALYPSO (in prep.) detect jet associated with the principal component with a PA of +205$^{\circ}$ and a tentative SiO and SO jet emission associated with a possible secondary with a PA = -65$^{\circ}$. In this study of the kinematics, we only used the first value (see Table \ref{table:sample}).

H$^{13}$CN observations from SMA with an angular resolution $<$1$\arcsec$ ($\sim$200~au) do not show organized velocity gradients or hints of Keplerian motions \citep{Brinch09}. CALYPSO methanol observations suggest the presence of a weak velocity gradient oriented in the direction perpendicular to the outflow axis, consistent with low rotational motions of the inner envelope \citep{Maret14}.

\begin{figure*}[!ht]
\centering
\includegraphics[scale=0.5,angle=0,trim=0cm 1.5cm 0cm 1.5cm,clip=true]{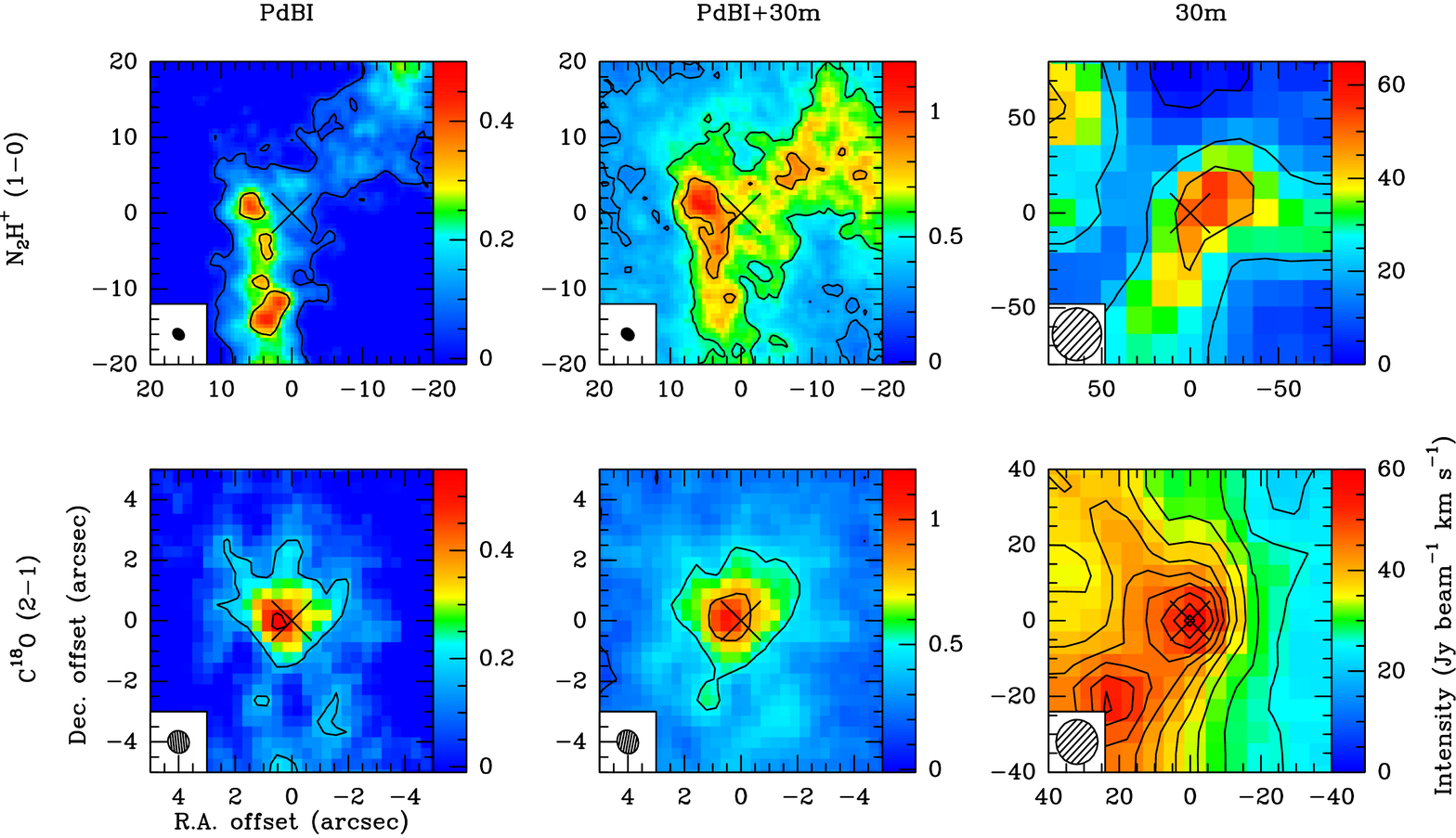}
\caption{Same as Figure \ref{fig:intensity-maps-L1448-2A}, but for IRAS2A. The black lines represent the integrated intensity contours of each tracer starting at 5$\sigma$ and increasing in steps of 25$\sigma$ for N$_{2}$H$^{+}$ and 10$\sigma$ for C$^{18}$O (see Tables \ref{table:details-obs-c18o} and \ref{table:details-obs-n2hp}.
}
\label{fig:intensity-maps-IRAS2A}
\end{figure*}

Figures \ref{fig:intensity-maps-IRAS2A} and \ref{fig:velocity-maps-IRAS2A} show the integrated intensity and centroid velocity maps obtained for IRAS2A from the PdBI, combined, and 30m CALYPSO datasets for the C$^{18}$O and N$_2$H$^+$ emission respectively. The gradients observed in the PdBI and combined velocity maps of the C$^{18}$O emission (see bottom left and middle panels on Figure \ref{fig:velocity-maps-IRAS2A}) have an angle difference of $\Delta \Theta\leq$20$^{\circ}$ with respect to the equatorial axis (see Table \ref{table:gradient-velocity-fit}). These gradients seem to be due to rotational motions of the envelope slightly contaminated by the outflows or the orbital motions of the multiple system. The other panels on the Figure \ref{fig:velocity-maps-IRAS2A} show velocity gradient with a PA along a northwest southeast axis and a angle difference of -20$^{\circ}$ < $\Theta$ <-20$^{\circ}$. These gradients are not due to rotational motions of the envelope but motions from larger scales. Moreover, the integrated intensity trace a filamentary structure which are consistent with the filament in which the source is embedded.

\begin{figure*}[!ht]
\centering
\includegraphics[width=10cm]{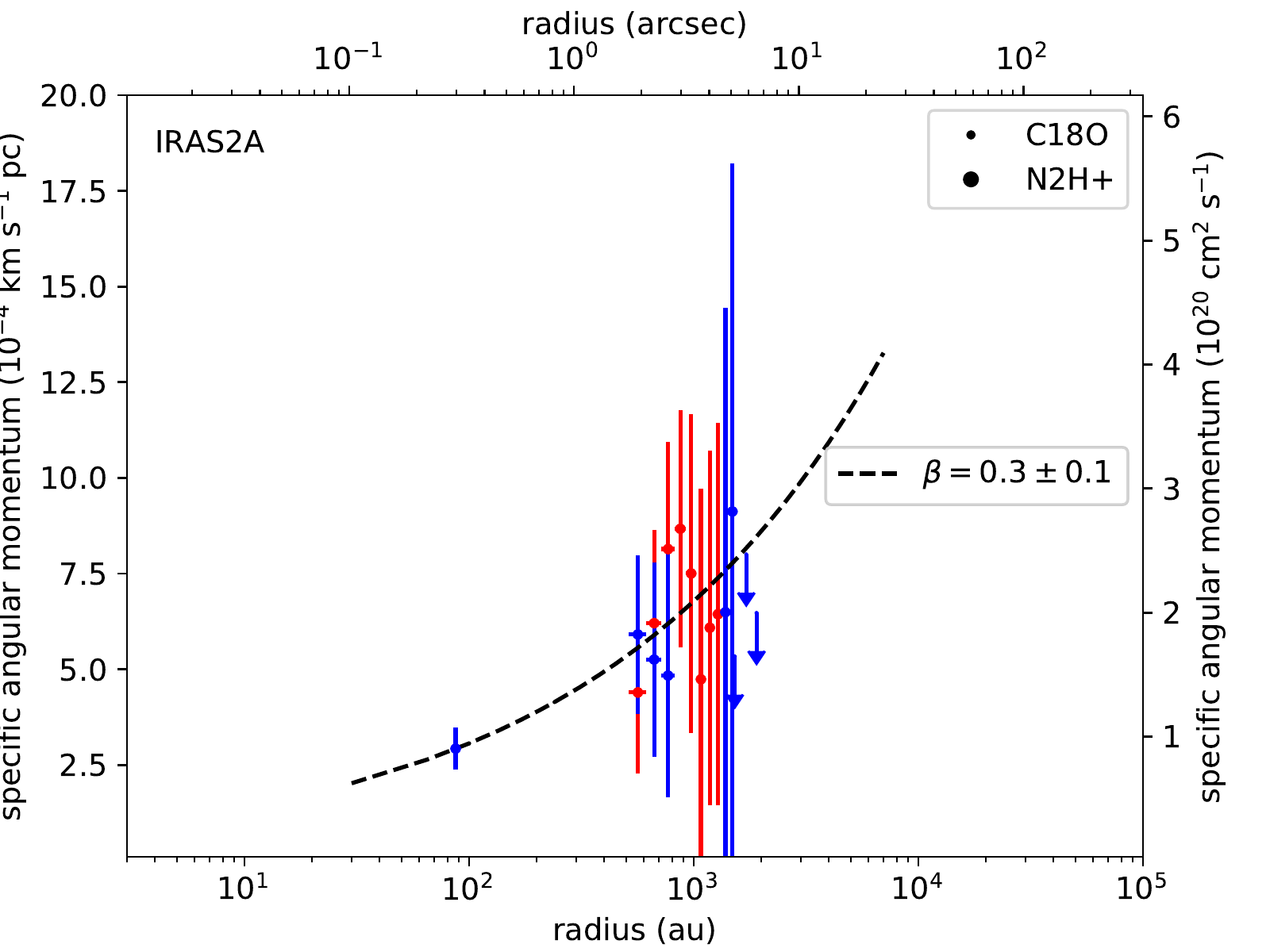}
\caption{Same as Figure \ref{fig:angular-momentum-profil-L1448-2A}, but for IRAS2A.}
\label{fig:angular-momentum-profil-IRAS2A}
\end{figure*}

The panel (d) of Fig. \ref{fig:PV-diagrams-1} shows the PV$_\mathrm{rot}$ diagram of IRAS2A built from the velocity gradients observed in C$^{18}$O emission at scales of $r <$1500~au. The index of the fitting by a power-law ($\alpha \sim$-0.7, see Table \ref{table:chi2-fit-profil-rotation}) is consistent with a Keplerian rotation. However, we also obtain a good reduced $\chi^2$ when we fit the PV$_\mathrm{rot}$ diagram by an infalling and rotating envelope by fixing the index of the power-law at -1 ($\sim$0.9, see Table \ref{table:chi2-fit-profil-rotation}). 
Thus, for this source, the CALYPSO dataset only allow us to estimate a range of the power-law index between -1 and -0.7 (see Table \ref{table:chi2-fit-profil-rotation}) at the scales of 80$-$1500~au probed in our analysis. Moreover, Keplerian rotation is not detected at the smaller envelope radii investigated by \cite{Maret20}. Thus, the Keplerian rotation due to a large disk is not a robust interpretation for this source.
Figure \ref{fig:angular-momentum-profil-IRAS2A} shows the radial distribution of the specific angular momentum of IRAS2A at radii of 80$-$1500~au.


%%%%%%%%%%%%%
\clearpage
\subsection{SVS13-B}

SVS13 is a multiple system located in the molecular cloud NGC1333 in the Perseus complex at a distance of (293 $\pm$ 20)~pc \citep{OrtizLeon18}. This system is composed of three main sources which are called SVS13-A, B, and C, aligned in the northeast southwest direction \citep{Looney03}. SVS13-A is a Class~I protostar and SVS13-B is a Class~0 protostar. These sources are gathered in a big filamentary structure observed at 450~$\mu$m and 1.3~mm \citep{Chandler00,Hull14}. Observations at 70~$\mu$m detected SVS13-A and C but not SVS13-B, which suggests that the latter is deeply embedded in the filamentary structure \citep{Chen09}. SVS13-A has a powerful outflow observed in the northwest southeast direction by \cite{Bachiller00, Plunkett13}. SVS13-B also harbors a very collimated outflows \citep{Bachiller98}. Podio \& CALYPSO (in prep.) estimate the PA of the outflows at 167$^{\circ}$ (see Table \ref{table:sample}).

A velocity gradients with a magnitude of 28~km~s$^{-1}$~pc$^{-1}$ and symmetric with respect to SVS13-A and B is detected in N$_2$H$^+$ emission from PdBI observations. This gradient suggests that the binary system is physically linked \citep{Chen09}. A study of dust continuum emission at 8~mm by \cite{Segura-Cox16} suggests a Keplerian disk with a radius $\lesssim$25~au. A similar study from CALYPSO observations at 1.3~mm also suggests an unresolved Keplerian disk with a radius $<$60~au \citep{Maury18}.

\begin{figure*}[!ht] 
\centering
\includegraphics[scale=0.5,angle=0,trim=0cm 1.5cm 0cm 1.5cm,clip=true]{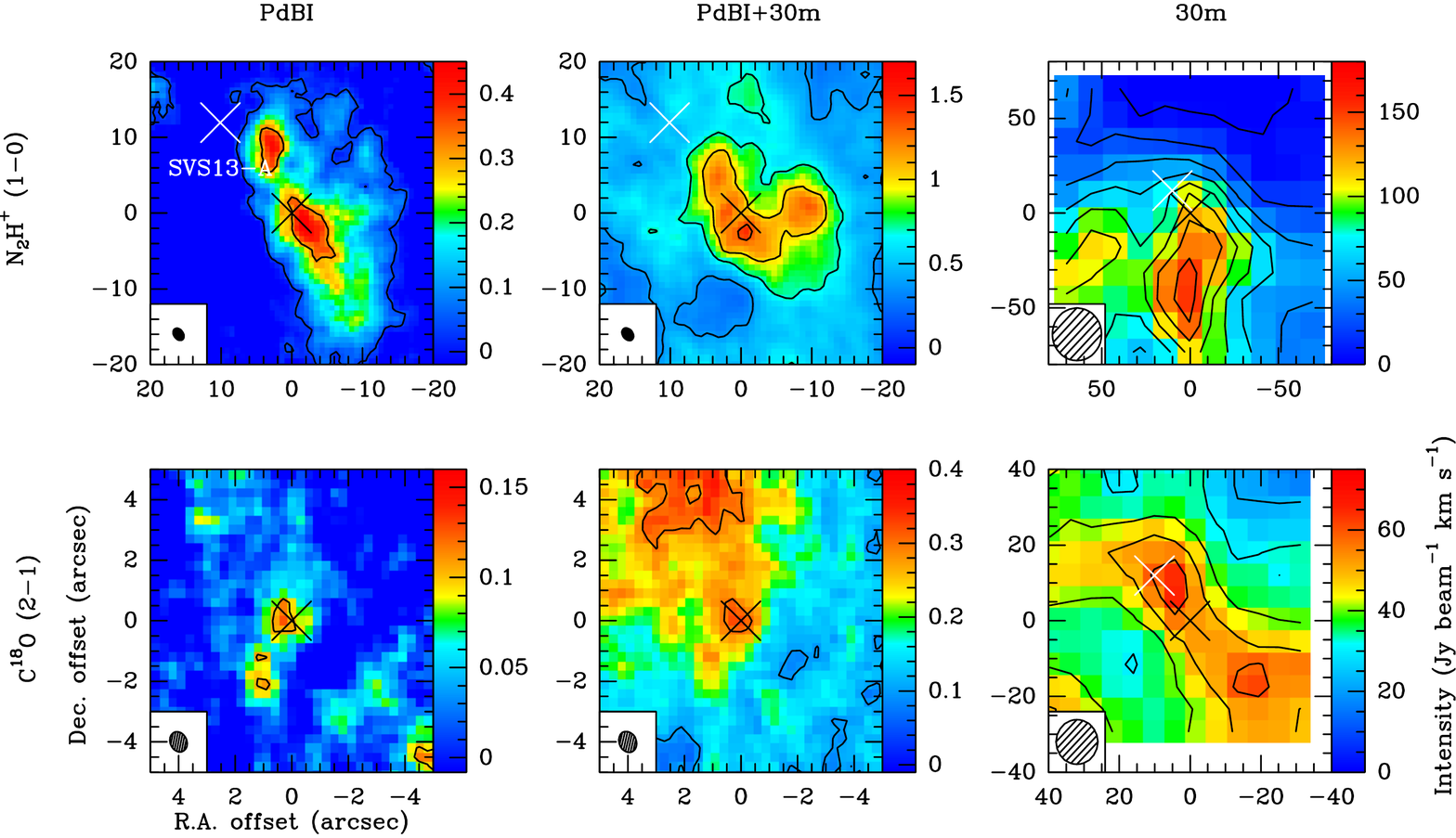}
\caption{Same as Figure \ref{fig:intensity-maps-L1448-2A}, but for SVS13-B. The white cross represents the position of the Class~I protostar SVS13-A determined from the 1.3~mm dust continuum emission \citep{Maury18}. The black lines represent the integrated intensity contours of each tracer starting at 5$\sigma$ and increasing in steps of 40$\sigma$ for N$_{2}$H$^{+}$ and 10$\sigma$ for C$^{18}$O (see Tables \ref{table:details-obs-c18o} and \ref{table:details-obs-n2hp}.
}
\label{fig:intensity-maps-SVS13B}
\end{figure*}

\begin{figure*}[!ht]
\centering
\includegraphics[scale=0.5,angle=0,trim=0cm 1.5cm 0cm 1.5cm,clip=true]{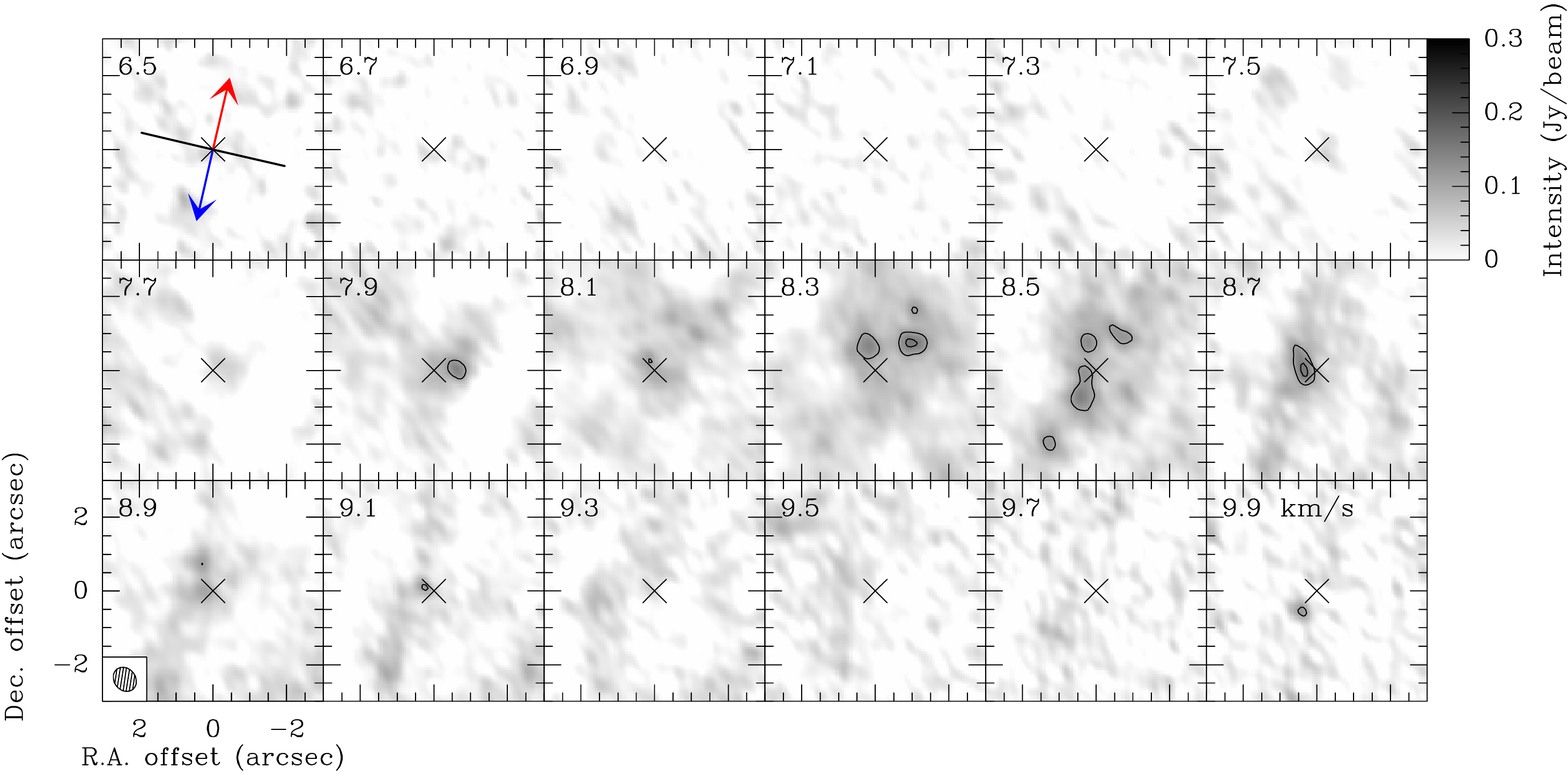}
\caption{Same as Figure \ref{fig:channel-maps-L1448-2A}, but for SVS13-B. The systemic velocity is estimated to be $\mathrm{v}_\mathrm{sys}=$8.3~km~s$^{-1}$ (see Table \ref{table:chi2-fit-lines}).
}
\label{fig:channel-maps-SVS13B}
\end{figure*}

Figures \ref{fig:intensity-maps-SVS13B} and \ref{fig:velocity-maps-SVS13B} show the integrated intensity and centroid velocity maps obtained for SVS13-B from the PdBI, combined, and 30m CALYPSO datasets for the C$^{18}$O and N$_2$H$^+$ emission respectively. The C$^{18}$O emission is weakly detected ($\sim$5$\sigma$) from PdBI observations for this source: the  emission  is  dominated  by  its  companion SVS13-A. The channel maps on Figure \ref{fig:channel-maps-SVS13B} allow us to constrain the PV$_\mathrm{rot}$ diagram with one point at $r\sim$100~au in the envelope (see panel (e) of Fig. \ref{fig:PV-diagrams-1}). We observed a velocity gradient along the equatorial axis ($\mathrm{v}-\mathrm{v}_\mathrm{sys}<$0.4~km~s$^{-1}$, see panel (e) of Fig. \ref{fig:PV-diagrams-1}) in the combined velocity maps of the C$^{18}$O emission ($r<$500~au). From the PdBI and combined velocity maps in N$_2$H$^+$ emission (see top left and middle panels on Figure \ref{fig:velocity-maps-SVS13B}), the velocity gradient is dominated by the intrinsic velocity of the protostar SVS13-A. At larger scales ($r>$4500~au), the velocity gradients have an angle difference $\Delta \Theta \geq$70$^{\circ}$ with respect to the equatorial axis (see Table \ref{table:gradient-velocity-fit}). These gradients are not due to rotational motions of the envelope but motions from larger scales. Moreover, the integrated intensity trace a filamentary structure which are consistent with the filament in which the source is embedded (see right panels on Figure \ref{fig:velocity-maps-SVS13B}).

\begin{figure*}[!ht]
\centering
\includegraphics[width=10cm]{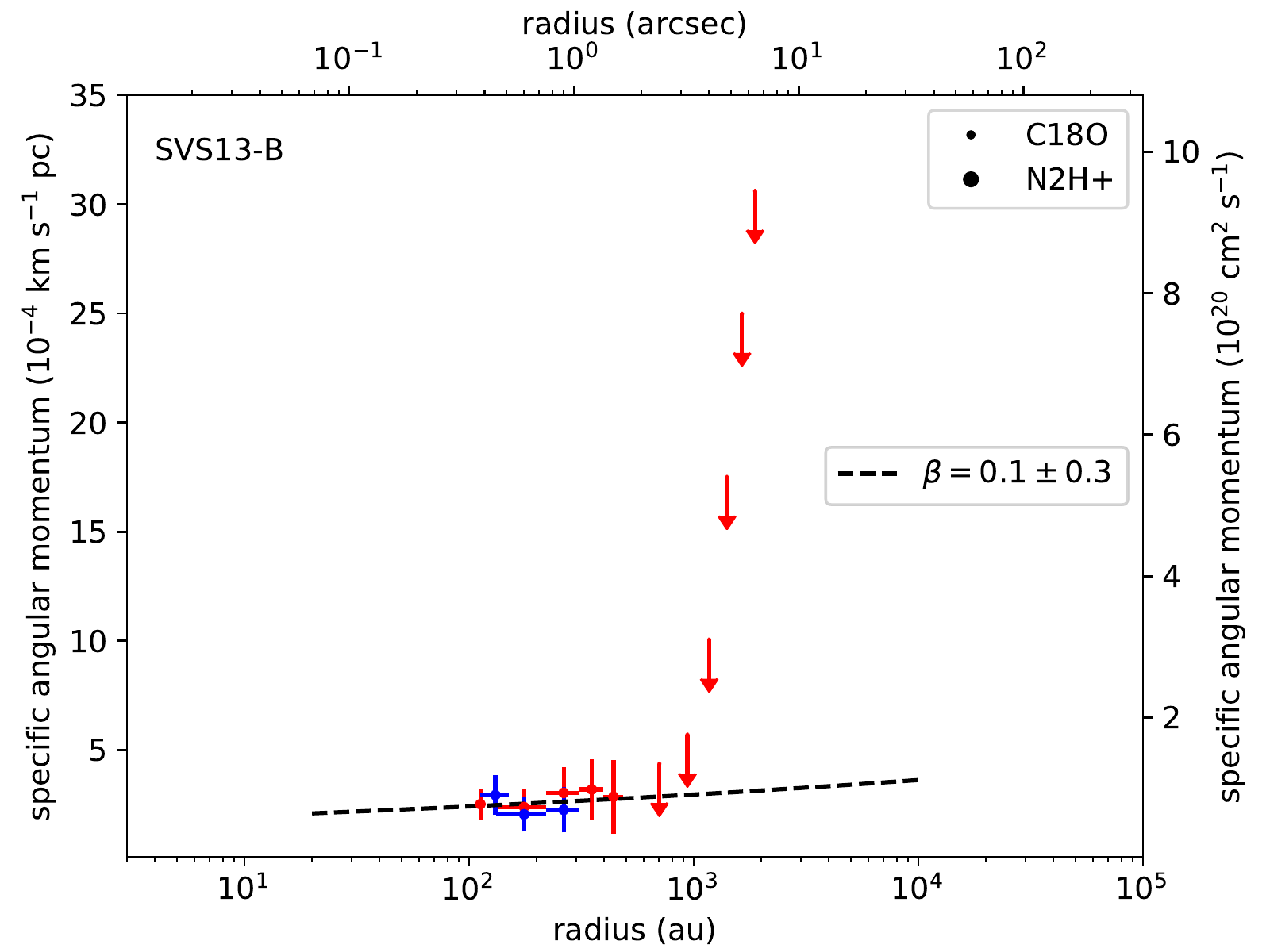}
\caption{Same as Figure \ref{fig:angular-momentum-profil-L1448-2A}, but for SVS13-B. }
\label{fig:angular-momentum-profil-SVS13B}
\end{figure*}

The panel (e) of Fig. \ref{fig:PV-diagrams-1} shows the PV$_\mathrm{rot}$ diagram of SVS13-B built from the velocity gradients observed in C$^{18}$O emission at scales of $r <$500~au. The index of the fitting by a power-law ($\alpha \sim$-0.9, see Table \ref{table:chi2-fit-profil-rotation}) is consistent with an infalling and rotating envelope. 
Figure \ref{fig:angular-momentum-profil-SVS13B} shows the radial distribution of the specific angular momentum of SVS13-B at radii of 100$-$500~au.


%%%%%%%%%%%%%
\clearpage
\subsection{IRAS4A} \label{sec:comments-IRAS4A}
The class~0 protostar IRAS4A is located in the molecular cloud NGC1333 in the Perseus complex at (293 $\pm$ 20)~pc \citep{OrtizLeon18}, in the vicinity of another young multiple system, IRAS4B \citep{Lay95}. It harbors a $\sim$2$\arcsec$ binary system \citep{Looney00, Girart06, Jorgensen07, Lopez-Sepulcre17, Maury18} and the dust continuum emission at (sub)millimeter wavelengths is dominated by the main protostar IRAS4A1. \cite{Santangelo15} and Podio \& CALYPSO (in prep.) estimate the PA of the outflows at 180$^{\circ}$ (see Table \ref{table:sample}).

\begin{figure*}[!ht]
\centering
\includegraphics[scale=0.3,angle=0,trim=0cm 3cm 0cm 3.5cm,clip=true]{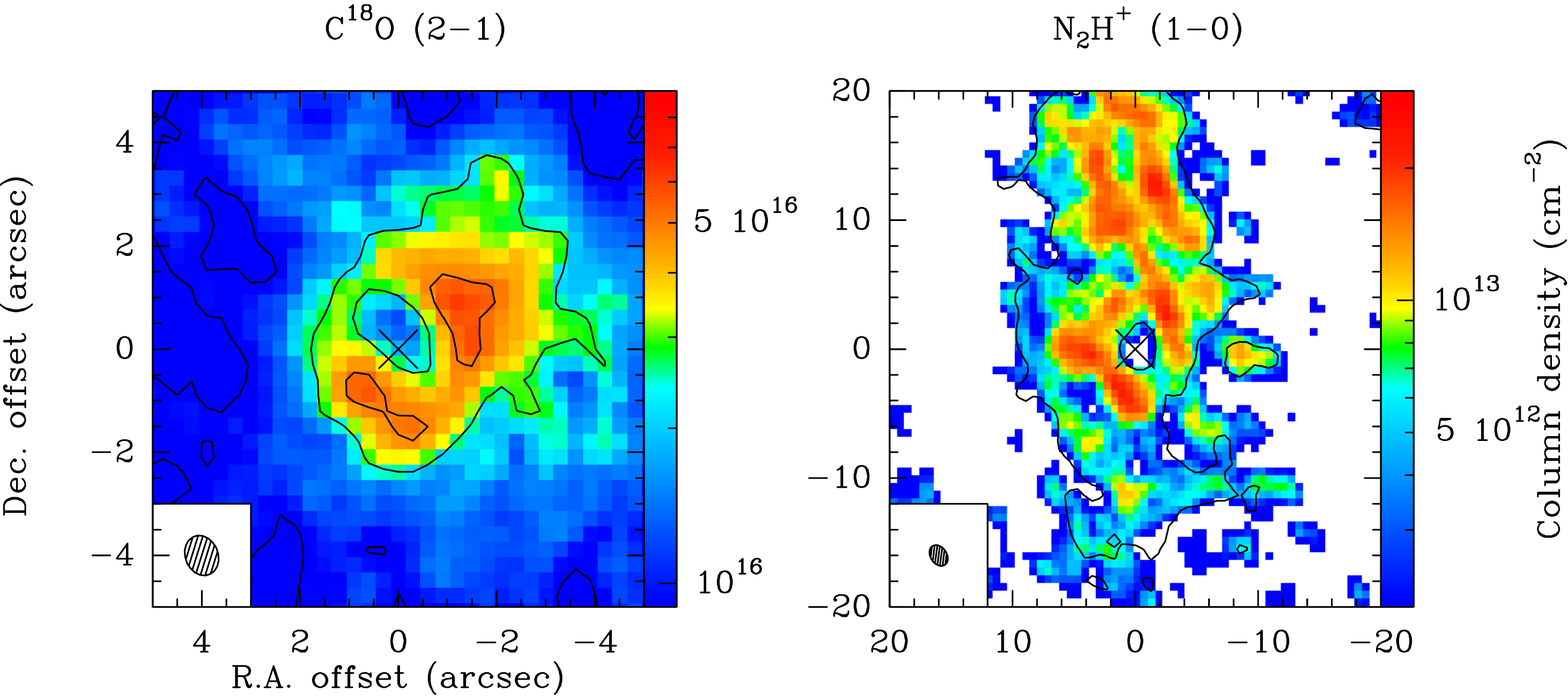}
\caption{Same Figure as \ref{fig:column-density-maps-L1448-2A} for IRAS4A. The N$_{2}$H$^{+}$ (1$-$0) column density maps is determined from the PdBI datasets. The integrated intensity overlaid in black solid line is the integrated intensity from the 7 hyperfine components of the transition (see top left panel on Figure \ref{fig:intensity-maps-IRAS4A}).
}
\label{fig:column-density-maps-IRAS4A}
\end{figure*}

\begin{figure*}[!ht]
\centering
\includegraphics[scale=0.5,angle=0,trim=0cm 1.5cm 0cm 1.5cm,clip=true]{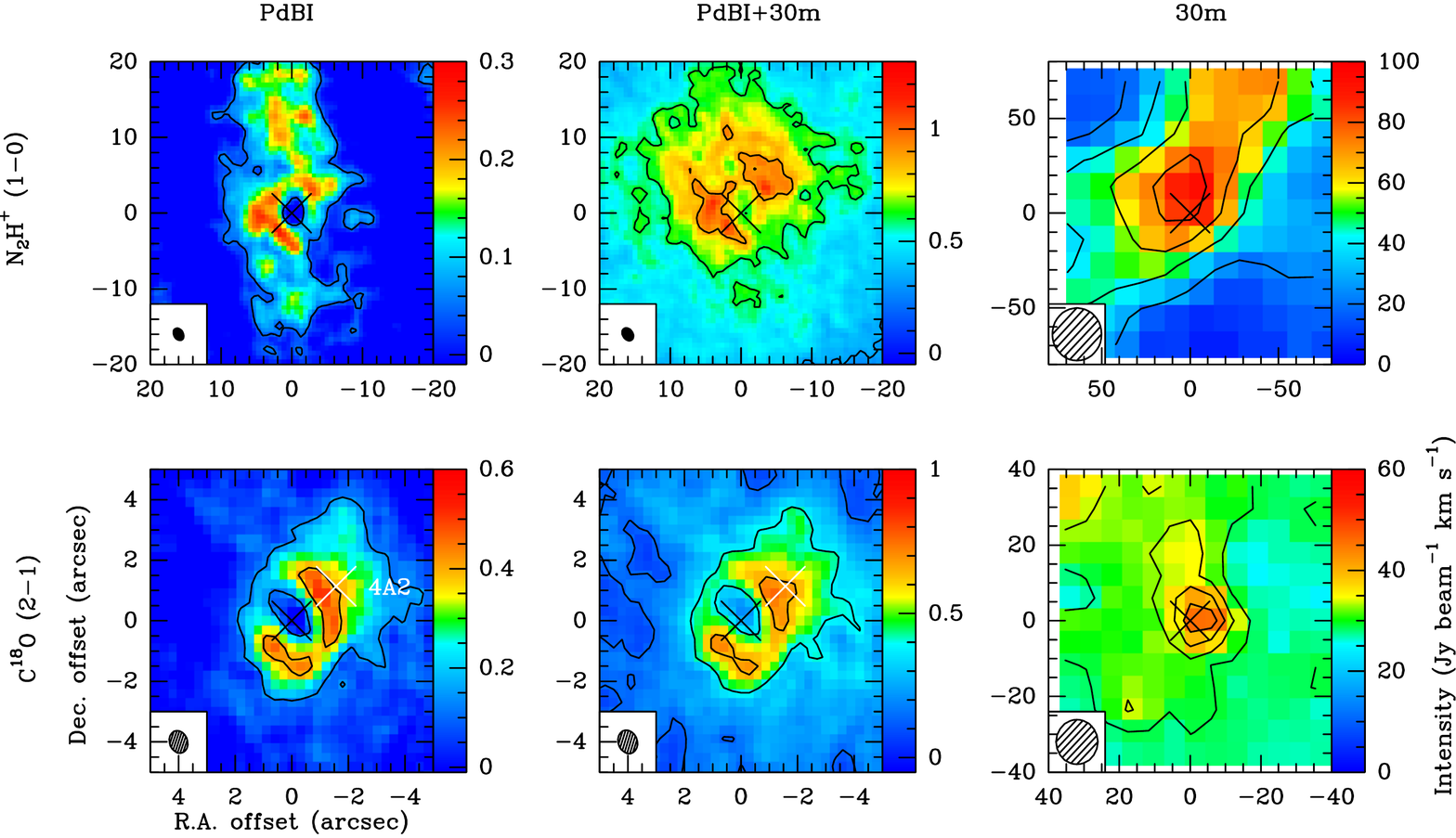}
\caption{Same as Figure \ref{fig:intensity-maps-L1448-2A}, but for IRAS4A. The white cross represents the position of secondary protostar IRAS4A2 determined from the 1.3~mm dust continuum emission (see Table \ref{table:sample}). The black cross represents the position of the main protostar IRAS4A1 of the multiple system. The black lines represent the integrated intensity contours of each tracer starting at 5$\sigma$ and increasing in steps of 30$\sigma$ for N$_{2}$H$^{+}$ and 10$\sigma$ for C$^{18}$O (see Tables \ref{table:details-obs-c18o} and \ref{table:details-obs-n2hp}.
}
\label{fig:intensity-maps-IRAS4A}
\end{figure*}

Figure \ref{fig:intensity-maps-IRAS4A} shows the integrated intensity obtained for IRAS4A from the PdBI, combined, and 30m CALYPSO datasets for the C$^{18}$O and N$_2$H$^+$ emission respectively. C$^{18}$O emission from the PdBI dataset is centered on the main protostar IRAS4A1 (RA: 03$^h$29$^m$10$^s$.537, Dec.: 31$^{\circ}$13$\arcmin$30$\arcsec$.98; see Table \ref{table:sample} and Figure \ref{fig:intensity-maps-IRAS4A}). 
This is why, the origin of the coordinate offsets is chosen to be the main protostar to study the kinematics in the protostellar envelope of IRAS4A. We also noticed a hole in the integrated intensity and column density (see Figure \ref{fig:column-density-maps-IRAS4A}) that suggests the C$^{18}$O emission may be optically thick at scales of $r \lesssim$100~au.
 
\cite{Belloche06} find two velocity components at $r \sim$6000~au by analyzing CS, C$_{34}$S and, N$_{2}$H$^{+}$ lines obtained with the 30m. This suggests that IRAS4A is collapsing with a velocity of $\mathrm{v}<$ 7.30~km~s$^{-1}$ triggered by a fast external compression with a velocity of $\mathrm{v} >$ 7.30~km~s$^{-1}$.
Figures \ref{fig:velocity-maps-SVS13B} show the centroid velocity maps obtained for SVS13-B from the PdBI, combined, and 30m CALYPSO datasets for the C$^{18}$O and N$_2$H$^+$ emission respectively.
To minimize contamination by external cloud compression, the velocity maps were constructed by fitting and removing a second gaussian component fixed at 7.7~km~s$^{-1}$.

The gradients observed in the PdBI and combined velocity maps of the C$^{18}$O emission (see bottom left and middle panels on Figure \ref{fig:velocity-maps-IRAS4A}) have an angle difference of $\Delta \Theta\leq$11$^{\circ}$ with respect to the equatorial axis (see Table \ref{table:gradient-velocity-fit}). These gradients are consistent with those observed in C$^{17}$O emission from SMA observations \citep{Ching16}. We notice that the gradient is aligned along and symmetric with respect to the binary system IRAS4A1 and IRAS4A2. This could suggest that the binary system is physically linked by orbital motions. 

\begin{figure*}[!ht]
\centering
\includegraphics[scale=0.5,angle=0,trim=0cm 1.5cm 0cm 1.5cm,clip=true]{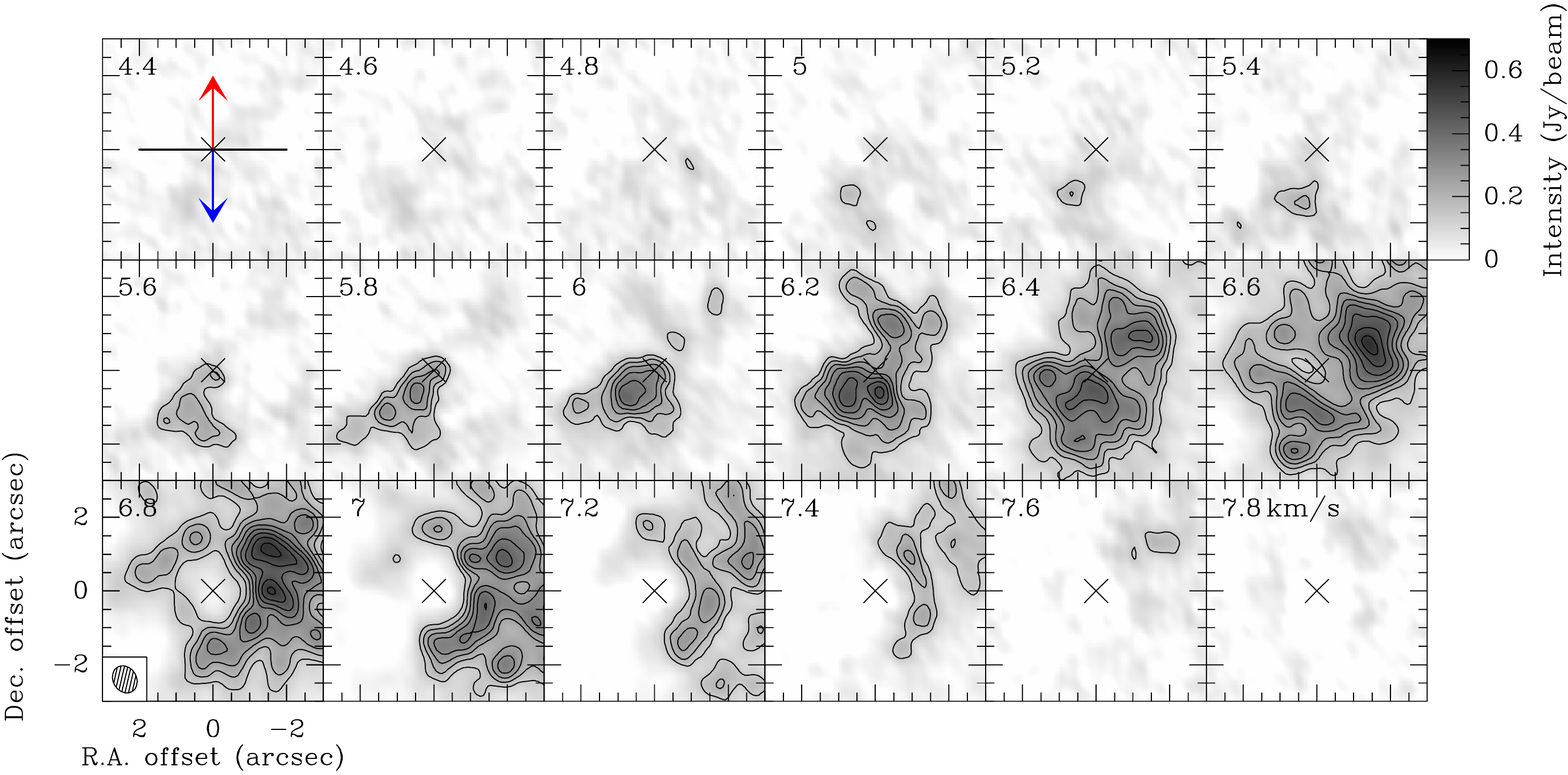}
\caption{Same as Figure \ref{fig:channel-maps-L1448-2A}, but for IRAS4A. The systemic velocity is estimated to be $\mathrm{v}_\mathrm{sys}=$6.6~km~s$^{-1}$ (see Table \ref{table:chi2-fit-lines}).
}
\label{fig:channel-maps-IRAS4A}
\end{figure*}

The channel maps do not exhibit an organized spatial distribution of velocities along the equatorial axis : the central emission fit show a position angle $>$|45$^{\circ}|$ with respect to the equatorial axis, suggesting a possible contamination by the outflow kinematics (see Figure \ref{fig:channel-maps-IRAS4A}). To minimize this contamination and probe rotational motions in the equatorial axis, we used the method in the image plane in the PdBI dataset instead of working in the visibilities to constrain in the inner envelope (200$-$800~au).

The velocity gradients observed in the PdBI velocity maps of the N$_2$H$^+$ emission (see top left panels on Figure \ref{fig:velocity-maps-IRAS4A}) have a direction of $\Theta \sim$-69$^{\circ}$ consistent with the equatorial axis. 
However, at larger scales ($r>$1000~au), the N$_2$H$^+$ emission is dominated by external cloud compression. The gradient have a direction of $\Theta \geq$32$^{\circ}$. Moreover, the integrated intensity of N$_2$H$^+$ emission from the 30m datasets (see right top panel on Figure \ref{fig:intensity-maps-IRAS4A}) trace the filamentary structure which are consistent with the filament in which the source is embedded.

The panel (l) of Fig. \ref{fig:PV-diagrams-1} shows the PV$_\mathrm{rot}$ diagram of IRAS4A built from the velocity gradients observed at scales of $r <$1600~au. This is the only source with increasing velocities from small to large scales. The index of the fitting by a power-law ($\alpha \sim$0.8, see Table \ref{table:chi2-fit-profil-rotation}) could be interpreted by solid-body rotation, but as the velocity gradient is not uniform as expected, this is not a robust interpretation for this source (see Sect. \ref{sec:results-PV-diagram}). 
Moreover, by considering only point at $r<$550~au, the index of the fitting by a power-law is consistent with an infalling and rotating envelope ($\alpha \sim$-1.3, see Table \ref{table:chi2-fit-profil-rotation}). We did not take this source into account to build the radial distribution of the specific angular momentum due to rotation motions.

%
%

%
%

%%%%%%%%%%%%%
\clearpage
\subsection{IRAS4B} \label{sec:comments-IRAS4B}
\begin{figure*}[!ht]
\centering
\includegraphics[scale=0.5,angle=0,trim=0cm 1.5cm 0cm 1.5cm,clip=true]{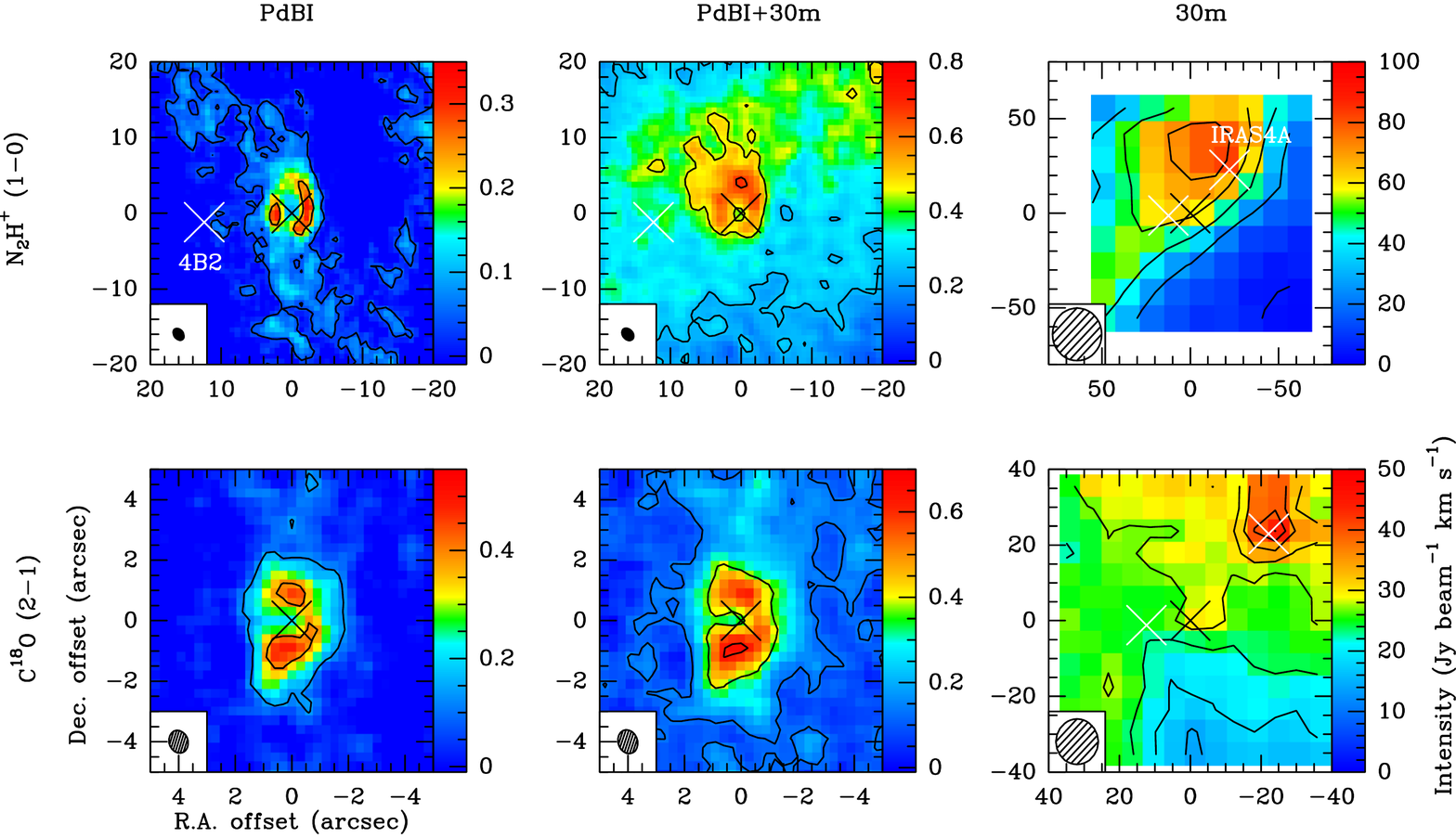}
\caption{Same as Figure \ref{fig:intensity-maps-L1448-2A}, but for IRAS4B. The white crosses represent the position of secondary protostar IRAS4B2 and the position of IRAS4A, respectively, determined from the 1.3~mm dust continuum emission (see Table \ref{table:sample}). The black cross represents the position of the secondary protostar L1448-NB2 of the multiple system. The black lines represent the integrated intensity contours of each tracer starting at 5$\sigma$ and increasing in steps of 25$\sigma$ for N$_{2}$H$^{+}$ and 10$\sigma$ for C$^{18}$O (see Tables \ref{table:details-obs-c18o} and \ref{table:details-obs-n2hp}.
}
\label{fig:intensity-maps-IRAS4B}
\end{figure*}

\begin{figure*}[!ht]
\centering
\includegraphics[scale=0.3,angle=0,trim=0cm 3cm 0cm 3.5cm,clip=true]{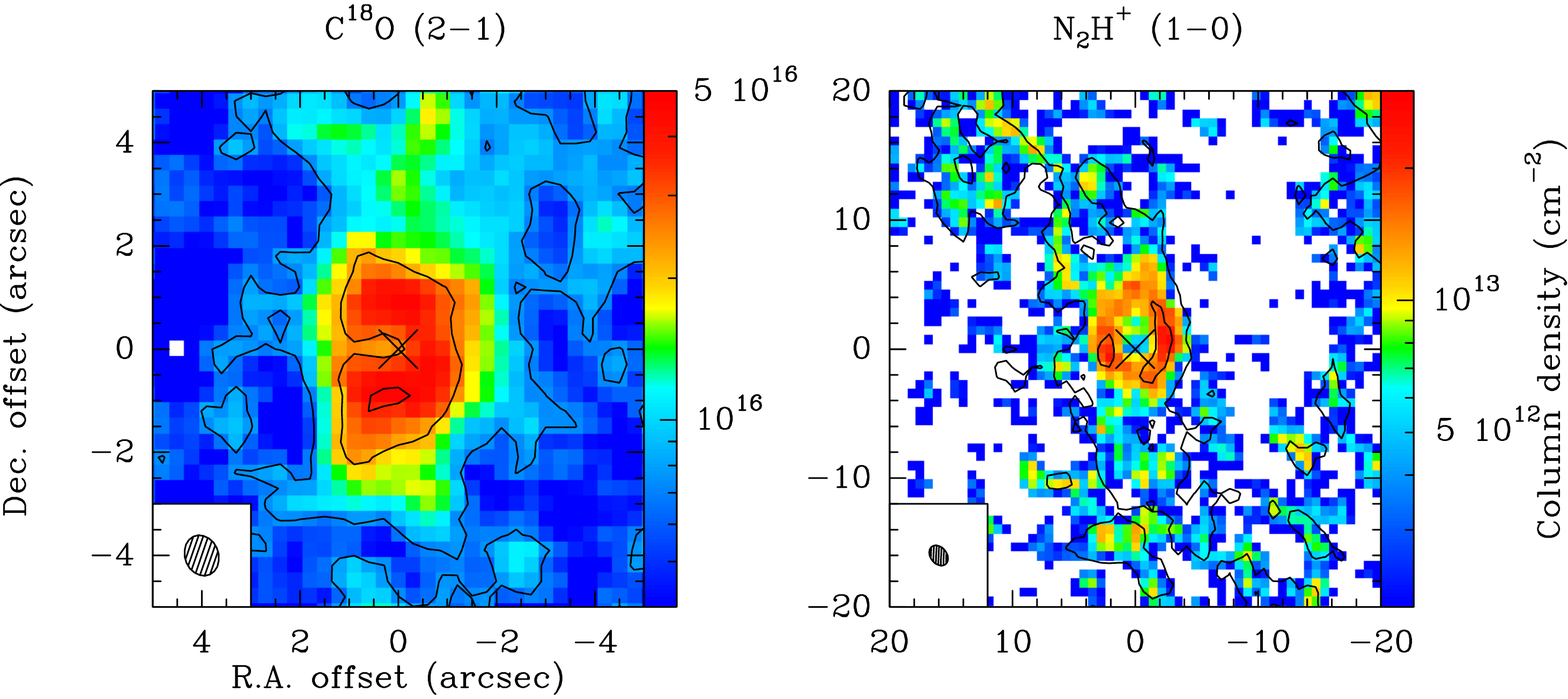}
\caption{Same Figure as \ref{fig:column-density-maps-IRAS4A} for IRAS4B. The N$_{2}$H$^{+}$ (1$-$0) column density maps is determined from the PdBI datasets. The integrated intensity overlaid in black solid line is the integrated intensity from the 7 hyperfine components of the transition (see top left panel on Figure \ref{fig:intensity-maps-IRAS4B})
}
\label{fig:column-density-maps-IRAS4B}
\end{figure*}

IRAS4B (or NGC1333-IRAS4B) is a Class~0 protobinary system in the Perseus molecular cloud at a distance of (293 $\pm$ 20)~pc \citep{OrtizLeon18}. The binary companions 4B1 and 4B2 are separated by $\sim$11$\arcsec$ \citep{Looney00,Jorgensen07}. The source harbors a young bipolar outflow almost aligned along a north-south direction \citep{Choi01}. Podio \& CALYPSO (in prep.) distinguish two outflows: a first associated with the main protostar 4B1 at 167$^{\circ}$ and a second with a PA of -99$^{\circ}$ (see Table \ref{table:sample}).
IRAS4B is in the vicinity of IRAS4A which undergo a collapse triggered by a fast external compression with a velocity of $\mathrm{v}>$ 7.30~km~s$^{-1}$ \citep{Belloche06}. Thus, collapse of IRAS4B could have triggered in the same way. It is why, to minimize contamination by external cloud compression, the velocity maps were constructed by fitting and removing a second Gaussian component fixed at 7.7~km~s$^{-1}$ (see Figure \ref{fig:intensity-maps-IRAS4B}). Previous studies of C$^{18}$O (2$-$1) emission from SMA observations are detected no organized velocity gradients consistent with rotational motions \citep{Yen13,Yen15}.

Figure \ref{fig:velocity-maps-IRAS4B} shows the centroid velocity maps obtained for IRAS4B from the PdBI, combined, and 30m CALYPSO datasets for the C$^{18}$O and N$_2$H$^+$ emission. The gradients observed in the PdBI and combined velocity maps of the C$^{18}$O emission (see bottom left and middle panels on Figure \ref{fig:velocity-maps-IRAS4B}) have an angle difference of $\Delta \Theta\leq$20$^{\circ}$ with respect to the equatorial axis (see Table \ref{table:gradient-velocity-fit}). 
The velocity gradients observed in the PdBI velocity maps of the N$_2$H$^+$ emission (see top left panels on Figure \ref{fig:velocity-maps-IRAS4B}) is also consistent with the equatorial axis.

\begin{figure*}[!ht]
\centering
\includegraphics[scale=0.5,angle=0,trim=0cm 1.5cm 0cm 1.5cm,clip=true]{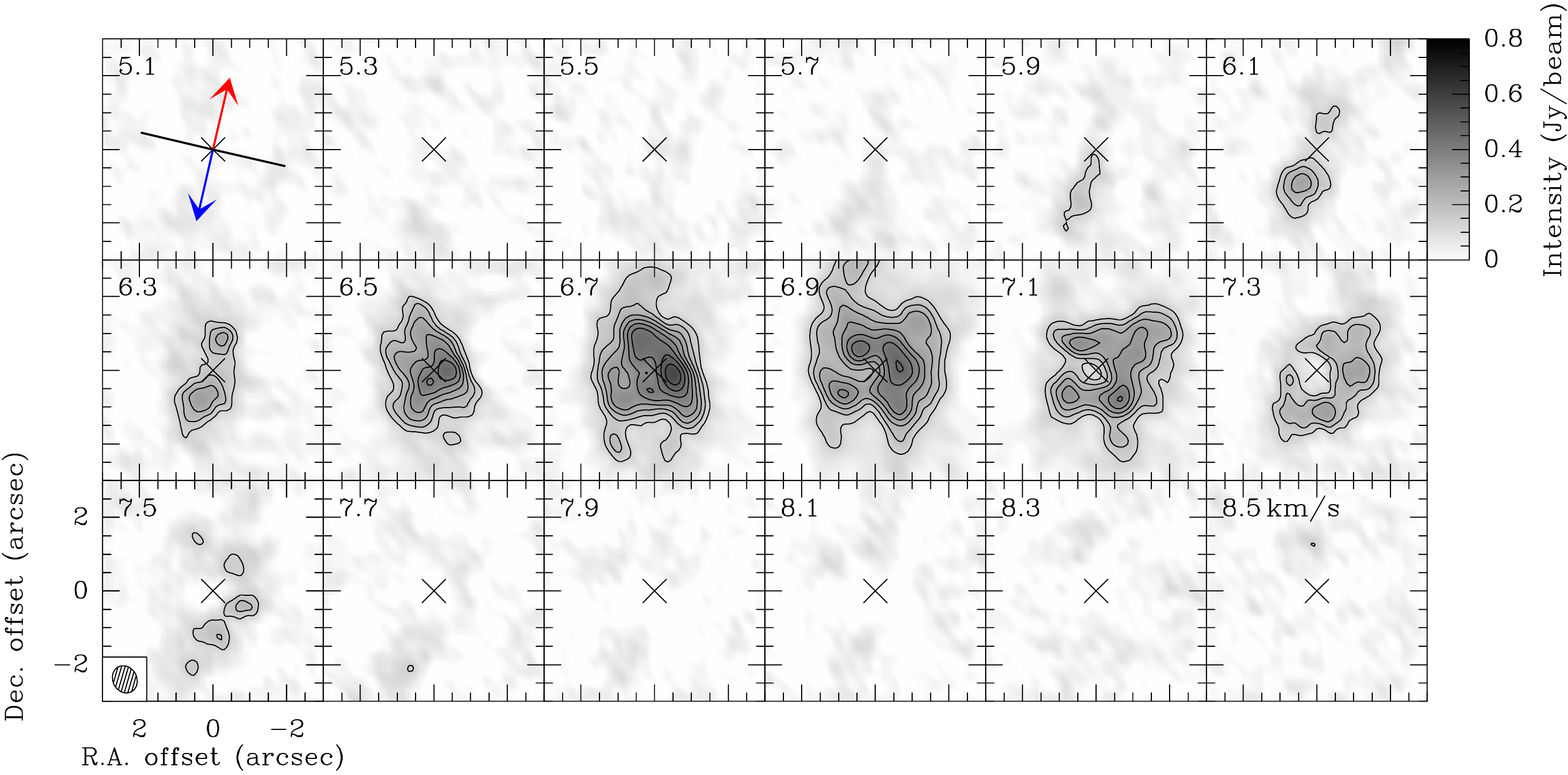}
\caption{Same as Figure \ref{fig:channel-maps-L1448-2A}, but for IRAS4B. The systemic velocity is estimated to be $\mathrm{v}_\mathrm{sys}=$6.9~km~s$^{-1}$ (see Table \ref{table:chi2-fit-lines}).
}
\label{fig:channel-maps-IRAS4B}
\end{figure*}

However, the channel maps do not exhibit an organized spatial distribution of velocities along the equatorial axis : the central emission fit show a position angle $>$|45$^{\circ}|$ with respect to the equatorial axis, suggesting a possible contamination by the outflow kinematics (see Figure \ref{fig:channel-maps-IRAS4B}). To minimize this contamination and probe rotational motions in the equatorial axis, we used the method in the image plane in the PdBI dataset instead of working in the visibilities to constrain in the inner envelope (200$-$500~au).

At outer envelope scales ($r>$1000~au), the N$_2$H$^+$ emission shows a reversed gradient dominated by external cloud compression. Moreover, the integrated intensity of N$_2$H$^+$ emission from the 30m datasets (see right top panel on Figure \ref{fig:intensity-maps-IRAS4B}) trace the a filamentary structure which are consistent with the filament in which the source is embedded.

\begin{figure*}[!ht]
\centering
\includegraphics[width=10cm]{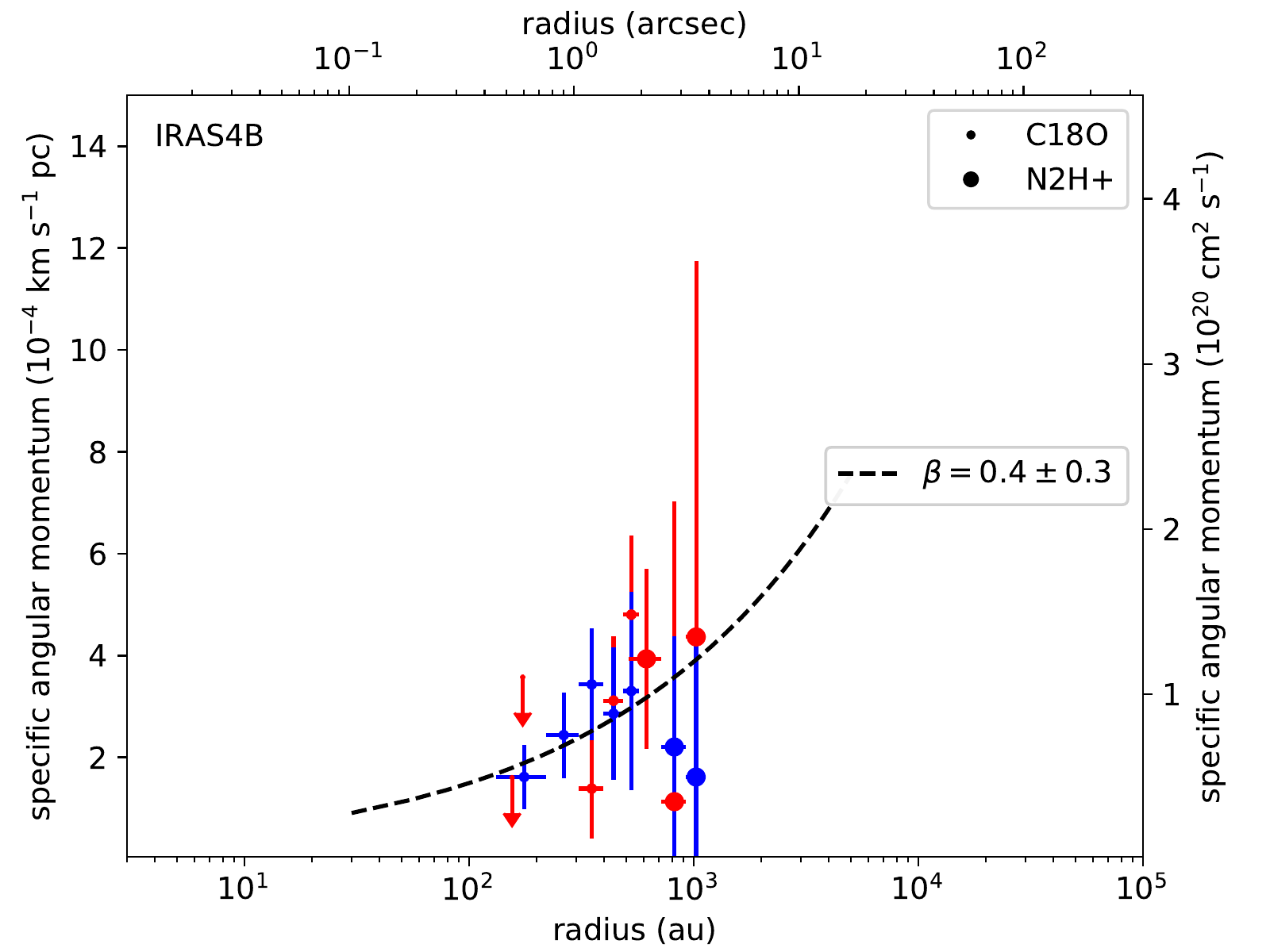}
\caption{Same as Figure \ref{fig:angular-momentum-profil-L1448-2A}, but for IRAS4B.}
\label{fig:angular-momentum-profil-IRAS4B}
\end{figure*}

The panel (g) of Fig. \ref{fig:PV-diagrams-1} shows the PV$_\mathrm{rot}$ diagram of IRAS4B built from the velocity gradients observed at scales of $r <$1100~au. The index of the fitting by a power-law ($\alpha \sim$-0.6, see Table \ref{table:chi2-fit-profil-rotation}) is consistent with a Keplerian rotation. However, we also obtain a good reduced $\chi^2$ when we fit the PV$_\mathrm{rot}$ diagram by an infalling and rotating envelope by fixing the index of the power-law at -1 ($\sim$0.6, see Table \ref{table:chi2-fit-profil-rotation}). 
Thus, for this source, the CALYPSO dataset only allow us to estimate a range of the power-law index between -1 and -0.6 (see Table \ref{table:chi2-fit-profil-rotation}) at the scales of 170$-$1100~au probed in our analysis. Moreover, Keplerian rotation is not detected at the smaller envelope radii investigated by \cite{Maret20}. Thus, the Keplerian rotation due to a large disk is not a robust interpretation for this source.
Figure \ref{fig:angular-momentum-profil-IRAS4B} shows the radial distribution of the specific angular momentum of IRAS4B at radii of 170$-$1100~au.


%%%%%%%%%%%
\clearpage
\subsection{IRAM04191}
At a distance of 140~pc \citep{Torres09}, IRAM04191+1522 (hereafter IRAM04191) is located in the southern part of the Taurus cloud in the viciny of the Class~I protostar IRAS04191. Its envelope mass (0.5~M$_{\odot}$, \citealt{Andre00}), its low luminosity (0.1~L$_{\odot}$, \citealt{Dunham06}), and its temperature suggest that is one of the youngest accreting and isolated Class 0 sources known in Taurus ($t \sim$1$-$3 $\times$ 10$^{4}$ yr, \citealt{Andre99}). This source harbors a strongly collimated outflow with a PA=28$^{\circ}$ \citep{Belloche02} and an inclination angle to the line of sight estimated at $\sim$40$^{\circ}$ \citep{Andre99,Maret14}. A new study by Podio \& CALYPSO (in prep.) estimates the PA of the outflows at 20$^{\circ}$ (see Table \ref{table:sample}). We used this later value for our study. From CALYPSO observations, \cite{Maury18} estimate an envelope mass of 0.45~M$_{\odot}$ and did not resolve candidate disk down to $<$50~au.

The 30m observations of several molecular lines suggests signatures of the envelope infalling motions with an estimated velocity of $\sim$0.2~km~s$^{-1}$ to $r \sim$1000~au and $\sim$0.1~km~s$^{-1}$ to $r \sim$10000~au \citep{Belloche02}. These observations associated with radiative transfer model allowed kinematics of the outer envelope to be constrained. A strong velocity gradient of $\sim$40~km~s$^{-1}$~pc$^{-1}$ was detected at scales of $\sim$3500~au along the northwest-southeast direction, namely in the equatorial axis. At $\sim$1500~au, \cite{Lee05} detect a velocity gradient of $\sim$100~km~s$^{-1}$~pc$^{-1}$. These gradient are interpreted as rotational motions of the protostellar envelope \citep{Belloche02, Takakuwa03}.

\begin{figure*}[!ht]
\centering
\includegraphics[scale=0.5,angle=0,trim=0cm 1.5cm 0cm 1.5cm,clip=true]{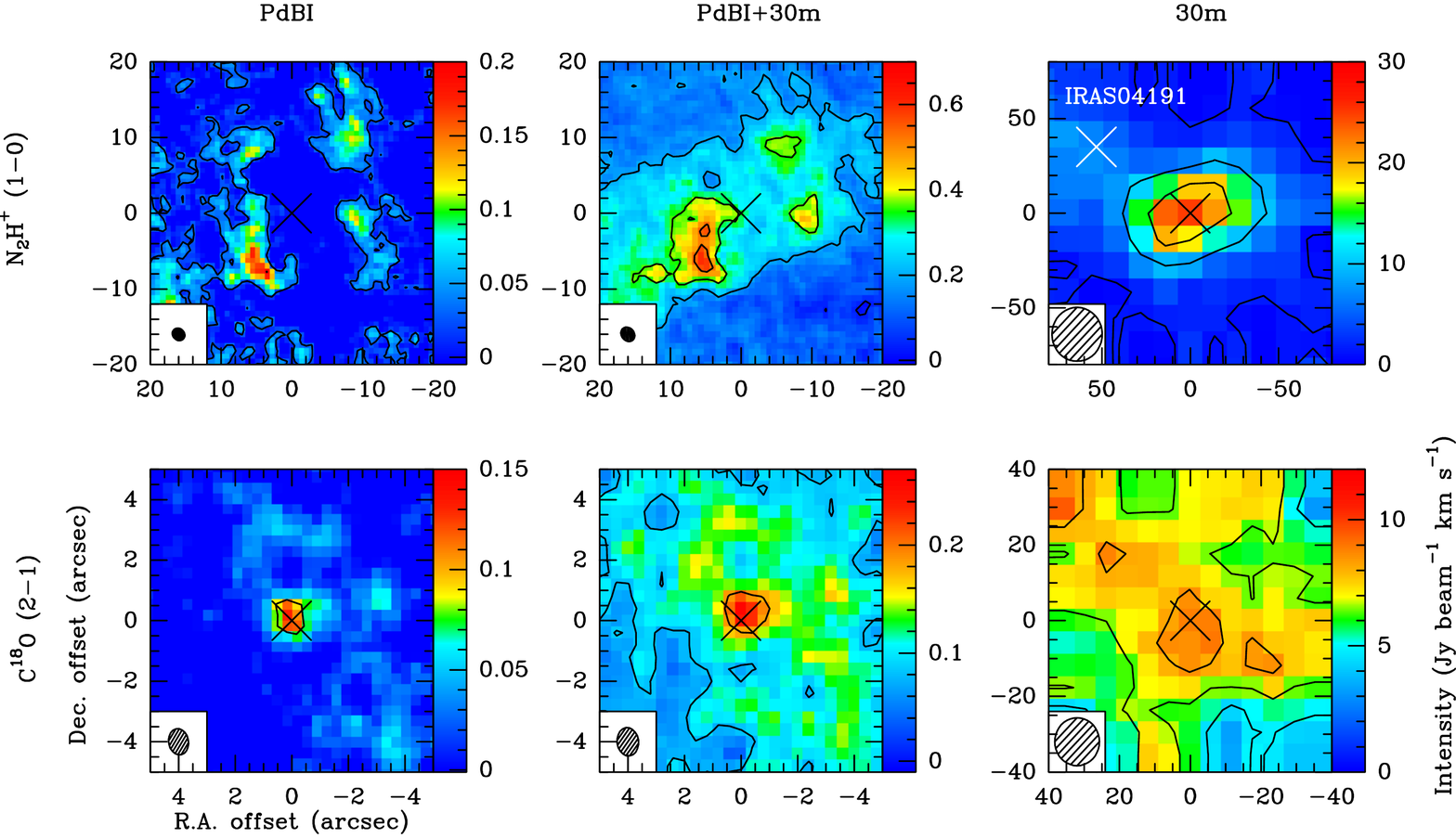}
\caption{Same as Figure \ref{fig:intensity-maps-L1448-2A}, but for IRAM04191. The white cross represents the position of the Class~I protostar IRAS04191. The black cross represents the position of IRAM04191 determined from the 1.3~mm dust continuum emission (see Table \ref{table:sample}). The black lines represent the integrated intensity contours of each tracer starting at 5$\sigma$ and increasing in steps of 30$\sigma$ for N$_{2}$H$^{+}$ and 6$\sigma$ for C$^{18}$O (see Tables \ref{table:details-obs-c18o} and \ref{table:details-obs-n2hp}).
}
\label{fig:intensity-maps-IRAM04191}
\end{figure*}

Figures \ref{fig:intensity-maps-IRAM04191} and \ref{fig:velocity-maps-IRAM04191} show the integrated intensity and centroid velocity maps obtained for IRAM04191 from the PdBI, combined, and 30m CALYPSO datasets for the C$^{18}$O and N$_2$H$^+$ emission, respectively. The C$^{18}$O emission is weakly detected ($\sim$5$\sigma$) from PdBI observations for this source which is one of the lowest  luminosity sources in our sample (see bottom left panel on Figure \ref{fig:velocity-maps-IRAM04191}). However, the combined map allows the kinematics of the protostellar envelope between 50 and 1000~au to be constrained. At those scales (see bottom middle panel on Figure \ref{fig:velocity-maps-IRAM04191}), we noticed a velocity gradients with a direction of $\Theta$= -83$^{\circ}$, consistent with the equatorial axis. At $r>$1000~au, the velocity gradients observed in the velocity maps of the N$_2$H$^+$ emission (see top panels on Figure \ref{fig:velocity-maps-IRAM04191}) have a direction of$\Theta \sim$100$^{\circ}$, consistent with those detected by \cite{Belloche02} at similar scales. Therefore, there is a reversal of the kinematics between inner ($\sim$500~au) and outer envelope scales ($>$1000~au).
 
\begin{figure*}[!ht]
\centering
\includegraphics[width=10cm]{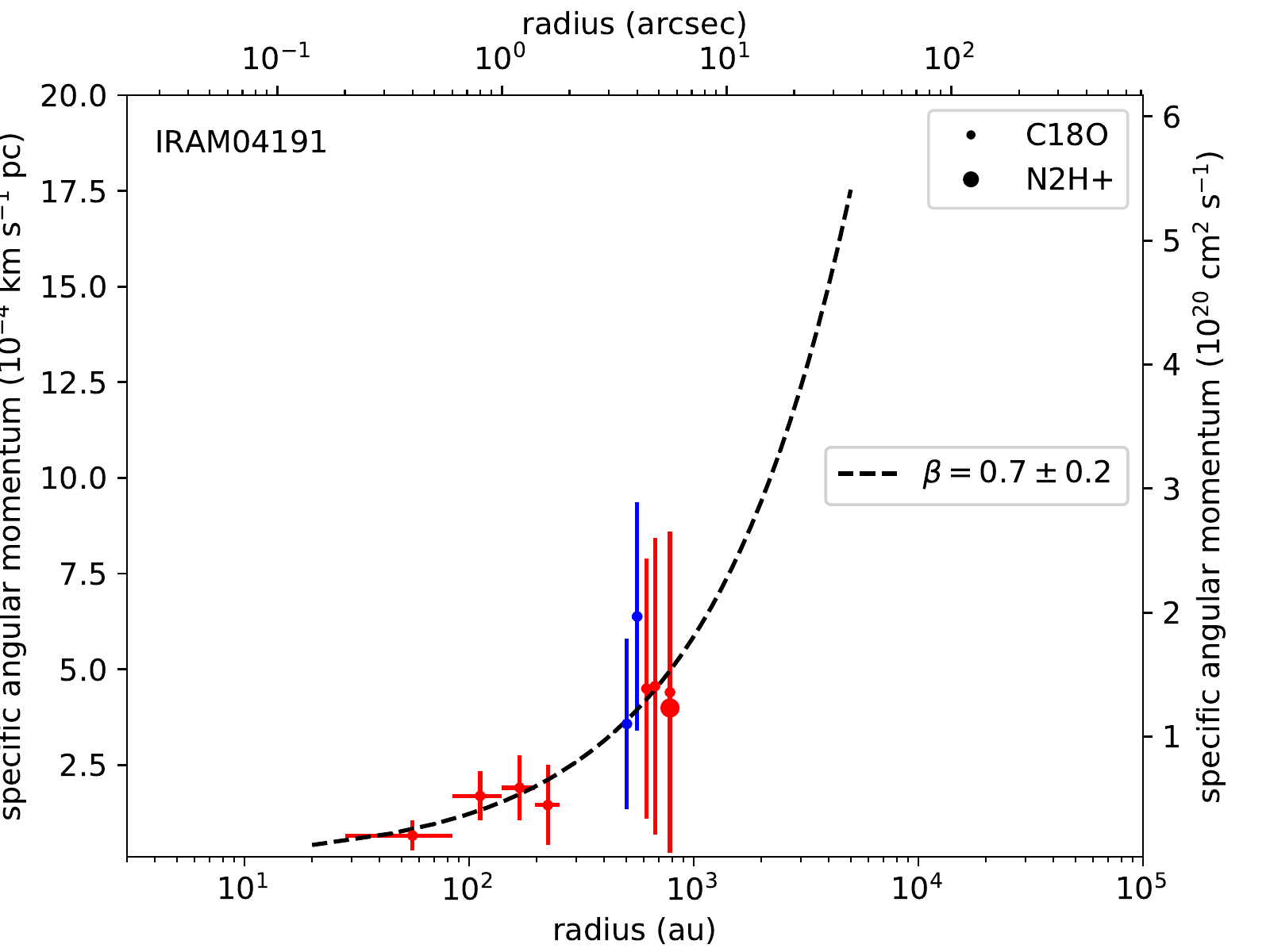}
\caption{Same as Figure \ref{fig:angular-momentum-profil-L1448-2A}, but for IRAM04191. 
}
\label{fig:angular-momentum-profil-IRAM04191}
\end{figure*}

The panel (h) of Fig. \ref{fig:PV-diagrams-1} shows the PV$_\mathrm{rot}$ diagram of IRAM04191 built from the velocity gradients observed at scales of 50$-$800~au. The index of the fitting by a power-law ($\alpha \sim$-0.3, see Table \ref{table:chi2-fit-profil-rotation}) is consistent with a Keplerian rotation. However, we also obtain a good reduced $\chi^2$ when we fit the PV$_\mathrm{rot}$ diagram by an infalling and rotating envelope by fixing the index of the power-law at -1 ($\sim$1.1, see Table \ref{table:chi2-fit-profil-rotation}). 
Thus, for this source, the CALYPSO dataset only allow us to estimate a range of the power-law index between -1 and -0.3 (see Table \ref{table:chi2-fit-profil-rotation}) at the scales of 50$-$800~au probed in our analysis. Moreover, Keplerian rotation is not detected at the smaller envelope radii investigated by \cite{Maret20}. Thus, the Keplerian rotation due to a large disk is not a robust interpretation for this source.
Figure \ref{fig:angular-momentum-profil-IRAM04191} shows the radial distribution of the specific angular momentum of IRAM04191 at radii of 50$-$800~au.


%%%%%%%%%%%
\clearpage
\subsection{L1521F}
At a distance of 140~pc \citep{Torres09}, L1521F (or MC27) is one of the densest cores in Taurus cloud. This source was firstly identified as a starless core \citep{Codella97,Onishi99,Crapsi04}. Its very low luminosity (L<0.004; \citealt{Maury18}) indicates that it is among the youngest known Class~0 protostars \citep{Andre00} and still may preserve the initial conditions of star formation \citep{Terebey09}. The emission of the molecular line HCO$^{+}$ showing a high central density and an asymmetric signature typical of the collapse suggests that the source is in the early stages of gravitational collapse \citep{Onishi99}. Its low luminosity of $L_\mathrm{bol}<$0.04 (\citealt{Maury18}; Ladjelate et al. in prep.) indicating that is one of the known Class~0 youngest prostars \citep{Andre00}, L1521F could still preserve its initial conditions \citep{Bourke06,Terebey09}. 
N$_2$H$^+$ and NH$_3$ emissions from VLA observations show a flattened structure perpendicular to the outflows \citep{Tobin10}. 
Podio \& CALYPSO (in prep.) do not detect clear jet signature in SiO and SO but only blue and red-shifted clumps in CO. In our study of the kinematics, we used the PA of the outflows at 240$^{\circ}$ observed in CO and HCO$^{+}$ by \cite{Tokuda14, Tokuda16} (see Table \ref{table:sample}).
Recent ALMA observations of $^{12}$CO emission showed a clear velocity gradient in the equatorial axis consistent with the presence of a disk of radius of $\sim$10~au surrounded a stellar embryo of $\sim$0.2~$M_{\odot}$ \citep{Tokuda17}.

\begin{figure*}[!ht]
\centering
\includegraphics[scale=0.5,angle=0,trim=0cm 1.5cm 0cm 1.5cm,clip=true]{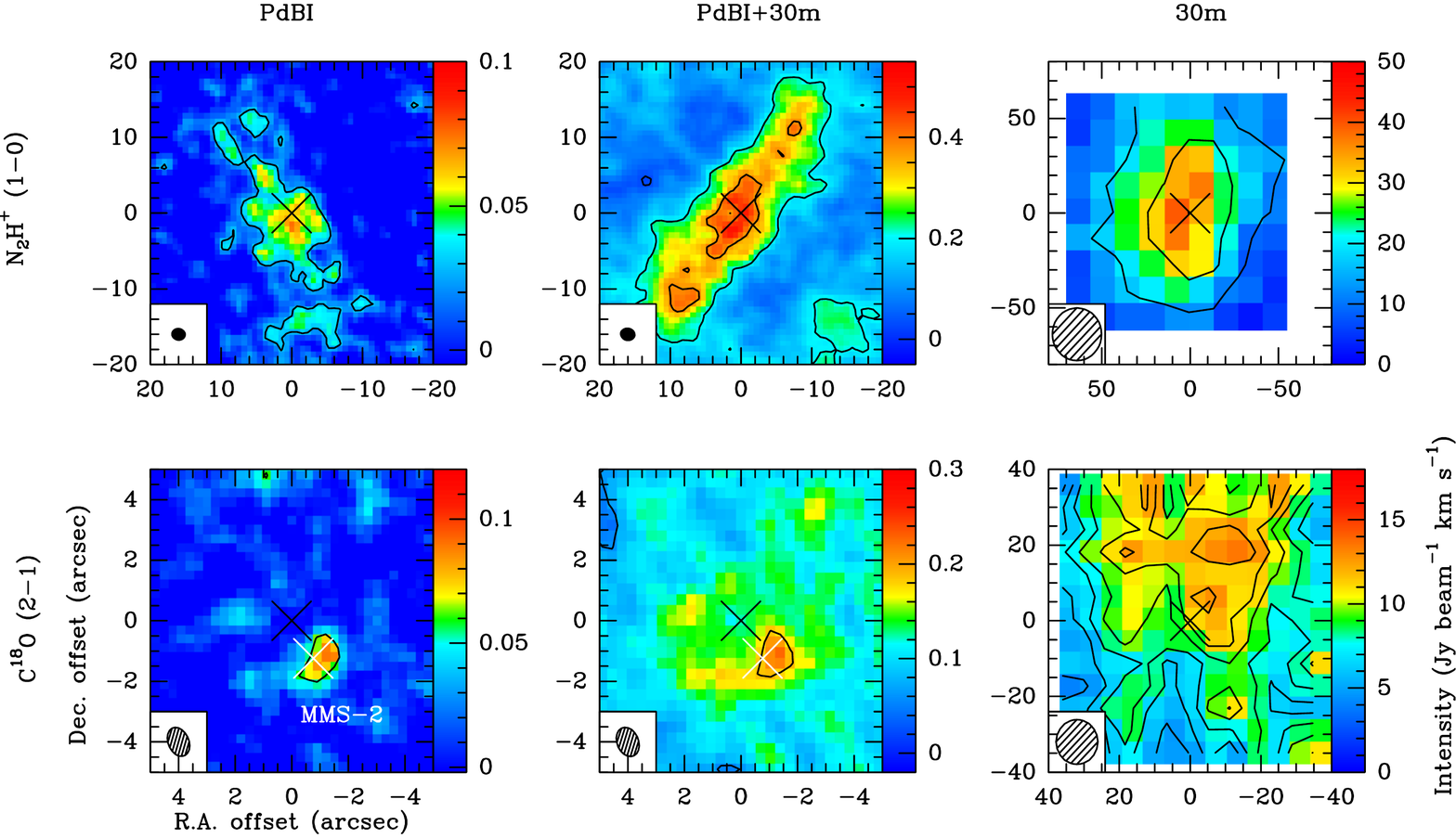}
\caption{Same as Figure \ref{fig:intensity-maps-L1448-2A}, but for L1521F. The white cross represents the position of the starless dense core MMS-2. The black lines represent the integrated intensity contours of each tracer starting at 5$\sigma$ and increasing in steps of 30$\sigma$ for N$_{2}$H$^{+}$ and 10$\sigma$ for C$^{18}$O (see Tables \ref{table:details-obs-c18o} and \ref{table:details-obs-n2hp}.
}
\label{fig:intensity-maps-L1521F}
\end{figure*}

Figure \ref{fig:intensity-maps-L1521F} shows the integrated intensity maps obtained for L1521F from the PdBI, combined, and 30m CALYPSO datasets for the C$^{18}$O and N$_2$H$^+$ emission. The C$^{18}$O emission is weakly detected ($\sim$5$\sigma$) from the PdBI observations for this source which is one of the lowest luminosity sources in our sample (see bottom left panel on Figure \ref{fig:velocity-maps-L1521F}). The emission peak does not correspond to the continuum emission peak at 1.3~mm determined from the PdBI dataset by \cite{Maury18} (see Table \ref{table:sample}). This is consistent with the existence of a starless high-density core, MMS-2 (RA:4$^h$28$^m$38$^s$.89, Dec.:+26$^{\circ}$51$\arcmin$33$\arcsec$.9) detected by ALMA Cycle~0 observations in dust continuum and H$^{13}$CO(3–2) emissions \citep{Tokuda14}. The protostar MMS-1 (RA:4$^h$28$^m$38$^s$.96, Dec.: +26$^{\circ}$51$\arcmin$35$\arcsec$) is also not spatially resolved in H$^{13}$CO(3–2) emission. The L1521F protostar is not spatially resolved by the emission H$^{13}$CO (3$-$2) but a clear velocity gradient is observed along the northwest southeast axis at outer envelope scales consistent with the equatorial axis \citep{Tokuda14}. This velocity gradient is in the same direction as the elongated structure we observed in combined N$_2$H$^+$ emission (see top middle panel on Figure \ref{fig:velocity-maps-L1521F}) and also observed by \cite{Tobin10}. The absence of velocity gradient at small scales ($r<$1500~au) is interpreted as due to low dust temperature of $T \sim$10K \citep{Tokuda14,Tokuda16} which is not sufficient to excite high density molecules ($E_\mathrm{upper}$=15.8~K for the molecular transition C$^{18}$O (2$-$1); \href{http://www.astro.uni-koeln.de/cdms/catalog}{CDMS}).

Figures \ref{fig:velocity-maps-L1521F} show the centroid velocity maps obtained for L1521F from the PdBI, combined, and 30m CALYPSO datasets for the C$^{18}$O and N$_2$H$^+$ emission. Only the 30m velocity map of the C$^{18}$O emission (see bottom right panel on Figure \ref{fig:velocity-maps-L1521F}) shows a gradient with a blue- and a red-shifted velocity components along the equatorial axis at radii $\sim$3000~au. The direction of this gradient ($\Theta \sim$ -8$^{\circ}$) is consistent with the equatorial axis axis and the gradients previously observed by \cite{Tokuda14}. However, we noted that the integrated intensity of the C$^{18}$O emission from the 30m datasets does not trace the envelope but a more complex and diffuse structure.

\begin{figure*}[!ht]
\centering
\includegraphics[width=10cm]{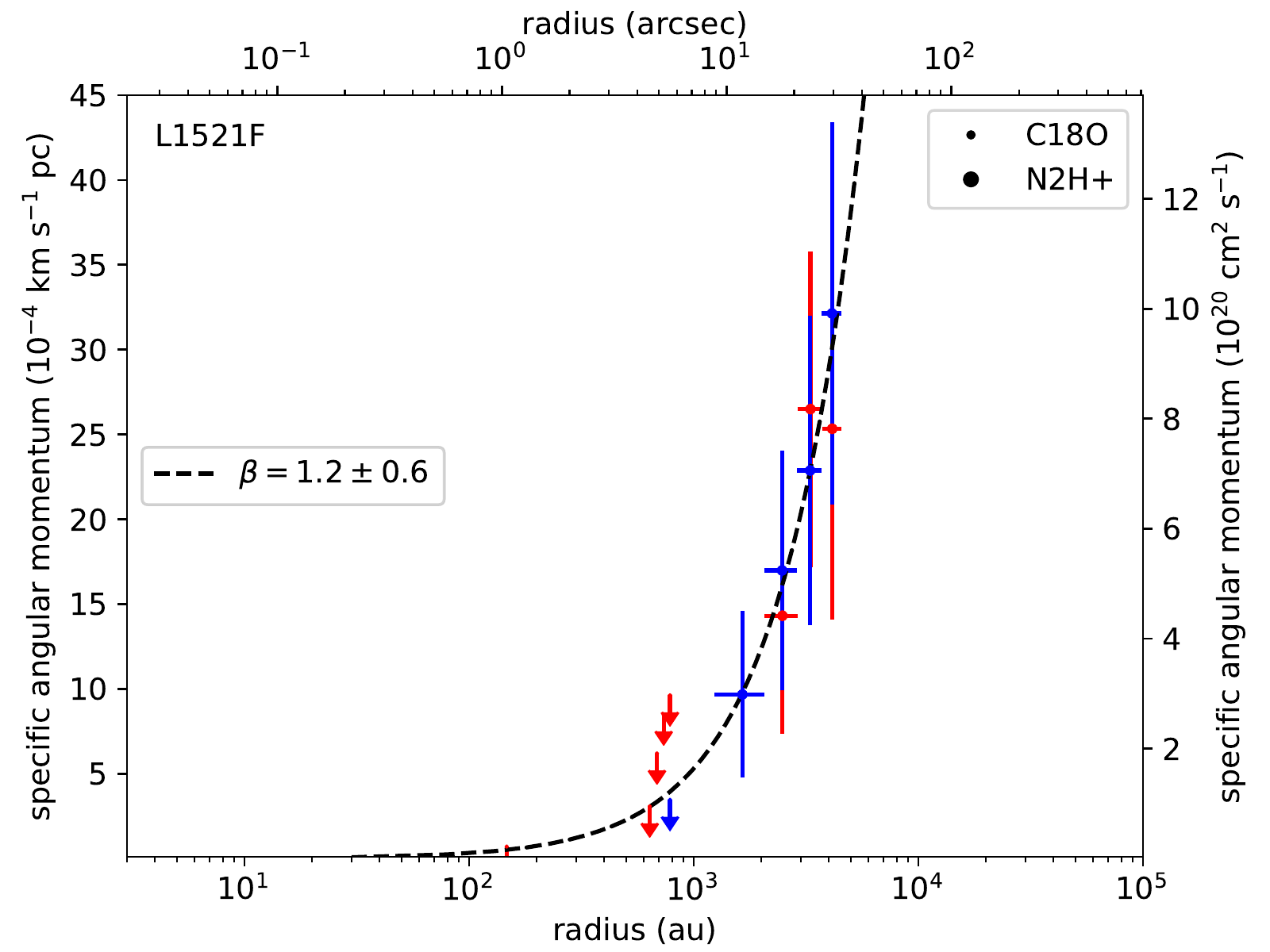}
\caption{Same as Figure \ref{fig:angular-momentum-profil-L1448-2A}, but for L1521F. 
}
\label{fig:angular-momentum-profil-L1521F}
\end{figure*}

The panel (i) of Fig. \ref{fig:PV-diagrams-1} shows the PV$_\mathrm{rot}$ diagram of L1521F built from the velocity gradients observed at scales of 1500$-$4200~au. The index of the fitting by a power-law is close to 0 (see Table \ref{table:chi2-fit-profil-rotation}). However, we also obtain a quite good reduced $\chi^2$ when we fit the PV$_\mathrm{rot}$ diagram by an infalling and rotating envelope by fixing the index of the power-law at -1 ($\sim$1, see Table \ref{table:chi2-fit-profil-rotation}). Thus, for this source, the CALYPSO dataset only allow us to estimate a range of the power-law index between -1 and 0 (see Table \ref{table:chi2-fit-profil-rotation}) at the scales of 1500$-$4200~au probed in our analysis. Figure \ref{fig:angular-momentum-profil-L1521F} shows the radial distribution of the specific angular momentum of L1521F at radii of 1500$-$4200~au.


%%%%%%%%%%%%%
\clearpage
\subsection{L1527}
At a distance of 140~pc \citep{Torres09} in the Taurus molecular cloud, L1527 is the closest Class~0/I source which exhibits a candidate proto-planetary disk with an outer radius of 50$-$90~au around a 0.2$-$0.3 M$_{\odot}$ central object \citep{Tobin12,Ohashi14}. A recent study by \cite{Aso17} find an outer radius of 74~au and a central mass of 0.45~M$_{\odot}$. The embryo mass is close of the envelope mass ($\sim$1~M$_{\odot}$), which sets the source at the boundary between the phases Class~0 and Class~I.
This source is an ideal target for studying the kinematics of the disk and the envelope at high angular resolution because it is seen almost edge-on, with an angle $\sim$90$^{\circ}$ with respect to the sky plane \citep{Ohashi97}. L1527 would also be embedded in a filament with an angle of PA$\sim$135$^{\circ}$ (Marsch \& HGBS, in prep).

\begin{figure*}[!ht]
\centering
\includegraphics[scale=0.5,angle=0,trim=0cm 1.5cm 0cm 1.5cm,clip=true]{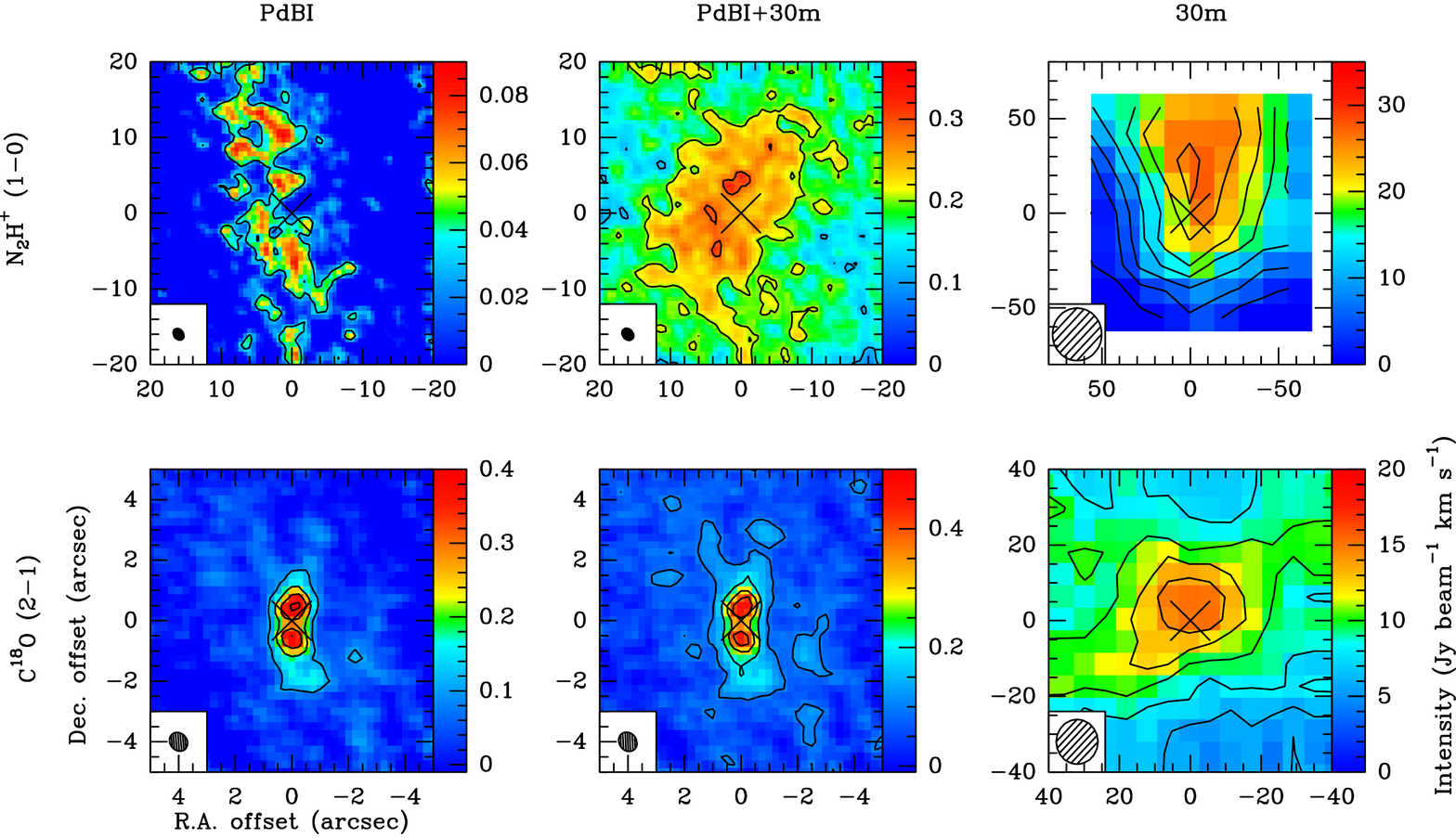}
\caption{Same as Figure \ref{fig:intensity-maps-L1448-2A}, but for L1527. The black lines represent the integrated intensity contours of each tracer starting at 5$\sigma$ and increasing in steps of 13$\sigma$ for N$_{2}$H$^{+}$ and 5$\sigma$ for C$^{18}$O (see Tables \ref{table:details-obs-c18o} and \ref{table:details-obs-n2hp}.
}
\label{fig:intensity-maps-L1527}
\end{figure*}

Figures \ref{fig:intensity-maps-L1527} and \ref{fig:velocity-maps-L1527} show the integrated intensity and centroid velocity maps obtained for L1527 from the PdBI, combined, and 30m CALYPSO datasets for the C$^{18}$O and N$_2$H$^+$ emission, respectively. At scales of $r<$2000~au (see left and middle panels on Figure \ref{fig:velocity-maps-L1527}), show a clear velocity gradient along the north-south axis. These velocity gradients are consistent with those observed in C$^{18}$O (2$-$1) emission with CARMA, SMA and ALMA \citep{Tobin12,Yen13,Yen15,Ohashi14}. The linear gradient fits give orientations of -9$^{\circ}$ $< \Theta <$26$^{\circ}$ which is consistent with the equatorial axis direction assumed perpendicular to the outflows (see Table \ref{table:gradient-velocity-fit}). Velocity maps from the 30m datasets at scales $\sim$3000$-$5000~au (see right panels on Figure \ref{fig:velocity-maps-L1527}) show gradients along the east-west axis opposite to the outflow direction, with an orientation $\sim$113$-$123$^{\circ}$ (see Table \ref{table:gradient-velocity-fit}).

\begin{figure*}[!ht]
\centering
\includegraphics[width=10cm]{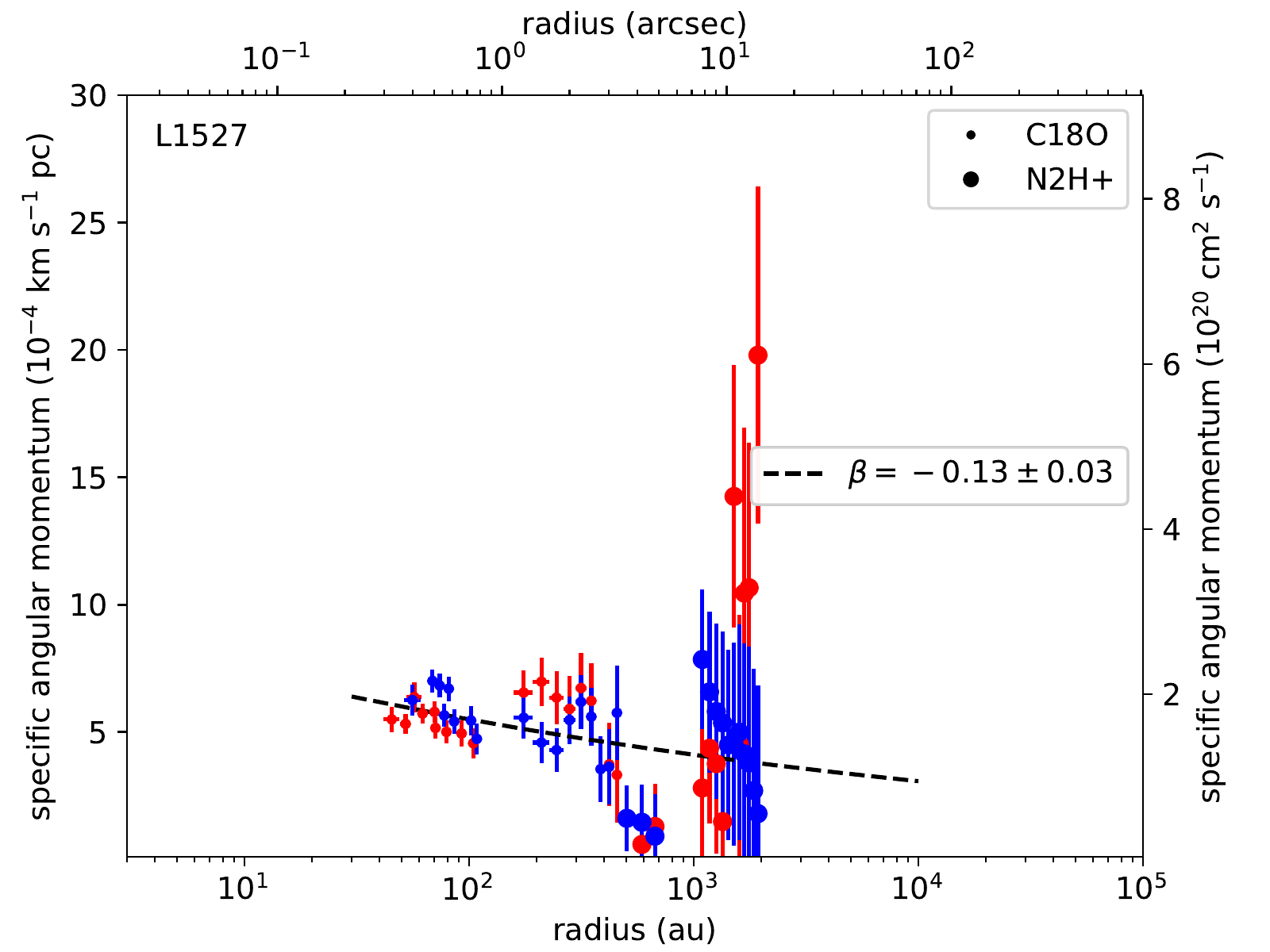}
\caption{Same as Figure \ref{fig:angular-momentum-profil-L1448-2A}, but for L1527. 
}
\label{fig:angular-momentum-profil-L1527}
\end{figure*}

The panel (j) of Fig. \ref{fig:PV-diagrams-1} shows the PV$_\mathrm{rot}$ diagram of L1527 built from the velocity gradients observed at scales of $r<$2000~au. The index of the fitting by a power-law is consistent with a rotating and infalling envelope ($\alpha \sim$-1.1, see Table \ref{table:chi2-fit-profil-rotation}). This result is consistent with previous studies of \cite{Yen13} and \cite{Ohashi14} which are found indices of $\sim$-1 and $\sim$-1.2 on scales of 40$-$500~au and 100$-$1500~au, respectively. \cite{Ohashi14} constrain the specific angular momentum at $j \sim$6 $\times$10$^{-4}$~km~s$^{-1}$~pc at radii of 100$-$1600~au. This value is consistent with the value we found on the Fig. \ref{fig:angular-momentum-profil-L1527} which shows the radial distribution of the specific angular momentum of L1527 at radii of 40$-$2000~au.


%%%%%%%%%%%%%
\clearpage
\subsection{L1157}
In the Cepheus molecular cloud at a distance previously estimated to be 250~pc \citep{Looney07} but determined at (352 $\pm$ 18)~pc by recent Gaia parallax measurements \citep{Zucker19}, L1157 is a Class~0 protostar deeply embedded within a large circumstellar envelope \citep{Gueth03,Beltran04}. The source harbors a powerful asymmetrical bipolar outflows at large scales \citep{Tafalla95,Zhang95,Gueth96}. \cite{Podio16} and Podio \& CALYPSO (in prep.) estimate the PA of the outflows at 163$^{\circ}$ (see Table \ref{table:sample}). Previous study of C$^{18}$O (2$-$1) observations with SMA showed no clear sign of rotational motions but kinematic models predict L1157 exhibit a possible disk with an outer radius $<$5~au \citep{Yen15}.

\begin{figure*}[!ht]
\centering
\includegraphics[scale=0.5,angle=0,trim=0cm 1.5cm 0cm 1.5cm,clip=true]{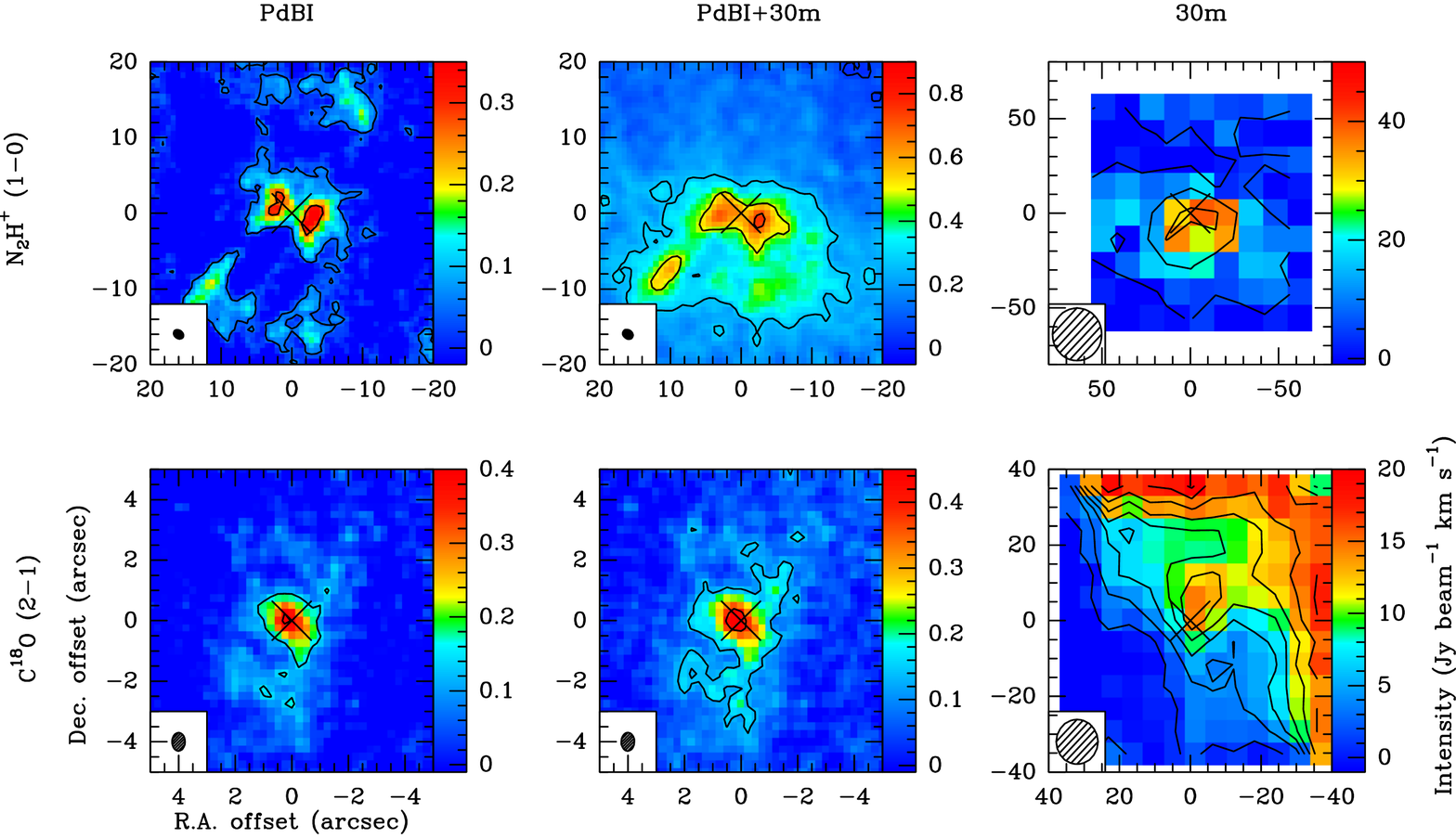}
\caption{Same as Figure \ref{fig:intensity-maps-L1448-2A}, but for L1157. The black lines represent the integrated intensity contours of each tracer starting at 5$\sigma$ and increasing in steps of 25$\sigma$ for N$_{2}$H$^{+}$ and 10$\sigma$ for C$^{18}$O (see Tables \ref{table:details-obs-c18o} and \ref{table:details-obs-n2hp}.
}
\label{fig:intensity-maps-L1157}
\end{figure*}

Figures \ref{fig:intensity-maps-L1157} and \ref{fig:velocity-maps-L1157} show the integrated intensity and centroid velocity maps obtained for L1157 from the PdBI, combined, and 30m CALYPSO datasets for the C$^{18}$O and N$_2$H$^+$ emission, respectively. Velocity maps of C$^{18}$O emission from the PdBI and combined datasets (see bottom left and middle panels on Figure \ref{fig:velocity-maps-L1157}) show gradients with a direction of $\Theta \sim$13$-$35$^{\circ}$ (see Table \ref{table:gradient-velocity-fit}). They are dominated by the kinematics of the outflows but a weak velocity gradient is observed along the equatorial axis at radii $r<$2000~au. Velocity maps of N$_{2}$H$^{+}$ emission from PdBI and combined datasets (see top left and middle panels on Figure \ref{fig:velocity-maps-L1157}) trace cavities along the outflow axis. These gradients are consistent with those detected in CARMA observations by \cite{Chiang10} and \cite{Tobin11}. At scales of $r>$2500~au, velocity maps from the 30m datasets (right panels on Figure \ref{fig:velocity-maps-L1157}) show a gradient with a direction of $\Theta \sim$ -130$^{\circ}$, in the opposite direction with respect to small scales observed along the equatorial axis.

\begin{figure*}[!ht]
\centering
\includegraphics[scale=0.5,angle=0,trim=0cm 1.5cm 0cm 1.5cm,clip=true]{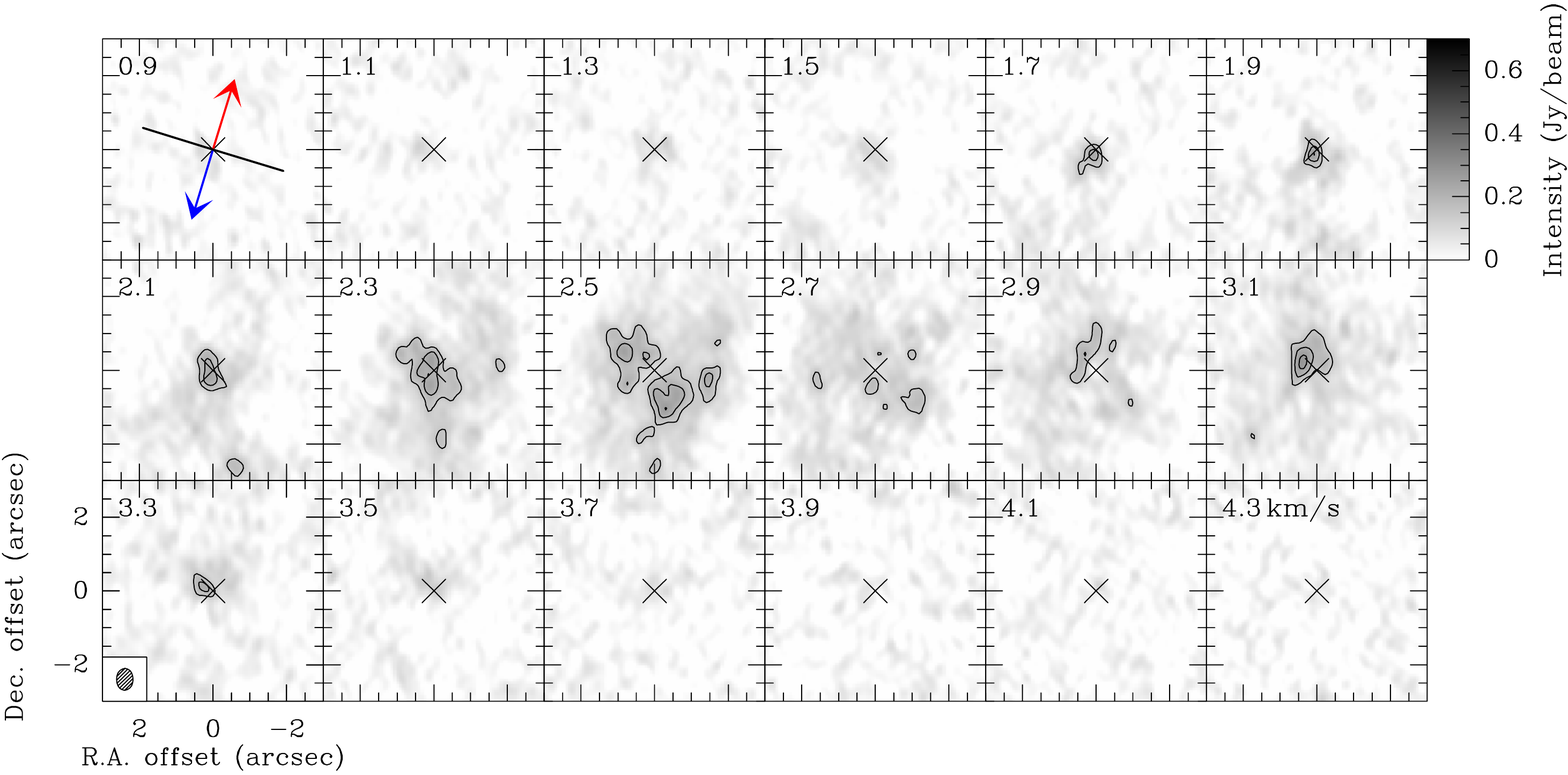}
\caption{Same as Figure \ref{fig:channel-maps-L1448-2A}, but for L1157. The systemic velocity is estimated to be $\mathrm{v}_\mathrm{sys}=$2.65~km~s$^{-1}$ (see Table \ref{table:chi2-fit-lines}).
}
\label{fig:channel-maps-L1157}
\end{figure*}

\begin{figure*}[!ht]
\centering
\includegraphics[width=10cm]{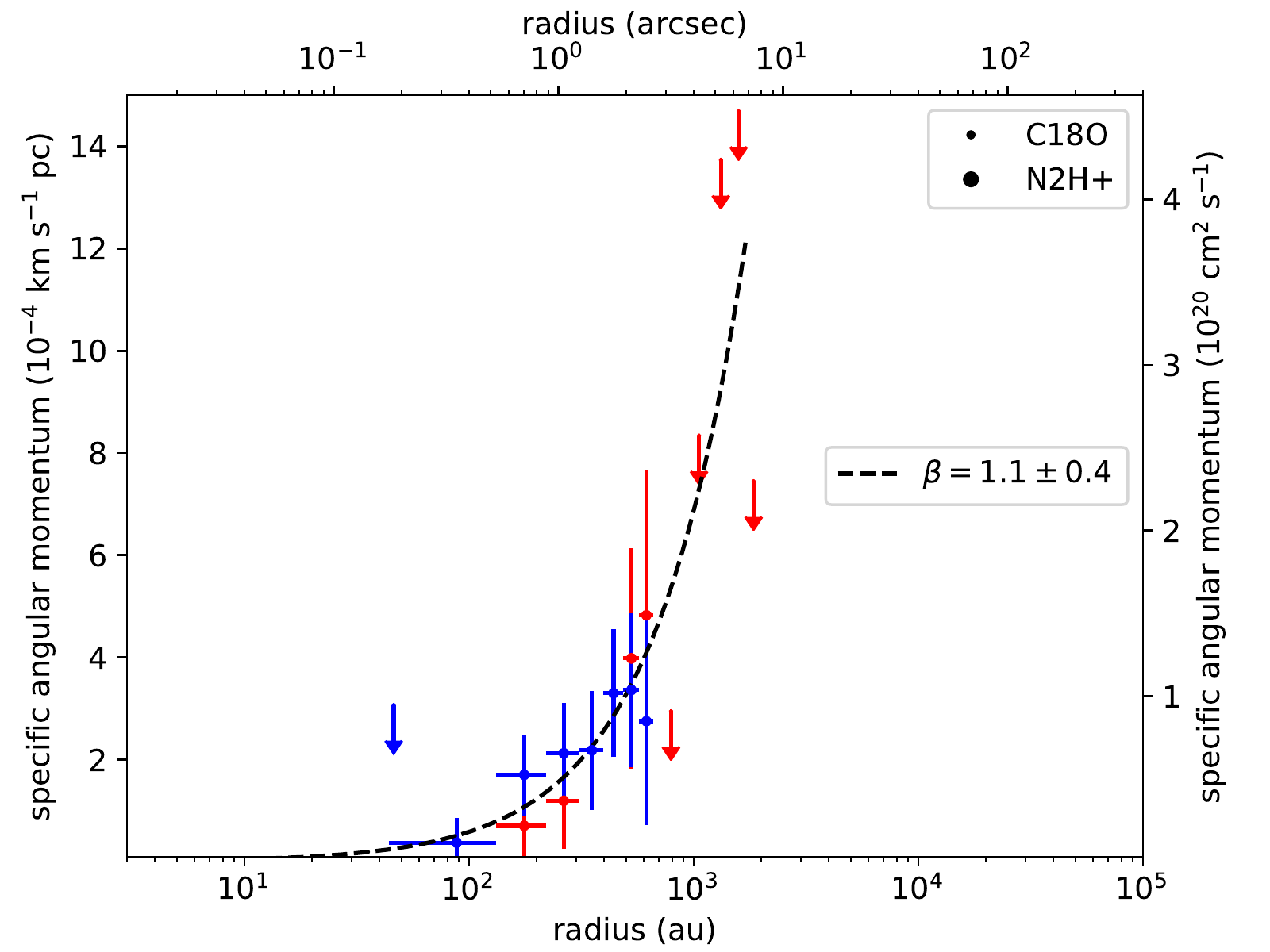}
\caption{Same as Figure \ref{fig:angular-momentum-profil-L1448-2A}, but for L1157. }
\label{fig:angular-momentum-profil-L1157}
\end{figure*}

The panel (k) of Fig. \ref{fig:PV-diagrams-1} shows the PV$_\mathrm{rot}$ diagram of L1157 built from the velocity gradients observed at scales of $r<$700~au. The index of the fitting by a power-law is close to 0 (see Table \ref{table:chi2-fit-profil-rotation}), consistent with differential rotation of the envelope. At each PdBI emission channel, the central emission fit show a position angle $>$|45$^{\circ}|$ with respect to the equatorial axis, suggesting a contamination by the outflow kinematics (see Figure \ref{fig:channel-maps-L1157}). To minimize this contamination and probe rotational motions in the equatorial axis, we used the method in the image plane in the PdBI dataset instead of working in the visibilities to constrain in the inner envelope (80$-$200~au). Figure \ref{fig:angular-momentum-profil-L1157} shows the radial distribution of the specific angular momentum of L1157 at radii of 80$-$700~au.


%%%%%%%%%%%
\clearpage
\subsection{GF9-2}
GF9-2 (L1082C \citealt{Benson89} or GF9-Core \citealt{Ciardi00}) is located in the filamentary cloud GF9-2 (or LDN 1082) at a distance of 200~pc \citep{Wiesemeyer97, Wiesemeyer98}. In this study, we adopt this distance but it is very uncertain and some studies estimated a higher distance between 440$-$470~pc (\citealt{Viotti69}, C. Zucker, priv. comm.) and 900~pc \citep{Reid16}. Due to its low luminosity (0.3~L$_{\odot}$, \citealt{Wiesemeyer97}), GF9-2 is has been cataloged as a transitional object between the prestellar and the Class~0 phases \citep{Wiesemeyer99}. GF9-2 would be the youngest source in our sample and would therefore retain the initial conditions of its gravitational collapse because the central embryo does not have a powerful bipolar flow \citep{Furuya06}. Podio \& CALYPSO (in prep.) detect for the first time the outflows with a PA of 0$^{\circ}$ (see Table \ref{table:sample}).

\begin{figure*}[!ht]
\centering
\includegraphics[scale=0.5,angle=0,trim=0cm 1.5cm 0cm 1.5cm,clip=true]{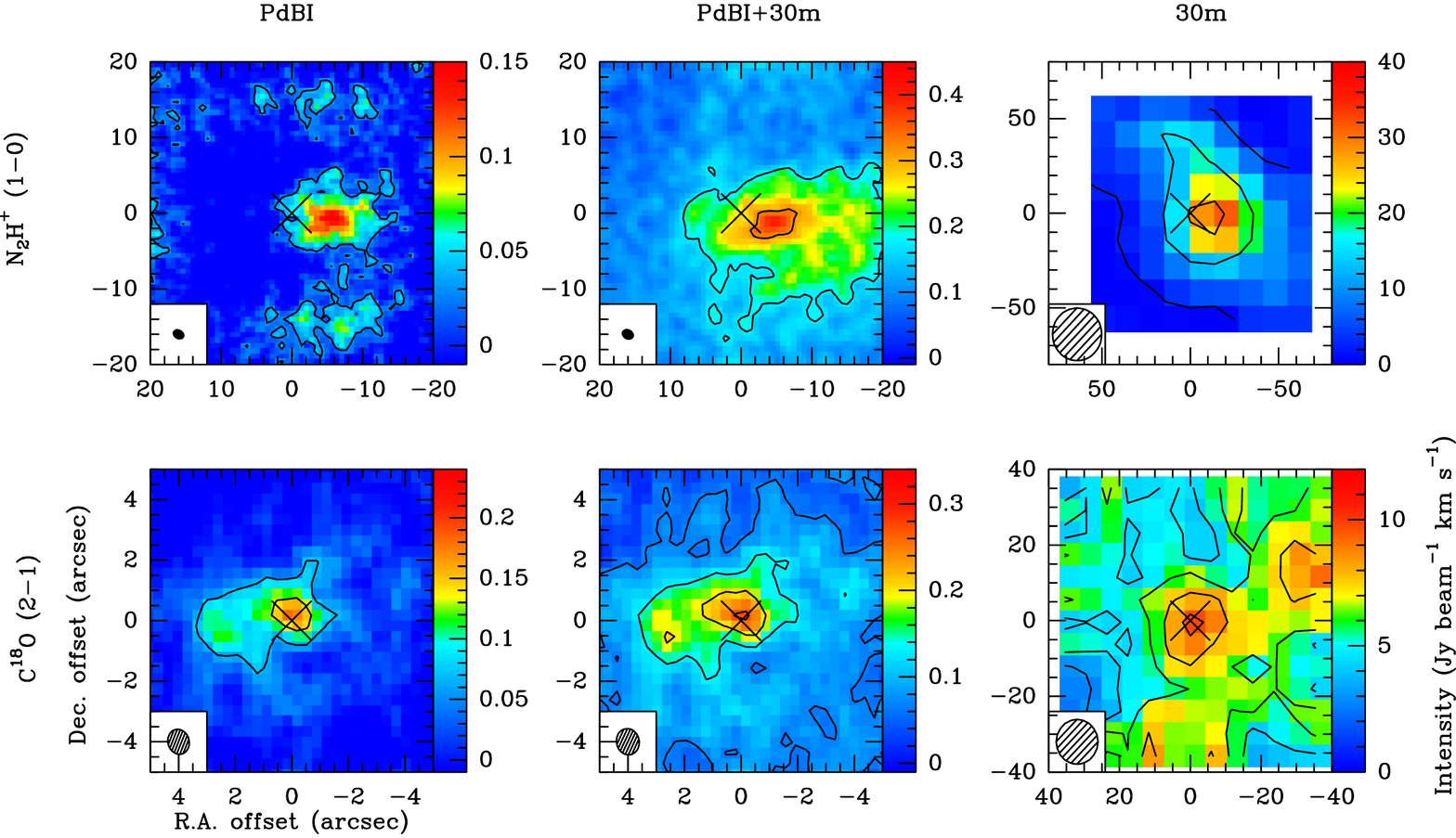}
\caption{Same as Figure \ref{fig:intensity-maps-L1448-2A}, but for GF9-2. The black lines represent the integrated intensity contours of each tracer starting at 5$\sigma$ and increasing in steps of 25$\sigma$ for N$_{2}$H$^{+}$ and 5$\sigma$ for C$^{18}$O (see Tables \ref{table:details-obs-c18o} and \ref{table:details-obs-n2hp}.
}
\label{fig:intensity-maps-GF92}
\end{figure*}

Figures \ref{fig:intensity-maps-GF92} and \ref{fig:velocity-maps-GF92} show the integrated intensity and centroid velocity maps obtained for GF9-2 from the PdBI, combined, and 30m CALYPSO datasets for the C$^{18}$O and N$_2$H$^+$ emission, respectively. The CALYPSO datasets allow us to constrain the kinematics of this source at the envelope scales for the first time.
Only the PdBI and combined maps of the C$^{18}$O emission (see bottom left and middle panels on Figure \ref{fig:velocity-maps-GF92}) show a velocity gradient associated with the source. The gradient is consistent with the equatorial axis, with a direction of -160$^{\circ} < \Theta < $-130$^{\circ}$ (see Table \ref{table:gradient-velocity-fit}). Moreover, the N$_2$H$^+$ integrated intensity from the 30m datasets at scales of $r>$3000~au traces the filament structure in which the source is embedded.

\begin{figure*}[!ht]
\centering
\includegraphics[width=10cm]{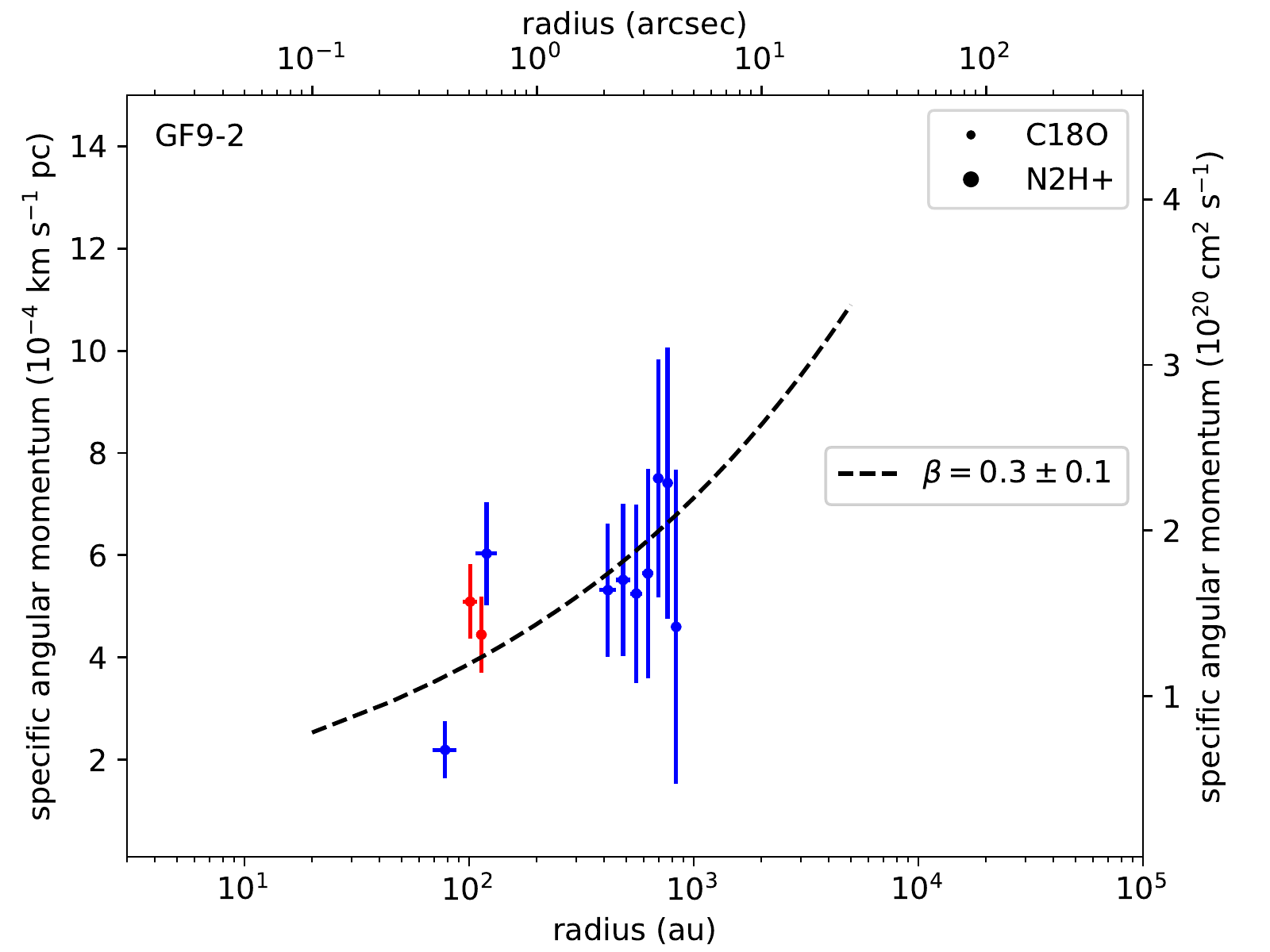}
\caption{Same as Figure \ref{fig:angular-momentum-profil-L1448-2A}, but for GF9-2.}
\label{fig:angular-momentum-profil-GF92}
\end{figure*}

The panel (l) of Fig. \ref{fig:PV-diagrams-1} shows the PV$_\mathrm{rot}$ diagram of GF9-2 built from the velocity gradients observed at scales of $r<$900~au. The index of the fitting by a power-law is consistent with an infalling and rotating envelope ($\sim$-0.8, see Table \ref{table:chi2-fit-profil-rotation}). Figure \ref{fig:angular-momentum-profil-GF92} shows the radial distribution of the specific angular momentum of GF9-2 at radii of 70$-$900~au.


\end{appendix}
\end{document}